%% file: BulletCluster.tex
\documentclass[preprintnumbers,amsmath,amssymb,rmp]{revtex4}

\usepackage{graphicx}
\usepackage{pstricks}
\newcounter{kpc}[figure]
\usepackage{color}
\definecolor{LightMagenta}{cmyk}{0.1,0.8,0,0.1}
\definecolor{DarkBlue}{rgb}{0,0.08,0.45}
\definecolor{chocolate}{rgb}{0.2784,0.1216,0.1020}
\definecolor{orange}{rgb}{1.0,0.75,0}
\definecolor{DarkOrange}{rgb}{0.8,0.4,0}
\definecolor{DarkGreen}{rgb}{0.0,0.5,0.0}
\definecolor{sand}{rgb}{0.82,0.73,0.5}
\definecolor{cayenne}{rgb}{0.62,0.15,0.06}
\usepackage{bm}
\usepackage[pdftitle={The Bullet Cluster 1E0657-558 evidence shows Modified Gravity in the absence of Dark Matter},pdfauthor={J. R. Brownstein and J. W. Moffat}, pdfsubject={modified gravity, dark matter, bullet cluster, 1E0657-558, X-ray imaging, gravitational lensing}, bookmarks, bookmarksopen, colorlinks=true, linkcolor=blue, citecolor=blue, anchorcolor=blue, urlcolor=blue]{hyperref}
\usepackage{multirow}
\usepackage[outercaption]{sidecap}
\usepackage{subfig}
\usepackage{caption}
\input{macros}

\begin{document}


\title{ The {\bc} evidence shows Modified Gravity \\
in the absence of Dark Matter}

\author{J.\,R.\,Brownstein} \email{jbrownstein@perimeterinsitute.ca}
\author{J.\,W\,Moffat}  \email{john.moffat@utoronto.ca}
\affiliation{Perimeter Institute for Theoretical Physics, Waterloo, Ontario N2L 2Y5, Canada}
\affiliation{Department of Physics, University of Waterloo, Waterloo, Ontario N2L 3G1, Canada}

\accepted{for publication in the {\mnras} -- July 26, 2007. In press.}{}

\begin{abstract}
A detailed analysis of the November 15, 2006 data release~\protect\citep{Clowe:ApJL:2006,Clowe:ApJ:2006,Clowe:astro-ph.0611496,Clowe:dataProduct} X-ray surface density \map{\Sigma} and the strong and weak gravitational lensing convergence \map{\kappa} for the {\bc} is
performed and the results are compared with the predictions of a modified gravity (MOG) and dark matter. Our surface
density $\Sigma$-model is computed using a King $\beta$-model density, and a mass profile of the main cluster and an isothermal
temperature profile are determined by the MOG. We find that the main cluster thermal profile is nearly isothermal. The
MOG prediction of the  isothermal temperature of the main cluster is $T=15.5 \pm 3.9\ \mbox{keV}$, in good agreement with the
experimental value $T=14.8^{+2.0}_{-1.7}\ \mbox{keV}$.  Excellent fits to the two-dimensional convergence $\kappa$-map data are
obtained without non-baryonic dark matter, accounting for the $8\sigma$ spatial offset between the \map{\Sigma}
and the \map{\kappa} reported in \protect\citet{Clowe:ApJL:2006}.   The MOG prediction for the 
\map{\kappa} results in two baryonic components distributed across the {\bc} with averaged mass-fraction of  83\%
intracluster medium (ICM) gas and 17\% galaxies.
Conversely, the Newtonian dark matter $\kappa$-model has on average 76\% dark matter (neglecting the indeterminant contribution
due to the galaxies) and 
24\% ICM gas for a baryon to dark matter mass-fraction of $0.32$, a statistically significant result when compared to the predicted $\Lambda$-CDM cosmological baryon mass-fraction of $0.176^{+0.019}_{-0.012}$~\protect\citep{Spergel:2006}.
\end{abstract}

\maketitle

\tableofcontents
\section{\label{section.Introduction}Introduction}
\subsection{\label{section.MissingMass.1}The Question of Missing Mass}
Galaxy cluster masses have been known to require some form of energy density that makes its presence felt
only by its gravitational effects since \protect\citet{Zwicky:1933} analysed the velocity dispersion for the Coma cluster. 
Application of the Newtonian $1/r^{2}$ gravitational force law inevitably points to the question of missing mass, and may be
explained by dark matter~\protect\citep{Oort:1932}.  The amount of non-baryonic dark matter required to maintain consistency with Newtonian physics increases as
the mass scale increases so that the mass to light ratio of clusters of galaxies exceeds the mass to light ratio for individual
galaxies by as much as a factor of $\sim 6$, which exceeds the mass to light ratio for the luminous inner core of
galaxies by as much as a factor of $\sim 10$.  In clusters of galaxies, the dark matter paradigm leads to a mass to light
ratio as much as $300\ M_{\sun}/L_{\sun}$.  In this scenario, non-baryonic dark matter dominates over baryons outside the cores of galaxies (by $\approx
80 - 90\%$).   

Dark matter has dominated cosmology for the last five decades, although the search for dark matter has to this point come up empty.  Regardless, one of the triumphs
of cosmology has been the precise determination of the standard (power-law flat, $\Lambda$-CDM) cosmological model
parameters. The highly anticipated third year results from the WMAP team have determined the cosmological baryon mass-fraction (to non-baryonic
dark matter) to be $0.176^{+0.019}_{-0.012}$~\protect\citep{Spergel:2006}.  This ratio may be inverted, so that for every gram
of baryonic matter, there are 5.68 grams of non-baryonic dark matter -- at least on cosmological scales.  There seems to
be no evidence of dark matter on the scale of the solar system, and the cores of galaxies also seem to be devoid of dark
matter.

Galaxy clusters and superclusters are the largest virialized (gravitationally bound) objects in the Universe and make 
ideal laboratories for gravitational physicists.  The data come from three sources: 
\begin{enumerate} \item X-ray imaging of the hot
intracluster medium (ICM), \item Hubble, Spitzer and Magellan telescope images of the galaxies comprising the
clusters, \item strong and weak gravitational
lensing surveys which may be used to calculate the mass distribution projected onto the sky (within a particular theory
of gravity).
\end{enumerate}

The alternative to the dark matter paradigm is to modify the Newtonian $1/r^{2}$ gravitational force law so that the
ordinary (visible) baryonic matter accounts for the observed gravitational effect.  An analysis of the {\bc} surface density \map{\Sigma} and convergence \map{\kappa} data by
\protect\citet{Angus:MNRAS:2006,Angus:ApJL:2007} based on MilgromÕs Modified Newtonian Dynamics (MOND) model~\protect\citep{Milgrom:ApJ:1983,Sanders:ARAA:2002} and
Bekenstein's relativistic version of MOND~\protect\citep{Bekenstein:PRD:2004} failed to fit the data without
dark matter. More recently, further evidence that MOND needs dark matter in weak lensing of clusters has been obtained
by \protect\citet{Takahashi:astro-ph.0701365}. Problems with fitting X-ray temperature profiles with MilgromÕs MOND model
without dark matter were shown in ~\protect\citet{Aguirre:ApJ:2001,Sanders:2006,Pointecouteau:MNRAS:2005,Brownstein:MNRAS:2006}.
Neutrino matter with an electron neutrino mass $m_{\nu}\sim 2\ \mbox{eV}$ can fit the bullet cluster data~\protect\citep{Angus:MNRAS:2006,Angus:ApJL:2007,Sanders:2006}. This mass is
at the upper bound obtained from observations. The Karlsruhe Tritium Neutrino (KATRIN) experiment will be able to
falsify 2 eV electron neutrinos at 95\% confidence level within months of taking data in 2009.

The theory of modified gravity -- or MOG model -- based upon a covariant
generalization of Einstein's theory with auxiliary (gravitational) fields in addition to the metric was proposed in 
\protect\citet{Moffat:JCAP:2005,Moffat:JCAP:2006} including metric skew-tensor gravity (MSTG) theory and scalar-tensor-vector
gravity (STVG) theory. Both versions of MOG, MSTG and STVG, modify the Newtonian $1/r^{2}$ gravitational force
law in the same way so that it is valid at small distances, say at terrestrial scales.

\protect\citet{Brownstein:ApJ:2006} applied MOG to the question of galaxy rotation curves, and presented the
fits to a large sample of over 100 low surface brightness (LSB), high surface brightness (HSB) and dwarf galaxies.
Each galaxy rotation curve was fit without dark 
matter using only the available photometric data (stellar matter and visible gas) and alternatively
a two-parameter mass distribution model which made no assumption regarding the mass to light ratio.  The results were
compared to MOND and were nearly indistinguishably right out to the edge of the rotation curve data, where MOND predicts a forever
flat rotation curve, but MOG predicts an eventual return to the familiar  $1/r^{2}$ gravitational force law.  The mass
to light ratio varied between $2 - 5\ M_{\sun}/L_{\sun}$ across the sample of 101 galaxies in contradiction to the dark
matter paradigm which predicts a mass to light ratio typically as high as $50\ M_{\sun}/L_{\sun}$.

In a sequel, \protect\citet{Brownstein:MNRAS:2006} applied MOG to the question of galaxy cluster masses, and presented the
fits to a large sample of over 100 X-ray galaxy clusters of temperatures ranging from 0.52 keV (6 million kelvin) to
13.29 keV (150 million kelvin).  For each of the 106 galaxy clusters, the MOG provided a parameter-free
prediction for the ICM gas mass profile, which reasonably matched the X-ray observations (King $\beta$-model) for the
same sample compiled by \protect\citet{Reiprich:2001} and \protect\citet{Reiprich:2002}.  The MOND predictions were presented for each galaxy
cluster, but failed to fit the data.  The Newtonian dark matter result outweighed the visible ICM gas mass
profiles by an order of magnitude.  

In the solar system, the Doppler data from the Pioneer 10 and 11 spacecraft suggest a deviation from the Newtonian
$1/r^{2}$ gravitational force law beyond Saturn's orbit.  \protect\citet{Brownstein:CQG:2006} applied MOG to fit the
available anomalous acceleration 
data~\protect\citep{Nieto:2005} for the Pioneer 10/11 spacecraft.  The solution showed a remarkably low variance of residuals
corresponding to a reduced $\chi^{2}$ per degree of freedom of 0.42 signalling a good fit.  The magnitude of the
satellite acceleration exceeds the MOND critical acceleration, negating the MOND solution~\protect\citep{Sanders:2006}.  The
dark matter paradigm is severely limited 
within the solar system by stability issues of the sun, and precision gravitational experiments including satellite,
lunar laser ranging, and measurements of the Gaussian gravitational constant and Kepler's law of planetary motion. 
Without an actual theory of dark matter, no attempt to fit the Pioneer anomaly with dark 
matter has been suggested.  Remarkably, MOG provides a closely fit solution to the Pioneer 10/11 anomaly and is
consistent with the accurate equivalence principle, all 
current satellite, laser ranging observations for the inner planets, and the precession of perihelion for all of the planets.

A fit to the acoustical wave peaks observed in the cosmic microwave background (CMB) data using MOG has been achieved 
without dark matter.  Moreover, a possible explanation for the accelerated expansion of the Universe 
has been obtained in MOG~\protect\citep{Moffat:JMPD:2007}.

Presently, on both an empirical and theoretical level, MOG is the most
successful alternative to dark matter.
The successful application of MOG across scales ranging from clusters of galaxies (Megaparsecs) to HSB, LSB and dwarf
galaxies (kiloparsecs), to the solar system (AU's) provides a clue to the question of missing mass.  The apparent
necessity of the dark matter paradigm may be an artifact of applying the Newtonian $1/r^{2}$ gravitational force law to
scales where it is not valid, where a theory such as MOG takes over. The ``excess gravity'' that MOG accounts for may
have nothing to do with the hypothesized missing mass of dark matter.  But how can we distinguish the two?  In most
observable systems, gravity creates a central potential, where the baryon density is naturally
the highest.  So in most situations, the matter which is creating the gravity potential occupies the same volume as the
visible matter. \protect\citet{Clowe:astro-ph.0611496} describes this as a degeneracy between whether gravity comes from dark matter, or
from the observed baryonic mass of the hot ICM and visible galaxies where the excess gravity is due to MOG.  This
degeneracy may be split by examining a system that is out of steady state, where there is spatial separation between the
hot ICM and visible galaxies.  This is precisely the case in galaxy cluster mergers: the galaxies will experience a different gravitational potential created by the hot ICM
than if they were concentrated at the center of the ICM.  \protect\citet{Moffat:astro-ph.0608675} considered the possibility
 that MOG may provide the explanation of the recently reported ``extra
gravity'' without non-baryonic dark matter which has so far been interpreted as direct evidence of dark matter. 
The research presented here addresses the full-sky data product for the \bc, recently released to the public~\protect\citep{Clowe:dataProduct}.
\begin{SCfigure}[0.9][h]
\includegraphics[width=10cm]{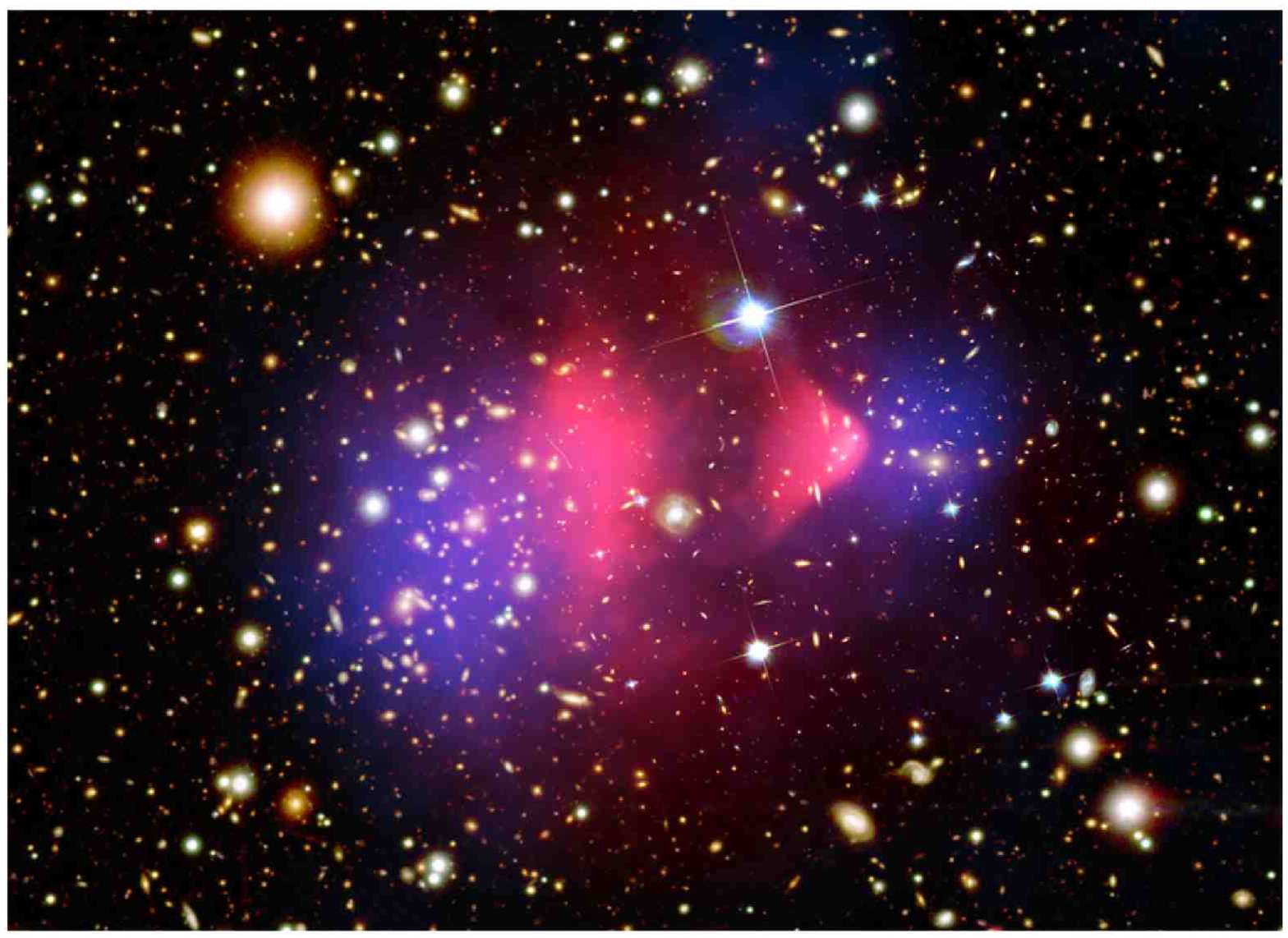}
\caption[False colour image of {\bc}]{\label{figure.1e0657} {{\sf\small False colour
image of {\bc}.}}\break  \small The surface density \map{\Sigma} reconstructed from X-ray imaging observations is shown
in red
and the convergence \map{\kappa} as reconstructed from strong and weak gravitational lensing observations is shown in
blue.  Image provided courtesy of \href{http://chandra.harvard.edu/photo/2006/1e0657/}{Chandra
X-ray Observatory}.}
\end{SCfigure}

\subsection{\label{section.LatestResults}The Latest Results from the \bc}

\begin{itemize}
\item The {\em Chandra} Peer Review has declared the {\bc} to be the most interesting cluster in the sky. 
This system, located at a redshift $z=0.296$ has the highest X-ray luminosity and temperature ($T = 14.1 \pm 0.2\
\mbox{keV} \sim 1.65 \times 10^{8}\ \mbox{K}$), and demonstrates a spectacular merger in the plane of the sky exhibiting
a supersonic shock front, with Mach number as high as $3.0 \pm 0.4$~\protect\citep{Markevitch:2006}.  The {\bc} has provided a rich dataset in the X-ray spectrum which has been
modeled to high precision.  From the extra-long $5.2
\times 10^{5}\ \mbox{s}$ {\em Chandra} space satellite X-ray image, the surface mass density, $\Sigma(x,y)$, was
reconstructed providing a high resolution map of the ICM gas~\protect\citep{Clowe:astro-ph.0611496}.  The \map{\Sigma}, shown in
a false colour composite map (in red) in \fref{figure.1e0657} is the result of a normalized geometric mass model
 based upon a $16^{\prime} \times 16^{\prime}$ field in the plane of the sky that covers the entire cluster and is
composed of a square grid of $185 \times 185$ pixels ($\sim 8000$ data-points)$^{1}$\protect\footnotetext[1]{Technical details in
Markevitch, M. et al. 2007, in preparation.}. Our analysis of the \map{\Sigma}
provides first published results for the King $\beta$-model density, as shown in \tref{table.king}, and mass profiles of 
the main cluster and the isothermal temperature profile as determined
by MOG, as shown in \fref{figure.MassProfile}.
\item Based on observations made with the NASA/ESA Hubble Space Telescope, the Spitzer Space Telescope and with the 6.5 meter
Magellan Telescopes, \protect\citet{Clowe:ApJL:2006,Clowe:ApJ:2006,Clowe:astro-ph.0611496} report on a combined strong and weak
gravitational lensing survey used to reconstruct a high-resolution, absolutely calibrated convergence \map{\kappa} of the
region of sky surrounding \bc, without assumptions on the underlying gravitational potential.  The
\map{\kappa} is shown in the false colour composite map (in blue) in \fref{figure.1e0657}.  The gravitational lensing reconstruction of the convergence map is a remarkable result, considering it is based on a
catalog of strong and weak lensing events and relies upon a thorough understanding of the distances
involved  -- ranging from the redshift of the {\bc} ($z = 0.296 $) which puts it at a distance of the
order of one million parsecs away.  Additionally, the typical angular diameter distances to the lensing event sources
($z \sim 0.8$ to $z\sim 1.0$) are several million parsecs distant!  This is perhaps the greatest source of error in the
\map{\kappa} which limits its precision. Regardless, we are able to learn much about the convergence map and its peaks. 
\end{itemize}

Both the \map{\Sigma} and the \map{\kappa} are two-dimensional distributions based upon line-of-sight integration.  We are fortunate,
indeed, that the {\bc} is not only one of the hottest, most supersonic, most massive cluster mergers seen, but the plane of
the merger is aligned with our sky!  As exhibited in \fref{figure.1e0657}, the
latest results from the {\bc} show, beyond a shadow of doubt, that the
\map{\Sigma}, which is a direct measure of the hot ICM gas, is offset from the \map{\kappa}, which is a direct measure
of the curvature (convergence) of space-time.  The fact that the \map{\kappa} is centered on the galaxies, and not on the
ICM gas mass is certainly either evidence of ``missing mass'', as in the case of the dark matter paradigm, or ``extra
gravity'', as in the case of MOG.   \protect\citet{Clowe:astro-ph.0611496} states
\begin{quotation}
One would expect that this (the offset $\Sigma$- and $\kappa$-peaks) indicates that dark matter must be present
regardless of the gravitational force law, but in some alternative gravity models, the multiple peaks can alter the lensing surface potential so that the strength of the peaks is no longer directly related to the matter density in them. As such, all of the alternative gravity models have to be tested individually against the
observations.
\end{quotation}
  The data from the {\bc} provides a laboratory of the greatest scale, where
the degeneracy between   ``missing mass'' and  ``extra gravity'' may be distinguished.  We demonstrate that in MOG, the
convergence \map{\kappa} correctly accounts for all of the baryons in each of the main and subclusters, including all
of the galaxies in the regions near the main central
dominant (cD) and the subcluster's brightest central galaxy (BCG), without non-baryonic dark matter.

\subsection{\label{section.Overview}{How Modified Gravity may account for the {\bc} evidence}} 

{We will show how MOG may account for the {\bc} evidence, without dominant dark matter, by deriving the modifications to the gravitational lensing equations from MOG.  Concurrently, we will provide comparisons to the equivalent Einstein-Newton results utilizing dark matter to explain the missing mass.  The paper is divided as follows:}

{\sref{section.Theory} is dedicated to the theory used to perform all of the derivations and numerical computations and is separated into three pieces:   \sref{section.MOG} presents the Poisson equations in MOG for a non-spherical distribution of matter and the corresponding
derivation of the acceleration law and the dimensionless gravitational coupling, $\mathcal{G}$.   \sref{section.Model} presents the King $\beta$-model density profile, $\rho$.
\sref{section.WeightedSurfaceMassDensity}  presents the derivation of the weighted surface mass density, ${\bar \Sigma} = \int \mathcal{G} \rho$, from the convergence $\kappa = {\bar \Sigma}/\Sigma_c$.  The effect of the dimensionless gravitational coupling, $\mathcal{G}$, is to carry more weight away from the center of the system.  If the galaxies occur away from the center of the ICM gas, as in the {\bc}, their contribution to the \map{\kappa} will be weighted by as much as a factor of 6 as shown in \fref{figure.G}.}

{\sref{section.Sigma} is dedicated to the surface density {\map{\Sigma} from the X-ray imaging observations of the {\bc} from the November 15, 2006 data release  \map{\Sigma}~\protect\citep{Clowe:dataProduct}.  \sref{section.SigmaMap}} presents a visualization of the \map{\Sigma} and our low $\chi^2$ best-fit King $\beta$-model (neglecting the subcluster).  \sref{section.MOGMain} presents a determination of $\mathcal{G}$ for the {\bc} based upon the ($>100$) galaxy cluster survey of~\protect\citet{Brownstein:MNRAS:2006}.  \sref{section.CylindricalMassProfile} presents the cylindrical mass profile, $\int \Sigma$, about the main cluster \map{\Sigma} peak.  \sref{section.T} presents the isothermal spherical mass profile from which we have derived
a parameter-free (unique) prediction  for the X-ray temperature of the {\bc} of $T=15.5 \pm 3.9\ \mbox{keV}$ which agrees with the experimental value of $T=14.8^{+2.0}_{-1.7}\ \mbox{keV}$, within the uncertainty. \sref{section.Subcluster} presents the details of the separation of the \map{\Sigma} into the main cluster and subcluster components.}

{\sref{section.Convergence} is dedicated to the convergence {\map{\kappa} from the weak and strong gravitational lensing survey of the {\bc} from the November 15, 2006 data release  \map{\kappa}~\protect\citep{Clowe:dataProduct}.  \sref{section.KappaMap} presents a visualization of the \map{\kappa} and some remarks on the evidence for dark matter or conversely, extra gravity.  \sref{section.MOG.solution}} presents the MOG solution which uses a projective approximation for the density profile to facilitate numerical integration, although the full non-spherically symmetric expressions are provided in \sref{section.MOG}.  To a zeroth order approximation, we present the spherically symmetric solution, which does not fit the {\bc}.  In our next approximation, the density profile of the {\bc}, $\rho$, is taken as the best-fit King $\beta$-model (spherically symmetric), but the dimensionless gravitational coupling, $\mathcal{G}$, also assumed to be spherically symmetric, has a different center -- in the direction closer to the subcluster.  We determined that our best-fit $\kappa$ model corresponds to a location of the MOG center 140 kpc away from the main cluster \map{\Sigma} toward the sub-cluster \map{\Sigma} peak.  \sref{section.Galaxies} presents the MOG prediction of the galaxy surface mass density, computed by taking the difference between the \map{\kappa} data and our $\kappa$-model of the \map{\Sigma} (ICM gas) data.  \sref{section.DM} presents the mass profile of dark matter computed by taking the difference between the \map{\kappa} data and scaled \map{\Sigma} (ICM gas) data.  This corresponds to the amount of dark matter (not a falsifiable prediction) necessary to explain the {\bc} data using Einstein/Newton gravity theory.}

\section{\label{section.Theory}The Theory}
\subsection{\label{section.MOG}Modified Gravity (MOG)}
Analysis of the recent X-ray data from the \bc~\protect\citep{Markevitch:2006}, probed by our computation in Modified Gravity (or MOG), provides direct
evidence that the  convergence \map{\kappa} reconstructed from strong and weak gravitational lensing observations~\protect\citep{Clowe:ApJL:2006,Clowe:ApJ:2006,Clowe:astro-ph.0611496} correctly accounts for
all of the baryons in each of the main and subclusters including all of the galaxies in the regions near the main
central dominant (cD) and the subcluster's brightest central galaxy (BCG).  The available baryonic mass, in addition
to a second-rank skew symmetric tensor field (in MSTG), or massive vector field (in STVG), are the only properties of the system which
contribute to the running gravitational coupling, $G(r)$.  It is precisely this effect which allows MOG to fulfill its
requirement
as a relativistic theory of gravitation to correctly describe astrophysical phenomena without the necessity of dark
matter~\protect\citep{Moffat:JCAP:2005,Moffat:JCAP:2006,Moffat:astro-ph.0608675}.  MOG contains a running gravitational coupling -- in the infrared
(IR) at astrophysical scales --
which has successfully been applied to galaxy rotation curves~\protect\citep{Brownstein:ApJ:2006}, X-ray cluster
masses~\protect\citep{Brownstein:MNRAS:2006}, and is
within limits set by solar system observations~\protect\citep{Brownstein:CQG:2006}.

{The weak field, point particle spherically symmetric acceleration
law in MOG is obtained from the action principle for the
relativistic equations of motion of a test particle in \protect\citet{Moffat:JCAP:2005,Moffat:JCAP:2006}.
The weak field point particle gravitational
potential for a static spherically symmetric system consists of
two parts:
\begin{equation}
\label{phipotential} \Phi(r)=\Phi_N(r)+\Phi_Y(r),
\end{equation}
where
\begin{equation}
\label{Newton} \Phi_N(r)=-\frac{G_{\infty}M}{r},
\end{equation}
and
\begin{equation}
\label{Yukawa} \Phi_Y(r)=\sigma \frac{\exp(-\mu r)}{r}
\end{equation}
denote the Newtonian and Yukawa potentials, respectively. $M$ denotes the total constant mass of the system and $\mu$ denotes the effective mass of the vector particle in STVG. The Poisson equations
for $\Phi_N({\bf r})$ and $\Phi_Y({\bf r})$ are given by
\begin{equation}
\label{NewtonPoiss} \nabla^2\Phi_N({\bf r})=-G_{\infty}\rho({\bf r}),
\end{equation}
and
\begin{equation}
\label{YukawaPoiss} (\nabla^2-\mu^2)\Phi_Y({\bf
r})=\frac{\sigma}{M}\rho({\bf r}),
\end{equation}
respectively. For sufficiently weak fields, we can assume that the
Poisson \erefs{NewtonPoiss}{YukawaPoiss} are
uncoupled and determine the potentials $\Phi_N({\bf r})$ and
$\Phi_Y({\bf r})$ for non-spherically symmetric systems, which can
be solved analytically and numerically. The Green's function for
the Yukawa Poisson equation is given by
\begin{equation}
(\nabla^2-\mu^2)\Delta_Y({\bf r})=-\delta^3({\bf r}).
\end{equation}
The full solutions to the potentials are given by
\begin{equation}
\label{fullNewton} \Phi_N({\bf r})=- G_{\infty}\int d^3{\bf
r'}\frac{\rho({\bf r'})}{\vert {\bf r}-{\bf r'}\vert}
\end{equation}
and
\begin{equation}
\label{fullYukawa} \Phi_Y({\bf r})=\frac{\sigma}{\int d^3\mathbf{r}'\rho(\mathbf{r}')} {\int d^3{\bf
r'}\frac{\exp(-\mu \vert{\bf r}-{\bf r'}\vert)\rho({\bf
r'})}{\vert {\bf r}-{\bf r'}\vert}}.
\end{equation}
For a delta function source density
\begin{equation}
\label{delta} \rho({\bf r})=M \delta^3({\bf r}),
\end{equation}
we obtain the point particle solutions of \erefs{Newton}{Yukawa}.}

{The modified acceleration law is obtained from
\begin{equation} \label{moda}
{\bf a}({\bf r})=-{\mathbf\nabla}\Phi=-\bigl({\mathbf\nabla}\Phi_N({\bf
r})+{\mathbf\nabla}\Phi_Y({\bf r})\bigr).
\end{equation}
Let us set
\begin{equation} \label{Gconst}
G_{\infty}=G_N\biggl[1+\biggl(\frac{M_0}{\int d^3{\bf
r}'\rho({\bf r}')}\biggr)^{1/2}\biggr],
\end{equation}
\begin{equation}
\label{sigmaconst} \sigma=G_N[M_0\int d^3{\bf r'}\rho({\bf
r}')]^{1/2},
\end{equation}
where $M_0$ is a constant  and $G_N$ denotes Newton's gravitational constant. From \erefss{fullNewton}{fullYukawa}{moda}, we obtain
\begin{eqnarray} \nonumber
{\bf a}({\bf r})&=&- G_N\int d^3{\bf r}'\frac{({\bf r}-{\bf r'})
\rho({\bf r'})}{\vert {\bf r}-{\bf
r}'\vert^3}\\ \nonumber
&& \times\biggl\{1+\biggl(\frac{M_0}{\int d^3{\bf r}'\rho({\bf
r}')}\biggr)^{1/2}\\
&& \times \Bigl[1-\exp(-\mu\vert{\bf r}-{\bf
r}'\vert)(1+\mu\vert{\bf r}-{\bf r}'\vert)\Bigr]\biggr\}.
\end{eqnarray}
We can write this equation in the form:
\begin{equation}
\label{fullGaccel} {\bf a}({\bf r})=-\int d^3{\bf
r}'\frac{({\bf r}-{\bf r'}) \rho({\bf r'})}{\vert {\bf r}-{\bf
r}'\vert^3}G({\bf r}-{\bf r}'),
\end{equation}
where
\begin{eqnarray}
 \label{fullG} G({\bf r}-{\bf
r}')&=&G_N\biggl\{1+\biggl(\frac{M_0}{\int d^3{\bf r}'\rho({\bf
r}')}\biggr)^{1/2}\\
\nonumber &&  \times \Bigl[1-\exp(-\mu\vert{\bf r}-{\bf
r}'\vert)(1+\mu\vert{\bf r}-{\bf r}'\vert)\Bigr]\biggr\}.
\end{eqnarray}}

{For a static spherically symmetric point particle system, we
obtain using \eref{delta} the effective modified acceleration
law:
\begin{equation} \label{Gforce}
a(r)=-\frac{G(r)M(r)}{r^2},
\end{equation}
where
\begin{equation}
\label{fullGspherical}
G(r)=G_N\biggl\{1+\sqrt{\frac{M_0}{M}}\biggl[1-\exp(-r/r_0)
\biggl(1+\frac{r}{r_0}\biggr)\biggr]\biggr\}.
\end{equation}
Here, $M$ is the total baryonic mass of the system and we have set
$\mu=1/r_0$ and $r_0$ is a distance range parameter. We observe
that $G(r)\rightarrow G_N$ as $r\rightarrow 0$.    }

\begin{SCfigure}[0.9][hb]
\scalebox{0.825}{\input{figure/gnuplot/M0r0}} 
\caption{\label{figure.M0r0} {{\sf\small Running Scales in Modified Gravity.}}\break
\small A plot of the
MOG mass scale, $M_{0}$, vs.\,the MOG range parameter, $r_{0}$ compiled from published fits to galaxy rotation curves and X-ray
cluster masses~\protect\citep{Brownstein:ApJ:2006, Brownstein:MNRAS:2006} The values of $M_{0}$
vs.\,$r_{0}$ are plotted for galaxies with green squares, dwarf galaxies with black circles, clusters of galaxies with blue crosses, and
dwarf clusters of galaxies with red triangles.}
\end{SCfigure}
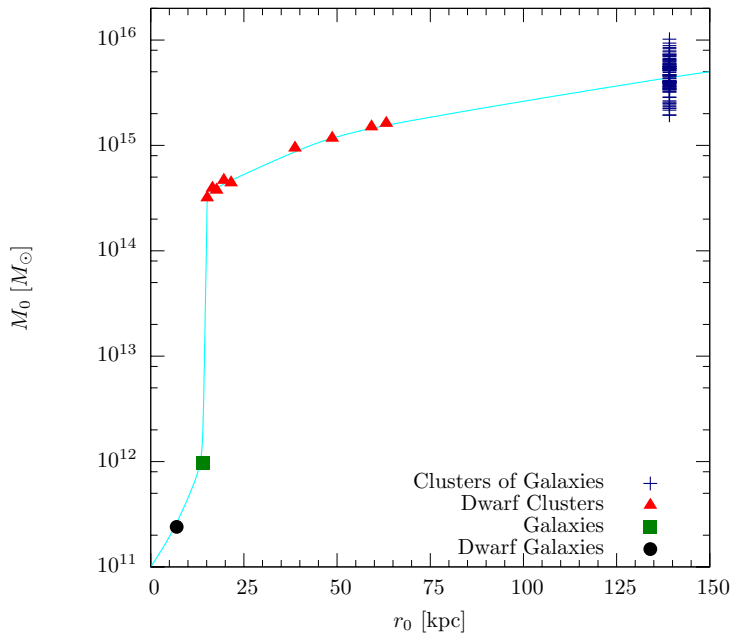
For the spherically symmetric static solution in MOG, the modified acceleration law  \eref{Gforce} is determined by the
baryon density and the parameters $G_N$, $M_0$ and $r_0$. However, the parameters $M_0$ and $r_0$ are scaling parameters that vary
with distance, $r$, according to the field equations for the scalar fields  $\omega(r) \propto M_0(r)$
and $\mu(r) =1/r_0(r)$, obtained from the action principle~\protect\citep{Moffat:JCAP:2006}.  In
principle, solutions of the effective field equations for the variations of $\omega(r)$ and $\mu(r)$ can be derived given
the potentials $V(\omega)$ and $V(\mu)$ in Equations (24) and (26) of \protect\citet{Moffat:JCAP:2006} 
However, at present the variations of $\omega(r)$ and $\mu(r)$ are empirically determined from
the galaxy rotation curve and X-ray cluster mass data~\protect\citep{Brownstein:ApJ:2006, Brownstein:MNRAS:2006}. 

In \fref{figure.M0r0}, we display the values of $M_0$ and $r_0$ for spherically symmetric systems obtained from
the published fits to the galaxy rotation curves for dwarf galaxies, elliptical galaxies and spiral galaxies, and the
fits to (normal and dwarf) X -ray cluster masses. A complete continuous relation between $M_0$ and $r_0$ at all mass
scales needs to be determined from the MOG that fits the empirical mass scales and distance scales shown in \fref{figure.M0r0}.
This will be a subject of future investigations.

{The modifications to gravity leading to \eref{fullGspherical} would be negated by the vanishing of either $M_{0} \rightarrow 0$ or the
$r_{0} \rightarrow \infty$ infrared limit.  These scale parameters are not to be taken as universal constants, but
are source dependent and scale according to the system.  This is precisely the opposite case in MOND, where the
Milgrom acceleration, $a_{0{\rm Milgrom}} = 1.2 \times 10^{-8}\ \mbox{cm s}^{-2}$\citep{Milgrom:ApJ:1983}, is a
phenomenologically derived universal constant -- arising from a classical modification to the Newtonian potential. 
Additionally, MOND has an arbitrary function, $\mu(x)$, which should be counted among the degrees of freedom of that
theory.  Conversely, MOG does not arise from a classical modification, but from the equations of motion of a relativistic
modification to general relativity.}

{The cases we have
examined until now have been modeled assuming spherical symmetry.
These include 101 galaxy rotation curves~\protect\citep{Brownstein:ApJ:2006}, 106 X-ray
cluster masses~\protect\citep{Brownstein:MNRAS:2006}, and the gravitational solution to the Pioneer 10/11 anomaly
in the solar system~\protect\citep{Brownstein:CQG:2006}.  In applications of MOG, we vary the gravitational coupling $G$, the vector field $\phi_\mu$ coupling to matter and the effective mass $\mu$ of the vector field according to a renormalization group (RG) flow description of quantum gravity theory formulated in terms of an
effective classical action~\protect\citep{Moffat:JCAP:2005,Moffat:JCAP:2006}. Large infrared
renormalization effects may cause the effective coupling constants
to run with momentum. A cutoff leads to spatially varying values
of $G, M_0$ and $r_0$ and these values increase in size according
to the mass scale and distance scale of a physical system~\protect\citep{Reuter:IJMPD:2006}.}

The spatially varying dimensionless gravitational coupling is given by
\begin{equation} \label{G}
\mathcal{G}(r) \equiv \frac{G(r)}{G_{N}} = 1 + \sqrt{\frac{M_{0}}{M(r)}}\left\{1-\exp\left(-\frac{r}{r_{0}}\right)\left(1+\frac{r}{r_{0}}\right)\right\},
\end{equation}
where 
\begin{itemize}
\item $G_{N} = 6.6742 \times 10^{-11}\ \mbox{m}^{3}/\mbox{kg s}^{2}$
is the ordinary (terrestrial) Newtonian gravitational constant measured experimentally$^{2}$\protect\footnotetext[2]{\href{http://www.physics.nist.gov/cgi-bin/cuu/Value?bg}{NIST
2002 CODATA value.}}, 
\item $M(r)$ is the total (ordinary) mass enclosed in a sphere or radius, r.  This may include all of the visible
(X-ray) ICM
gas and all of the galactic (baryonic) matter, but none of the non-baryonic dark matter.
\item $M_{0}$ is the MOG mass scale (usually measured in units of [$M_{\sun}$]),
\item $r_{0}$ is the MOG range parameter (usually measured in units of [kpc]).
\end{itemize}

The dimensionless gravitational coupling, $\mathcal{G}(r)$, of \eref{G}, 
approaches an asymptotic value as $r\rightarrow \infty$:
\begin{equation} \label{Ginf}
\mathcal{G}_{\infty}  = 1 + \sqrt{\frac{M_{0}}{M}},
\end{equation}
where $M$ is the total baryonic mass of the system. Conversely, taken in the limit of $r \ll r_{0}$, $G(r)\rightarrow
G_{N}$ or $\mathcal{G}(r) \rightarrow 1$ down to the Planck length.

In order to apply the MOG model of \eref{G} to the \bc, we must
first generalize the spherically symmetric case.  Our approach follows a sequence of approximations:
\begin{enumerate}
\item Treat the subcluster as a perturbation, and neglect it as a zeroth order approximation.
\item Treat the subcluster as a perturbation, and shift the origin of the gravitational coupling, $\mathcal{G}(r)$ toward the subcluster (toward the center-of-mass of the system).
\item Use the concentric cylinder mass $M(R)$ as an approximation for $M(r)$, and shift that toward the subcluster (toward the MOG center).
\item Treat the subcluster as a perturbation, and utilize the isothermal sphere model to approximate $M(R)$ and shift that toward the subcluster (toward the center-of-mass of the system -- where the origin of $\mathcal{G}(r)$ is located).
\end{enumerate}
\subsection{\label{section.Model}The King $\beta$-Model for the \map{\Sigma}}
Starting with the zeroth order approximation, we neglect the subcluster, and perform a best-fit to determine a spherically symmetric King $\beta$-model density of the main cluster.  
We assume the
main cluster gas (neglecting the subcluster) is in nearly hydrostatic equilibrium with the gravitational potential of the galaxy cluster. Within a
few core radii, the distribution of gas within a galaxy cluster may be fit by a King ``$\beta$-model''.  The
observed surface brightness of the X-ray cluster can be fit to a radial distribution profile~\protect\citep{Chandrasekhar:1960,King:AJ:1966}:
\begin{equation}
\label{betaModel} I(r)= I_{0}\left[  1+\left(\frac{r}{r_{c}}\right)^{2}\right]^{-3 \beta + 1/2},
\end{equation}
resulting in best-fit parameters, $\beta$ and $r_{c}$. A deprojection of the $\beta$-model of \eref{betaModel}
assuming a nearly isothermal gas sphere then results in a physical gas density distribution~\protect\citep{Cavaliere:AAP:1976}:
\begin{equation}
\label{betaRhoModel} \rho(r)= \rho_{0}\left[  1+\left(\frac{r}{r_{c}}\right)^{2}\right]^{-3 \beta /2},
\end{equation}
where $\rho(r)$ is the ICM mass density profile, and $\rho_{0}$ denotes the central density.  The mass profile associated with this density is given by
\begin{equation}
\label{massProfile}
M(r)  = 4\pi\int_{0}^{r} \rho({r^{\prime}}) {r^{\prime}}^{2} d{r^{\prime}},
\end{equation}
where $M(r)$ is the total mass contained within a sphere of radius r.  Galaxy clusters are observed to have finite 
spatial extent.  This allows an approximate determination of the total mass of the galaxy cluster by first solving
equation (\ref{betaRhoModel})
for the position, $r_{\rm out}$, at which the density, $\rho(r_{\rm out})$, drops to $\approx 10^{-28}\,\mbox{g/cm}^{3}$, or 250
times the mean cosmological density of baryons:
\begin{equation}
\label{rout.0}
r_{\rm out} = r_{c} \left[\left(
\frac{\rho_{0}}{10^{-28}\,\mbox{g/cm}^{3}}\right)^{2/3\beta}-1\right]^{1/2}.
\end{equation}
Then, the total mass of the ICM gas may be taken as $M_{\rm gas} \approx M(r_{\rm out})$:
\begin{equation}
\label{Mgas.0}
M_{\rm gas}  = 4\pi\int_{0}^{r_{\rm out}}\rho_{0} \left[  1+\left(\frac{{r^{\prime}}}{r_{c}}\right)^{2}\right]^{-3 \beta /2} {r^{\prime}}^{2}
d{r^{\prime}}.
\end{equation}
To make contact with the
experimental data, we must calculate the surface mass density by integrating $\rho(r)$ of \eref{betaRhoModel} along the
line-of-sight:
\begin{equation}
\label{surfaceMassDensity.1}
\Sigma(x,y)  = \int_{-z_{\rm out}}^{z_{\rm out}} \rho(x,y,z) dz,
\end{equation} where
\begin{equation}
\label{zout}
z_{\rm out} = \sqrt{r_{\rm out}^{2} - x^{2} - y^{2}}.
\end{equation}
Substituting \eref{betaRhoModel} into \eref{surfaceMassDensity.1}, we obtain
\begin{equation}
\label{surfaceMassDensity.2}
\Sigma(x,y)  = \rho_{0} \int_{-z_{\rm out}}^{z_{\rm out}} \left[  1+\frac{{x^{2}+y^{2}+z^{2}}}{r_{c}^{2}}\right]^{-3
\beta /2} dz.
\end{equation}
This integral becomes tractable by making a substitution of variables:
\begin{equation}
\label{substitutedVariable}
u^{2} = 1+\frac{x^{2}+y^{2}}{r_{c}^{2}},
\end{equation}
so that
\begin{eqnarray} \nonumber
\Sigma(x,y)  &=& \rho_{0} \int_{-z_{\rm out}}^{z_{\rm out}} \left[ u^{2} + \left(\frac{z}{r_{c}}\right)^{2}\right]^{-3
\beta /2} dz\\
\nonumber &=& \frac{\rho_{0}}{u^{3\beta}}\int_{-z_{\rm out}}^{z_{\rm out}} \left[1 + \left(\frac{z}{u
r_{c}}\right)^{2}\right]^{-3 \beta /2} dz\\
\label{surfaceMassDensity.3} &=& 2 \frac{\rho_{0}}{u^{3\beta}} z_{\rm out}
F\left(\left[\frac{1}{2},\frac{3}{2}\beta\right],\left[\frac{3}{2}\right],-\left(\frac{z_{\rm out}}{u
r_{c}}\right)^{2}\right),
\end{eqnarray}
where we have made use of the hypergeometric function, $F([a,b],[c],z)$.  Substituting \eref{substitutedVariable} into
\eref{surfaceMassDensity.3} gives
\begin{equation}
\label{surfaceMassDensity.4}
\Sigma(x,y)  =  2 \rho_{0} z_{\rm out}\left(1+\frac{x^{2}+y^{2}}{r_{c}^{2}}\right)^{-3\beta/2} 
F\left(\left[\frac{1}{2},\frac{3}{2}\beta\right],\left[\frac{3}{2}\right],-\frac{z_{\rm
out}^{2}}{x^{2}+y^{2}+r_{c}^{2}}\right).
\end{equation}
We next define
\begin{equation}
\label{surfaceMassDensity0.1}
\Sigma_{0} \equiv \Sigma(0,0)  =  2 \rho_{0} z_{\rm out} 
F\left(\left[\frac{1}{2},\frac{3}{2}\beta\right],\left[\frac{3}{2}\right],-\left(\frac{z_{\rm
out}}{r_{c}}\right)^{2}\right),
\end{equation}
which we substitute into \eref{surfaceMassDensity.4}, yielding
\begin{equation}
\label{surfaceMassDensity.5}
\Sigma(x,y)  = \Sigma_{0} \left(1+\frac{x^{2}+y^{2}}{r_{c}^{2}}\right)^{-3\beta/2} 
\frac{F\left(\left[\frac{1}{2},\frac{3}{2}\beta\right],\left[\frac{3}{2}\right],-\frac{z_{\rm
out}^{2}}{x^{2}+y^{2}+r_{c}^{2}}\right)}{F\left(\left[\frac{1}{2},\frac{3}{2}\beta\right],\left[\frac{3}{2}\right],-\frac{z_{\rm
out}^{2}}{r_{c}^{2}}\right)}.
\end{equation}
In the limit $z_{\rm out} \gg r_{c}$, the Hypergeometric functions simplify to $\Gamma$ functions, and
\erefs{surfaceMassDensity0.1}{surfaceMassDensity.5} result in the simple, approximate solutions:
\begin{equation}
\label{surfaceMassDensity0}
\Sigma_{0} =  \sqrt{\pi} \rho_{0} r_{c}  \frac{\Gamma\left(\frac{3\beta-1}{2}\right)}{\Gamma\left(\frac{3}{2}\beta\right)}
\end{equation}
and
\begin{equation}
\label{surfaceMassDensity}
\Sigma(x,y)  = \Sigma_{0} \left(1+\frac{x^{2}+y^{2}}{r_{c}^{2}}\right)^{-(3\beta-1)/2 } .
\end{equation}
which we may, in principle, fit to the \map{\Sigma} data to determine the King $\beta$-model parameters, $\beta$,
$r_{c}$ and $\rho_{0}$.

\subsection{\label{section.WeightedSurfaceMassDensity}Deriving the Weighted Surface Mass Density from the Convergence \map{\kappa}}
The goal
of the strong and weak lensing survey of ~\protect\citet{Clowe:ApJL:2006,Clowe:ApJ:2006,Clowe:astro-ph.0611496} was to obtain a convergence
\map{\kappa} by measuring the distortion of images of background galaxies (sources) caused 
by the deflection of light as it passes the {\bc} (lens).  The distortions in image ellipticity are only measurable
statistically with large numbers of sources.  The data were first corrected for smearing by the point spread function in
the image, resulting in a noisy, but direct, measurement of the reduced shear $g=\gamma/(1-\kappa)$.  The shear,
$\gamma$, is the anisotropic stretching of the galaxy image, and the convergence, $\kappa$, is the shape-independent
change in the size of the image.  By recovering the \map{\kappa} from the measured reduced shear field, a measure of the
local curvature is obtained. In Einstein's general relativity, the local curvature is related to the distribution of mass/energy,
as it is in MOG.  In Newtonian gravity theory the relationship between the \map{\kappa} and the surface mass density becomes
very simple, allowing one to refer to $\kappa$ as the scaled surface mass density (see, for example, Chapter 4 of \protect\citet{Peacock:2003} for a derivation):
\begin{equation}
\label{scaledSurfaceMassDensity.Newton}
\kappa(x,y) =  \int \frac{4\pi G_{N}}{c^{2}} \frac{D_{\rm l}D_{\rm ls}}{D_{\rm s}} \rho(x,y,z) dz  \equiv \frac{\Sigma}{\SigmaN},
\end{equation}
where
\begin{equation}
\label{Sigma.Newton} \Sigma(x,y) = \int \rho(x,y,z) dz,
\end{equation}
is the surface mass density, and
\begin{equation}
\label{SigmaC.Newton}
\SigmaN = \frac{c^{2}}{4\pi G_{N}} \frac{D_{\rm s}}{D_{\rm l}D_{\rm ls}} \approx 3.1 \times 10^{9}\
M_{\sun}/\mbox{kpc}^{2}
\end{equation}
is the Newtonian critical surface mass density (with vanishing shear), $D_{\rm s}$  is the angular distance to a source
(background) galaxy, $D_{\rm l}$ is the angular distance to the lens 
(\bc), and $D_{\rm ls}$ is the angular distance from the {\bc} to a source galaxy.  The result of \eref{SigmaC.Newton} is equivalent to
\begin{equation}
\label{SigmaC.Deff}
\frac{D_{\rm l}D_{\rm ls}}{D_{\rm s}} \approx 540\ \mbox{kpc}.
\end{equation}
Since there is a multitude of
source galaxies, these distances become ``effective'', as is the numeric value presented in \eref{SigmaC.Newton}, quoted
from \protect\citep{Clowe:ApJ:2004} without estimate of the uncertainty.  In fact, due to the multitude of sources in the
lensing survey, both $D_{\rm s}$ and $D_{\rm ls}$ are distributions in $(x,y)$.  However, it is common practice  to move
them outside the integral, as a necessary approximation.  

We may obtain a similar result in MOG, as was shown in \protect\citet{Moffat:astro-ph.0608675}, by promoting the Newtonian
gravitational constant, $G_{N}$ to the running gravitational coupling, $G(r)$, but approximating $G(r)$ as sufficiently slow-varying to
allow it to be removed from the integral.  We have in general,
\begin{equation}
\label{scaledSurfaceMassDensity.MOG.0}
\kappa(x,y) =  \int \frac{4\pi G(r)}{c^{2}} \frac{D_{\rm l}D_{\rm ls}}{D_{\rm s}} \rho(x,y,z) dz  \equiv \frac{\Sigma(x,y)}{\Sigma_{c}(r)},
\end{equation}
and we assume that
\begin{equation}
\label{SigmaC.MOG.0}
\Sigma_{c}(r) = \frac{c^{2}}{4\pi G(r)} \frac{D_{\rm s}}{D_{\rm l}D_{\rm ls}}  = \frac{G_{N}}{G(r)}
\SigmaN = \frac{\SigmaN}{\mathcal{G}(r)},
\end{equation}
where we applied \eref{G}.  However, \erefs{scaledSurfaceMassDensity.MOG.0}{SigmaC.MOG.0} are only valid in the thin lens approximation. For the \bc, the thin lens approximation is inappropriate, and we must
use the
correct relationship between $\kappa$ and $\Sigma$:
\begin{equation}
\label{scaledSurfaceMassDensity.MOG}
\kappa(x,y) =  \int \frac{4\pi G(r)}{c^{2}} \frac{D_{\rm l}D_{\rm ls}}{D_{\rm s}} \rho(x,y,z) dz  \equiv \frac{\bar \Sigma(x,y)}{\SigmaN},
\end{equation}
where
\begin{equation}
\label{Sigma.MOG} {\bar \Sigma}(x,y) = \int \mathcal{G}(r)\rho(x,y,z) dz,
\end{equation}
is the weighted surface mass density (weighted by the dimensionless gravitational coupling $\mathcal{G}(r)$ of \eref{G}), and $\SigmaN$ is the usual Newtonian critical surface mass density \eref{SigmaC.Newton}.

For the remainder of this paper, we will use \erefsss{G}{SigmaC.Newton}{scaledSurfaceMassDensity.MOG}{Sigma.MOG} to
reconcile the experimental observations of the gravitational lensing \map{\kappa} with the X-ray imaging \map{\Sigma}.
We can already see from these equations, how in MOG the convergence, $\kappa$, is now related to the weighted surface
mass density, $\bar \Sigma$, so
\begin{quotation}
\ldots $\kappa$ is no longer a measurement 
of the surface density, but is a non-local function 
whose overall level is still tied to the amount of 
mass. For complicated system geometries, such 
as a merging cluster, the multiple peaks can deflect, suppress, or enhance some of the peaks~\protect\citep{Clowe:astro-ph.0611496}.
\end{quotation}

\section{\label{section.Sigma}The Surface Density Map from X-ray Image Observations}
\subsection{\label{section.SigmaMap}The \map{\Sigma}}
\begin{figure}[h]
\psset{unit=1pt}
\begin{pspicture}(500,325)(0,0)
\psset{linecolor=DarkGreen,linewidth=1pt}
{\thicklines \multiput(113,303)(90,0){4}{\line(0,1){9}} \put(113,303){\line(1,0){284.4}}}
{\thinlines \multiput(113,303)(4.5,0){64}{\line(0,1){3}} \multiput(113,303)(18,0){16}{\line(0,1){6}}}
\put(403,317){\makebox(0,0){\sf [kpc]}}
\multiput(103,317)(90,0){4}{\makebox(20,0){\sf\arabic{kpc}}\addtocounter{kpc}{500}}
\put(110,0){\includegraphics[width=290pt]{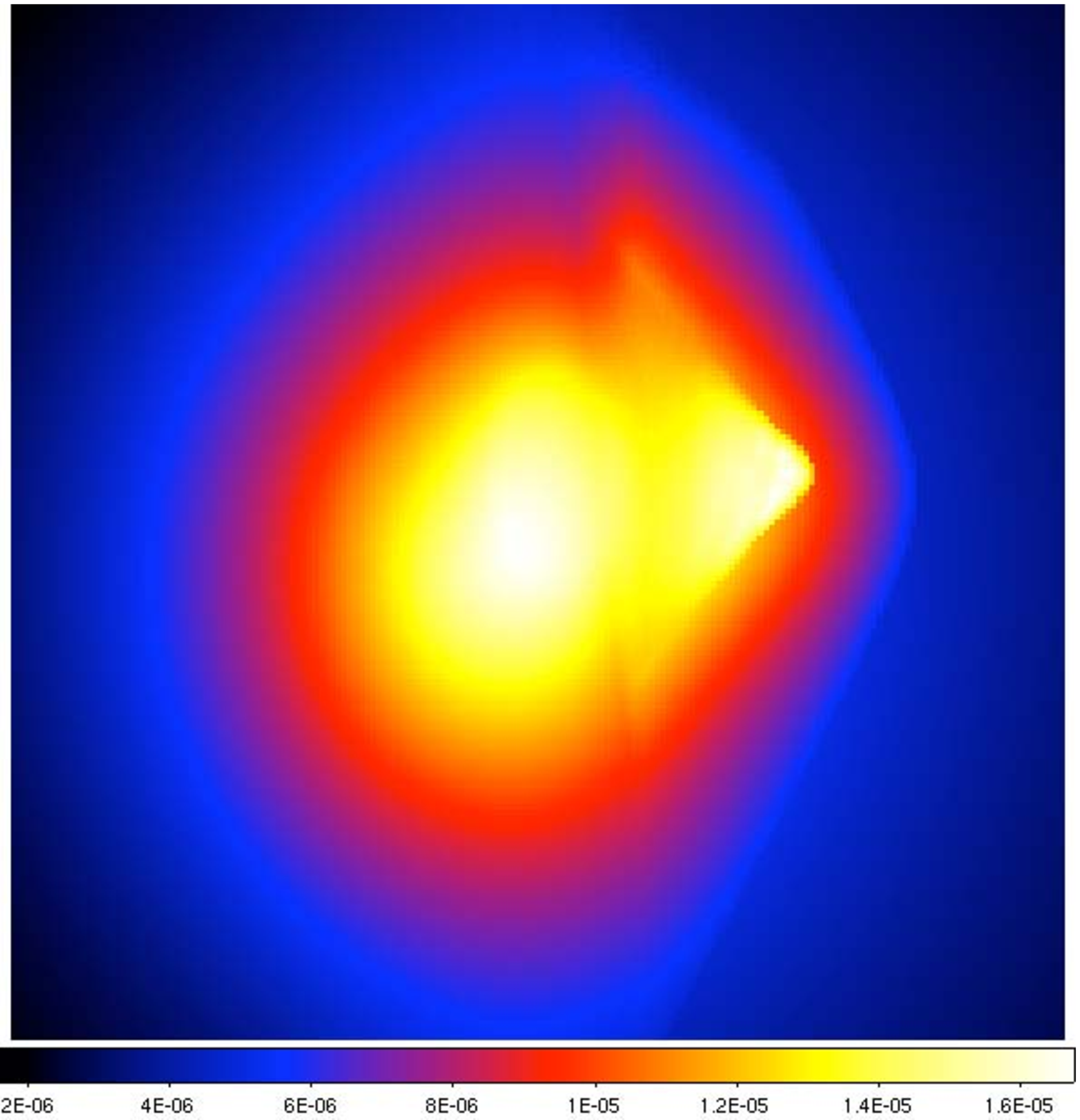}} 
\psline{->}(105,247)(249,155) \psline{->}(105,187)(221,153) \psline{->}(105,127)(219,141) \psline{->}(105,67)(273,159)
\psline{->}(403,247)(320,168) \psline{->}(403,187)(341,174) \psline{->}(403,127)(349,168) 
\put(0, 240){\fcolorbox{cayenne}{sand}{\makebox(100,14){\color{chocolate}\small\sf Main cluster \map{\Sigma} peak}}} 
\put(0, 180){\fcolorbox{cayenne}{sand}{\makebox(100,14){\color{chocolate}\small\sf Main cluster cD galaxy}}} 
\put(0, 120){\fcolorbox{cayenne}{sand}{\makebox(100,14){\color{chocolate}\small\sf Main cluster \map{\kappa} peak}}} 
\put(0, 60){\fcolorbox{cayenne}{sand}{\makebox(100,14){\color{chocolate}\small\sf MOG Center}}} 
\put(403, 240){\fcolorbox{cayenne}{sand}{\makebox(100,14){\color{chocolate}\small\sf Subcluster \map{\Sigma} peak}}} 
\put(403, 180){\fcolorbox{cayenne}{sand}{\makebox(100,14){\color{chocolate}\small\sf Subcluster \map{\kappa} peak}}} 
\put(403, 120){\fcolorbox{cayenne}{sand}{\makebox(100,14){\color{chocolate}\small\sf Subcluster BCG}}} 
\end{pspicture}
\caption{\label{figure.sigma} {{\sf\small Surface density \map{\Sigma}.}} \break \small Data
reconstructed from X-ray imaging observations of the
\bc, November 15, 2006 data release~\protect\citep{Clowe:dataProduct}. 
\map{\Sigma} observed peaks (local maxima) and
\map{\kappa} observed peaks are shown for comparison.  The
central dominant (cD) galaxy of the main cluster, the brightest cluster galaxy (BCG) of the subcluster, and the
MOG predicted gravitational center are shown.  The colourscale is shown at the bottom, in units of $10^{15}\ M_{\sun}/\mbox{pixel}^{2}$.  The resolution of the \map{\Sigma}  
is 8.5 kpc/pixel, based upon the measured redshift distance $\sim$ 260.0
kpc/arcminute of the \bc~\protect\citep{Clowe:ApJ:2006}.  The scale in kpc is shown at the top of the figure.  J2000 and map (x,y) coordinates are listed in \tref{table.coords}.}
\end{figure}

With an advance of the November 15, 2006 data release~\protect\citep{Clowe:dataProduct},  we began a precision analysis 
to model the gross features of the surface density \map{\Sigma} data in order to gain insight into the three-dimensional
matter distribution, $\rho(r)$, and to separate the components into a model representing the main
cluster and the subcluster -- the remainder after subtraction. 

The \map{\Sigma} is shown in false colour in \fref{figure.sigma}.  There are two distinct peaks in the surface density
\map{\Sigma} - the primary peak centered at the main cluster, and the secondary peak centered at the subcluster. The main
cluster gas is the brightly glowing (yellow) region to the left of the subcluster gas, which is the nearly equally bright
shockwave region (arrowhead shape to the right).
The \map{\kappa} observed peaks, the
central dominant (cD) galaxy of the main cluster, the brightest cluster galaxy (BCG) of the subcluster, and the
MOG predicted gravitational center are shown in \fref{figure.sigma} for comparison.  J2000 and map (x,y) coordinates are listed in
\tref{table.coords}.

\begin{table}[ht]
\begin{tabular}{|c||cc|c|c|} \hline
{\sf Observation} & \multicolumn{2}{c|}{{\sf J2000 Coordinates}} & {\sf\map{\Sigma}} & {\sf\map{\kappa}}\\
& {\sf RA} & {\sf Dec} & { $(x,\,y)$} & {$(x,\,y)$} \\ \hline \hline
Main cluster \map{\Sigma} peak & 06 : 58 : 31.1\quad & \quad-55 : 56 : 53.6 & $(89,\,89)$ & $(340,\,321)$ \\
Subcluster \map{\Sigma} peak & 06 : 58 : 20.4\quad & \quad-55 : 56 : 35.9 & $(135,\,98)$ & $(365,\,326)$ \\
Main cluster \map{\kappa} peak & 06 : 58 : 35.6\quad & \quad-55 : 57 : 10.8 & (70,\,80) & (329,\,317)\\
Subcluster \map{\kappa} peak & 06 : 58 : 17.0\quad & \quad-55 : 56 : 27.6 & (149,\,102) & (374,\,327) \\
Main cluster cD & 06 : 58 : 35.3\quad & \quad-55 : 56 : 56.3 & (71,\,88) & (330,\,320) \\
Subcluster BCG & 06 : 58 : 16.0\quad & \quad-55 : 56 : 35.1 & (154,\,98) & (375,\,326) \\
MOG Center & 06 : 58 : 27.6\quad & \quad-55 : 56 : 49.4 & (105,\,92) & (348,\,322) \\ \hline
\end{tabular}
 \caption{\label{table.coords} {{\sf\small The J2000 sky coordinates  of the \bc,  November 15, 2006 data release~\protect\citep{Clowe:dataProduct}:}}  \break \small Main  cluster and subcluster
\map{\Sigma} and
\map{\kappa} observed peaks, the
central dominant (cD) galaxy of the main cluster, the brightest cluster galaxy (BCG) of the subcluster, and the
MOG predicted gravitational center.  The resolution of the \map{\Sigma} and \map{\kappa} 
are 8.5 kpc/pixel, and 15.4 kpc/pixel, respectively, based upon the measured redshift distance $\sim$ 260.0
kpc/arcminute of the \bc~\protect\citep{Clowe:ApJ:2006}.}
 \end{table}
\begin{SCfigure}[1.0][hb]
\scalebox{0.7}{\input{figure/gnuplot/SigmaFit}} 
\caption{\label{figure.SigmaFit} {{\sf\small King $\beta$-model fit to scaled \map{\Sigma}.}}\break
\small A cross-section of the scaled
surface density \map{\Sigma} data reconstructed from X-ray
imaging observations of the \bc, on a straight-line connecting the
main X-ray cluster peak ($R \equiv 0\ \mbox{kpc}$) to the main central dominant (cD) galaxy ($R \approx -150\
\mbox{kpc}$). The 
Main cluster \map{\Sigma} data, taken from the November 15, 2006 data release~\protect\citep{Clowe:dataProduct}, is shown in
solid red, and the surface density \map{\Sigma} according to the best-fit King $\beta$-model (neglecting the subcluster)
of
\eref{surfaceMassDensity} is shown in
short-dashed blue.  The unmodeled peak (at $R \sim 300\ \mbox{kpc}$) is due to the subcluster.  We used the \protect\citet{Clowe:ApJ:2004} value 
for the Newtonian critical surface mass density (with vanishing shear), $\SigmaN = 3.1 \times 10^{9}\
M_{\sun}/\mbox{kpc}^{2}$.  J2000 and map (x,y) coordinates are listed in \tref{table.coords}.
  The best-fitting King $\beta$-model parameters are listed in \tref{table.king}.}
\end{SCfigure}

We may now proceed to calculate the best-fit parameters, $\beta$, $r_{c}$ and $\rho_{0}$ of the King $\beta$-model of
\erefs{surfaceMassDensity0}{surfaceMassDensity}, by applying
a nonlinear least-squares fitting routine (including estimated errors) to the entire \map{\Sigma}, or alternatively by
performing the fit to a subset of the {\map{\Sigma}} on a straight-line connecting the main cluster \map{\Sigma} peak
($R\equiv 0$)
to the main cD, and then extrapolating the fit to the entire map.  This reduces the complexity of the calculation
to a simple algorithm, but is not guaranteed to yield a global best-fit.  However, our approximate
best-fit will prove to agree with the \map{\Sigma} everywhere, except at the subcluster (which is neglected for the best-fit).

The scaled surface density \map{\Sigma} data is shown in solid red in \fref{figure.SigmaFit}. The unmodeled peak (at $R
\sim 300\ \mbox{kpc}$) is due to the subcluster.  The best-fit to the King $\beta$-model of \eref{surfaceMassDensity} is
shown in \fref{figure.SigmaFit} in short-dashed blue, and corresponds to
\begin{eqnarray}
\label{beta} \beta &=& 0.803\pm0.013,\\
\label{rc} r_{c} &=& 278.0\pm6.8\ \mbox{kpc},
\end{eqnarray}
where the value of the \map{\Sigma} at the main cluster peak is constrained to the observed value, 
\begin{equation} 
\label{Sigma0} \Sigma_{0} = 1.6859\times 10^{10} \frac{M_{\sun}}{\mbox{pixel}^{2}} \left(\frac{1\ \mbox{pixel}}{8.528\
\mbox{kpc}}\right)^{2}=2.3181\times 10^{8} M_{\sun}/\mbox{kpc}^{2}.
\end{equation}

\begin{figure}[ht]
\begin{center}
\subfloat[ {{\sf\small\map{\Sigma}}}]{\label{figure.SigmaSurface}\includegraphics[width=8cm]{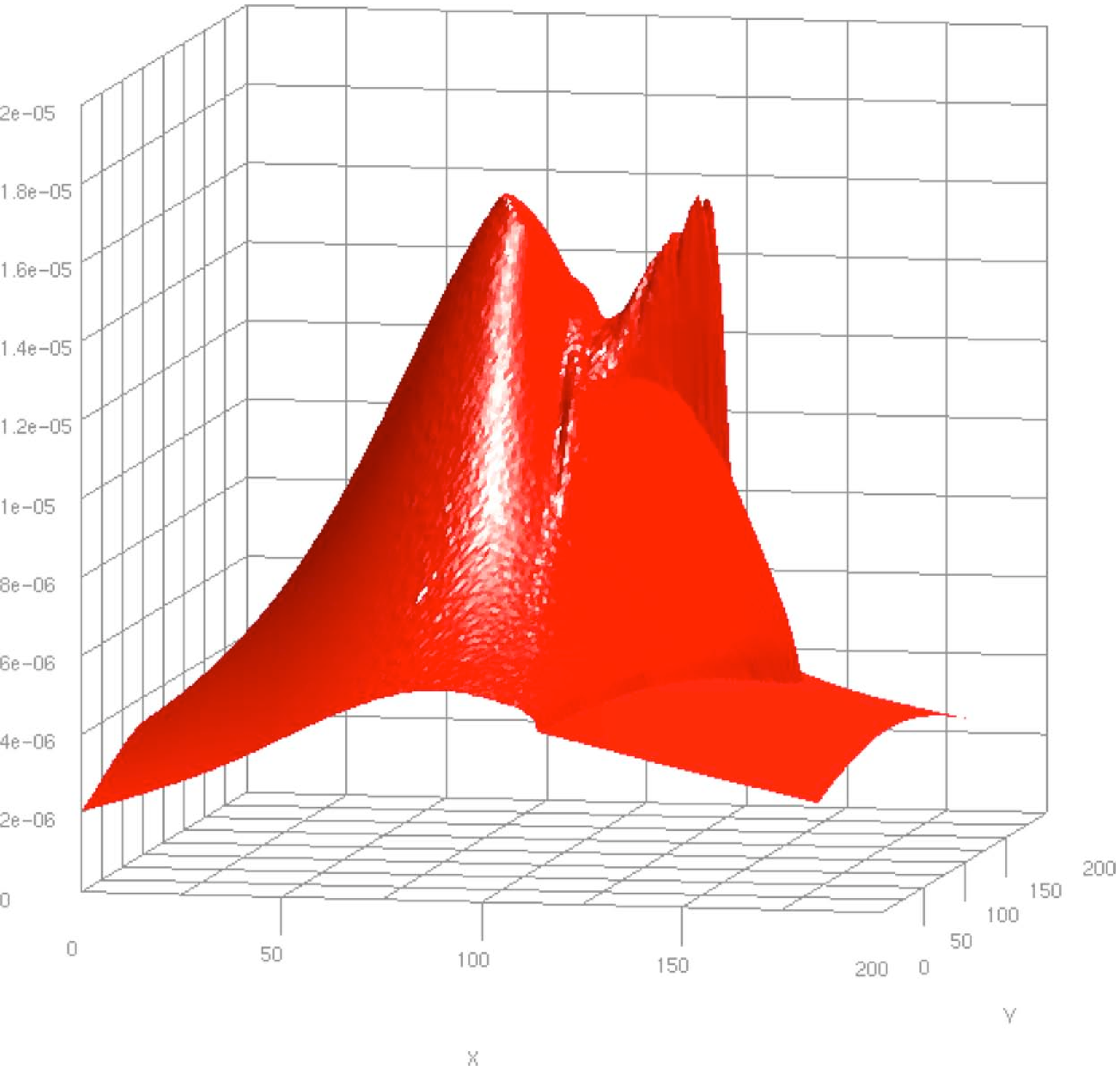}}\quad\quad 
\subfloat[ {{\sf\small King $\beta$-model}}]{\label{figure.SigmaModel}\includegraphics[width=8cm]{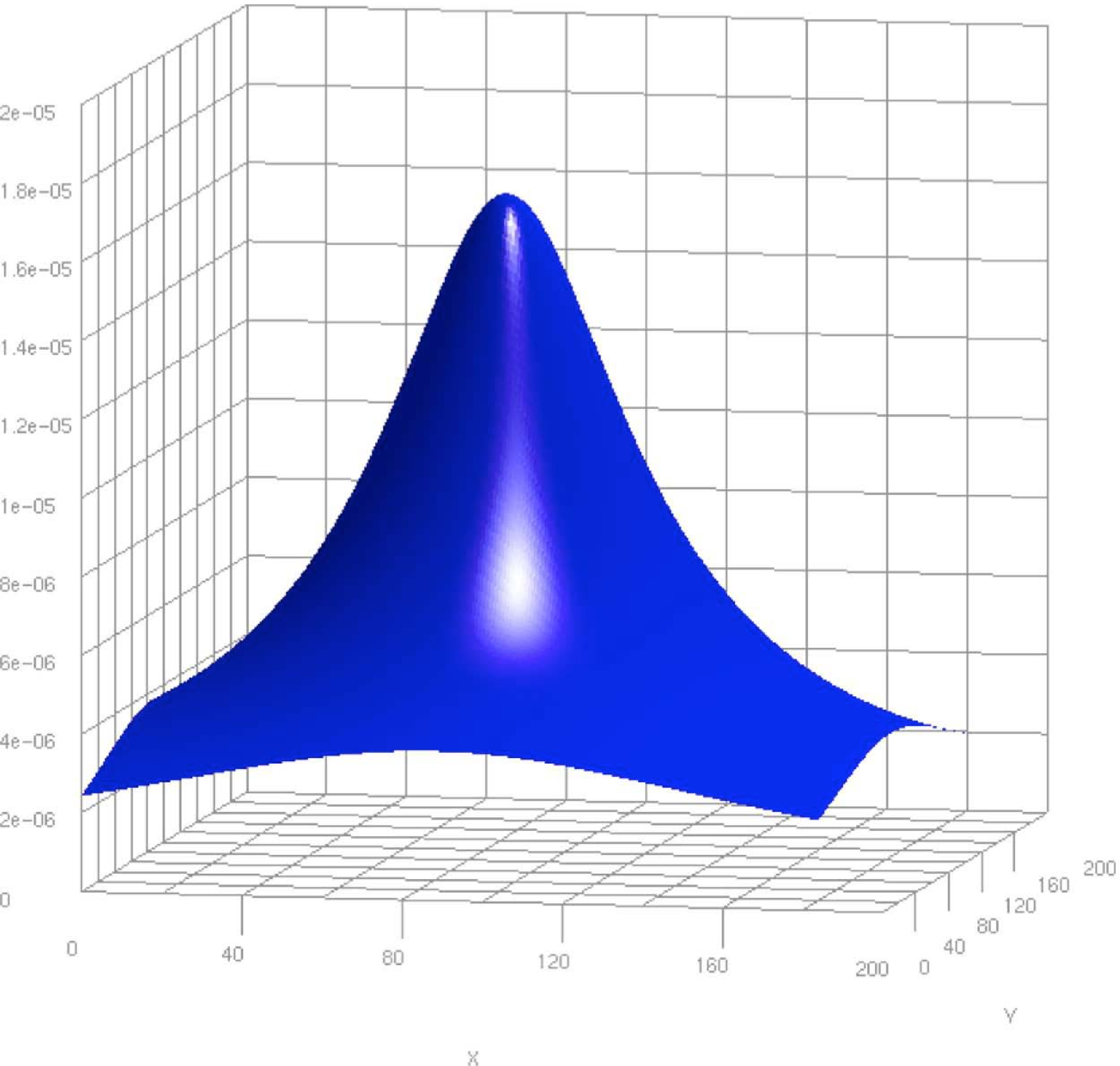}} 
\end{center}
\caption{\label{figure.SigmaSurfModel}  {{\sf\small Comparison of the \map{\Sigma} data
and the best-fit King $\beta$-model.}}\break \small The surface density
\map{\Sigma} data reconstructed from X-ray imaging observations of the
\bc, November 15, 2006 data
release~\protect\citep{Clowe:dataProduct}, is shown in red (a).  The surface density \map{\Sigma} according to the best-fit
King $\beta$-model (neglecting the subcluster) of \eref{surfaceMassDensity} is shown in
blue (b).  The best-fitting King $\beta$-model parameters are listed in \tref{table.king}.}
\end{figure} 
\begin{SCtable}[1.00][hb]
\begin{tabular}{|c||c|c|} \hline 
{\sf King Model} & {\sf Main Cluster} &{\sf Subcluster} \\ \hline
{$\beta$} & $0.803\pm0.013$ & \ldots \\
{$r_{c}$} & $278.0\pm6.8\ \mbox{kpc}$ & \ldots \\
{$\rho_{0}$} & $3.34 \times 10^{5}\ M_{\sun}/\mbox{kpc}^{2}$ & \ldots \\ \hline
{$M_{\rm gas}$} & $3.87  \times 10^{14}\ M_{\sun}$ & $2.58  \times 10^{13}\ M_{\sun}$ \\
{$r_{\rm out}$} & $2620\ \mbox{kpc}$ & \ldots \\ \hline \end{tabular} 
 \caption{\label{table.king} {\sf\small  The best fit King $\beta$-model parameters.}\break  Using a nonlinear least-squares fitting routine, \erefs{surfaceMassDensity0}{surfaceMassDensity} are fit to the \map{\Sigma} data for the main
cluster (neglecting the subcluster).  Results of our calculation for the main and subcluster mass, $M_{\rm gas}$, and cluster
outer radial extent, $r_{\rm out}$ are listed below.  The best-fit is shown in  \frefs{figure.SigmaFit}{figure.SigmaModel} in blue, and may be compared to the data, in red.}
 \end{SCtable}

We may now solve \eref{surfaceMassDensity0} for the central density of the main cluster,
\begin{equation}
\label{rho0}
\rho_{0} =  \frac{\Sigma_{0}}{\sqrt{\pi}  r_{c}} 
\frac{\Gamma\left(\frac{3}{2}\beta\right)}{\Gamma\left(\frac{3\beta-1}{2}\right)} = 3.34 \times 10^{5}\
M_{\sun}/\mbox{kpc}^{2}.
\end{equation}

 The values of the parameters, $\beta$, $r_{c}$ and $\rho_{0}$  completely
determine the King $\beta$-model for the density, $\rho(r)$, of
\eref{betaRhoModel} of the main cluster X-ray gas. We provide a comparison of the full
\map{\Sigma} data (\fref{figure.SigmaSurface}, in red) with the result of the 
surface density \map{\Sigma} derived from the best-fit King $\beta$-model to the main
cluster (\fref{figure.SigmaModel}, in
blue).
%
%
%
%
\begin{SCfigure}[0.9][h]
\caption{ \label{figure.runningM} {{\sf\small The MOG mass scale, $M_{0}$, vs. $M_{\rm gas}$.}}\break
\small A
graphical calculation based on a power-law running of the MOG mass scale, $M_{0}$,
depending on the total ICM gas mass, $M_{\rm gas}$, determined by the large ($>100$) galaxy cluster survey
of~\protect\citet{Brownstein:MNRAS:2006}.  The computed values of $M_{\rm gas}$ and $M_{0}$ for the main and subcluster are
listed in \tref{table.MOG}.}
\scalebox{0.825}{\input{figure/gnuplot/runningM}}
\end{SCfigure} 
\begin{SCtable}[1.0][h]
\begin{tabular}{|c||c|c||c|} \hline 
{\sf MOG} & {\sf Main Cluster} & {\sf Subcluster} & {\sf Best-fit to \map{\kappa}} \\ \hline
{\color{chocolate}{$M_{0}$}} & $1.02 \times 10^{16}\ M_{\sun}$ & $3.56 \times 10^{15}\ M_{\sun}$ & $3.07 \times 10^{15}\ M_{\sun}$\\
{\color{chocolate}{$r_{0}$}} & $139.2\ \mbox{kpc}$ & \ldots & 208.8\ \mbox{kpc}\\ \hline
{\color{chocolate}{$\mathcal{G}_{\infty}$}} & $6.14$ & \ldots & 3.82\\
{\color{chocolate}{$T$}} & $15.5\pm3.9\ \mbox{keV}$ & \ldots & \ldots \\ \hline
\end{tabular}
\caption{\label{table.MOG} \small {\sf MOG parameters.}\break  Results of calculation of the  power-law
running of the MOG mass scale, $M_{0}$,
depending on the computed total cluster mass, $M_{\rm gas}$, according to the large ($>100$) galaxy cluster survey of~\protect\citet{Brownstein:MNRAS:2006}
for the main and subcluster (shown graphically in \fref{figure.runningM}), and the overall best-fit
to the \map{\kappa}; and the results for the MOG range parameter, $r_{0}$.  Results of the calculations for the
asymptotic value of the dimensionless gravitational coupling, $\mathcal{G}_{\infty}$,
and the calculated X-ray temperature of the unperturbed isothermal sphere for the main cluster, $T$, are listed.}
 \end{SCtable}
Substituting \erefss{beta}{rc}{rho0} into \eref{rout.0}, we obtain the main cluster outer radial extent,
\begin{equation}
\label{rout}
r_{\rm out} = 2620\ \mbox{kpc},
\end{equation}
the distance at which the density, $\rho(r_{\rm out})$, drops to $\approx 10^{-28}\,\mbox{g/cm}^{3}$, or 250
times the mean cosmological density of baryons.  The total mass of the main cluster may be calculated by 
substituting \erefss{beta}{rc}{rho0} into \eref{Mgas.0}:
\begin{equation}
\label{Mgas}
M_{\rm gas}  = 3.87  \times 10^{14}\ M_{\sun}.
\end{equation}

\subsection{\label{section.MOGMain}The Gravitational Coupling for the Main cluster}
As discussed in \sref{section.MOG},  in order to apply the MOG model of \eref{G} to the \bc, we must first generalize the
spherically symmetric case by treating the subcluster as a perturbation.  In the zeroth order approximation, we begin by
neglecting the subcluster.  According to the large ($>100$) galaxy cluster survey of~\protect\citet{Brownstein:MNRAS:2006}, the MOG mass
scale, $M_{0}$, is
determined by a power-law, depending only on the computed total cluster mass, $M_{\rm gas}$:
\begin{equation}
\label{M0.0} M_{0} = (60.4 \pm 4.1) \times 10^{14}\ M_{\sun} \left(\frac{M_{\rm gas}}{10^{14} M_{\sun}}\right)^{0.39 \pm
0.10}.
\end{equation}
Substituting the result of \eref{Mgas} for the main cluster into \eref{M0.0}, we obtain
\begin{equation}
\label{M0} M_{0} = 1.02 \times 10^{16}\ M_{\sun},
\end{equation}
and substituting the result of \eref{Mgas} for $M$ and \eref{M0} for the main cluster of {\bc} into \eref{Ginf}, we
obtain
\begin{equation}
\label{Ginfty} \mathcal{G}_{\infty}=6.14.
\end{equation}
From~\protect\citet{Brownstein:MNRAS:2006}, the MOG range parameter, $r_{0}$, depends only on the computed outer radial
extent, $r_{\rm out}$:
\begin{eqnarray}
\label{r0.0} r_0 =& r_{\rm out}/10, &r_{\rm out} \leq 650\,\mbox{kpc},\\
\nonumber r_0 =& 139.2\,{\rm kpc}, &r_{\rm out} > 650\,\mbox{kpc},
\end{eqnarray}
from which we obtain
\begin{equation}
\label{r0} r_{0} = 139.2\ \mbox{kpc},
\end{equation}
for the main cluster.

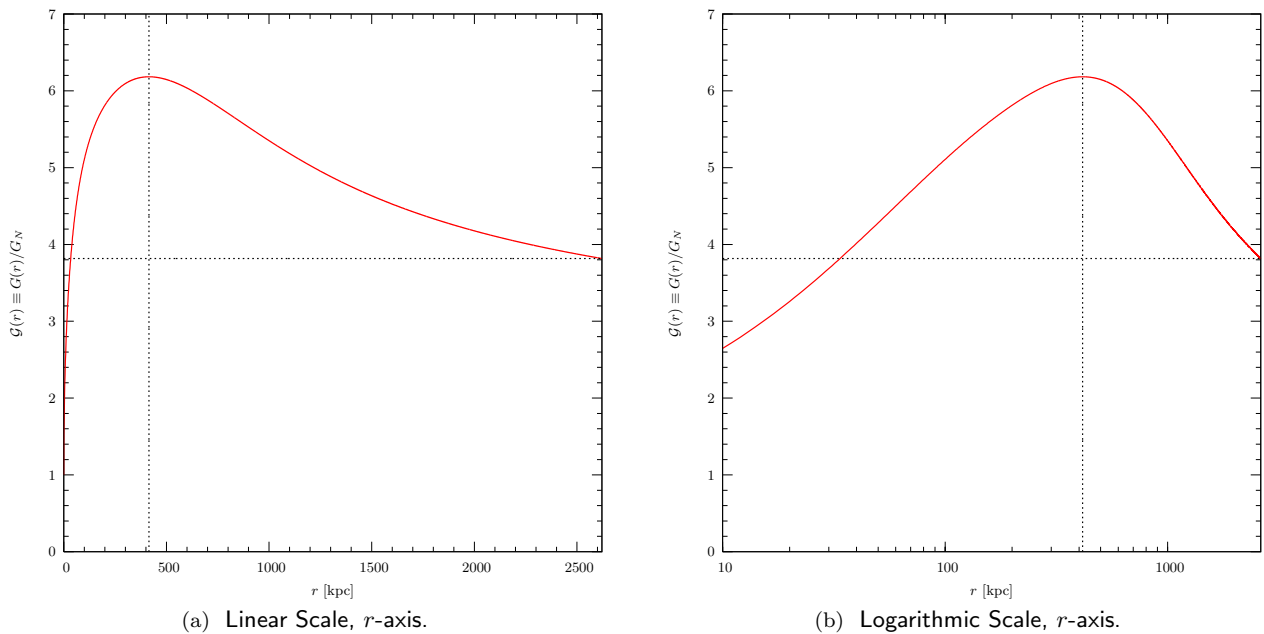
\begin{figure}[ht]
\begin{center}
\subfloat[ {\sf\small Linear Scale, $r$-axis.}]{\label{figure.G_MainXray.linear} \scalebox{0.6}{\input{figure/gnuplot/G_MainXrayLinear}}}  
\quad\quad
\subfloat[ {\sf\small Logarithmic Scale, $r$-axis.}]{\label{figure.G_MainXray.log}\scalebox{0.6}{\input{figure/gnuplot/G_MainXrayLog}}}
\end{center}
\caption{\label{figure.G} {\sf\small Plot of the dimensionless gravitational coupling, $\mathcal{G}(r) \equiv
G(r)/G_{N}$, of \eref{G} vs. the
distance, $r$, in kpc.}\break \small Shown in linear scale for the $r$-axis, (a), and in logarithmic scale, (b).  J2000
and map (x,y) coordinates of the MOG center ($R \equiv 0\ \mbox{kpc}$) are listed in
\tref{table.coords}, and located with respect to the \map{\Sigma} and \map{\kappa} in
\frefs{figure.sigma}{figure.kappa}, respectively. 
The running gravitational coupling, $\mathcal{G}(r)$, shown here, corresponds to our best-fit to the
\map{\kappa}, listed in \tref{table.MOG}.}
\end{figure}

The best-fitting King $\beta$-model parameters for the main cluster are listed in
\tref{table.king}, and the MOG parameters for the main cluster are listed in \tref{table.MOG}.  A plot of the
dimensionless gravitational coupling of \eref{G} using the MOG
parameter results of \erefs{M0}{r0} for the main cluster (neglecting the subcluster)  is plotted in
\fref{figure.G_MainXray.linear} using a linear scale for the $r$-axis, and in
\fref{figure.G_MainXray.log} on a logarithmic scale for the $r$-axis.

 \subsection{\label{section.CylindricalMassProfile}The Cylindrical Mass Profile}
\begin{SCfigure}[0.9][ht]
\scalebox{0.7}{\input{figure/gnuplot/MainXray}} 
\caption{\label{figure.MainXray} \small{\sf Cylindrical mass profile}\break 
The surface density
\map{\Sigma} reconstructed from X-ray image observations of the
\bc, November 15, 2006 data release~\protect\citep{Clowe:dataProduct} is integrated about the main cluster \map{\Sigma} peak
($R\equiv 0$), shown in solid red.  The cylindrical mass profile
derived from the best-fit King $\beta$-model is shown in short-dashed blue, demonstrating a good agreement with the
data.  The long-dash magenta line is the result of the galaxy subtraction of \erefs{sigma.galaxies}{mass.galaxies}. 
The lack of galactic mass in the vicinity of the main cluster \map{\Sigma} peak at $R = 0$ is due to the merger.  The
dash-dot black line is the result of the dark matter subtraction of \eref{mass.darkmatter}.  
The dark matter dominates the ICM cluster gas by a factor $>3$.
J2000 and map (x,y) coordinates are listed in \tref{table.coords}. The best-fitting King $\beta$-model parameters are listed in
\tref{table.king}.}
\end{SCfigure} 
$\Sigma(x,y)$ is an integrated density along the line-of-sight, $z$, as shown in \eref{Sigma.Newton}.  By summing the
\map{\Sigma} pixel-by-pixel, starting from the center of the main cluster \map{\Sigma} peak, one is performing an
integration of the surface density, yielding the mass,
\begin{equation}
\label{cylindricalmass} M(R) = \int_{0}^{R} \Sigma(R^{\prime}) R^{\prime} dR^{\prime},
\end{equation}
enclosed by concentric cylinders of radius, $R=\sqrt{x^{2}+y^{2}}$.  We performed such a sum over the \map{\Sigma}
data, and compared the result to the best-fit King $\beta$-model derived in \sref{section.SigmaMap}, with the parameters  listed
in \tref{table.king}.  The results are shown in
\fref{figure.MainXray}.  The fact that the data and model are in good agreement provides evidence that the
King
$\beta$-model is valid in all directions from the main cluster \map{\Sigma} peak, and not just on the straight line
connecting the peak to the main cD, where the fit was performed.  The model deviates slightly from the data,
underpredicting $M(R)$ for $R > 350\ \mbox{kpc}$, which may be explained by the subcluster (which is included in the
data, but not the model).  The integrated mass profile arising from the
dark matter analysis of the \map{\kappa} of \eref{mass.darkmatter} is shown for comparison on the same figure. 

The ratio of dark matter to ICM gas for the main cluster is $> 3$, which is significantly less than the
average cosmological ratio of 5.68 discussed in \sref{section.MissingMass.1}, whereas one should expect the highest mass to light ratios
from large, hot, luminous galaxy clusters, and the {\bc} is certainly one of the largest and hottest. 

\subsection{\label{section.T}The Isothermal Spherical Mass Profile}
For a spherical system in hydrostatic equilibrium, the structure equation can be derived from the collisionless Boltzmann
equation
\begin{equation}
\label{CBE} \frac{d}{dr}(\rho(r) \sigma_{r}^{2}) + \frac{2\rho(r)}{r}\left(\sigma_{r}^{2} - \sigma_{\theta,\phi}^{2}\right) =
-\rho(r) \frac{d\Phi(r)}{dr},
\end{equation}
where $\Phi(r)$ is the gravitational potential for a point source, $\sigma_{r}$ and $\sigma_{\theta,\phi}$ are mass-weighted
velocity dispersions in the radial ($r$) and tangential ($\theta, \phi$) directions, respectively.  For an isotropic
system,
\begin{equation}
\label{isotropicSystem}
\sigma_{r} = \sigma_{\theta,\phi}.
\end{equation}
The pressure profile, $P(r)$, can be related to these quantities by
\begin{equation}
\label{pressureProfile}
P(r) = \sigma_{r}^{2} \rho(r).
\end{equation}
Combining equations (\ref{CBE}), (\ref{isotropicSystem}) and (\ref{pressureProfile}), the result for the isotropic
sphere is
\begin{equation}
\label{isotropicSphere}
\frac {dP(r)}{dr} = -\rho(r) \frac{d\Phi(r)}{dr}.
\end{equation}
For a gas sphere with temperature profile, $T(r)$, the velocity dispersion becomes
\begin{equation}
\label{velocityDispersion}
\sigma_{r}^{2} = \frac{kT(r)}{\mu_{A} m_{p}},
\end{equation}
where $k$ is Boltzmann's constant, $\mu_{A} \approx 0.609$ is the mean atomic weight and $m_{p}$ is the proton mass.  We
may now substitute equations (\ref{pressureProfile}) and (\ref{velocityDispersion}) into equation (\ref{isotropicSphere}) to obtain
\begin{equation}
\label{isotropicSphere2}
\frac {d}{dr}\left(\frac{kT(r)}{\mu_{A} m_{p}} \rho(r)\right) = -\rho(r) \frac{d\Phi(r)}{dr}.
\end{equation}
Performing the differentiation on the left-hand side of equation (\ref{isotropicSphere}), we may solve for the
gravitational acceleration:
\begin{eqnarray}
\nonumber
a(r) & \equiv  & - \frac{d\Phi(r)}{dr} \\
\label{accelerationProfile}
& = & \frac{kT(r)}{\mu_{A} m_{p} r} \left[ \frac{d \ln(\rho(r))}{d \ln(r)} + \frac{d \ln(T(r))}{d \ln(r)}\right].
\end{eqnarray}
For the isothermal isotropic gas sphere, the temperature derivative on the right-hand side of equation
(\ref{accelerationProfile}) vanishes and the remaining derivative can be evaluated using the $\beta$-model of equation (\ref{betaRhoModel}):
\begin{equation}
\label{isothermalAccelerationProfile}
a(r) = -\frac{3\beta kT}{\mu_{A} m_{p}} \left(\frac{r}{r^{2}+r_{c}^{2}}\right).
\end{equation}

\begin{SCfigure}[0.9][b]
\scalebox{0.7}{\input{figure/gnuplot/MassProfile}} 
\caption{\label{figure.MassProfile} {\sf\small The spherically integrated mass profile of the main cluster ICM
gas.}\break {\small The mass profile integrated King $\beta$-model, $M_{\rm gas}(r)$, in  from the
\map{\Sigma} peak, is shown in long-dashed red.
 derived from the surface density \map{\Sigma} data reconstructed from
X-ray imaging observations of the \bc, November 15, 2006 data release~\protect\citep{Clowe:dataProduct}.
The surface
density \map{\Sigma} according to the best-fit King $\beta$-model (neglecting the subcluster) of
\eref{surfaceMassDensity} is shown in black (with uncertainty). The Newtonian dynamical mass (the dark matter paradigm)
is shown in short-dashed blue.  All masses are presented in units of $M_{\sun}$, and distances, $r$,
in kpc. The strong correspondence between the data (red) and the MOG fit (black) across distance
scales in the cluster from tens to thousands of kpc. is significant.  This is the first evidence that the X-ray imaging
data for the cluster is behaving similar to the large ($>100$) galaxy cluster survey
of~\protect\citet{Brownstein:MNRAS:2006}.   J2000 and map (x,y) coordinates are listed in
\tref{table.coords}. The best-fitting King $\beta$-model parameters are listed in
\tref{table.king}.  The best-fitting MOG parameters are listed in \tref{table.MOG}.}}
\end{SCfigure} 

The dynamical mass in Newton's theory of gravitation can be obtained as a function of radial position by replacing the
gravitational acceleration with Newton's Law:
\begin{equation}
\label{newtonsLaw}
a_{\rm N}(r) = - \frac{G_{N} M(r)}{r^{2}},
\end{equation}
so that equation (\ref{accelerationProfile}) can be rewritten as
\begin{equation}
\label{newtonsMass}
M_{\rm N}(r) = - \frac{r}{G_{N}}\frac{kT}{\mu_{A} m_{p}} \left[ \frac{d \ln(\rho(r))}{d \ln(r)} + \frac{d \ln(T(r))}{d
\ln(r)}\right],
\end{equation}
and the isothermal $\beta$-model result of equation (\ref{isothermalAccelerationProfile}) can be
rewritten as
\begin{equation}
\label{isothermalNewtonsMass}
M_{\rm N}(r) = \frac{3\beta kT}{\mu_{A} m_{p}G_{N}} \left(\frac{r^{3}}{r^{2}+r_{c}^{2}}\right).
\end{equation}

Similarly, the dynamical mass in MOG can be obtained as a function of radial position by substituting the MOG
gravitational acceleration law~\protect\citep{Moffat:JCAP:2005,Moffat:JCAP:2006,Brownstein:ApJ:2006,Brownstein:MNRAS:2006}
\begin{equation}
\label{runG} a(r)=-\frac{G(r)M}{r^2},
\end{equation}
so that our result for the isothermal $\beta$-model becomes
\begin{equation}
\label{MOGMassProfile}
M_{\rm MOG}(r) = \frac{3\beta kT}{\mu_{A} m_{p}G(r)} \left(\frac{r^{3}}{r^{2}+r_{c}^{2}}\right).
\end{equation}
We can express this result as a scaled version of equation (\ref{newtonsMass}) or the isothermal case of equation
(\ref{isothermalNewtonsMass}):
\begin{eqnarray}
\nonumber M_{\rm MOG}(r)  &=& \frac{M_{\rm N}(r)}{\mathcal{G}(r)} \\
 \label{MOGMassScaled} &=& \biggl\{1+\sqrt\frac{M_0}{M_{\rm MOG}(r)}\biggl[1-\exp(-r/r_0)\biggl(1+\frac{r}{r_0}\biggr)
\biggr]\biggr\}^{-1} M_{\rm N}(r),
\end{eqnarray}
where we have substituted \eref{G} for $\mathcal{G}(r)$.  \eref{MOGMassScaled} may be solved explicitly for $M_{\rm
MOG}(r)$ by squaring both sides and determing the positive root of the quadratic equation.  

\begin{SCtable}[1.0][t]
\begin{tabular}{|c|l|c|r|} \hline 
{\sf Year} & {\sf Source - Theory or Experiment} & {\sf $T\ (\mbox{keV})$} & {\sf \% error} \\ \hline\hline
\colorbox{yellow}{\color{cyan}{2007}} & \colorbox{yellow}{\color{cyan}{\sf MOG Prediction\phantom{Experiment 012}}}  & $15.5\pm3.9$& \\
2002 & accepted experimental value  &$14.8^{+1.7}_{-1.2}$ & \colorbox{yellow}{\color{cyan}{$\phantom{X1}4.5$}} \\
1999 & ASCA+ROSAT fit  & $14.5^{+2.0}_{1.7}$ & \colorbox{yellow}{\color{cyan}{$\phantom{X1}6.5$}} \\
1998 & ASCA fit & $17.4\pm2.5$ & \colorbox{yellow}{\color{cyan}{$\phantom{X}12.3$}} \\ \hline
\end{tabular}
\caption{\label{table.temp} \small The MOG best-fit temperature
 is consistent with the experimental values for the main cluster isothermal
temperature, the 1999 ASCA+ROSAT fit, and the 1998 ASCA fit~\protect\citep{Markevitch:ApJL:2002}. }
 \end{SCtable}
The scaling of the Newtonian dynamical mass by $\mathcal{G}(r)$ according to \eref{MOGMassScaled}
solved the dark matter problem for the galaxy clusters of the $>100$ galaxy cluster survey
of~\protect\citet{Brownstein:MNRAS:2006}.  The unperturbed {\bc} is no exception!  In \fref{figure.MassProfile}, we plotted the MOG
and the Newtonian 
dynamical masses, $M_{\rm MOG}(r)$ and $M_{\rm N}(r)$, respectively, and compared it to the spherically integrated
best-fit King
$\beta$-model for the main cluster gas
mass of \erefs{betaRhoModel}{massProfile} using the parameters listed in \tref{table.king}.  The MOG temperature prediction due to the best-fit is listed in \tref{table.temp} and compared to experimental values.
 
As shown in \fref{figure.MassProfile}, across the range of the $r$-axis, and throughout the radial extent of the {\bc}, the $1\sigma$ correlation
between the gas mass, $M(r)$ and the MOG dynamical mass, $M_{\rm MOG}(r)$, provides excellent agreement between
theory and experiment. 
The same cannot be said of any theory of dark matter in which the X-ray surface density map is negligible in relation to the DM
density.  So dark matter makes no prediction for the isothermal temperature, which has been measured to reasonable precision for
many clusters, but simply ``accounts for missing mass.''  Since there is no mysterious missing mass in MOG, the
prediction should be taken seriously as direct confirmation of the theory, or at least as hard evidence for MOG.

\subsection{\label{section.Subcluster}The Subcluster Subtraction}
\begin{figure}[ht]
\psset{unit=1pt}
\begin{pspicture}(500,325)(0,0)
\psset{linecolor=DarkGreen,linewidth=1pt}
{\thicklines \multiput(113,303)(90,0){4}{\line(0,1){9}} \put(113,303){\line(1,0){284.4}}}
{\thinlines \multiput(113,303)(4.5,0){64}{\line(0,1){3}} \multiput(113,303)(18,0){16}{\line(0,1){6}}}
\put(403,317){\makebox(0,0){\sf [kpc]}}
\multiput(103,317)(90,0){4}{\makebox(20,0){\sf\arabic{kpc}}\addtocounter{kpc}{500}}
\put(110,0){\includegraphics[width=290pt]{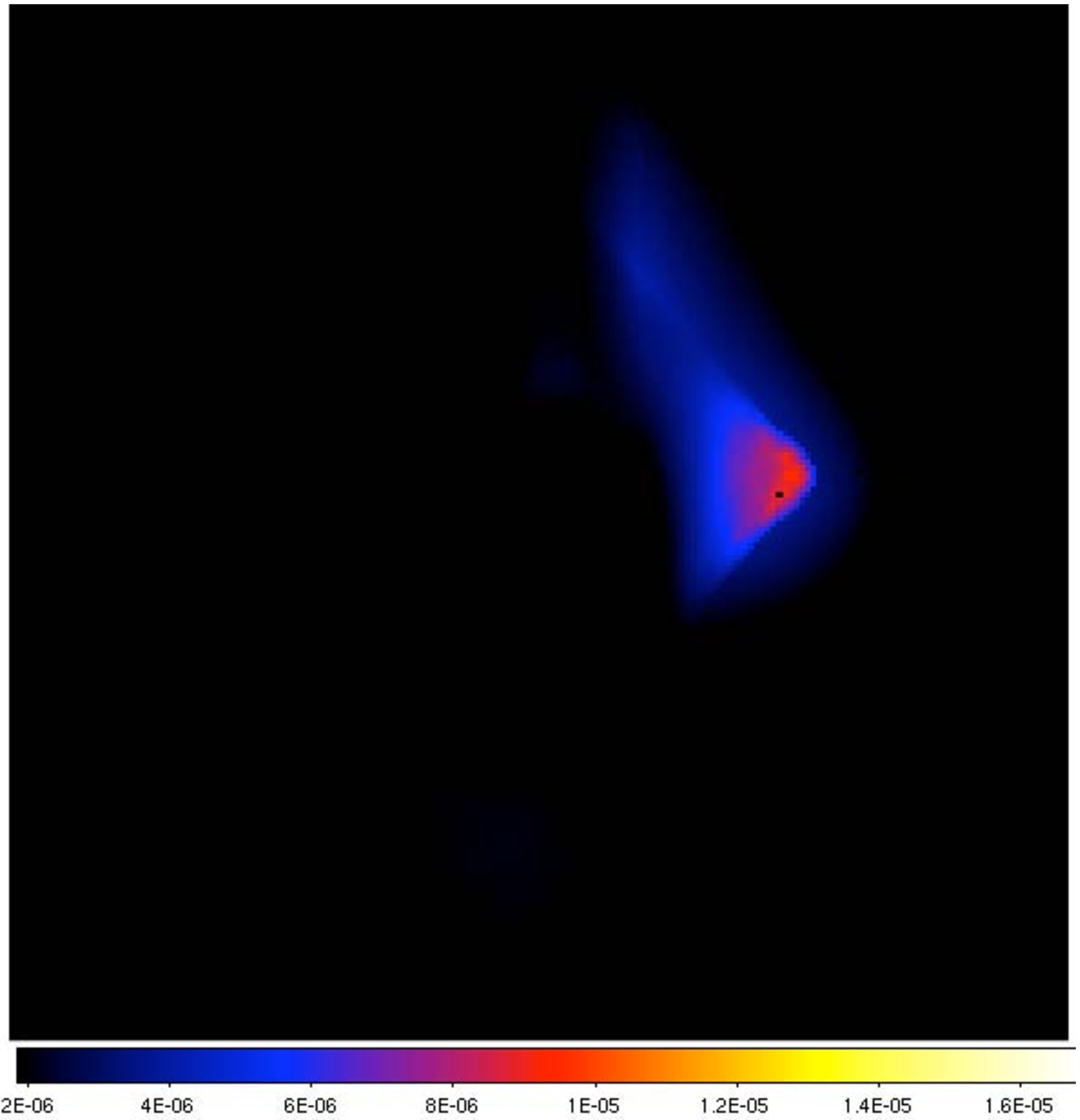}} 
\psline{->}(105,247)(249,155) \psline{->}(105,187)(221,153) \psline{->}(105,127)(219,141) \psline{->}(105,67)(273,159)
\psline{->}(403,247)(320,168) \psline{->}(403,187)(341,174) \psline{->}(403,127)(349,168) \psline{->}(403,67)(247,74)
\put(0, 240){\fcolorbox{cayenne}{sand}{\makebox(100,14){\color{chocolate}\small\sf Main cluster \map{\Sigma} peak}}} 
\put(0, 180){\fcolorbox{cayenne}{sand}{\makebox(100,14){\color{chocolate}\small\sf Main cluster cD galaxy}}} 
\put(0, 120){\fcolorbox{cayenne}{sand}{\makebox(100,14){\color{chocolate}\small\sf Main cluster \map{\kappa} peak}}} 
\put(0, 60){\fcolorbox{cayenne}{sand}{\makebox(100,14){\color{chocolate}\small\sf MOG Center}}} 
\put(403, 240){\fcolorbox{cayenne}{sand}{\makebox(100,14){\color{chocolate}\small\sf Subcluster \map{\Sigma} peak}}} 
\put(403, 180){\fcolorbox{cayenne}{sand}{\makebox(100,14){\color{chocolate}\small\sf Subcluster \map{\kappa} peak}}} 
\put(403, 120){\fcolorbox{cayenne}{sand}{\makebox(100,14){\color{chocolate}\small\sf Subcluster BCG}}} 
\put(403 , 60){\fcolorbox{cayenne}{sand}{\makebox(100,14){\color{chocolate}\small\sf X-ray Bulge}}} 
\end{pspicture}
\caption{\label{figure.SigmaSubcluster}  {{\sf\small
The subcluster subtracted surface density \map{\Sigma}}}\break \small
Main  cluster and subcluster \map{\kappa} observed peaks (local maxima) and
\map{\Sigma} observed peaks are shown for comparison.  The
central dominant (cD) galaxy of the main cluster, the brightest cluster galaxy (BCG) of the subcluster, and the
MOG predicted gravitational center are shown.  The colourscale is set to agree with \fref{figure.sigma} for
comparison. 
J2000 and map (x,y) coordinates are listed in \tref{table.coords}.}
\end{figure}

\begin{figure}[ht]
\begin{center} 
\subfloat[{\sf\small
The subcluster subtracted surface density \map{\Sigma}}]{\label{figure.SigmaSubtraction} \includegraphics[width=8cm]{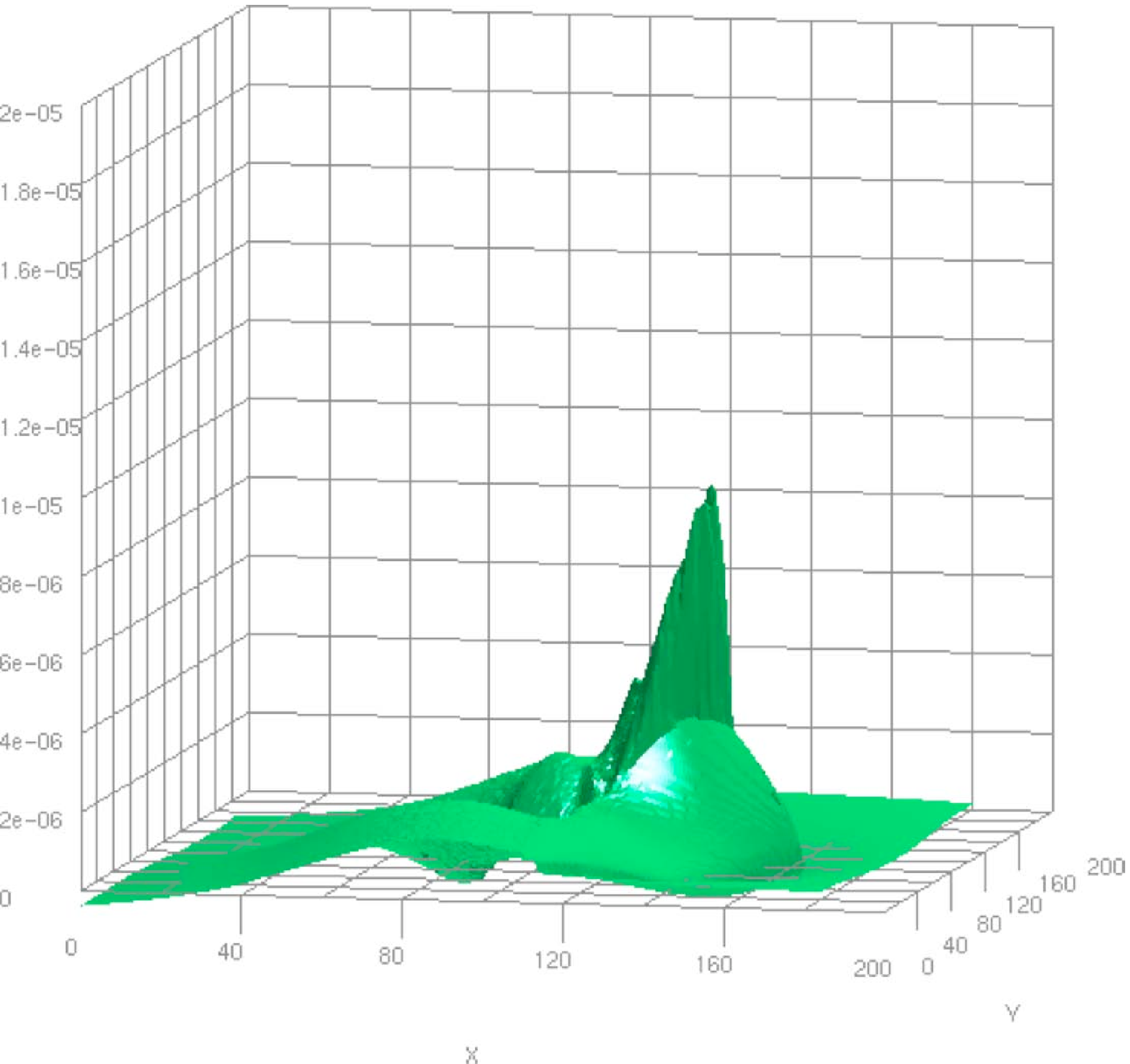}}
\quad\quad
\subfloat[ {\sf\small The subcluster subtracted \map{\Sigma} superposed onto the best-fit King $\beta$-model
of the main cluster.}  \small Note the odd X-ray Bulge, here, and in
\fref{figure.SigmaSubcluster}.]{\label{figure.SigmaSuperposed}\includegraphics[width=8cm]{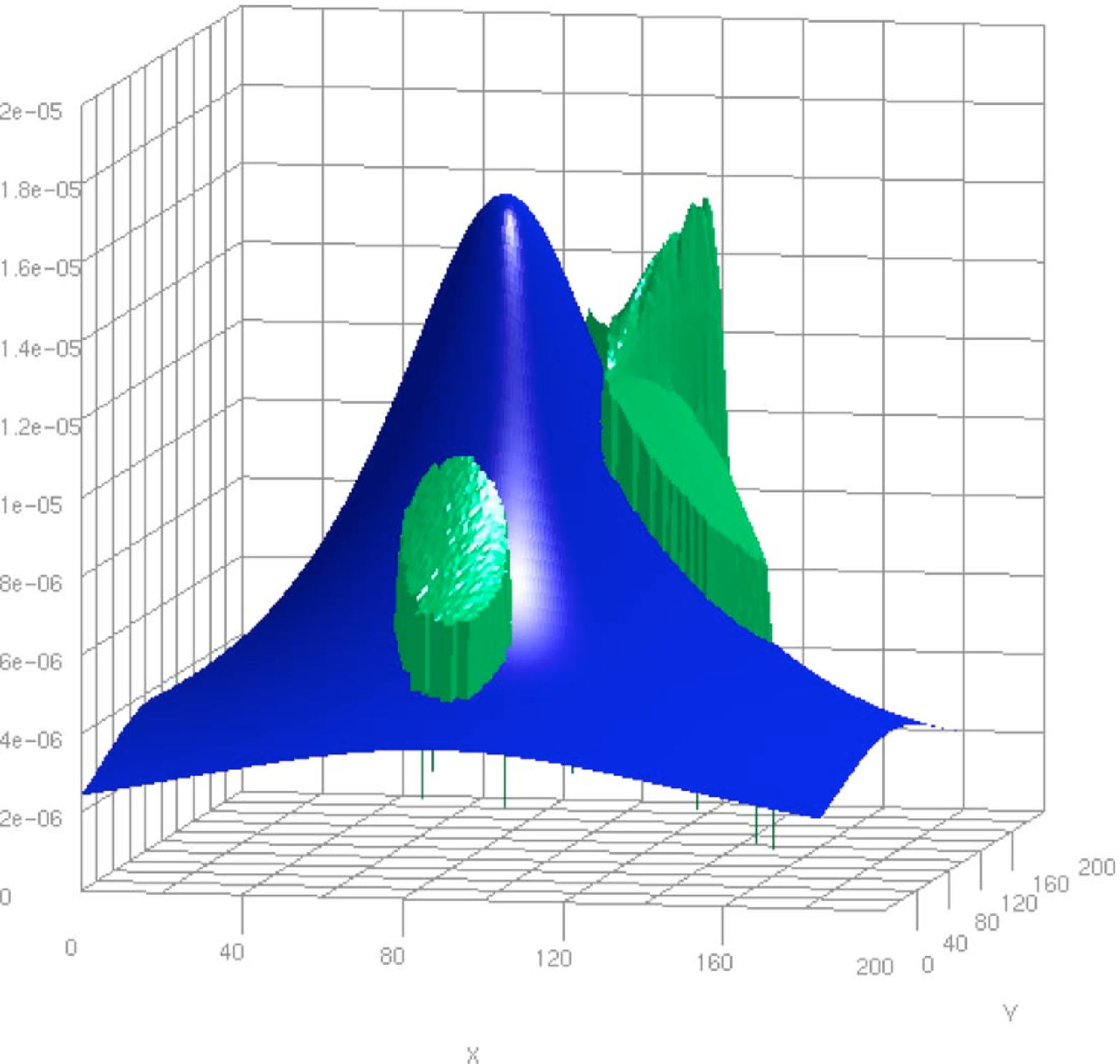}}
\end{center}
\caption{ \label{figure.SigmaStereo}{{\sf\small The subcluster subtracted surface density
\map{\Sigma}.}}\break  \small The blue
surface represents the \map{\Sigma} due to the
integrated (line-of-sight) King $\beta$-model fit to main cluster (as shown in blue in \fref{figure.SigmaModel}).  The
green surface is the contribution to the \map{\Sigma} from the subcluster (as calculated from our model).   J2000 and map (x,y) coordinates are listed in \tref{table.coords}. 
The best-fitting King $\beta$-model parameters are listed in \tref{table.king}.}
\end{figure}

Provided the surface density \map{\Sigma} derived from the best-fit 
King $\beta$-model of \eref{surfaceMassDensity} is sufficiently close to the \map{\Sigma} data (consider
\frefs{figure.SigmaFit}{figure.SigmaSurfModel}) then the difference between the data and the $\beta$-model is the
surface density \map{\Sigma} due to the subcluster. Our subcluster subtraction, shown in
\frefs{figure.SigmaSubcluster}{figure.SigmaSubtraction}, is based upon a high precision ($\chi^{2}<0.2$) best-fit King
$\beta$-model to the main cluster, which agrees with the main cluster surface mass \map{\Sigma} (data) within 1\%
everywhere and to a mean uncertainy of 0.8\%.  The subcluster subtraction is accurate down to $\rho = 10^{-28}\ \mbox{g}/\mbox{cm}^{3}\sim 563.2\ M_{\sun}/\mbox{pc}^{3}$ baryonic background
density.  After subtraction, the subcluster \map{\Sigma} peak takes a value of $1.30 \times 10^{8}\
M_{\sun}/\mbox{kpc}^{2}$, whereas the full \map{\Sigma} has a value of  $2.32\times 10^{8}\
M_{\sun}/\mbox{kpc}^{2}$ at the subcluster \map{\Sigma} peak.  Thus the subcluster (at its most dense position)
provides only $\approx 56\%$ of the X-ray ICM, the rest is due to the extended distribution of the main cluster.
 The subcluster subtraction surface density \map{\Sigma} shown in \fref{figure.SigmaSubcluster} uses the same colorscale
as the full \map{\Sigma} shown in \fref{figure.sigma}, for
comparison.  \fref{figure.SigmaStereo} is a stereogram of the subcluster subtracted surface density
\map{\Sigma} and the
subcluster superposed onto the surface density \map{\Sigma} of the best-fit King
$\beta$-model to the main cluster. There is an odd
X-ray bulge in \frefs{figure.SigmaSubcluster}{figure.SigmaSuperposed} which may be an artifact of the subtraction, or
perhaps evidence of an as yet unidentified component.

Since the outer radial extent of the subluster gas is less than 400 kpc, the \map{\Sigma} completely contains
all of the subcluster gas mass.  By summing the subcluster subtracted
\map{\Sigma} pixel-by-pixel over the entire \map{\Sigma} peak, one is
performing an integration of the surface density, yielding the total subcluster mass.  We performed such a sum over the
subcluster subtracted \map{\Sigma}
data, obtaining
\begin{equation}
\label{Mgas.subcluster}
M_{\rm gas}  = 2.58  \times 10^{13}\ M_{\sun},
\end{equation}
for the mass of the subcluster gas, which is less than 6.7\% of the mass of main cluster gas (the best-fitting King $\beta$-model parameters are listed in
\tref{table.king}.)  This justifies our initial assumption that the subcluster may be treated as a perturbation in
order to fit the main cluster to the King $\beta$-model.  Our subsequent analysis of the thermal profile confirms that
the main cluster X-ray temperature is nearly isothermal, lending further support to the validity of the King
$\beta$-model.  

We may now calculate the MOG mass scale, $M_{0}$, of the subcluster by substituting the subcluster gas mass, $M_{\rm
gas}$, of \eref{Mgas.subcluster} into the power-law relation of \eref{M0.0}, yielding,
\begin{equation}
\label{M0.subcluster} M_{0} = 3.56 \times 10^{15}\ M_{\sun},
\end{equation}
as shown in \fref{figure.runningM}.  The computed values of $M_{\rm gas}$ and $M_{0}$ for the main and subcluster are
listed in \tref{table.MOG}.

\section{\label{section.Convergence}The Convergence Map from Lensing Analysis}
\subsection{\label{section.KappaMap}The $\kappa$-Map}
\begin{figure}[h]
\psset{unit=1pt}
\begin{pspicture}(500,325)(0,0)
\psset{linecolor=DarkGreen,linewidth=1pt}
{\thicklines \multiput(113,306)(90,0){4}{\line(0,1){9}} \put(113,306){\line(1,0){284.4}}}
{\thinlines \multiput(113,306)(4.5,0){64}{\line(0,1){3}} \multiput(113,306)(18,0){16}{\line(0,1){6}}}
\put(403,320){\makebox(0,0){\sf [kpc]}}
\multiput(103,320)(90,0){4}{\makebox(20,0){\sf\arabic{kpc}}\addtocounter{kpc}{500}}
\put(110,0){\includegraphics[width=290pt]{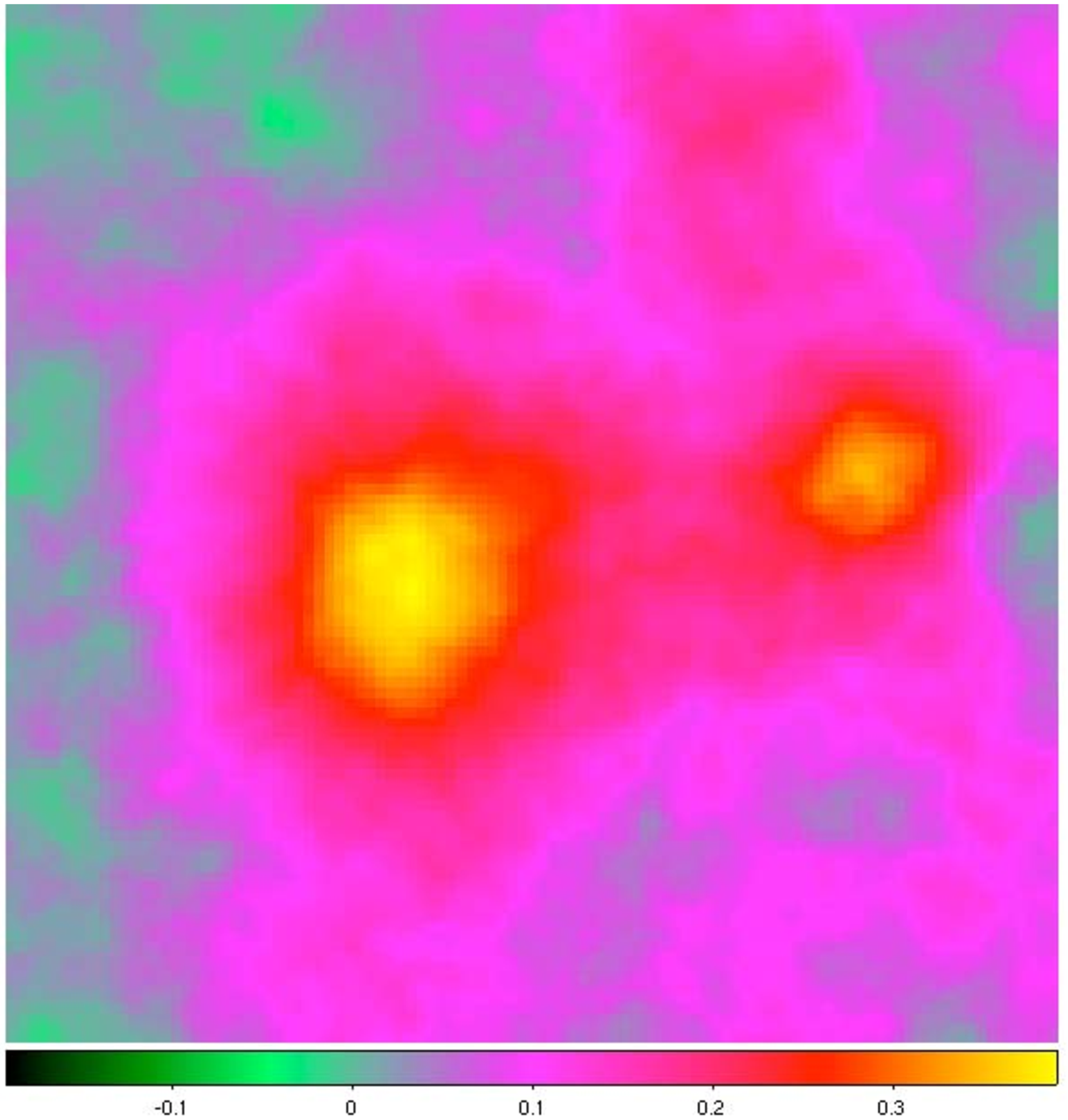}} 
\psline{->}(105,247)(249,155) \psline{->}(105,187)(221,153) \psline{->}(105,127)(219,141) \psline{->}(105,67)(273,159)
\psline{->}(403,247)(313,270) \psline{->}(403,187)(341,174) \psline{->}(403,127)(349,168) \psline{->}(403,67)(320,168)
\put(0, 240){\fcolorbox{cayenne}{sand}{\makebox(100,14){\color{chocolate}\small\sf Main cluster \map{\Sigma} peak}}} 
\put(0, 180){\fcolorbox{cayenne}{sand}{\makebox(100,14){\color{chocolate}\small\sf Main cluster cD galaxy}}} 
\put(0, 120){\fcolorbox{cayenne}{sand}{\makebox(100,14){\color{chocolate}\small\sf Main cluster \map{\kappa} peak}}} 
\put(0, 60){\fcolorbox{cayenne}{sand}{\makebox(100,14){\color{chocolate}\small\sf MOG Center}}} 
\put(403, 240){\fcolorbox{cayenne}{sand}{\makebox(100,14){\color{chocolate}\small\sf Mysterious Plateau Nearby}}} 
\put(403, 180){\fcolorbox{cayenne}{sand}{\makebox(100,14){\color{chocolate}\small\sf Subcluster \map{\kappa} peak}}} 
\put(403, 120){\fcolorbox{cayenne}{sand}{\makebox(100,14){\color{chocolate}\small\sf Subcluster BCG}}} 
\put(403 , 60){\fcolorbox{cayenne}{sand}{\makebox(100,14){\color{chocolate}\small\sf Subcluster \map{\Sigma} peak}}} 
\end{pspicture}
\caption{\label{figure.kappa}  {{\sf\small The surface density \map{\kappa} reconstructed from strong and weak gravitational lensing.}}
\break \small
Main  cluster and subcluster of the \bc, November 15, 2006 data release  \map{\kappa}~\protect\citep{Clowe:dataProduct} observed peaks (local maxima) and
\map{\Sigma} observed peaks are shown for comparison.  The
central dominant (cD) galaxy of the main cluster, the brightest cluster galaxy (BCG) of the subcluster, and the
MOG predicted gravitational center are shown.  J2000 and map (x,y) coordinates are listed in \tref{table.coords}.}
\end{figure}
As tempting as it is to see the convergence \map{\kappa} of \fref{figure.kappa} -- a false color image of the  strong
and weak gravitational lensing reconstruction~\protect\citep{Clowe:ApJL:2006,Clowe:ApJ:2006,Clowe:astro-ph.0611496} -- 
as a photograph of the ``curvature'' around the {\bc}, it is actually a
reconstruction of all of the bending of light over the entire distance from the lensing event source toward the Hubble
Space Telescope.  The source of the \map{\kappa} is  $\propto\int G_{N}\rho(r)$, along the line-of-site, as in
\eref{scaledSurfaceMassDensity.Newton},  $\propto G(r)\int\rho(r)$ as in \eref{scaledSurfaceMassDensity.MOG.0}, or
$\propto\int G(r)\rho(r)$ as in \eref{scaledSurfaceMassDensity.MOG}.  
For the {\bc}, we are looking along a line-of-sight which is at least as long as indicated by a redshift of $z=0.296$ (Gpc scale).  The sources of the lensing events are in a large neighbourhood of redshifts, an estimated $z=0.85\pm 0.15$.  This fantastic scale (several Gpc) is naturally far in excess of the distance
scales involved in the X-ray imaging surface density \map{\Sigma}.  It is an accumulated effect, but
only over the range of the
X-ray source -- as much as 2.2 Mpc.  A comparison of these two scales indicates that the distance scales within the
\map{\Sigma} are $10^{-3}$ below the Gpc's scale of the \map{\kappa}.

\begin{quote}\noindent{{\sf Preliminary comments on the November 15, 2006 data release~\protect\citep{Clowe:dataProduct}:}}\end{quote}
\begin{itemize}
\item The conclusion of \protect\citet{Clowe:ApJL:2006,Clowe:ApJ:2006}, that the \map{\kappa} shows direct
evidence
for the existence of dark matter may be premature.  Until dark matter has been detected in the lab, it remains an open
question whether a modified gravity theory, such as MOG, can account for the \map{\kappa} without nonbaryonic
dark 
matter.  MOG, due to the varying gravitational coupling, \eref{G}, gives the Newtonian $1/r^{2}$ gravitational force law a
considerable boost  --  ``extra gravity'' as much as $\mathcal{G}_{\infty}\approx 6$ for the {\bc}.
\item It may be feasible that the mysterious plateau in the north-east corner of the \map{\kappa} is from some distribution of mass unrelated to the {\bc}.  This ``background curvature'' contribution to the \map{\kappa} is one part of the uncertainty in the reconstruction.  The second, dominant source of uncertainty certainly must be the estimate of the angular diameter distances between the source of the lensing event and the {\bc}, as shown in \eref{SigmaC.Newton}.
\end{itemize}

We have completed a large ($>100$) galaxy cluster survey in~\protect\citet{Brownstein:MNRAS:2006} that provides a statistically
significant answer to the
question of how much dark matter is expected.  The relative abundance across the scales of the X-ray clusters would imply that the
\map{\kappa} should peak at $\sim 1.0$ as opposed to the November 15, 2006 data release~\protect\citep{Clowe:dataProduct}, which peaks at a value of
$\kappa \approx 0.38$ for the main cluster \map{\kappa} peak.

\begin{quote}\noindent{{\sf Where is the missing dark matter?}}\end{quote}
The dark matter paradigm cannot statistically account for such an observation without resorting to further ad-hoc
explanation, so the question of ``missing matter''  may be irrelevant.   The problem of ``extra gravity'' due
to MOG may be a step in the right direction, with our solution as the subject of the next section.
\subsection{\label{section.MOG.solution}The MOG Solution}
The MOG $\kappa$-model we have developed in \erefs{scaledSurfaceMassDensity.MOG}{Sigma.MOG} can now be applied
to the {\bc}.  In order to model the \map{\kappa}, shown in \frefs{figure.kappa}{figure.kappaMap}, in MOG, we must
integrate the product of the dimensionless gravitational coupling, $\mathcal{G}(r)$ of \eref{G} with the mass density, $\rho(r)$, over the volume of
the {\bc}. As discussed in \sref{section.CylindricalMassProfile}, we may integrate the surface density \map{\Sigma} data according to \eref{cylindricalmass} to obtain the integrated cluster gas mass about concentric cylinders centered about $R=0$.  However, what we require in the 
calculation of the $\mathcal{G}(r)$ of \eref{G} is the integrated mass about concentric spheres,
\begin{equation}
\label{sphericalmass} M(r) = \int_{0}^{2\pi} d\theta \int_{0}^{\pi} \sin\phi\,d\phi \int_{0}^{r} {r^{\prime}}^{2} d{r^{\prime}}
\rho({r^{\prime}},\theta,\phi).
\end{equation}
The calculation of $M(r)$ using \eref{sphericalmass} is non-trivial, and we proceed by making a suitable 
approximation
to the density, $\rho(r)$.  As discussed in \sref{section.MOG}, we may apply a sequence of approximations to develop a MOG solution.
\begin{description}
\item[{\sf\small $0^{\rm th}$-order approximation:}]  We have obtained the spherically integrated best-fit King $\beta$-model for the main cluster gas mass of
\erefs{betaRhoModel}{massProfile} using the parameters listed in \tref{table.king}.  By neglecting the subcluster, we
have generated a base-line solution similar to every other spherically symmetric galaxy cluster.  The result of
substituting the spherically symmetric King $\beta$-model mass profile of \eref{massProfile} into \eref{G} is shown in
\fref{figure.G} for the dimensionless gravitational coupling, $\mathcal{G}(r)$. We obtain a zeroth order, spherically
symmetric approximation to the \map{\kappa} by substituting $\rho(r)$ and $\mathcal{G}(r)$ into
\erefs{scaledSurfaceMassDensity.MOG}{Sigma.MOG} and integrating over the volume. 
\item[{\sf\small MOG Center:}] Treat the subcluster as a perturbation, and shift the origin of
$\mathcal{G}(r)$ toward the subcluster (toward the center-of-mass of the
system).  In this approximation, we continue to use the spherically integrated best-fit King $\beta$-model for the main
cluster gas mass of
\erefs{betaRhoModel}{massProfile} using the parameters listed in \tref{table.king}, but we allow the subcluster to
perturb (shift) the origin of $\mathcal{G}(r)$ toward the true center-of-mass of
the system.  The integrals of \erefs{scaledSurfaceMassDensity.MOG}{Sigma.MOG} become nontrivial as the integrand is no
longer spherically symmetric.  We were able to obtain a full numerical integration, but the computation proved to be too
time consuming  ($\sim 70,000$ numerical integrations to cover the $185\times185\ \mbox{pixel}^{2}$) to treat by means
of a nonlinear least-squares fitting routine.
\item[]  {\sf\small Projective approximation:} Approximate $M(r)$ with the concentric cylinder mass, $M(R)$ of
\eref{cylindricalmass}, calculated directly from the \map{\Sigma}, where the pixel-by-pixel sum proceeds from the MOG
center. 
This is a poor approximation for small $R$ (few pixels), but becomes very good for large $R$ (many pixels), as can 
be seen by comparing \frefs{figure.MainXray}{figure.MassProfile}.  The cylindrical mass, $M(R)$, can be computed directly
from the \map{\Sigma} data using \eref{cylindricalmass} without the need of a King $\beta$-model.
\item[ {\sf\small Isothermal $\beta$-model approximation:}]  Treat
the subcluster as a perturbation, and utilize the isothermal $\beta$-model of \eref{MOGMassScaled}
to approximate $M(r)$ and shift that toward the subcluster (toward the MOG center).  This is a fully analytic
expression, allowing ease of integration and an iterative fitting routine.
\end{description}

We will
base our analysis on our best-fit determined by the isothermal  $\beta$-model approximation of \eref{MOGMassScaled},
with $\mathcal{G}(r)$ located at the MOG center, a distance away from the main cluster \map{\Sigma} peak toward the
subcluster  \map{\Sigma} peak.  Since the dimensionless gravitational coupling, $\mathcal{G}(r)$, of  \eref{G} depends on $M(r)$, and the
isothermal $\beta$-model of \eref{MOGMassScaled} depends on $\mathcal{G}(r)$,
we must solve the system simultaneously.  Let us rewrite \eref{G}
\begin{equation}
\label{G.xi} \mathcal{G}(r) = 1 + \sqrt{\frac{\xi(r)}{M(r)}},
\end{equation}
where
\begin{equation}
 \xi(r) = M_{0} \left\{1-\exp\left(-\frac{r}{r_{0}}\right)\left(1+\frac{r}{r_{0}}\right)\right\}^{2}.
\end{equation}
We may solve \eref{G.xi} for
\begin{equation}
\label{G.mass} M(r) =  {\frac{\xi(r)}{\left(\mathcal{G}(r)-1\right)^{2}}},
\end{equation}
and equate this to the isothermal $\beta$-model of \eref{MOGMassScaled}:
\begin{equation}
\label{G.equation} {\frac{\xi(r)}{\left(\mathcal{G}(r)-1\right)^{2}}} = \frac{M_{\rm N}(r)}{\mathcal{G}(r)},
\end{equation}
and we may solve the quadratic equation for
\begin{equation}
\label{G.thermal} \mathcal{G}(r) = 1 + \frac{\xi(r)}{2 M_{\rm N}(r)}\left\{1+\sqrt{\frac{4 M_{\rm N}(r)}{\xi(r)}+1}\right\},
\end{equation}
where $M_{\rm N}$ is the isothermal $\beta$-model Newtonian dynamic mass of \eref{isothermalNewtonsMass}. The result of
\eref{G.thermal} is a fully analytic function, as opposed to a hypergeometric integral, and may be
easily computed across the full \map{\kappa}.  There was a
noticeable
parameter degeneracy in choosing the MOG center, the best-fit corresponded to a distance of 140 kpc away from the main cluster \map{\Sigma} peak toward the
subcluster  \map{\Sigma} peak.  This is reasonable since the 
\map{\Sigma} and \map{\kappa} data are two-dimensional ``surface projections'', due to the line-of-sight integral.  The
full simulation was run iteratively over a range of positions for the MOG center, while covarying the MOG parameters,
$M_{0}$ and $r_{0}$, the MOG mass scale and MOG range parameter, respectively.  This yielded a best-fit MOG model for
the dimensionless gravitational coupling, $\mathcal{G}(r)$ of \eref{G}, where $r$ is the distance from the MOG center,
as listed in \tref{table.MOG}.  Our best-fit to the \map{\kappa} corresponds to the MOG $\kappa$-model of
\erefs{scaledSurfaceMassDensity.MOG}{Sigma.MOG}.  The best-fit location of the MOG center is provided on the \map{\Sigma}
in \fref{figure.sigma}, and the coordinates are listed in \tref{table.coords}.
%

We show a 3D visualization of the convergence \map{\kappa} data in \fref{figure.kappaMap}.  The $0^{\rm th}$-order
approximation result, rescaled is shown in \fref{figure.KappaMainXray}.  The twin humps at the peaks of our prediction
will be a generic prediction for any
spherically symmetric galaxy cluster (non-mergers).  The best-fit MOG $\kappa$-model is shown in
\fref{figure.KappaMainXrayMainGalaxy} along the line connecting the MOG center to the main cluster \map{\kappa} peak. 
We show a 3D visualization  of the full convergence \map{\kappa} model in \fref{figure.kappaModel}.
\subsection{\label{section.Galaxies}Including the Galaxies}
\begin{figure}[ht] 
\psset{unit=1pt}
\begin{pspicture}(500,325)(0,0)
\psset{linecolor=DarkGreen,linewidth=1pt}
{\thicklines \multiput(113,303)(90,0){4}{\line(0,1){9}} \put(113,303){\line(1,0){284.4}}}
{\thinlines \multiput(113,303)(4.5,0){64}{\line(0,1){3}} \multiput(113,303)(18,0){16}{\line(0,1){6}}}
\put(403,317){\makebox(0,0){\sf [kpc]}}
\multiput(103,317)(90,0){4}{\makebox(20,0){\sf\arabic{kpc}}\addtocounter{kpc}{500}}
\put(110,0){\includegraphics[width=290pt]{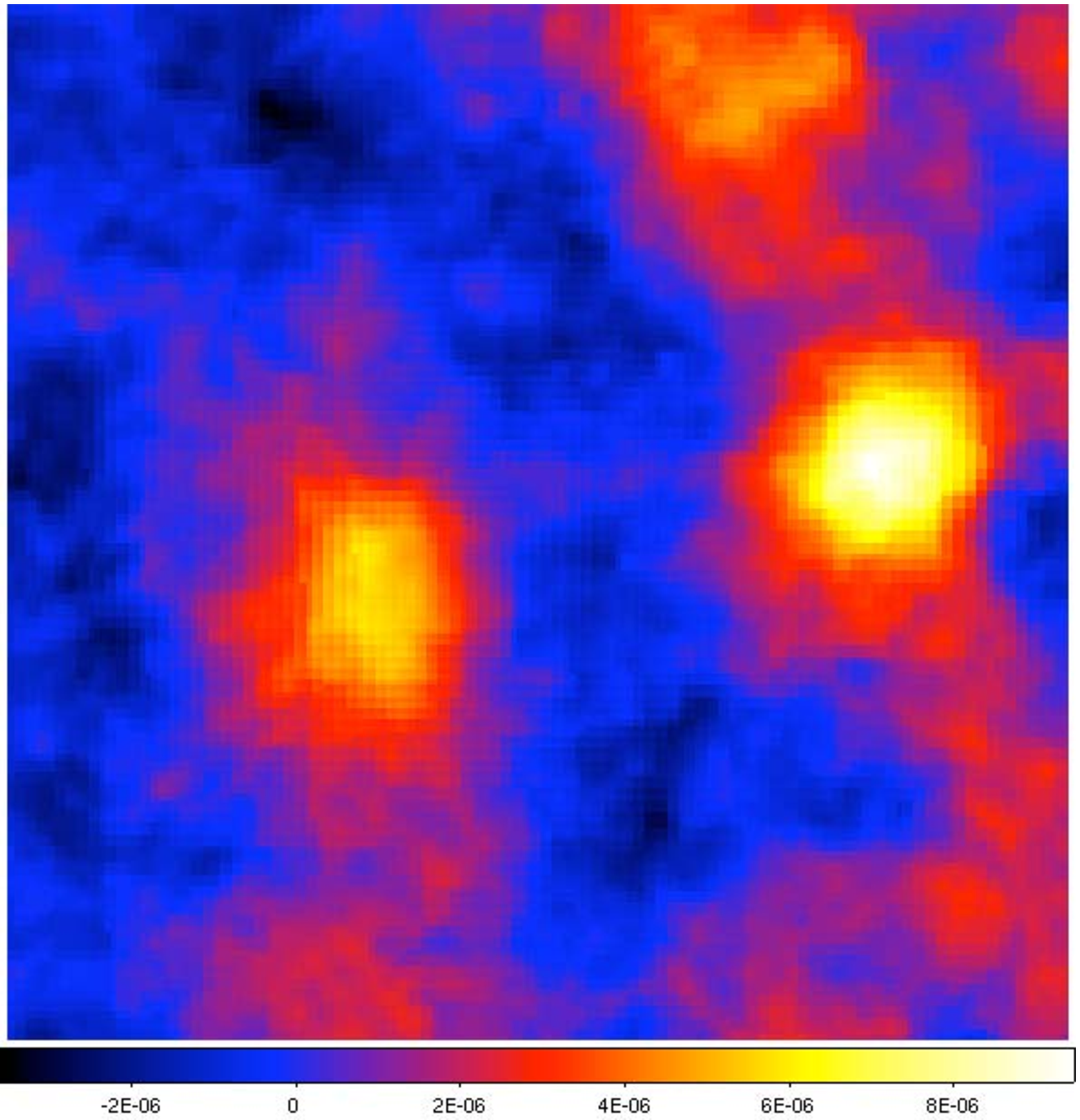}} 
\psline{->}(105,247)(249,155) \psline{->}(105,187)(221,153) \psline{->}(105,127)(219,141) \psline{->}(105,67)(273,159)
\psline{->}(403,247)(313,270) \psline{->}(403,187)(341,174) \psline{->}(403,127)(349,168) \psline{->}(403,67)(320,168)
\put(0, 240){\fcolorbox{cayenne}{sand}{\makebox(100,14){\color{chocolate}\small\sf Main cluster \map{\Sigma} peak}}} 
\put(0, 180){\fcolorbox{cayenne}{sand}{\makebox(100,14){\color{chocolate}\small\sf Main cluster cD galaxy}}} 
\put(0, 120){\fcolorbox{cayenne}{sand}{\makebox(100,14){\color{chocolate}\small\sf Main cluster \map{\kappa} peak}}} 
\put(0, 60){\fcolorbox{cayenne}{sand}{\makebox(100,14){\color{chocolate}\small\sf MOG Center}}} 
\put(403, 240){\fcolorbox{cayenne}{sand}{\makebox(100,14){\color{chocolate}\small\sf Mysterious Plateau Nearby}}} 
\put(403, 180){\fcolorbox{cayenne}{sand}{\makebox(100,14){\color{chocolate}\small\sf Subcluster \map{\kappa} peak}}} 
\put(403, 120){\fcolorbox{cayenne}{sand}{\makebox(100,14){\color{chocolate}\small\sf Subcluster BCG}}} 
\put(403 , 60){\fcolorbox{cayenne}{sand}{\makebox(100,14){\color{chocolate}\small\sf Subcluster \map{\Sigma} peak}}} 
\end{pspicture}
\caption{\label{figure.galaxySigma}  {{\sf\small
The galaxy surface density \map{\Sigma} prediction.}}\break \small
The prediction of the \map{\Sigma} due to the galaxies as computed by the difference between the \map{\kappa}
and our MOG $\kappa$-model,
scaled as surface mass density according to \eref{sigma.galaxies}.  \map{\Sigma} and \map{\kappa} observed peaks are shown for comparison.  The
central dominant (cD) galaxy of the main cluster, the brightest cluster galaxy (BCG) of the subcluster, and the
MOG predicted gravitational center are shown.  J2000 and map (x,y) coordinates are listed in \tref{table.coords}. Component masses (integrated within a 100 kpc radius aperture) for the main and
subcluster, the MOG center and the total predicted baryonic mass, $M_{\rm bary}$, for the {\bc} are shown in \tref{table.mass}.}
\end{figure}
\begin{figure}[ht]
\begin{center} 
\subfloat[ {\sf\small 3D visualization of the \map{\kappa} November 15, 2006 data release~\protect\citep{Clowe:dataProduct}}]{\label{figure.kappaMap}\includegraphics[width=8cm]{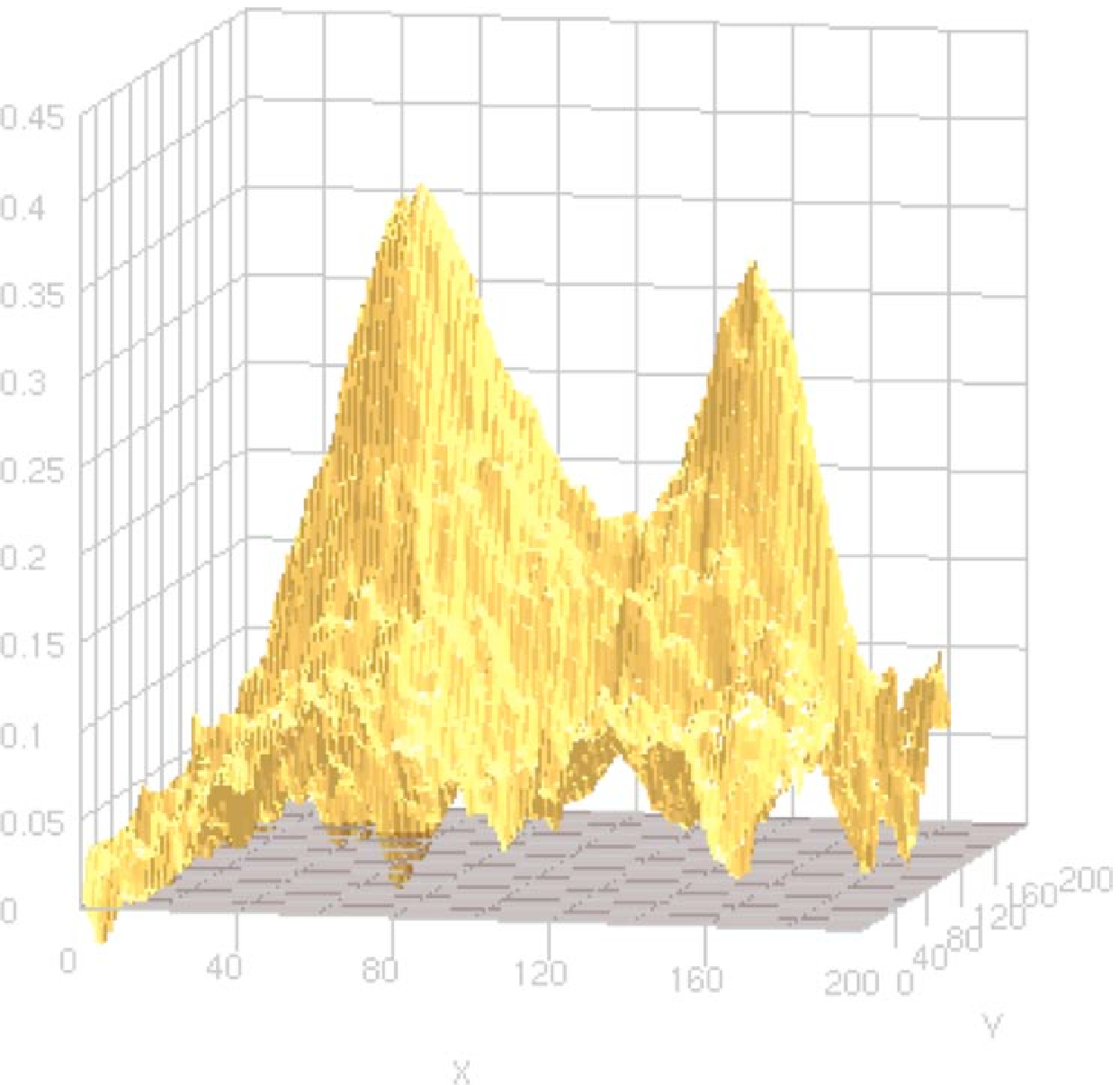}} 
\quad\quad
\subfloat[ {\sf\small $0^{\rm th}$-order approximation -- neglecting the subcluster}]{\label{figure.KappaMainXray}\scalebox{0.6}{\input{figure/gnuplot/Kappa_MainXray}}}
\\
\subfloat[ {\sf\small Best-fit $\kappa$-model}]{\label{figure.KappaMainXrayMainGalaxy}\scalebox{0.6}{\input{figure/gnuplot/Kappa_MainXray_MainGalaxy}}}
\quad\quad
\subfloat[
{\sf\small 3D visualization of the best-fit $\kappa$-model}]{\label{figure.kappaModel}\includegraphics[width=8cm]{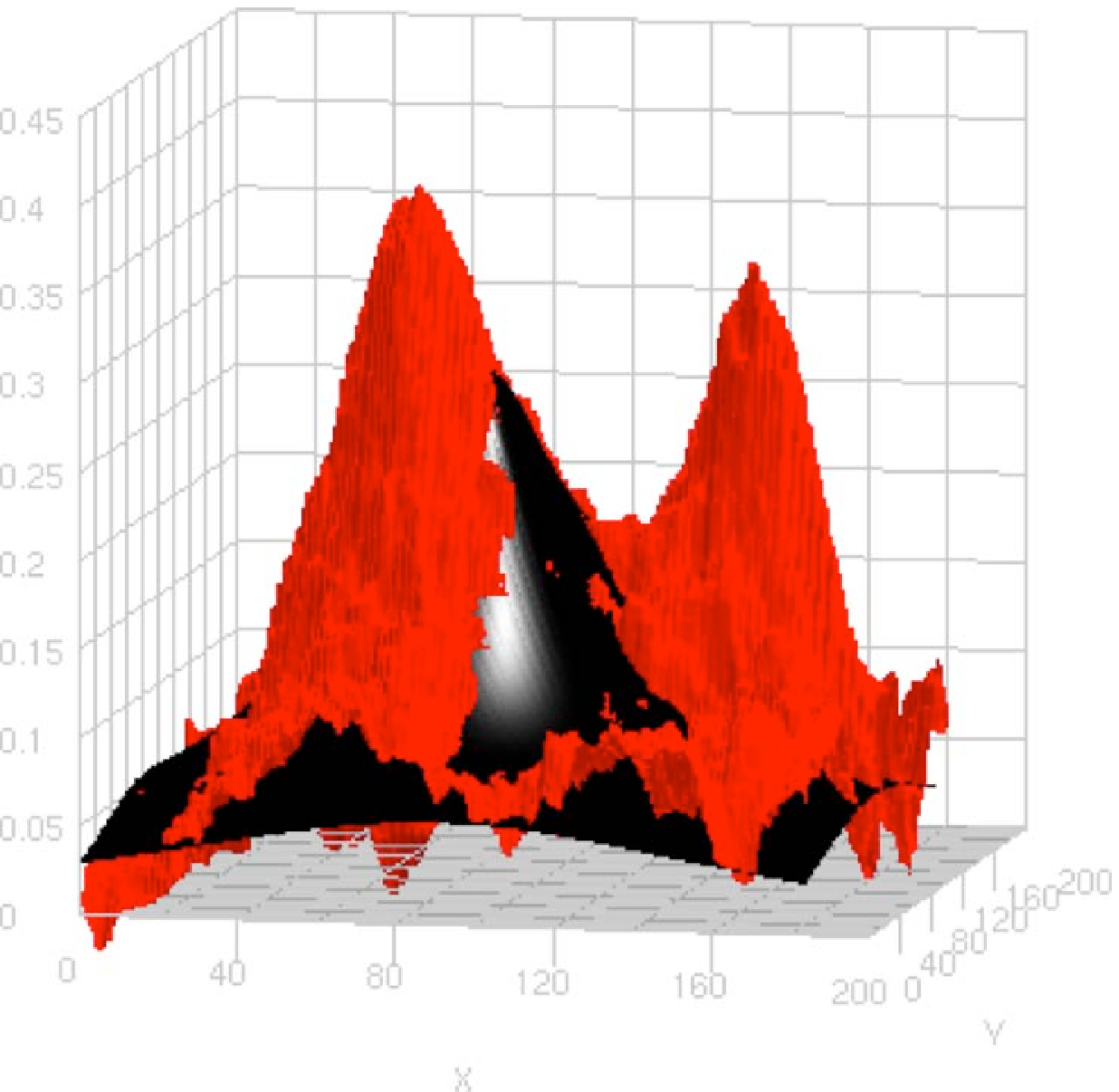}}
\caption{ {\sf The convergence \map{\kappa} November 15, 2006 data release~\protect\citep{Clowe:dataProduct}
and our $\kappa$-models.}\break \small
  The best-fit MOG $\kappa$-model is shown in solid black in
\frefss{figure.KappaMainXray}{figure.KappaMainXrayMainGalaxy}{figure.kappaModel}.  The convergence \map{\kappa}
November 15, 2006 data release~\protect\citep{Clowe:dataProduct} is shown as \fref{figure.kappaMap} and in red in
\frefss{figure.KappaMainXray}{figure.KappaMainXrayMainGalaxy}{figure.kappaModel}.  The scaled \map{\Sigma},
$\Sigma(x,y)/\Sigma_{c}$ data  is shown in short-dashed green, also shown in \fref{figure.SigmaFit}.}
\end{center}
\end{figure}

In considering the MOG $\kappa$-model resulting from the  $0^{\rm th}$-order approximation, as shown in
\fref{figure.KappaMainXray}, it is tempting to try to explain the entire convergence \map{\kappa} by the X-ray gas mass,
just by
shifting the solid-black line to the left by $\sim 200$ kpc, but then there would be no way to explain the subcluster
\map{\kappa} peak.  We next proceed to account for the effect of the subcluster on the
dimensionless gravitational coupling, $\mathcal{G}(r)$, of MOG, as shown in the best-fit $\kappa$-model of
\frefs{figure.KappaMainXrayMainGalaxy}{figure.kappaModel}.  Remarkably, as the MOG center is separated from the main
cluster \map{\Sigma} peak, say due to the gravitational effect of the subcluster,  the centroid naturally shifts
toward the \map{\kappa} peak, and the predicted height of the \map{\kappa} drops.  Let us  take the difference
between the \map{\kappa} data and our best-fit $\kappa$-model, to see if there is any ``missing mass''.   We 
hypothesize that the difference can be explained by including the galaxies, 
\begin{equation}
\label{kappa.galaxies}\kappa(x,y) = \frac{{\bar \Sigma}(x,y) + {\bar \Sigma}_{\rm galax}(x,y)}{\SigmaN},
\end{equation}
where $\bar \Sigma$ is the weighted surface mass density of \eref{Sigma.MOG}, and the best-fit $\kappa$-model of ${\bar \Sigma}/\SigmaN$ is
derived from \erefs{scaledSurfaceMassDensity.MOG}{Sigma.MOG}.  Therefore, the galaxies contribute a ``measurable'' surface mass density,
\begin{equation}
\label{sigma.galaxies}\Sigma_{\rm galax}(x,y) \approx \frac{\kappa(x,y) \SigmaN-{\bar \Sigma}(x,y)}{\mathcal{G}(x,y)},
\end{equation}
where $\mathcal{G}(x,y)$ corresponds to the best-fit model of \eref{G} listed in \tref{table.MOG}.  The result of the galaxy subtraction of  \eref{sigma.galaxies} is shown in
\fref{figure.galaxySigma}.  Now we may interpret \fref{figure.kappaModel} as the total convergence \map{\kappa} where the
black surface is the contribution from the weighted surface density of the ICM gas, ${\bar \Sigma}/\SigmaN$, and  the red surface is the remainder of the \map{\kappa} due to the contribution of the weighted surface density of the galaxies, ${\bar \Sigma}_{\rm galax}/\SigmaN$.
We may calculate the total mass of the galaxies, 
\begin{equation}
\label{mass.galaxies} M_{\rm galax} = \int \Sigma_{\rm galax}(x^{\prime},y^{\prime}) dx^{\prime}\,dy^{\prime}.
\end{equation}
\begin{SCtable}[0.9][t]
\begin{tabular}{|c||c|c|c||c|} \hline 
{\sf Component}& {\sf Main cluster} & {\sf Subcluster} & {\sf
Central ICM} &
{\sf Total} \\\hline\hline
{$M_{\rm gas}$} & $7.0  \times 10^{12}\ M_{\sun}$ & $5.8  \times 10^{12}\ M_{\sun}$ &$6.3  \times 10^{12}\ M_{\sun}$& $2.2 
\times 10^{14}\ M_{\sun}$ \\
{$M_{\rm galax}$} & $1.8  \times 10^{12}\ M_{\sun}$ & $3.1  \times 10^{12}\ M_{\sun}$ &$2.4  \times 10^{10}\ M_{\sun}$&  $3.8  \times 10^{13}\ M_{\sun}$
\\
\hline
{$M_{\rm bary}$} & \colorbox{yellow}{\color{cyan}{$8.8\times 10^{12}\ M_{\sun}$}} &
\colorbox{yellow}{\color{cyan}{$9.0\times 10^{12}\ M_{\sun}$}} &\colorbox{yellow}{\color{cyan}{$4.9 \times 10^{12}\ M_{\sun}$}}&  \colorbox{yellow}{\color{cyan}{$2.6 \times 10^{14}\ M_{\sun}$}} \\ \hline\hline
{$M_{\rm DM}$} & \colorbox{orange}{\color{cyan}{$2.1\times 10^{13}\ M_{\sun}$}} &
\colorbox{orange}{\color{cyan}{$1.7\times 10^{13}\ M_{\sun}$}} &\colorbox{orange}{\color{cyan}{$1.4\times 10^{13}\ M_{\sun}$}}&  \colorbox{orange}{\color{cyan}{$6.8\times 10^{14}\ M_{\sun}$}}\\ \hline\hline
{$M_{\rm galax}/M_{\rm gas}$} & $26\%$ & $53\%$ &$0.4\%$&  $17\%$ \\ 
{$M_{\rm gas}/M_{\rm DM}$} & $33\%$ & $34\%$ &$45\%$&  $32\%$ \\  \hline
\end{tabular}
\caption{\label{table.mass} {{\sf\small Summary of component mass predictions.}}\break \small
Component masses (integrated within a 100 kpc radius aperture) for the main and subcluster and the MOG center. 
The total predicted mass for the {\bc} is integrated over the full \map{\Sigma}.}
 \end{SCtable}
\pagebreak
We were able to perform the integration within a 100 kpc radius aperture about the main cluster cD and subcluster
BCG, separately, the results of which are listed in \tref{table.mass}, where they are compared with the upper limits on
galaxy masses set by HST observations.  If the hypothesis that the predicted $M_{\rm galax}$ is below the bound set by HST observations is true, then it follows that 
\begin{equation}
\label{mass.bary} M_{\rm bary} = M_{\rm gas} + M_{\rm galax},
\end{equation}
requires no addition of non-baryonic dark matter.  The results of our best-fit for $M_{\rm gas}$, $M_{\rm galax}$ and $M_{\rm
bary}$ of \eref{mass.bary} are listed in  \tref{table.mass}.

\subsection{\label{section.DM} Dark Matter}
From the alternative point-of-view, dark matter is hypothesized to account for all of the ``missing mass'' which results
in applying Newton/Einstein gravity.  This means, for the November 15, 2006 data
release~\protect\citep{Clowe:ApJL:2006,Clowe:ApJ:2006,Clowe:astro-ph.0611496,Clowe:dataProduct}, 
that the ``detected''  dark matter must contribute a surface mass density,
\begin{equation}
\label{sigma.darkmatter}\Sigma_{\rm DM}(x,y) \approx  \kappa(x,y)  \SigmaN - \Sigma(x,y), 
\end{equation}
with an associated total mass,
\begin{equation}
\label{mass.darkmatter.0} M_{\rm DM} = \int \Sigma_{\rm DM}(x^{\prime},y^{\prime})
dx^{\prime}\,dy^{\prime}.
\end{equation}
The integral of \eref{mass.darkmatter.0} becomes trivial upon substitution of \eref{sigma.darkmatter}:
\begin{equation}
\label{mass.darkmatter} M_{\rm DM} = \SigmaN \int \kappa(x^{\prime},y^{\prime}) dx^{\prime}\,dy^{\prime} - \int \Sigma(x^{\prime},y^{\prime}) dx^{\prime}\,dy^{\prime}
\end{equation}
where we have neglected $M_{\rm galax}$ in \eref{mass.darkmatter}, because as is usually argued, the contribution from
the
galaxies in the dark matter paradigm is $\le 1$ -- $4\%$ of $M_{\rm total}$.  The calculation of $M_{\rm DM}$ in
\eref{mass.darkmatter} was performed by a pixel-by-pixel sum over the convergence \map{\kappa} data and surface density
\map{\Sigma}
data, within a 100 kpc radius aperture around the main and subcluster \map{\kappa} peaks, respectively.  The
result of our computation is included in \tref{table.mass}.
We emphasize, here, that for each of the main cluster, subcluster and total \map{\Sigma}, our results of \tref{table.mass}
indicate that 
\begin{equation}
\label{mass.question} M_{\rm bary} = M_{\rm gas} + M_{\rm galax} \ll M_{\rm DM}
\end{equation}
implying that we have successfully put {\bc} on a lean diet! This seems to us to be a proper use of
\textcolor{DarkOrange}{\it Occam's razor.}  The mass ratios, $M_{\rm galax}/M_{\rm gas}$, for the
main and subcluster and central ICM are shown at the bottom of \tref{table.mass}.  The result of $M_{\rm galax}/M_{\rm
gas} \approx 0.4\%$ in the central ICM is due to the excellent fit in MOG across the hundreds of kpc separating the main
and subcluster.  The dark matter result of $M_{\rm gas}/M_{\rm
DM} \approx 45\%$ in the central ICM implies that the evolutionary scenario does not lead to a spatial dissociation
between the dark matter and the ICM gas, which indicates that the merger is ongoing.  In contrast, the MOG result shows a true
dissociation between the galaxies and the ICM gas as required by the evolutionary scenario.  The baryon to dark matter fraction over the full \map{\Sigma} is $32\%$, which is significantly higher than the $\Lambda$-CDM cosmological baryon mass-fraction of $17^{+1.9}_{-1.2}\%$~\protect\citep{Spergel:2006}.  The distribution of mass predicted by MOG vs.\,the dark matter paradigm is shown in \fref{figure.SigmaGalaxDM}.
\begin{figure}[ht]
\begin{center}
\subfloat[ {\sf\small  Scaled surface density for the MOG predicted galaxies, $\Sigma_{\rm
galax}/\SigmaN$, and the MOG predicted visible baryons, $\Sigma_{\rm
bary}/\SigmaN$, compared to the ICM gas.}]{\label{figure.SigmaGalax}\scalebox{0.6}{\input{figure/gnuplot/SigmaGalax}}}
\quad\quad
\subfloat[ {\sf\small Scaled surface density of dark matter, $\Sigma_{\rm DM}/\SigmaN$,
compared to ICM gas.}]{\label{figure.SigmaDM}\scalebox{0.6}{\input{figure/gnuplot/SigmaDM}}}
\end{center}
\caption{\label{figure.SigmaGalaxDM} {{\sf\small Plot of the scaled surface density
$\Sigma/\SigmaN$ along the line connecting the main cluster \map{\Sigma} peak with the main cD.}}\break {\small In
\fref{figure.SigmaGalax}, the prediction of
\eref{sigma.galaxies} for the galaxies is shown in long-dashed magenta, and the prediction of \eref{mass.bary} for 
the visible baryonic mass is shown in solid brown.  The calculation of \eref{sigma.darkmatter} for dark matter is shown in
\fref{figure.SigmaDM} 
in dash-dot black.  The ICM gas distribution inferred from the \map{\Sigma} data is shown in short-dashed green on each
plot.}}
\end{figure} 

\section{\label{section.Conclusions}Conclusions}

The modified gravity (MOG) theory provides a fit to the \map{\kappa} of the November 15, 2006 data
release~\protect\citep{Clowe:dataProduct}.  The model, derived purely from
the X-ray imaging \map{\Sigma} observations combined with the galaxy \map{\Sigma} predicted by MOG, accounts for the
\map{\kappa} peak amplitudes and their spatial dissociation without the introduction of nonbaryonic dark matter.

The question of the internal degrees of freedom in MOG has to be further investigated. It would be desirable to derive a
theoretical prediction from the MOG field equations that fits the empirically determined mass and distance scales in
\fref{figure.M0r0}.

{It could be argued that any modified theory designed to solve the dark matter conundrum, such as MOND or MOG, has
less freedom than dark matter.  So the important question to resolve is precisely how much freedom the MOG solution
has.  On one hand we said, definitely, that there was no freedom in choosing the pair of $M_{0}$ and $r_{0}$ for the
main cluster since it was well described by an isothermal sphere, to an excellent approximation.  We further argued
that the subcluster was (per mass) a small perturbation to the ICM.  But if MOG has more freedom than MOND, but less
freedom than Dark Matter, then what is the additional degree of freedom that enters the Bullet Cluster observations?}

{The question is resolved in that there is a physical degree of freedom due to a lack of spherical symmetry in the Bullet
Cluster, and whence the galaxies sped outward, beyond the ICM gas clouds which lagged behind -- effectively allowing
the galaxies to climb out of the spherical minimum of the Newtonian core where MOG effects are
small (inside the MOG range $r_{0}$) upwards along the divergence of the stress-energy tensor (Newtonian potential, if
you prefer a simple choice) towards the far infrared region of large gravitational coupling, ${G}_{\infty}$.}

{In fact, the Bullet Cluster data results describe, to a remarkable precision, a simple King $\beta$-Model. Our analysis, with the result to the best-fit shown in 
\tref{table.king},
 uniquely determines the mass profile $\rho(r)$ of \eref{betaRhoModel} used throughout our computations.  We permitted only
a single further degree of freedom to account for the fits of 
\fref{figure.KappaMainXrayMainGalaxy} and the predictions of \frefs{figure.galaxySigma}{figure.SigmaGalaxDM};
this was the location of the MOG center, where the gravitational coupling, $\mathcal{G}(0) \rightarrow 1$, is a minimum at the
Newtonian core.  Remarkably, the data did not permit a vanishing MOG center, with respect to the peak of the ICM gas
$\rho(0)$.  We have shown the location of the MOG center as determined by a numerical simulation of convergence map
according to \erefs{scaledSurfaceMassDensity.MOG}{Sigma.MOG} in each of \frefsss{figure.sigma}{figure.SigmaSubcluster}{figure.kappa}{figure.galaxySigma} and provided the coordinates in \tref{table.coords}.}

The surface density \map{\Sigma} derived from X-ray imaging observations  is separable into the main cluster and the subcluster subtracted surface density \map{\Sigma} through a low $\chi^{2}$-fitting King $\beta$-model.
Following the ($>100$) galaxy cluster survey of~\protect\citet{Brownstein:MNRAS:2006}, we have derived
a parameter-free (unique) prediction  for the X-ray temperature of the {\bc} which has already been experimentally
confirmed.  In \erefs{scaledSurfaceMassDensity.MOG}{Sigma.MOG}, we have derived a weighted surface mass density, $\bar
\Sigma$, from the
convergence \map{\kappa} which produced a best-fit model (\frefs{figure.KappaMainXrayMainGalaxy}{figure.kappaModel} and
\tref{table.MOG}).  We have computed the dark matter and the MOG predicted 
galaxies and baryons (\fref{figure.SigmaGalaxDM}), and noted the tremendous predictive power of MOG as a means of
utilizing strong
and weak gravitational lensing to do galactic photometry -- a powerful tool simply not provided by any candidate dark
matter (\fref{figure.galaxySigma}).  The predictions for galaxy photometry will be the subject of future investigations
in MOG, and the availability of weak and strong gravitational lensing surveys will prove invaluable in the future.

Although dark matter allows us to continue to use Einstein (weak-field) and Newtonian gravity theory, these theories may be misleading at astrophysical scales.  By searching for dark matter, we may have
arrived at a means to answer one of the most fundamental questions remaining in astrophysics and cosmology:  How much matter (energy) is
there in the Universe and how is it distributed?  For the {\bc}, dark matter dominance is a ready answer, but in MOG we must answer the 
question with only the visible contribution of galaxies, ICM gas and gravity.
\begin{acknowledgments}
This work was supported by the Natural Sciences and Engineering Research Council of Canada (NSERC). We thank Douglas Clowe and
Scott Randall for providing early releases of the gravitational lensing convergence data and X-ray 
surface mass density data, respectively, and for stimulating and helpful discussions. Research at Perimeter Institute for Theoretical Physics is supported in part by the Government of Canada through NSERC and by the Province of Ontario through the Ministry of Research and Innovation (MRI).
\end{acknowledgments}

\bibliography{BulletCluster}
\end{document}

%% file: macros.tex
\newcommand{\SigmaN}{\Sigma_{c}}
\newcommand{\bc}{Bullet Cluster 1E0657-558}
\newcommand{\map}[1]{$#1$-map}

\newcommand{\frefsss}[4]{Figures \ref{#1}, \ref{#2}, \ref{#3} and \ref{#4}}
\newcommand{\frefss}[3]{Figures \ref{#1}, \ref{#2} and \ref{#3}}
\newcommand{\frefs}[2]{Figures \ref{#1} and \ref{#2}}
\newcommand{\fref}[1]{Figure \ref{#1}}
\newcommand{\tref}[1]{Table \ref{#1}}
\newcommand{\sref}[1]{Section \ref{#1}}

\newcommand{\eref}[1]{Equation (\ref{#1})}
\newcommand{\erefs}[2]{Equations (\ref{#1}) and (\ref{#2})}
\newcommand{\erefss}[3]{Equations (\ref{#1}), (\ref{#2}) and (\ref{#3})}
\newcommand{\erefsss}[4]{Equations (\ref{#1}), (\ref{#2}), (\ref{#3}) and (\ref{#4})}

\newcommand{\sun}{\odot}

\newcommand{\mnras}{Mon.\,Not.\,Roy.\,Astron.\,Soc.}


%% file: figure/gnuplot/M0r0.tex
\begingroup%
\makeatletter%
\newcommand{\GNUPLOTspecial}{%
  \@sanitize\catcode`\%=14\relax\special}%
\setlength{\unitlength}{0.0500bp}%
\begin{picture}(8496,5947)(0,0)%
  {\GNUPLOTspecial{"
/gnudict 256 dict def
gnudict begin
%
%
/Color true def
/Blacktext false def
/Solid true def
/Dashlength 1 def
/Landscape false def
/Level1 false def
/Rounded false def
/TransparentPatterns false def
/gnulinewidth 5.000 def
/userlinewidth gnulinewidth def
/vshift -66 def
/dl1 {
  10.0 Dashlength mul mul
  Rounded { currentlinewidth 0.75 mul sub dup 0 le { pop 0.01 } if } if
} def
/dl2 {
  10.0 Dashlength mul mul
  Rounded { currentlinewidth 0.75 mul add } if
} def
/hpt_ 31.5 def
/vpt_ 31.5 def
/hpt hpt_ def
/vpt vpt_ def
Level1 {} {
/SDict 10 dict def
systemdict /pdfmark known not {
  userdict /pdfmark systemdict /cleartomark get put
} if
SDict begin [
  /Title (M0r0_out.tex)
  /Subject (gnuplot plot)
  /Creator (gnuplot 4.2 patchlevel 0)
  /Author (Joel Brownstein)
  /CreationDate (Tue Aug  7 19:04:11 2007)
  /DOCINFO pdfmark
end
} ifelse
%
%
/M {moveto} bind def
/L {lineto} bind def
/R {rmoveto} bind def
/V {rlineto} bind def
/N {newpath moveto} bind def
/Z {closepath} bind def
/C {setrgbcolor} bind def
/f {rlineto fill} bind def
/vpt2 vpt 2 mul def
/hpt2 hpt 2 mul def
/Lshow {currentpoint stroke M 0 vshift R 
	Blacktext {gsave 0 setgray show grestore} {show} ifelse} def
/Rshow {currentpoint stroke M dup stringwidth pop neg vshift R
	Blacktext {gsave 0 setgray show grestore} {show} ifelse} def
/Cshow {currentpoint stroke M dup stringwidth pop -2 div vshift R 
	Blacktext {gsave 0 setgray show grestore} {show} ifelse} def
/UP {dup vpt_ mul /vpt exch def hpt_ mul /hpt exch def
  /hpt2 hpt 2 mul def /vpt2 vpt 2 mul def} def
/DL {Color {setrgbcolor Solid {pop []} if 0 setdash}
 {pop pop pop 0 setgray Solid {pop []} if 0 setdash} ifelse} def
/BL {stroke userlinewidth 2 mul setlinewidth
	Rounded {1 setlinejoin 1 setlinecap} if} def
/AL {stroke userlinewidth 2 div setlinewidth
	Rounded {1 setlinejoin 1 setlinecap} if} def
/UL {dup gnulinewidth mul /userlinewidth exch def
	dup 1 lt {pop 1} if 10 mul /udl exch def} def
/PL {stroke userlinewidth setlinewidth
	Rounded {1 setlinejoin 1 setlinecap} if} def
/LCw {1 1 1} def
/LCb {0 0 0} def
/LCa {0 0 0} def
/LC0 {1 0 0} def
/LC1 {0 1 0} def
/LC2 {0 0 1} def
/LC3 {1 0 1} def
/LC4 {0 1 1} def
/LC5 {1 1 0} def
/LC6 {0 0 0} def
/LC7 {1 0.3 0} def
/LC8 {0.5 0.5 0.5} def
/LTw {PL [] 1 setgray} def
/LTb {BL [] LCb DL} def
/LTa {AL [1 udl mul 2 udl mul] 0 setdash LCa setrgbcolor} def
/CCg {0 0.5 0} def
/CCb {0 0 0.5} def
/fatlinewidth 7.500 def
/gatlinewidth 10.000 def
/GL { stroke gatlinewidth setlinewidth Rounded { 1 setlinejoin 1 setlinecap } if } def
/FL { stroke fatlinewidth setlinewidth Rounded { 1 setlinejoin 1 setlinecap } if } def
/LT0 {GL [] LC4 DL} def
/LT1 {GL [] CCb DL} def
/LT2 {GL [] LC0 DL} def
/LT3 {GL [] CCg DL} def
/LT4 {GL [] LC6 DL} def                                                                                                             
/LT5 {PL [3 dl1 3 dl2 1 dl1 3 dl2] LC5 DL} def
/LT6 {PL [2 dl1 2 dl2 2 dl1 6 dl2] LC6 DL} def
/LT7 {PL [1 dl1 2 dl2 6 dl1 2 dl2 1 dl1 2 dl2] LC7 DL} def
/LT8 {PL [2 dl1 2 dl2 2 dl1 2 dl2 2 dl1 2 dl2 2 dl1 4 dl2] LC8 DL} def
/Pnt {stroke [] 0 setdash gsave 1 setlinecap M 0 0 V stroke grestore} def
/Dia {stroke [] 0 setdash 2 copy vpt add M
  hpt neg vpt neg V hpt vpt neg V
  hpt vpt V hpt neg vpt V closepath stroke
  Pnt} def
/Pls {stroke [] 0 setdash vpt sub M 0 vpt2 V
  currentpoint stroke M
  hpt neg vpt neg R hpt2 0 V stroke
 } def
/Box {stroke [] 0 setdash 2 copy exch hpt sub exch vpt add M
  0 vpt2 neg V hpt2 0 V 0 vpt2 V
  hpt2 neg 0 V closepath stroke
  Pnt} def
/Crs {stroke [] 0 setdash exch hpt sub exch vpt add M
  hpt2 vpt2 neg V currentpoint stroke M
  hpt2 neg 0 R hpt2 vpt2 V stroke} def
/TriU {stroke [] 0 setdash 2 copy vpt 1.12 mul add M
  hpt neg vpt -1.62 mul V
  hpt 2 mul 0 V
  hpt neg vpt 1.62 mul V closepath stroke
  Pnt} def
/Star {2 copy Pls Crs} def
/BoxF {stroke [] 0 setdash exch hpt sub exch vpt add M
  0 vpt2 neg V hpt2 0 V 0 vpt2 V
  hpt2 neg 0 V closepath fill} def
/TriUF {stroke [] 0 setdash vpt 1.12 mul add M
  hpt neg vpt -1.62 mul V
  hpt 2 mul 0 V
  hpt neg vpt 1.62 mul V closepath fill} def
/TriD {stroke [] 0 setdash 2 copy vpt 1.12 mul sub M
  hpt neg vpt 1.62 mul V
  hpt 2 mul 0 V
  hpt neg vpt -1.62 mul V closepath stroke
  Pnt} def
/TriDF {stroke [] 0 setdash vpt 1.12 mul sub M
  hpt neg vpt 1.62 mul V
  hpt 2 mul 0 V
  hpt neg vpt -1.62 mul V closepath fill} def
/DiaF {stroke [] 0 setdash vpt add M
  hpt neg vpt neg V hpt vpt neg V
  hpt vpt V hpt neg vpt V closepath fill} def
/Pent {stroke [] 0 setdash 2 copy gsave
  translate 0 hpt M 4 {72 rotate 0 hpt L} repeat
  closepath stroke grestore Pnt} def
/PentF {stroke [] 0 setdash gsave
  translate 0 hpt M 4 {72 rotate 0 hpt L} repeat
  closepath fill grestore} def
/Circle {stroke [] 0 setdash 2 copy
  hpt 0 360 arc stroke Pnt} def
/CircleF {stroke [] 0 setdash hpt 0 360 arc fill} def
/C0 {BL [] 0 setdash 2 copy moveto vpt 90 450 arc} bind def
/C1 {BL [] 0 setdash 2 copy moveto
	2 copy vpt 0 90 arc closepath fill
	vpt 0 360 arc closepath} bind def
/C2 {BL [] 0 setdash 2 copy moveto
	2 copy vpt 90 180 arc closepath fill
	vpt 0 360 arc closepath} bind def
/C3 {BL [] 0 setdash 2 copy moveto
	2 copy vpt 0 180 arc closepath fill
	vpt 0 360 arc closepath} bind def
/C4 {BL [] 0 setdash 2 copy moveto
	2 copy vpt 180 270 arc closepath fill
	vpt 0 360 arc closepath} bind def
/C5 {BL [] 0 setdash 2 copy moveto
	2 copy vpt 0 90 arc
	2 copy moveto
	2 copy vpt 180 270 arc closepath fill
	vpt 0 360 arc} bind def
/C6 {BL [] 0 setdash 2 copy moveto
	2 copy vpt 90 270 arc closepath fill
	vpt 0 360 arc closepath} bind def
/C7 {BL [] 0 setdash 2 copy moveto
	2 copy vpt 0 270 arc closepath fill
	vpt 0 360 arc closepath} bind def
/C8 {BL [] 0 setdash 2 copy moveto
	2 copy vpt 270 360 arc closepath fill
	vpt 0 360 arc closepath} bind def
/C9 {BL [] 0 setdash 2 copy moveto
	2 copy vpt 270 450 arc closepath fill
	vpt 0 360 arc closepath} bind def
/C10 {BL [] 0 setdash 2 copy 2 copy moveto vpt 270 360 arc closepath fill
	2 copy moveto
	2 copy vpt 90 180 arc closepath fill
	vpt 0 360 arc closepath} bind def
/C11 {BL [] 0 setdash 2 copy moveto
	2 copy vpt 0 180 arc closepath fill
	2 copy moveto
	2 copy vpt 270 360 arc closepath fill
	vpt 0 360 arc closepath} bind def
/C12 {BL [] 0 setdash 2 copy moveto
	2 copy vpt 180 360 arc closepath fill
	vpt 0 360 arc closepath} bind def
/C13 {BL [] 0 setdash 2 copy moveto
	2 copy vpt 0 90 arc closepath fill
	2 copy moveto
	2 copy vpt 180 360 arc closepath fill
	vpt 0 360 arc closepath} bind def
/C14 {BL [] 0 setdash 2 copy moveto
	2 copy vpt 90 360 arc closepath fill
	vpt 0 360 arc} bind def
/C15 {BL [] 0 setdash 2 copy vpt 0 360 arc closepath fill
	vpt 0 360 arc closepath} bind def
/Rec {newpath 4 2 roll moveto 1 index 0 rlineto 0 exch rlineto
	neg 0 rlineto closepath} bind def
/Square {dup Rec} bind def
/Bsquare {vpt sub exch vpt sub exch vpt2 Square} bind def
/S0 {BL [] 0 setdash 2 copy moveto 0 vpt rlineto BL Bsquare} bind def
/S1 {BL [] 0 setdash 2 copy vpt Square fill Bsquare} bind def
/S2 {BL [] 0 setdash 2 copy exch vpt sub exch vpt Square fill Bsquare} bind def
/S3 {BL [] 0 setdash 2 copy exch vpt sub exch vpt2 vpt Rec fill Bsquare} bind def
/S4 {BL [] 0 setdash 2 copy exch vpt sub exch vpt sub vpt Square fill Bsquare} bind def
/S5 {BL [] 0 setdash 2 copy 2 copy vpt Square fill
	exch vpt sub exch vpt sub vpt Square fill Bsquare} bind def
/S6 {BL [] 0 setdash 2 copy exch vpt sub exch vpt sub vpt vpt2 Rec fill Bsquare} bind def
/S7 {BL [] 0 setdash 2 copy exch vpt sub exch vpt sub vpt vpt2 Rec fill
	2 copy vpt Square fill Bsquare} bind def
/S8 {BL [] 0 setdash 2 copy vpt sub vpt Square fill Bsquare} bind def
/S9 {BL [] 0 setdash 2 copy vpt sub vpt vpt2 Rec fill Bsquare} bind def
/S10 {BL [] 0 setdash 2 copy vpt sub vpt Square fill 2 copy exch vpt sub exch vpt Square fill
	Bsquare} bind def
/S11 {BL [] 0 setdash 2 copy vpt sub vpt Square fill 2 copy exch vpt sub exch vpt2 vpt Rec fill
	Bsquare} bind def
/S12 {BL [] 0 setdash 2 copy exch vpt sub exch vpt sub vpt2 vpt Rec fill Bsquare} bind def
/S13 {BL [] 0 setdash 2 copy exch vpt sub exch vpt sub vpt2 vpt Rec fill
	2 copy vpt Square fill Bsquare} bind def
/S14 {BL [] 0 setdash 2 copy exch vpt sub exch vpt sub vpt2 vpt Rec fill
	2 copy exch vpt sub exch vpt Square fill Bsquare} bind def
/S15 {BL [] 0 setdash 2 copy Bsquare fill Bsquare} bind def
/D0 {gsave translate 45 rotate 0 0 S0 stroke grestore} bind def
/D1 {gsave translate 45 rotate 0 0 S1 stroke grestore} bind def
/D2 {gsave translate 45 rotate 0 0 S2 stroke grestore} bind def
/D3 {gsave translate 45 rotate 0 0 S3 stroke grestore} bind def
/D4 {gsave translate 45 rotate 0 0 S4 stroke grestore} bind def
/D5 {gsave translate 45 rotate 0 0 S5 stroke grestore} bind def
/D6 {gsave translate 45 rotate 0 0 S6 stroke grestore} bind def
/D7 {gsave translate 45 rotate 0 0 S7 stroke grestore} bind def
/D8 {gsave translate 45 rotate 0 0 S8 stroke grestore} bind def
/D9 {gsave translate 45 rotate 0 0 S9 stroke grestore} bind def
/D10 {gsave translate 45 rotate 0 0 S10 stroke grestore} bind def
/D11 {gsave translate 45 rotate 0 0 S11 stroke grestore} bind def
/D12 {gsave translate 45 rotate 0 0 S12 stroke grestore} bind def
/D13 {gsave translate 45 rotate 0 0 S13 stroke grestore} bind def
/D14 {gsave translate 45 rotate 0 0 S14 stroke grestore} bind def
/D15 {gsave translate 45 rotate 0 0 S15 stroke grestore} bind def
/DiaE {stroke [] 0 setdash vpt add M
  hpt neg vpt neg V hpt vpt neg V
  hpt vpt V hpt neg vpt V closepath stroke} def
/BoxE {stroke [] 0 setdash exch hpt sub exch vpt add M
  0 vpt2 neg V hpt2 0 V 0 vpt2 V
  hpt2 neg 0 V closepath stroke} def
/TriUE {stroke [] 0 setdash vpt 1.12 mul add M
  hpt neg vpt -1.62 mul V
  hpt 2 mul 0 V
  hpt neg vpt 1.62 mul V closepath stroke} def
/TriDE {stroke [] 0 setdash vpt 1.12 mul sub M
  hpt neg vpt 1.62 mul V
  hpt 2 mul 0 V
  hpt neg vpt -1.62 mul V closepath stroke} def
/PentE {stroke [] 0 setdash gsave
  translate 0 hpt M 4 {72 rotate 0 hpt L} repeat
  closepath stroke grestore} def
/CircE {stroke [] 0 setdash 
  hpt 0 360 arc stroke} def
/Opaque {gsave closepath 1 setgray fill grestore 0 setgray closepath} def
/DiaW {stroke [] 0 setdash vpt add M
  hpt neg vpt neg V hpt vpt neg V
  hpt vpt V hpt neg vpt V Opaque stroke} def
/BoxW {stroke [] 0 setdash exch hpt sub exch vpt add M
  0 vpt2 neg V hpt2 0 V 0 vpt2 V
  hpt2 neg 0 V Opaque stroke} def
/TriUW {stroke [] 0 setdash vpt 1.12 mul add M
  hpt neg vpt -1.62 mul V
  hpt 2 mul 0 V
  hpt neg vpt 1.62 mul V Opaque stroke} def
/TriDW {stroke [] 0 setdash vpt 1.12 mul sub M
  hpt neg vpt 1.62 mul V
  hpt 2 mul 0 V
  hpt neg vpt -1.62 mul V Opaque stroke} def
/PentW {stroke [] 0 setdash gsave
  translate 0 hpt M 4 {72 rotate 0 hpt L} repeat
  Opaque stroke grestore} def
/CircW {stroke [] 0 setdash 
  hpt 0 360 arc Opaque stroke} def
/BoxFill {gsave Rec 1 setgray fill grestore} def
/Density {
  /Fillden exch def
  currentrgbcolor
  /ColB exch def /ColG exch def /ColR exch def
  /ColR ColR Fillden mul Fillden sub 1 add def
  /ColG ColG Fillden mul Fillden sub 1 add def
  /ColB ColB Fillden mul Fillden sub 1 add def
  ColR ColG ColB setrgbcolor} def
/BoxColFill {gsave Rec PolyFill} def
/PolyFill {gsave Density fill grestore grestore} def
/h {rlineto rlineto rlineto gsave fill grestore} bind def
%
%
/PatternFill {gsave /PFa [ 9 2 roll ] def
  PFa 0 get PFa 2 get 2 div add PFa 1 get PFa 3 get 2 div add translate
  PFa 2 get -2 div PFa 3 get -2 div PFa 2 get PFa 3 get Rec
  gsave 1 setgray fill grestore clip
  currentlinewidth 0.5 mul setlinewidth
  /PFs PFa 2 get dup mul PFa 3 get dup mul add sqrt def
  0 0 M PFa 5 get rotate PFs -2 div dup translate
  0 1 PFs PFa 4 get div 1 add floor cvi
	{PFa 4 get mul 0 M 0 PFs V} for
  0 PFa 6 get ne {
	0 1 PFs PFa 4 get div 1 add floor cvi
	{PFa 4 get mul 0 2 1 roll M PFs 0 V} for
 } if
  stroke grestore} def
/languagelevel where
 {pop languagelevel} {1} ifelse
 2 lt
	{/InterpretLevel1 true def}
	{/InterpretLevel1 Level1 def}
 ifelse
%
%
/Level2PatternFill {
/Tile8x8 {/PaintType 2 /PatternType 1 /TilingType 1 /BBox [0 0 8 8] /XStep 8 /YStep 8}
	bind def
/KeepColor {currentrgbcolor [/Pattern /DeviceRGB] setcolorspace} bind def
<< Tile8x8
 /PaintProc {0.5 setlinewidth pop 0 0 M 8 8 L 0 8 M 8 0 L stroke} 
>> matrix makepattern
/Pat1 exch def
<< Tile8x8
 /PaintProc {0.5 setlinewidth pop 0 0 M 8 8 L 0 8 M 8 0 L stroke
	0 4 M 4 8 L 8 4 L 4 0 L 0 4 L stroke}
>> matrix makepattern
/Pat2 exch def
<< Tile8x8
 /PaintProc {0.5 setlinewidth pop 0 0 M 0 8 L
	8 8 L 8 0 L 0 0 L fill}
>> matrix makepattern
/Pat3 exch def
<< Tile8x8
 /PaintProc {0.5 setlinewidth pop -4 8 M 8 -4 L
	0 12 M 12 0 L stroke}
>> matrix makepattern
/Pat4 exch def
<< Tile8x8
 /PaintProc {0.5 setlinewidth pop -4 0 M 8 12 L
	0 -4 M 12 8 L stroke}
>> matrix makepattern
/Pat5 exch def
<< Tile8x8
 /PaintProc {0.5 setlinewidth pop -2 8 M 4 -4 L
	0 12 M 8 -4 L 4 12 M 10 0 L stroke}
>> matrix makepattern
/Pat6 exch def
<< Tile8x8
 /PaintProc {0.5 setlinewidth pop -2 0 M 4 12 L
	0 -4 M 8 12 L 4 -4 M 10 8 L stroke}
>> matrix makepattern
/Pat7 exch def
<< Tile8x8
 /PaintProc {0.5 setlinewidth pop 8 -2 M -4 4 L
	12 0 M -4 8 L 12 4 M 0 10 L stroke}
>> matrix makepattern
/Pat8 exch def
<< Tile8x8
 /PaintProc {0.5 setlinewidth pop 0 -2 M 12 4 L
	-4 0 M 12 8 L -4 4 M 8 10 L stroke}
>> matrix makepattern
/Pat9 exch def
/Pattern1 {PatternBgnd KeepColor Pat1 setpattern} bind def
/Pattern2 {PatternBgnd KeepColor Pat2 setpattern} bind def
/Pattern3 {PatternBgnd KeepColor Pat3 setpattern} bind def
/Pattern4 {PatternBgnd KeepColor Landscape {Pat5} {Pat4} ifelse setpattern} bind def
/Pattern5 {PatternBgnd KeepColor Landscape {Pat4} {Pat5} ifelse setpattern} bind def
/Pattern6 {PatternBgnd KeepColor Landscape {Pat9} {Pat6} ifelse setpattern} bind def
/Pattern7 {PatternBgnd KeepColor Landscape {Pat8} {Pat7} ifelse setpattern} bind def
} def
%
%
%
/PatternBgnd {
  TransparentPatterns {} {gsave 1 setgray fill grestore} ifelse
} def
%
%
/Level1PatternFill {
/Pattern1 {0.250 Density} bind def
/Pattern2 {0.500 Density} bind def
/Pattern3 {0.750 Density} bind def
/Pattern4 {0.125 Density} bind def
/Pattern5 {0.375 Density} bind def
/Pattern6 {0.625 Density} bind def
/Pattern7 {0.875 Density} bind def
} def
%
%
Level1 {Level1PatternFill} {Level2PatternFill} ifelse
/Symbol-Oblique /Symbol findfont [1 0 .167 1 0 0] makefont
dup length dict begin {1 index /FID eq {pop pop} {def} ifelse} forall
currentdict end definefont pop
end
gnudict begin
gsave
0 0 translate
0.050 0.050 scale
0 setgray
newpath
0.500 UL
LTb
2234 600 M
126 0 V
4981 0 R
-126 0 V
2234 890 M
63 0 V
5044 0 R
-63 0 V
2234 1273 M
63 0 V
5044 0 R
-63 0 V
2234 1470 M
63 0 V
5044 0 R
-63 0 V
-5044 93 R
126 0 V
4981 0 R
-126 0 V
2234 1853 M
63 0 V
5044 0 R
-63 0 V
2234 2237 M
63 0 V
5044 0 R
-63 0 V
2234 2433 M
63 0 V
5044 0 R
-63 0 V
-5044 94 R
126 0 V
4981 0 R
-126 0 V
2234 2817 M
63 0 V
5044 0 R
-63 0 V
2234 3200 M
63 0 V
5044 0 R
-63 0 V
2234 3397 M
63 0 V
5044 0 R
-63 0 V
-5044 93 R
126 0 V
4981 0 R
-126 0 V
2234 3780 M
63 0 V
5044 0 R
-63 0 V
2234 4164 M
63 0 V
5044 0 R
-63 0 V
2234 4360 M
63 0 V
5044 0 R
-63 0 V
-5044 94 R
126 0 V
4981 0 R
-126 0 V
2234 4744 M
63 0 V
5044 0 R
-63 0 V
2234 5127 M
63 0 V
5044 0 R
-63 0 V
2234 5324 M
63 0 V
5044 0 R
-63 0 V
-5044 93 R
126 0 V
4981 0 R
-126 0 V
2234 5707 M
63 0 V
5044 0 R
-63 0 V
2234 600 M
0 126 V
0 4981 R
0 -126 V
2404 600 M
0 63 V
0 5044 R
0 -63 V
2574 600 M
0 63 V
0 5044 R
0 -63 V
2745 600 M
0 63 V
0 5044 R
0 -63 V
2915 600 M
2915 663 L
0 5044 R
0 -63 V
3085 600 M
0 126 V
0 4981 R
0 -126 V
3255 600 M
0 63 V
0 5044 R
0 -63 V
3426 600 M
0 63 V
0 5044 R
0 -63 V
3596 600 M
0 63 V
0 5044 R
0 -63 V
3766 600 M
0 63 V
0 5044 R
0 -63 V
3936 600 M
0 126 V
0 4981 R
0 -126 V
4107 600 M
0 63 V
0 5044 R
0 -63 V
4277 600 M
0 63 V
0 5044 R
0 -63 V
4447 600 M
0 63 V
0 5044 R
0 -63 V
4617 600 M
0 63 V
0 5044 R
0 -63 V
4788 600 M
0 126 V
0 4981 R
0 -126 V
4958 600 M
0 63 V
0 5044 R
0 -63 V
5128 600 M
0 63 V
0 5044 R
0 -63 V
5298 600 M
0 63 V
0 5044 R
0 -63 V
5468 600 M
0 63 V
0 5044 R
0 -63 V
5639 600 M
0 126 V
0 4981 R
0 -126 V
5809 600 M
0 63 V
0 5044 R
0 -63 V
5979 600 M
0 63 V
0 5044 R
0 -63 V
6149 600 M
0 63 V
0 5044 R
0 -63 V
6320 600 M
0 63 V
0 5044 R
0 -63 V
6490 600 M
0 126 V
0 4981 R
0 -126 V
6660 600 M
0 63 V
0 5044 R
0 -63 V
6830 600 M
0 63 V
0 5044 R
0 -63 V
7001 600 M
0 63 V
0 5044 R
0 -63 V
7171 600 M
0 63 V
0 5044 R
0 -63 V
7341 600 M
0 126 V
0 4981 R
7341 5581 L
2234 5707 M
0 -5107 V
5107 0 V
0 5107 V
-5107 0 V
stroke
LCb setrgbcolor
LTb
LCb setrgbcolor
LTb
2.000 UP
0.500 UL
LTb
1.000 UL
LT0
7341 5127 M
-16 -2 V
-19 -2 V
-21 -3 V
-24 -4 V
-26 -3 V
-28 -4 V
-30 -5 V
-32 -5 V
-34 -5 V
-36 -5 V
-37 -6 V
-39 -6 V
-41 -6 V
-41 -7 V
-43 -6 V
-45 -7 V
-45 -7 V
-46 -8 V
-48 -7 V
-48 -8 V
-49 -8 V
-49 -8 V
-50 -8 V
-51 -8 V
-52 -8 V
-51 -8 V
-52 -9 V
-52 -8 V
-53 -9 V
-52 -9 V
-53 -8 V
-53 -9 V
-52 -8 V
-53 -9 V
-52 -9 V
-53 -8 V
-52 -9 V
-51 -9 V
-51 -8 V
-51 -9 V
-51 -8 V
-50 -8 V
-49 -9 V
-49 -8 V
-49 -8 V
-47 -8 V
-47 -8 V
-47 -8 V
-45 -8 V
-45 -8 V
-44 -7 V
-44 -8 V
-42 -7 V
-42 -7 V
-41 -8 V
-40 -7 V
-39 -7 V
-38 -6 V
-38 -7 V
-36 -7 V
-36 -6 V
-34 -6 V
-34 -6 V
-33 -6 V
-32 -6 V
-31 -6 V
-30 -6 V
-29 -5 V
-28 -6 V
-27 -5 V
-27 -5 V
-25 -5 V
-25 -5 V
-24 -5 V
-23 -5 V
-23 -5 V
-21 -4 V
-21 -5 V
-20 -4 V
-20 -4 V
-19 -4 V
-18 -5 V
-17 -4 V
-18 -4 V
-16 -4 V
-16 -4 V
-16 -4 V
-15 -3 V
-15 -4 V
-15 -4 V
-14 -4 V
-14 -4 V
-14 -4 V
-14 -4 V
-14 -4 V
-14 -3 V
-14 -4 V
-14 -5 V
-15 -4 V
-21 -6 V
-21 -6 V
-22 -7 V
-23 -7 V
-23 -8 V
3758 4476 L
-23 -8 V
-24 -9 V
-24 -9 V
-24 -10 V
-24 -9 V
-24 -10 V
-24 -10 V
-24 -10 V
-24 -10 V
-24 -10 V
-23 -10 V
-23 -9 V
-23 -10 V
-22 -10 V
-22 -9 V
-22 -10 V
-21 -9 V
-21 -10 V
-20 -9 V
-20 -9 V
-20 -8 V
-19 -9 V
-19 -8 V
-18 -8 V
-17 -8 V
-18 -8 V
-16 -8 V
-16 -7 V
-16 -7 V
-15 -7 V
-15 -7 V
-14 -7 V
-14 -6 V
-14 -6 V
-13 -6 V
-12 -6 V
-12 -6 V
-11 -5 V
-12 -5 V
-10 -5 V
-11 -5 V
-9 -5 V
-10 -5 V
-9 -4 V
-9 -4 V
-8 -5 V
-8 -4 V
-8 -3 V
-8 -4 V
-7 -4 V
-7 -3 V
-6 -4 V
-6 -3 V
-6 -3 V
-6 -4 V
-6 -3 V
-5 -3 V
-5 -2 V
-5 -3 V
-5 -3 V
-4 -3 V
-5 -2 V
-4 -3 V
-4 -3 V
-4 -2 V
-4 -3 V
-4 -2 V
-3 -2 V
-4 -3 V
-3 -2 V
-4 -3 V
-3 -2 V
-3 -3 V
-3 -2 V
-3 -3 V
-3 -2 V
-3 -3 V
-3 -3 V
-3 -2 V
-3 -3 V
-3 -3 V
-3 -3 V
-3 -3 V
-2 -3 V
-3 -4 V
-3 -3 V
-3 -4 V
-3 -4 V
-2 -4 V
-3 -4 V
-3 -5 V
-3 -5 V
-3 -5 V
-3 -166 V
-2 -157 V
-3 -147 V
-2 -139 V
-3 -129 V
-2 -122 V
-2 -115 V
-2 -107 V
-1 -100 V
-2 -94 V
-2 -87 V
2725 2529 L
-1 -77 V
-2 -71 V
-1 -67 V
-1 -62 V
-1 -58 V
-1 -54 V
-1 -50 V
-2 -46 V
-1 -44 V
-1 -40 V
-1 -38 V
-1 -34 V
-1 -33 V
-1 -30 V
-1 -28 V
-1 -27 V
-2 -24 V
-1 -23 V
-1 -22 V
-2 -20 V
-1 -19 V
-2 -18 V
-2 -17 V
-2 -16 V
-2 -16 V
-2 -15 V
-3 -14 V
-2 -14 V
-3 -13 V
-3 -13 V
-3 -13 V
-3 -13 V
-3 -12 V
-4 -13 V
-3 -13 V
-4 -12 V
-4 -13 V
-4 -12 V
-5 -13 V
-4 -13 V
-5 -13 V
-5 -14 V
-5 -13 V
-6 -14 V
-5 -14 V
-6 -14 V
-6 -15 V
-6 -14 V
-6 -15 V
-6 -15 V
-7 -16 V
-6 -15 V
-7 -16 V
-7 -16 V
-7 -16 V
-7 -16 V
-8 -17 V
-7 -16 V
-8 -17 V
-7 -16 V
-8 -17 V
-8 -17 V
-8 -17 V
-8 -17 V
-8 -17 V
-9 -16 V
-8 -17 V
-9 -17 V
-8 -17 V
-9 -17 V
-9 -17 V
-9 -17 V
-9 -17 V
-10 -17 V
-10 -17 V
-10 -18 V
-10 -17 V
-11 -18 V
-11 -18 V
-11 -19 V
-12 -19 V
-13 -20 V
-13 -21 V
-14 -22 V
-14 -22 V
-16 -24 V
-16 -25 V
2.000 UP
stroke
LT1
LTb
LT1
6973 5027 Pls
6973 5098 Pls
6973 5270 Pls
6973 5155 Pls
6973 5045 Pls
6973 4988 Pls
6973 5189 Pls
6973 4896 Pls
6973 5288 Pls
6973 5249 Pls
6973 5156 Pls
6973 4963 Pls
6973 4992 Pls
6973 4949 Pls
6973 5138 Pls
6973 5079 Pls
6973 4776 Pls
6973 4992 Pls
6973 5198 Pls
6973 5173 Pls
6973 5144 Pls
6973 5172 Pls
6973 5239 Pls
6973 5174 Pls
6973 5064 Pls
6973 5177 Pls
6973 5032 Pls
6973 5219 Pls
6973 5278 Pls
6973 5028 Pls
6973 5025 Pls
6973 5088 Pls
6973 5348 Pls
6973 5265 Pls
6973 4861 Pls
6973 5389 Pls
6973 5038 Pls
6973 5288 Pls
6973 5177 Pls
6973 5179 Pls
6973 5198 Pls
6973 5366 Pls
6973 4981 Pls
6973 5038 Pls
6973 4977 Pls
6973 5009 Pls
6973 5041 Pls
6973 4840 Pls
6973 5134 Pls
6973 5155 Pls
6973 5238 Pls
6973 5034 Pls
6973 5096 Pls
6973 5012 Pls
6973 5020 Pls
6973 4942 Pls
6973 4973 Pls
6973 5159 Pls
6973 4969 Pls
6973 5048 Pls
6973 5228 Pls
6973 5037 Pls
6973 5176 Pls
6973 5209 Pls
6973 4793 Pls
6973 5166 Pls
6973 5338 Pls
6973 5200 Pls
6973 5141 Pls
6973 5156 Pls
6973 5150 Pls
6973 5100 Pls
6973 4995 Pls
6973 5026 Pls
6973 4962 Pls
6973 5226 Pls
6973 4892 Pls
6973 4971 Pls
6973 5088 Pls
6973 4808 Pls
6973 5024 Pls
6973 5425 Pls
6973 5002 Pls
6973 4822 Pls
6973 4850 Pls
6973 4729 Pls
6973 5191 Pls
6973 5309 Pls
6973 5281 Pls
6973 5032 Pls
6973 4942 Pls
6973 5102 Pls
6973 4834 Pls
6973 4733 Pls
6973 5318 Pls
6973 4793 Pls
6973 5068 Pls
6973 5121 Pls
6798 1363 Pls
2.000 UP
1.000 UL
LT2
LTb
LT2
4386 4655 TriUF
4251 4623 TriUF
3892 4519 TriUF
3552 4429 TriUF
2968 4111 TriUF
2902 4134 TriUF
2836 4044 TriUF
2798 4062 TriUF
2750 3974 TriUF
6798 1163 TriUF
2.000 UP
1.000 UL
LT3
LTb
LT3
2708 1546 BoxF
6798 963 BoxF
2.000 UP
1.000 UL
LT4
LTb
LT4
2471 966 CircleF
6798 763 CircleF
0.500 UL
LTb
2234 5707 M
0 -5107 V
5107 0 V
0 5107 V
-5107 0 V
2.000 UP
stroke
grestore
end
showpage
  }}%
  \put(6375,763){\makebox(0,0)[r]{\strut{}Dwarf Galaxies}}%
  \put(6375,963){\makebox(0,0)[r]{\strut{}Galaxies}}%
  \put(6375,1163){\makebox(0,0)[r]{\strut{}Dwarf Clusters}}%
  \put(6375,1363){\makebox(0,0)[r]{\strut{}Clusters of Galaxies}}%
  \put(4787,100){\makebox(0,0){\strut{}$r_{0}\ [\mbox{kpc}]$}}%
  \put(1054,3153){%
  \special{ps: gsave currentpoint currentpoint translate
270 rotate neg exch neg exch translate}%
  \makebox(0,0){\strut{}$M_{0}\ [M_{\odot}]$}%
  \special{ps: currentpoint grestore moveto}%
  }%
  \put(7341,400){\makebox(0,0){\strut{} 150}}%
  \put(6490,400){\makebox(0,0){\strut{} 125}}%
  \put(5639,400){\makebox(0,0){\strut{} 100}}%
  \put(4788,400){\makebox(0,0){\strut{} 75}}%
  \put(3936,400){\makebox(0,0){\strut{} 50}}%
  \put(3085,400){\makebox(0,0){\strut{} 25}}%
  \put(2234,400){\makebox(0,0){\strut{} 0}}%
  \put(2114,5417){\makebox(0,0)[r]{\strut{}$10^{16}$}}%
  \put(2114,4454){\makebox(0,0)[r]{\strut{}$10^{15}$}}%
  \put(2114,3490){\makebox(0,0)[r]{\strut{}$10^{14}$}}%
  \put(2114,2527){\makebox(0,0)[r]{\strut{}$10^{13}$}}%
  \put(2114,1563){\makebox(0,0)[r]{\strut{}$10^{12}$}}%
  \put(2114,600){\makebox(0,0)[r]{\strut{}$10^{11}$}}%
\end{picture}%
\endgroup
 

%% file: figure/gnuplot/SigmaFit.tex
\begingroup%
  \makeatletter%
  \newcommand{\GNUPLOTspecial}{%
    \@sanitize\catcode`\%=14\relax\special}%
  \setlength{\unitlength}{0.1bp}%
\begin{picture}(4050,3780)(0,0)%
{\GNUPLOTspecial{"
/gnudict 256 dict def
gnudict begin
/Color true def
/Solid false def
/gnulinewidth 5.000 def
/userlinewidth gnulinewidth def
/vshift -33 def
/dl {10.0 mul} def
/hpt_ 31.5 def
/vpt_ 31.5 def
/hpt hpt_ def
/vpt vpt_ def
/Rounded false def
/M {moveto} bind def
/L {lineto} bind def
/R {rmoveto} bind def
/V {rlineto} bind def
/N {newpath moveto} bind def
/C {setrgbcolor} bind def
/f {rlineto fill} bind def
/vpt2 vpt 2 mul def
/hpt2 hpt 2 mul def
/Lshow { currentpoint stroke M
  0 vshift R show } def
/Rshow { currentpoint stroke M
  dup stringwidth pop neg vshift R show } def
/Cshow { currentpoint stroke M
  dup stringwidth pop -2 div vshift R show } def
/UP { dup vpt_ mul /vpt exch def hpt_ mul /hpt exch def
  /hpt2 hpt 2 mul def /vpt2 vpt 2 mul def } def
/DL { Color {setrgbcolor Solid {pop []} if 0 setdash }
 {pop pop pop 0 setgray Solid {pop []} if 0 setdash} ifelse } def
/BL { stroke userlinewidth 2 mul setlinewidth
      Rounded { 1 setlinejoin 1 setlinecap } if } def
/AL { stroke userlinewidth 2 div setlinewidth
      Rounded { 1 setlinejoin 1 setlinecap } if } def
/UL { dup gnulinewidth mul /userlinewidth exch def
      dup 1 lt {pop 1} if 10 mul /udl exch def } def
/PL { stroke userlinewidth setlinewidth
      Rounded { 1 setlinejoin 1 setlinecap } if } def
/LTw { PL [] 1 setgray } def
/LTb { BL [] 0 0 0 DL } def
/LTa { AL [1 udl mul 2 udl mul] 0 setdash 0 0 0 setrgbcolor } def
/fatlinewidth 7.500 def
/gatlinewidth 10.000 def
/FL { stroke fatlinewidth setlinewidth Rounded { 1 setlinejoin 1 setlinecap } if } def
/GL { stroke gatlinewidth setlinewidth Rounded { 1 setlinejoin 1 setlinecap } if } def
/LT0 { FL [] 1 0 0 DL } def
/LT1 { GL [2 dl 3 dl] 0 0 1 DL } def/LT2 { PL [2 dl 3 dl] 0 0 1 DL } def
/LT3 { PL [1 dl 1.5 dl] 1 0 1 DL } def
/LT4 { PL [5 dl 2 dl 1 dl 2 dl] 0 1 1 DL } def
/LT5 { PL [4 dl 3 dl 1 dl 3 dl] 1 1 0 DL } def
/LT6 { PL [2 dl 2 dl 2 dl 4 dl] 0 0 0 DL } def
/LT7 { PL [2 dl 2 dl 2 dl 2 dl 2 dl 4 dl] 1 0.3 0 DL } def
/LT8 { PL [2 dl 2 dl 2 dl 2 dl 2 dl 2 dl 2 dl 4 dl] 0.5 0.5 0.5 DL } def
/Pnt { stroke [] 0 setdash
   gsave 1 setlinecap M 0 0 V stroke grestore } def
/Dia { stroke [] 0 setdash 2 copy vpt add M
  hpt neg vpt neg V hpt vpt neg V
  hpt vpt V hpt neg vpt V closepath stroke
  Pnt } def
/Pls { stroke [] 0 setdash vpt sub M 0 vpt2 V
  currentpoint stroke M
  hpt neg vpt neg R hpt2 0 V stroke
  } def
/Box { stroke [] 0 setdash 2 copy exch hpt sub exch vpt add M
  0 vpt2 neg V hpt2 0 V 0 vpt2 V
  hpt2 neg 0 V closepath stroke
  Pnt } def
/Crs { stroke [] 0 setdash exch hpt sub exch vpt add M
  hpt2 vpt2 neg V currentpoint stroke M
  hpt2 neg 0 R hpt2 vpt2 V stroke } def
/TriU { stroke [] 0 setdash 2 copy vpt 1.12 mul add M
  hpt neg vpt -1.62 mul V
  hpt 2 mul 0 V
  hpt neg vpt 1.62 mul V closepath stroke
  Pnt  } def
/Star { 2 copy Pls Crs } def
/BoxF { stroke [] 0 setdash exch hpt sub exch vpt add M
  0 vpt2 neg V  hpt2 0 V  0 vpt2 V
  hpt2 neg 0 V  closepath fill } def
/TriUF { stroke [] 0 setdash vpt 1.12 mul add M
  hpt neg vpt -1.62 mul V
  hpt 2 mul 0 V
  hpt neg vpt 1.62 mul V closepath fill } def
/TriD { stroke [] 0 setdash 2 copy vpt 1.12 mul sub M
  hpt neg vpt 1.62 mul V
  hpt 2 mul 0 V
  hpt neg vpt -1.62 mul V closepath stroke
  Pnt  } def
/TriDF { stroke [] 0 setdash vpt 1.12 mul sub M
  hpt neg vpt 1.62 mul V
  hpt 2 mul 0 V
  hpt neg vpt -1.62 mul V closepath fill} def
/DiaF { stroke [] 0 setdash vpt add M
  hpt neg vpt neg V hpt vpt neg V
  hpt vpt V hpt neg vpt V closepath fill } def
/Pent { stroke [] 0 setdash 2 copy gsave
  translate 0 hpt M 4 {72 rotate 0 hpt L} repeat
  closepath stroke grestore Pnt } def
/PentF { stroke [] 0 setdash gsave
  translate 0 hpt M 4 {72 rotate 0 hpt L} repeat
  closepath fill grestore } def
/Circle { stroke [] 0 setdash 2 copy
  hpt 0 360 arc stroke Pnt } def
/CircleF { stroke [] 0 setdash hpt 0 360 arc fill } def
/C0 { BL [] 0 setdash 2 copy moveto vpt 90 450  arc } bind def
/C1 { BL [] 0 setdash 2 copy        moveto
       2 copy  vpt 0 90 arc closepath fill
               vpt 0 360 arc closepath } bind def
/C2 { BL [] 0 setdash 2 copy moveto
       2 copy  vpt 90 180 arc closepath fill
               vpt 0 360 arc closepath } bind def
/C3 { BL [] 0 setdash 2 copy moveto
       2 copy  vpt 0 180 arc closepath fill
               vpt 0 360 arc closepath } bind def
/C4 { BL [] 0 setdash 2 copy moveto
       2 copy  vpt 180 270 arc closepath fill
               vpt 0 360 arc closepath } bind def
/C5 { BL [] 0 setdash 2 copy moveto
       2 copy  vpt 0 90 arc
       2 copy moveto
       2 copy  vpt 180 270 arc closepath fill
               vpt 0 360 arc } bind def
/C6 { BL [] 0 setdash 2 copy moveto
      2 copy  vpt 90 270 arc closepath fill
              vpt 0 360 arc closepath } bind def
/C7 { BL [] 0 setdash 2 copy moveto
      2 copy  vpt 0 270 arc closepath fill
              vpt 0 360 arc closepath } bind def
/C8 { BL [] 0 setdash 2 copy moveto
      2 copy vpt 270 360 arc closepath fill
              vpt 0 360 arc closepath } bind def
/C9 { BL [] 0 setdash 2 copy moveto
      2 copy  vpt 270 450 arc closepath fill
              vpt 0 360 arc closepath } bind def
/C10 { BL [] 0 setdash 2 copy 2 copy moveto vpt 270 360 arc closepath fill
       2 copy moveto
       2 copy vpt 90 180 arc closepath fill
               vpt 0 360 arc closepath } bind def
/C11 { BL [] 0 setdash 2 copy moveto
       2 copy  vpt 0 180 arc closepath fill
       2 copy moveto
       2 copy  vpt 270 360 arc closepath fill
               vpt 0 360 arc closepath } bind def
/C12 { BL [] 0 setdash 2 copy moveto
       2 copy  vpt 180 360 arc closepath fill
               vpt 0 360 arc closepath } bind def
/C13 { BL [] 0 setdash  2 copy moveto
       2 copy  vpt 0 90 arc closepath fill
       2 copy moveto
       2 copy  vpt 180 360 arc closepath fill
               vpt 0 360 arc closepath } bind def
/C14 { BL [] 0 setdash 2 copy moveto
       2 copy  vpt 90 360 arc closepath fill
               vpt 0 360 arc } bind def
/C15 { BL [] 0 setdash 2 copy vpt 0 360 arc closepath fill
               vpt 0 360 arc closepath } bind def
/Rec   { newpath 4 2 roll moveto 1 index 0 rlineto 0 exch rlineto
       neg 0 rlineto closepath } bind def
/Square { dup Rec } bind def
/Bsquare { vpt sub exch vpt sub exch vpt2 Square } bind def
/S0 { BL [] 0 setdash 2 copy moveto 0 vpt rlineto BL Bsquare } bind def
/S1 { BL [] 0 setdash 2 copy vpt Square fill Bsquare } bind def
/S2 { BL [] 0 setdash 2 copy exch vpt sub exch vpt Square fill Bsquare } bind def
/S3 { BL [] 0 setdash 2 copy exch vpt sub exch vpt2 vpt Rec fill Bsquare } bind def
/S4 { BL [] 0 setdash 2 copy exch vpt sub exch vpt sub vpt Square fill Bsquare } bind def
/S5 { BL [] 0 setdash 2 copy 2 copy vpt Square fill
       exch vpt sub exch vpt sub vpt Square fill Bsquare } bind def
/S6 { BL [] 0 setdash 2 copy exch vpt sub exch vpt sub vpt vpt2 Rec fill Bsquare } bind def
/S7 { BL [] 0 setdash 2 copy exch vpt sub exch vpt sub vpt vpt2 Rec fill
       2 copy vpt Square fill
       Bsquare } bind def
/S8 { BL [] 0 setdash 2 copy vpt sub vpt Square fill Bsquare } bind def
/S9 { BL [] 0 setdash 2 copy vpt sub vpt vpt2 Rec fill Bsquare } bind def
/S10 { BL [] 0 setdash 2 copy vpt sub vpt Square fill 2 copy exch vpt sub exch vpt Square fill
       Bsquare } bind def
/S11 { BL [] 0 setdash 2 copy vpt sub vpt Square fill 2 copy exch vpt sub exch vpt2 vpt Rec fill
       Bsquare } bind def
/S12 { BL [] 0 setdash 2 copy exch vpt sub exch vpt sub vpt2 vpt Rec fill Bsquare } bind def
/S13 { BL [] 0 setdash 2 copy exch vpt sub exch vpt sub vpt2 vpt Rec fill
       2 copy vpt Square fill Bsquare } bind def
/S14 { BL [] 0 setdash 2 copy exch vpt sub exch vpt sub vpt2 vpt Rec fill
       2 copy exch vpt sub exch vpt Square fill Bsquare } bind def
/S15 { BL [] 0 setdash 2 copy Bsquare fill Bsquare } bind def
/D0 { gsave translate 45 rotate 0 0 S0 stroke grestore } bind def
/D1 { gsave translate 45 rotate 0 0 S1 stroke grestore } bind def
/D2 { gsave translate 45 rotate 0 0 S2 stroke grestore } bind def
/D3 { gsave translate 45 rotate 0 0 S3 stroke grestore } bind def
/D4 { gsave translate 45 rotate 0 0 S4 stroke grestore } bind def
/D5 { gsave translate 45 rotate 0 0 S5 stroke grestore } bind def
/D6 { gsave translate 45 rotate 0 0 S6 stroke grestore } bind def
/D7 { gsave translate 45 rotate 0 0 S7 stroke grestore } bind def
/D8 { gsave translate 45 rotate 0 0 S8 stroke grestore } bind def
/D9 { gsave translate 45 rotate 0 0 S9 stroke grestore } bind def
/D10 { gsave translate 45 rotate 0 0 S10 stroke grestore } bind def
/D11 { gsave translate 45 rotate 0 0 S11 stroke grestore } bind def
/D12 { gsave translate 45 rotate 0 0 S12 stroke grestore } bind def
/D13 { gsave translate 45 rotate 0 0 S13 stroke grestore } bind def
/D14 { gsave translate 45 rotate 0 0 S14 stroke grestore } bind def
/D15 { gsave translate 45 rotate 0 0 S15 stroke grestore } bind def
/DiaE { stroke [] 0 setdash vpt add M
  hpt neg vpt neg V hpt vpt neg V
  hpt vpt V hpt neg vpt V closepath stroke } def
/BoxE { stroke [] 0 setdash exch hpt sub exch vpt add M
  0 vpt2 neg V hpt2 0 V 0 vpt2 V
  hpt2 neg 0 V closepath stroke } def
/TriUE { stroke [] 0 setdash vpt 1.12 mul add M
  hpt neg vpt -1.62 mul V
  hpt 2 mul 0 V
  hpt neg vpt 1.62 mul V closepath stroke } def
/TriDE { stroke [] 0 setdash vpt 1.12 mul sub M
  hpt neg vpt 1.62 mul V
  hpt 2 mul 0 V
  hpt neg vpt -1.62 mul V closepath stroke } def
/PentE { stroke [] 0 setdash gsave
  translate 0 hpt M 4 {72 rotate 0 hpt L} repeat
  closepath stroke grestore } def
/CircE { stroke [] 0 setdash 
  hpt 0 360 arc stroke } def
/Opaque { gsave closepath 1 setgray fill grestore 0 setgray closepath } def
/DiaW { stroke [] 0 setdash vpt add M
  hpt neg vpt neg V hpt vpt neg V
  hpt vpt V hpt neg vpt V Opaque stroke } def
/BoxW { stroke [] 0 setdash exch hpt sub exch vpt add M
  0 vpt2 neg V hpt2 0 V 0 vpt2 V
  hpt2 neg 0 V Opaque stroke } def
/TriUW { stroke [] 0 setdash vpt 1.12 mul add M
  hpt neg vpt -1.62 mul V
  hpt 2 mul 0 V
  hpt neg vpt 1.62 mul V Opaque stroke } def
/TriDW { stroke [] 0 setdash vpt 1.12 mul sub M
  hpt neg vpt 1.62 mul V
  hpt 2 mul 0 V
  hpt neg vpt -1.62 mul V Opaque stroke } def
/PentW { stroke [] 0 setdash gsave
  translate 0 hpt M 4 {72 rotate 0 hpt L} repeat
  Opaque stroke grestore } def
/CircW { stroke [] 0 setdash 
  hpt 0 360 arc Opaque stroke } def
/BoxFill { gsave Rec 1 setgray fill grestore } def
/BoxColFill {
  gsave Rec
  /Fillden exch def
  currentrgbcolor
  /ColB exch def /ColG exch def /ColR exch def
  /ColR ColR Fillden mul Fillden sub 1 add def
  /ColG ColG Fillden mul Fillden sub 1 add def
  /ColB ColB Fillden mul Fillden sub 1 add def
  ColR ColG ColB setrgbcolor
  fill grestore } def
%
%
/PatternFill { gsave /PFa [ 9 2 roll ] def
    PFa 0 get PFa 2 get 2 div add PFa 1 get PFa 3 get 2 div add translate
    PFa 2 get -2 div PFa 3 get -2 div PFa 2 get PFa 3 get Rec
    gsave 1 setgray fill grestore clip
    currentlinewidth 0.5 mul setlinewidth
    /PFs PFa 2 get dup mul PFa 3 get dup mul add sqrt def
    0 0 M PFa 5 get rotate PFs -2 div dup translate
	0 1 PFs PFa 4 get div 1 add floor cvi
	{ PFa 4 get mul 0 M 0 PFs V } for
    0 PFa 6 get ne {
	0 1 PFs PFa 4 get div 1 add floor cvi
	{ PFa 4 get mul 0 2 1 roll M PFs 0 V } for
    } if
    stroke grestore } def
/Symbol-Oblique /Symbol findfont [1 0 .167 1 0 0] makefont
dup length dict begin {1 index /FID eq {pop pop} {def} ifelse} forall
currentdict end definefont pop
end
gnudict begin
gsave
0 0 translate
0.100 0.100 scale
0 setgray
newpath
0.500 UL
LTb
500 300 M
63 0 V
3317 0 R
-63 0 V
0.500 UL
LTb
500 469 M
31 0 V
3349 0 R
-31 0 V
500 638 M
31 0 V
3349 0 R
-31 0 V
500 807 M
31 0 V
3349 0 R
-31 0 V
500 976 M
63 0 V
3317 0 R
-63 0 V
0.500 UL
LTb
500 1145 M
31 0 V
3349 0 R
-31 0 V
500 1314 M
31 0 V
3349 0 R
-31 0 V
500 1483 M
31 0 V
3349 0 R
-31 0 V
500 1652 M
63 0 V
3317 0 R
-63 0 V
0.500 UL
LTb
500 1821 M
31 0 V
3349 0 R
-31 0 V
500 1990 M
31 0 V
3349 0 R
-31 0 V
500 2159 M
31 0 V
3349 0 R
-31 0 V
500 2328 M
63 0 V
3317 0 R
-63 0 V
0.500 UL
LTb
500 2497 M
31 0 V
3349 0 R
-31 0 V
500 2666 M
31 0 V
3349 0 R
-31 0 V
500 2835 M
31 0 V
3349 0 R
-31 0 V
500 3004 M
63 0 V
3317 0 R
-63 0 V
0.500 UL
LTb
500 3173 M
31 0 V
3349 0 R
-31 0 V
500 3342 M
31 0 V
3349 0 R
-31 0 V
500 3511 M
31 0 V
3349 0 R
-31 0 V
500 3680 M
63 0 V
3317 0 R
-63 0 V
0.500 UL
LTb
500 300 M
0 63 V
0 3317 R
0 -63 V
0.500 UL
LTb
669 300 M
0 31 V
0 3349 R
0 -31 V
838 300 M
0 31 V
0 3349 R
0 -31 V
1007 300 M
0 31 V
0 3349 R
0 -31 V
1176 300 M
0 31 V
0 3349 R
0 -31 V
1345 300 M
0 63 V
0 3317 R
0 -63 V
0.500 UL
LTb
1514 300 M
0 31 V
0 3349 R
0 -31 V
1683 300 M
0 31 V
0 3349 R
0 -31 V
1852 300 M
0 31 V
0 3349 R
0 -31 V
2021 300 M
0 31 V
0 3349 R
0 -31 V
2190 300 M
0 63 V
0 3317 R
0 -63 V
0.500 UL
LTb
2359 300 M
0 31 V
0 3349 R
0 -31 V
2528 300 M
0 31 V
0 3349 R
0 -31 V
2697 300 M
0 31 V
0 3349 R
0 -31 V
2866 300 M
0 31 V
0 3349 R
0 -31 V
3035 300 M
0 63 V
0 3317 R
0 -63 V
0.500 UL
LTb
3204 300 M
0 31 V
0 3349 R
0 -31 V
3373 300 M
0 31 V
0 3349 R
0 -31 V
3542 300 M
0 31 V
0 3349 R
0 -31 V
3711 300 M
0 31 V
0 3349 R
0 -31 V
3880 300 M
0 63 V
0 3317 R
0 -63 V
0.500 UL
LTb
0.500 UL
LTb
500 300 M
3380 0 V
0 3380 V
-3380 0 V
500 300 L
LTb
LTb
1.000 UP
1.000 UL
LT0
918 811 M
14 10 V
14 11 V
15 12 V
14 11 V
15 12 V
14 12 V
15 12 V
15 12 V
14 13 V
15 13 V
14 13 V
14 14 V
15 13 V
14 14 V
14 14 V
15 15 V
14 14 V
14 15 V
15 16 V
15 16 V
15 15 V
14 17 V
14 16 V
15 17 V
14 17 V
15 17 V
14 18 V
14 18 V
15 19 V
14 18 V
14 20 V
16 20 V
14 20 V
15 21 V
14 21 V
14 21 V
15 23 V
14 22 V
14 23 V
15 24 V
14 24 V
15 23 V
14 25 V
14 26 V
16 25 V
14 27 V
15 27 V
14 27 V
14 28 V
15 28 V
14 28 V
14 29 V
15 29 V
14 30 V
14 30 V
15 30 V
15 30 V
15 31 V
14 30 V
14 32 V
15 31 V
14 33 V
15 32 V
14 32 V
14 32 V
15 33 V
14 32 V
14 32 V
15 31 V
15 29 V
15 32 V
14 29 V
14 32 V
15 30 V
14 32 V
14 30 V
15 31 V
14 33 V
14 29 V
15 32 V
14 25 V
16 30 V
14 27 V
14 24 V
15 28 V
14 27 V
15 18 V
14 3 V
14 -2 V
15 -7 V
14 -11 V
15 -13 V
14 -14 V
14 -18 V
16 -21 V
14 -20 V
15 -24 V
14 -26 V
14 -27 V
15 -28 V
14 -27 V
14 -33 V
15 -38 V
14 -25 V
stroke
2421 2494 M
14 -24 V
15 -30 V
15 -11 V
15 -27 V
14 -15 V
14 -28 V
15 -21 V
14 6 V
14 0 V
15 11 V
14 11 V
15 18 V
14 15 V
14 16 V
15 27 V
15 39 V
15 25 V
14 30 V
14 11 V
15 -2 V
14 20 V
14 34 V
15 38 V
14 8 V
14 -26 V
15 -81 V
14 -50 V
16 40 V
14 -168 V
14 -352 V
15 -38 V
14 -33 V
15 -37 V
14 -36 V
14 -36 V
15 -37 V
14 -36 V
14 -36 V
15 -36 V
14 -36 V
16 -28 V
14 -36 V
14 -36 V
15 -37 V
14 -36 V
14 -37 V
15 -45 V
14 -38 V
15 -38 V
14 -38 V
14 -51 V
15 -51 V
15 -37 V
15 -65 V
14 -96 V
14 -7 V
15 -7 V
14 -7 V
14 -7 V
15 -6 V
14 -7 V
14 -6 V
15 -7 V
14 -6 V
15 -6 V
15 -7 V
14 -6 V
15 -5 V
14 -6 V
15 -6 V
14 -6 V
14 -5 V
15 -6 V
14 -5 V
14 -6 V
15 -5 V
14 -5 V
16 -6 V
14 -5 V
14 -5 V
1.000 UL
LT1
500 694 M
7 2 V
7 2 V
6 3 V
7 2 V
7 2 V
7 2 V
6 2 V
7 2 V
7 3 V
7 2 V
7 2 V
6 2 V
7 3 V
7 2 V
7 2 V
6 3 V
7 2 V
7 3 V
7 2 V
6 3 V
7 2 V
7 3 V
7 2 V
7 3 V
6 2 V
7 3 V
7 2 V
7 3 V
6 3 V
7 2 V
7 3 V
7 3 V
7 3 V
6 3 V
7 2 V
7 3 V
7 3 V
6 3 V
7 3 V
7 3 V
7 3 V
6 3 V
7 3 V
7 3 V
7 3 V
7 4 V
6 3 V
7 3 V
7 3 V
7 4 V
6 3 V
7 3 V
7 4 V
7 3 V
7 4 V
6 3 V
7 4 V
7 3 V
7 4 V
6 3 V
7 4 V
7 4 V
7 4 V
7 3 V
6 4 V
7 4 V
7 4 V
7 4 V
6 4 V
7 4 V
7 4 V
7 5 V
6 4 V
7 4 V
7 4 V
7 5 V
7 4 V
6 5 V
7 4 V
7 5 V
7 4 V
6 5 V
7 5 V
7 4 V
7 5 V
7 5 V
6 5 V
7 5 V
7 5 V
7 5 V
6 5 V
7 6 V
7 5 V
7 5 V
6 6 V
7 5 V
7 6 V
7 5 V
7 6 V
6 6 V
7 6 V
7 5 V
7 6 V
6 6 V
stroke
1204 1069 M
7 7 V
7 6 V
7 6 V
7 6 V
6 7 V
7 6 V
7 7 V
7 7 V
6 6 V
7 7 V
7 7 V
7 7 V
7 7 V
6 7 V
7 8 V
7 7 V
7 7 V
6 8 V
7 8 V
7 7 V
7 8 V
6 8 V
7 8 V
7 9 V
7 8 V
7 8 V
6 9 V
7 8 V
7 9 V
7 9 V
6 9 V
7 9 V
7 9 V
7 9 V
7 10 V
6 9 V
7 10 V
7 10 V
7 10 V
6 10 V
7 10 V
7 10 V
7 11 V
6 11 V
7 10 V
7 11 V
7 11 V
7 11 V
6 12 V
7 11 V
7 12 V
7 12 V
6 11 V
7 13 V
7 12 V
7 12 V
7 13 V
6 12 V
7 13 V
7 13 V
7 13 V
6 14 V
7 13 V
7 14 V
7 14 V
7 14 V
6 14 V
7 14 V
7 15 V
7 15 V
6 14 V
7 15 V
7 16 V
7 15 V
6 15 V
7 16 V
7 16 V
7 16 V
7 16 V
6 16 V
7 17 V
7 16 V
7 17 V
6 17 V
7 17 V
7 17 V
7 17 V
7 18 V
6 17 V
7 18 V
7 18 V
7 18 V
6 17 V
7 18 V
7 19 V
7 18 V
6 18 V
7 18 V
7 18 V
7 19 V
7 18 V
6 18 V
7 19 V
7 18 V
stroke
1909 2337 M
7 18 V
6 18 V
7 18 V
7 18 V
7 18 V
7 18 V
6 18 V
7 17 V
7 17 V
7 17 V
6 17 V
7 17 V
7 16 V
7 16 V
7 16 V
6 15 V
7 15 V
7 15 V
7 14 V
6 14 V
7 14 V
7 13 V
7 12 V
6 12 V
7 11 V
7 11 V
7 10 V
7 10 V
6 9 V
7 9 V
7 7 V
7 8 V
6 6 V
7 6 V
7 5 V
7 4 V
7 4 V
6 3 V
7 2 V
7 2 V
7 0 V
6 0 V
7 0 V
7 -2 V
7 -2 V
6 -3 V
7 -4 V
7 -4 V
7 -5 V
7 -6 V
6 -6 V
7 -8 V
7 -7 V
7 -9 V
6 -9 V
7 -10 V
7 -10 V
7 -11 V
7 -11 V
6 -12 V
7 -12 V
7 -13 V
7 -14 V
6 -14 V
7 -14 V
7 -15 V
7 -15 V
6 -15 V
7 -16 V
7 -16 V
7 -16 V
7 -17 V
6 -17 V
7 -17 V
7 -17 V
7 -17 V
6 -18 V
7 -18 V
7 -18 V
7 -18 V
7 -18 V
6 -18 V
7 -18 V
7 -18 V
7 -19 V
6 -18 V
7 -18 V
7 -19 V
7 -18 V
7 -18 V
6 -18 V
7 -18 V
7 -19 V
7 -18 V
6 -17 V
7 -18 V
7 -18 V
7 -18 V
6 -17 V
7 -18 V
7 -17 V
7 -17 V
7 -17 V
6 -17 V
stroke
2613 1962 M
7 -17 V
7 -16 V
7 -17 V
6 -16 V
7 -16 V
7 -16 V
7 -16 V
7 -16 V
6 -15 V
7 -15 V
7 -16 V
7 -15 V
6 -14 V
7 -15 V
7 -15 V
7 -14 V
6 -14 V
7 -14 V
7 -14 V
7 -14 V
7 -13 V
6 -14 V
7 -13 V
7 -13 V
7 -13 V
6 -12 V
7 -13 V
7 -12 V
7 -12 V
7 -13 V
6 -11 V
7 -12 V
7 -12 V
7 -11 V
6 -12 V
7 -11 V
7 -11 V
7 -11 V
7 -10 V
6 -11 V
7 -11 V
7 -10 V
7 -10 V
6 -10 V
7 -10 V
7 -10 V
7 -10 V
6 -9 V
7 -10 V
7 -9 V
7 -9 V
7 -9 V
6 -9 V
7 -9 V
7 -9 V
7 -8 V
6 -9 V
7 -8 V
7 -8 V
7 -9 V
7 -8 V
6 -8 V
7 -8 V
7 -7 V
7 -8 V
6 -8 V
7 -7 V
7 -7 V
7 -8 V
6 -7 V
7 -7 V
7 -7 V
7 -7 V
7 -7 V
6 -6 V
7 -7 V
7 -7 V
7 -6 V
6 -7 V
7 -6 V
7 -6 V
7 -6 V
7 -7 V
6 -6 V
7 -6 V
7 -5 V
7 -6 V
6 -6 V
7 -6 V
7 -5 V
7 -6 V
7 -5 V
6 -6 V
7 -5 V
7 -5 V
7 -6 V
6 -5 V
7 -5 V
7 -5 V
7 -5 V
6 -5 V
7 -5 V
7 -5 V
7 -4 V
stroke
3318 957 M
7 -5 V
6 -5 V
7 -4 V
7 -5 V
7 -4 V
6 -5 V
7 -4 V
7 -5 V
7 -4 V
7 -4 V
6 -4 V
7 -5 V
7 -4 V
7 -4 V
6 -4 V
7 -4 V
7 -4 V
7 -4 V
6 -4 V
7 -3 V
7 -4 V
7 -4 V
7 -4 V
6 -3 V
7 -4 V
7 -3 V
7 -4 V
6 -3 V
7 -4 V
7 -3 V
7 -4 V
7 -3 V
6 -3 V
7 -4 V
7 -3 V
7 -3 V
6 -3 V
7 -4 V
7 -3 V
7 -3 V
7 -3 V
6 -3 V
7 -3 V
7 -3 V
7 -3 V
6 -3 V
7 -3 V
7 -3 V
7 -2 V
6 -3 V
7 -3 V
7 -3 V
7 -3 V
7 -2 V
6 -3 V
7 -3 V
7 -2 V
7 -3 V
6 -2 V
7 -3 V
7 -2 V
7 -3 V
7 -2 V
6 -3 V
7 -2 V
7 -3 V
7 -2 V
6 -3 V
7 -2 V
7 -2 V
7 -3 V
6 -2 V
7 -2 V
7 -2 V
7 -3 V
7 -2 V
6 -2 V
7 -2 V
7 -2 V
7 -2 V
6 -3 V
7 -2 V
7 -2 V
0.500 UL
LTb
500 300 M
3380 0 V
0 3380 V
-3380 0 V
500 300 L
1.000 UP
stroke
grestore
end
showpage
}}%
\put(2190,50){\makebox(0,0){$R\ [\mbox{kpc}]$}}%
\put(100,1990){%
\special{ps: gsave currentpoint currentpoint translate
270 rotate neg exch neg exch translate}%
\makebox(0,0)[b]{\shortstack{${\Sigma(R)}/{\Sigma_{c}}$}}%
\special{ps: currentpoint grestore moveto}%
}%
\put(3880,200){\makebox(0,0){ 1000}}%
\put(3035,200){\makebox(0,0){ 500}}%
\put(2190,200){\makebox(0,0){ 0}}%
\put(1345,200){\makebox(0,0){-500}}%
\put(500,200){\makebox(0,0){-1000}}%
\put(450,3680){\makebox(0,0)[r]{ 0.1}}%
\put(450,3004){\makebox(0,0)[r]{ 0.08}}%
\put(450,2328){\makebox(0,0)[r]{ 0.06}}%
\put(450,1652){\makebox(0,0)[r]{ 0.04}}%
\put(450,976){\makebox(0,0)[r]{ 0.02}}%
\put(450,300){\makebox(0,0)[r]{ 0}}%
\end{picture}%
\endgroup
 

%% file: figure/gnuplot/runningM.tex
\begingroup%
  \makeatletter%
  \newcommand{\GNUPLOTspecial}{%
    \@sanitize\catcode`\%=14\relax\special}%
  \setlength{\unitlength}{0.1bp}%
\begin{picture}(3440,3240)(100,0)%
{\GNUPLOTspecial{"
/gnudict 256 dict def
gnudict begin
/Color true def
/Solid true def
/gnulinewidth 5.000 def
/userlinewidth gnulinewidth def
/vshift -33 def
/dl {10.0 mul} def
/hpt_ 31.5 def
/vpt_ 31.5 def
/hpt hpt_ def
/vpt vpt_ def
/Rounded false def
/M {moveto} bind def
/L {lineto} bind def
/R {rmoveto} bind def
/V {rlineto} bind def
/N {newpath moveto} bind def
/C {setrgbcolor} bind def
/f {rlineto fill} bind def
/vpt2 vpt 2 mul def
/hpt2 hpt 2 mul def
/Lshow { currentpoint stroke M
  0 vshift R show } def
/Rshow { currentpoint stroke M
  dup stringwidth pop neg vshift R show } def
/Cshow { currentpoint stroke M
  dup stringwidth pop -2 div vshift R show } def
/UP { dup vpt_ mul /vpt exch def hpt_ mul /hpt exch def
  /hpt2 hpt 2 mul def /vpt2 vpt 2 mul def } def
/DL { Color {setrgbcolor Solid {pop []} if 0 setdash }
 {pop pop pop 0 setgray Solid {pop []} if 0 setdash} ifelse } def
/BL { stroke userlinewidth 2 mul setlinewidth
      Rounded { 1 setlinejoin 1 setlinecap } if } def
/AL { stroke userlinewidth 2 div setlinewidth
      Rounded { 1 setlinejoin 1 setlinecap } if } def
/UL { dup gnulinewidth mul /userlinewidth exch def
      dup 1 lt {pop 1} if 10 mul /udl exch def } def
/PL { stroke userlinewidth setlinewidth
      Rounded { 1 setlinejoin 1 setlinecap } if } def
/LTw { PL [] 1 setgray } def
/LTb { BL [] 0 0 0 DL } def
/fatlinewidth 7.500 def
/FL { stroke fatlinewidth setlinewidth Rounded { 1 setlinejoin 1 setlinecap } if } def
/LTa { FL [1 udl mul 2 udl mul] 0 setdash 0 0 0 setrgbcolor } def
/LT0 { FL [] 1 0 0 DL } def
/LT1 { PL [4 dl 2 dl] 0 1 0 DL } def
/LT2 { PL [2 dl 3 dl] 0 0 1 DL } def
/LT3 { PL [1 dl 1.5 dl] 1 0 1 DL } def
/LT4 { PL [5 dl 2 dl 1 dl 2 dl] 0 1 1 DL } def
/LT5 { PL [4 dl 3 dl 1 dl 3 dl] 1 1 0 DL } def
/LT6 { PL [2 dl 2 dl 2 dl 4 dl] 0 0 0 DL } def
/LT7 { PL [2 dl 2 dl 2 dl 2 dl 2 dl 4 dl] 1 0.3 0 DL } def
/LT8 { PL [2 dl 2 dl 2 dl 2 dl 2 dl 2 dl 2 dl 4 dl] 0.5 0.5 0.5 DL } def
/Pnt { stroke [] 0 setdash
   gsave 1 setlinecap M 0 0 V stroke grestore } def
/Dia { stroke [] 0 setdash 2 copy vpt add M
  hpt neg vpt neg V hpt vpt neg V
  hpt vpt V hpt neg vpt V closepath stroke
  Pnt } def
/Pls { stroke [] 0 setdash vpt sub M 0 vpt2 V
  currentpoint stroke M
  hpt neg vpt neg R hpt2 0 V stroke
  } def
/Box { stroke [] 0 setdash 2 copy exch hpt sub exch vpt add M
  0 vpt2 neg V hpt2 0 V 0 vpt2 V
  hpt2 neg 0 V closepath stroke
  Pnt } def
/Crs { stroke [] 0 setdash exch hpt sub exch vpt add M
  hpt2 vpt2 neg V currentpoint stroke M
  hpt2 neg 0 R hpt2 vpt2 V stroke } def
/TriU { stroke [] 0 setdash 2 copy vpt 1.12 mul add M
  hpt neg vpt -1.62 mul V
  hpt 2 mul 0 V
  hpt neg vpt 1.62 mul V closepath stroke
  Pnt  } def
/Star { 2 copy Pls Crs } def
/BoxF { stroke [] 0 setdash exch hpt sub exch vpt add M
  0 vpt2 neg V  hpt2 0 V  0 vpt2 V
  hpt2 neg 0 V  closepath fill } def
/TriUF { stroke [] 0 setdash vpt 1.12 mul add M
  hpt neg vpt -1.62 mul V
  hpt 2 mul 0 V
  hpt neg vpt 1.62 mul V closepath fill } def
/TriD { stroke [] 0 setdash 2 copy vpt 1.12 mul sub M
  hpt neg vpt 1.62 mul V
  hpt 2 mul 0 V
  hpt neg vpt -1.62 mul V closepath stroke
  Pnt  } def
/TriDF { stroke [] 0 setdash vpt 1.12 mul sub M
  hpt neg vpt 1.62 mul V
  hpt 2 mul 0 V
  hpt neg vpt -1.62 mul V closepath fill} def
/DiaF { stroke [] 0 setdash vpt add M
  hpt neg vpt neg V hpt vpt neg V
  hpt vpt V hpt neg vpt V closepath fill } def
/Pent { stroke [] 0 setdash 2 copy gsave
  translate 0 hpt M 4 {72 rotate 0 hpt L} repeat
  closepath stroke grestore Pnt } def
/PentF { stroke [] 0 setdash gsave
  translate 0 hpt M 4 {72 rotate 0 hpt L} repeat
  closepath fill grestore } def
/Circle { stroke [] 0 setdash 2 copy
  hpt 0 360 arc stroke Pnt } def
/CircleF { stroke [] 0 setdash hpt 0 360 arc fill } def
/C0 { BL [] 0 setdash 2 copy moveto vpt 90 450  arc } bind def
/C1 { BL [] 0 setdash 2 copy        moveto
       2 copy  vpt 0 90 arc closepath fill
               vpt 0 360 arc closepath } bind def
/C2 { BL [] 0 setdash 2 copy moveto
       2 copy  vpt 90 180 arc closepath fill
               vpt 0 360 arc closepath } bind def
/C3 { BL [] 0 setdash 2 copy moveto
       2 copy  vpt 0 180 arc closepath fill
               vpt 0 360 arc closepath } bind def
/C4 { BL [] 0 setdash 2 copy moveto
       2 copy  vpt 180 270 arc closepath fill
               vpt 0 360 arc closepath } bind def
/C5 { BL [] 0 setdash 2 copy moveto
       2 copy  vpt 0 90 arc
       2 copy moveto
       2 copy  vpt 180 270 arc closepath fill
               vpt 0 360 arc } bind def
/C6 { BL [] 0 setdash 2 copy moveto
      2 copy  vpt 90 270 arc closepath fill
              vpt 0 360 arc closepath } bind def
/C7 { BL [] 0 setdash 2 copy moveto
      2 copy  vpt 0 270 arc closepath fill
              vpt 0 360 arc closepath } bind def
/C8 { BL [] 0 setdash 2 copy moveto
      2 copy vpt 270 360 arc closepath fill
              vpt 0 360 arc closepath } bind def
/C9 { BL [] 0 setdash 2 copy moveto
      2 copy  vpt 270 450 arc closepath fill
              vpt 0 360 arc closepath } bind def
/C10 { BL [] 0 setdash 2 copy 2 copy moveto vpt 270 360 arc closepath fill
       2 copy moveto
       2 copy vpt 90 180 arc closepath fill
               vpt 0 360 arc closepath } bind def
/C11 { BL [] 0 setdash 2 copy moveto
       2 copy  vpt 0 180 arc closepath fill
       2 copy moveto
       2 copy  vpt 270 360 arc closepath fill
               vpt 0 360 arc closepath } bind def
/C12 { BL [] 0 setdash 2 copy moveto
       2 copy  vpt 180 360 arc closepath fill
               vpt 0 360 arc closepath } bind def
/C13 { BL [] 0 setdash  2 copy moveto
       2 copy  vpt 0 90 arc closepath fill
       2 copy moveto
       2 copy  vpt 180 360 arc closepath fill
               vpt 0 360 arc closepath } bind def
/C14 { BL [] 0 setdash 2 copy moveto
       2 copy  vpt 90 360 arc closepath fill
               vpt 0 360 arc } bind def
/C15 { BL [] 0 setdash 2 copy vpt 0 360 arc closepath fill
               vpt 0 360 arc closepath } bind def
/Rec   { newpath 4 2 roll moveto 1 index 0 rlineto 0 exch rlineto
       neg 0 rlineto closepath } bind def
/Square { dup Rec } bind def
/Bsquare { vpt sub exch vpt sub exch vpt2 Square } bind def
/S0 { BL [] 0 setdash 2 copy moveto 0 vpt rlineto BL Bsquare } bind def
/S1 { BL [] 0 setdash 2 copy vpt Square fill Bsquare } bind def
/S2 { BL [] 0 setdash 2 copy exch vpt sub exch vpt Square fill Bsquare } bind def
/S3 { BL [] 0 setdash 2 copy exch vpt sub exch vpt2 vpt Rec fill Bsquare } bind def
/S4 { BL [] 0 setdash 2 copy exch vpt sub exch vpt sub vpt Square fill Bsquare } bind def
/S5 { BL [] 0 setdash 2 copy 2 copy vpt Square fill
       exch vpt sub exch vpt sub vpt Square fill Bsquare } bind def
/S6 { BL [] 0 setdash 2 copy exch vpt sub exch vpt sub vpt vpt2 Rec fill Bsquare } bind def
/S7 { BL [] 0 setdash 2 copy exch vpt sub exch vpt sub vpt vpt2 Rec fill
       2 copy vpt Square fill
       Bsquare } bind def
/S8 { BL [] 0 setdash 2 copy vpt sub vpt Square fill Bsquare } bind def
/S9 { BL [] 0 setdash 2 copy vpt sub vpt vpt2 Rec fill Bsquare } bind def
/S10 { BL [] 0 setdash 2 copy vpt sub vpt Square fill 2 copy exch vpt sub exch vpt Square fill
       Bsquare } bind def
/S11 { BL [] 0 setdash 2 copy vpt sub vpt Square fill 2 copy exch vpt sub exch vpt2 vpt Rec fill
       Bsquare } bind def
/S12 { BL [] 0 setdash 2 copy exch vpt sub exch vpt sub vpt2 vpt Rec fill Bsquare } bind def
/S13 { BL [] 0 setdash 2 copy exch vpt sub exch vpt sub vpt2 vpt Rec fill
       2 copy vpt Square fill Bsquare } bind def
/S14 { BL [] 0 setdash 2 copy exch vpt sub exch vpt sub vpt2 vpt Rec fill
       2 copy exch vpt sub exch vpt Square fill Bsquare } bind def
/S15 { BL [] 0 setdash 2 copy Bsquare fill Bsquare } bind def
/D0 { gsave translate 45 rotate 0 0 S0 stroke grestore } bind def
/D1 { gsave translate 45 rotate 0 0 S1 stroke grestore } bind def
/D2 { gsave translate 45 rotate 0 0 S2 stroke grestore } bind def
/D3 { gsave translate 45 rotate 0 0 S3 stroke grestore } bind def
/D4 { gsave translate 45 rotate 0 0 S4 stroke grestore } bind def
/D5 { gsave translate 45 rotate 0 0 S5 stroke grestore } bind def
/D6 { gsave translate 45 rotate 0 0 S6 stroke grestore } bind def
/D7 { gsave translate 45 rotate 0 0 S7 stroke grestore } bind def
/D8 { gsave translate 45 rotate 0 0 S8 stroke grestore } bind def
/D9 { gsave translate 45 rotate 0 0 S9 stroke grestore } bind def
/D10 { gsave translate 45 rotate 0 0 S10 stroke grestore } bind def
/D11 { gsave translate 45 rotate 0 0 S11 stroke grestore } bind def
/D12 { gsave translate 45 rotate 0 0 S12 stroke grestore } bind def
/D13 { gsave translate 45 rotate 0 0 S13 stroke grestore } bind def
/D14 { gsave translate 45 rotate 0 0 S14 stroke grestore } bind def
/D15 { gsave translate 45 rotate 0 0 S15 stroke grestore } bind def
/DiaE { stroke [] 0 setdash vpt add M
  hpt neg vpt neg V hpt vpt neg V
  hpt vpt V hpt neg vpt V closepath stroke } def
/BoxE { stroke [] 0 setdash exch hpt sub exch vpt add M
  0 vpt2 neg V hpt2 0 V 0 vpt2 V
  hpt2 neg 0 V closepath stroke } def
/TriUE { stroke [] 0 setdash vpt 1.12 mul add M
  hpt neg vpt -1.62 mul V
  hpt 2 mul 0 V
  hpt neg vpt 1.62 mul V closepath stroke } def
/TriDE { stroke [] 0 setdash vpt 1.12 mul sub M
  hpt neg vpt 1.62 mul V
  hpt 2 mul 0 V
  hpt neg vpt -1.62 mul V closepath stroke } def
/PentE { stroke [] 0 setdash gsave
  translate 0 hpt M 4 {72 rotate 0 hpt L} repeat
  closepath stroke grestore } def
/CircE { stroke [] 0 setdash 
  hpt 0 360 arc stroke } def
/Opaque { gsave closepath 1 setgray fill grestore 0 setgray closepath } def
/DiaW { stroke [] 0 setdash vpt add M
  hpt neg vpt neg V hpt vpt neg V
  hpt vpt V hpt neg vpt V Opaque stroke } def
/BoxW { stroke [] 0 setdash exch hpt sub exch vpt add M
  0 vpt2 neg V hpt2 0 V 0 vpt2 V
  hpt2 neg 0 V Opaque stroke } def
/TriUW { stroke [] 0 setdash vpt 1.12 mul add M
  hpt neg vpt -1.62 mul V
  hpt 2 mul 0 V
  hpt neg vpt 1.62 mul V Opaque stroke } def
/TriDW { stroke [] 0 setdash vpt 1.12 mul sub M
  hpt neg vpt 1.62 mul V
  hpt 2 mul 0 V
  hpt neg vpt -1.62 mul V Opaque stroke } def
/PentW { stroke [] 0 setdash gsave
  translate 0 hpt M 4 {72 rotate 0 hpt L} repeat
  Opaque stroke grestore } def
/CircW { stroke [] 0 setdash 
  hpt 0 360 arc Opaque stroke } def
/BoxFill { gsave Rec 1 setgray fill grestore } def
/BoxColFill {
  gsave Rec
  /Fillden exch def
  currentrgbcolor
  /ColB exch def /ColG exch def /ColR exch def
  /ColR ColR Fillden mul Fillden sub 1 add def
  /ColG ColG Fillden mul Fillden sub 1 add def
  /ColB ColB Fillden mul Fillden sub 1 add def
  ColR ColG ColB setrgbcolor
  fill grestore } def
%
%
/PatternFill { gsave /PFa [ 9 2 roll ] def
    PFa 0 get PFa 2 get 2 div add PFa 1 get PFa 3 get 2 div add translate
    PFa 2 get -2 div PFa 3 get -2 div PFa 2 get PFa 3 get Rec
    gsave 1 setgray fill grestore clip
    currentlinewidth 0.5 mul setlinewidth
    /PFs PFa 2 get dup mul PFa 3 get dup mul add sqrt def
    0 0 M PFa 5 get rotate PFs -2 div dup translate
	0 1 PFs PFa 4 get div 1 add floor cvi
	{ PFa 4 get mul 0 M 0 PFs V } for
    0 PFa 6 get ne {
	0 1 PFs PFa 4 get div 1 add floor cvi
	{ PFa 4 get mul 0 2 1 roll M PFs 0 V } for
    } if
    stroke grestore } def
/Symbol-Oblique /Symbol findfont [1 0 .167 1 0 0] makefont
dup length dict begin {1 index /FID eq {pop pop} {def} ifelse} forall
currentdict end definefont pop
end
gnudict begin
gsave
0 0 translate
0.100 0.100 scale
0 setgray
newpath
0.500 UL
LTb
600 300 M
63 0 V
2777 0 R
-63 0 V
0.500 UL
LTb
600 957 M
31 0 V
2809 0 R
-31 0 V
600 1342 M
31 0 V
2809 0 R
-31 0 V
600 1614 M
31 0 V
2809 0 R
-31 0 V
600 1826 M
31 0 V
2809 0 R
-31 0 V
600 1999 M
31 0 V
2809 0 R
-31 0 V
600 2145 M
31 0 V
2809 0 R
-31 0 V
600 2271 M
31 0 V
2809 0 R
-31 0 V
600 2383 M
31 0 V
2809 0 R
-31 0 V
600 2483 M
63 0 V
2777 0 R
-63 0 V
0.500 UL
LTb
600 3140 M
31 0 V
2809 0 R
-31 0 V
600 300 M
0 63 V
0 2777 R
0 -63 V
0.500 UL
LTb
1027 300 M
0 31 V
0 2809 R
0 -31 V
1278 300 M
0 31 V
0 2809 R
0 -31 V
1455 300 M
0 31 V
0 2809 R
0 -31 V
1593 300 M
0 31 V
0 2809 R
0 -31 V
1705 300 M
0 31 V
0 2809 R
0 -31 V
1800 300 M
0 31 V
0 2809 R
0 -31 V
1882 300 M
0 31 V
0 2809 R
0 -31 V
1955 300 M
0 31 V
0 2809 R
0 -31 V
2020 300 M
0 63 V
0 2777 R
0 -63 V
0.500 UL
LTb
2447 300 M
0 31 V
0 2809 R
0 -31 V
2698 300 M
0 31 V
0 2809 R
0 -31 V
2875 300 M
0 31 V
0 2809 R
0 -31 V
3013 300 M
0 31 V
0 2809 R
0 -31 V
3125 300 M
0 31 V
0 2809 R
0 -31 V
3220 300 M
0 31 V
0 2809 R
0 -31 V
3302 300 M
0 31 V
0 2809 R
0 -31 V
3375 300 M
0 31 V
0 2809 R
0 -31 V
3440 300 M
0 63 V
0 2777 R
0 -63 V
0.500 UL
LTb
0.500 UL
LTb
600 300 M
2840 0 V
0 2840 V
-2840 0 V
600 300 L
LTb
LTb
1.000 UP
LTb
LTb
1.000 UL
LT0
600 1154 M
427 256 V
251 150 V
177 106 V
138 83 V
112 67 V
95 57 V
82 49 V
73 44 V
65 39 V
59 35 V
53 32 V
50 30 V
46 27 V
42 26 V
40 24 V
37 22 V
35 21 V
34 20 V
31 19 V
31 18 V
28 17 V
28 17 V
26 16 V
25 15 V
24 14 V
24 14 V
22 14 V
22 13 V
21 12 V
20 12 V
19 12 V
19 11 V
19 11 V
18 11 V
17 11 V
17 10 V
16 9 V
16 10 V
16 9 V
15 10 V
15 9 V
15 8 V
14 9 V
14 8 V
13 8 V
13 8 V
13 8 V
13 7 V
13 8 V
12 7 V
12 7 V
11 8 V
12 6 V
11 7 V
11 7 V
11 6 V
11 7 V
11 6 V
10 6 V
10 6 V
10 6 V
10 6 V
10 6 V
9 6 V
10 6 V
9 5 V
9 6 V
9 5 V
9 5 V
9 6 V
8 5 V
9 5 V
8 5 V
9 5 V
8 5 V
8 5 V
8 4 V
8 5 V
7 5 V
8 4 V
8 5 V
7 4 V
7 5 V
8 4 V
7 4 V
7 5 V
7 4 V
7 4 V
7 4 V
7 4 V
7 4 V
6 4 V
7 4 V
6 4 V
7 4 V
6 4 V
7 4 V
6 4 V
6 3 V
1.000 UL
LTa
1184 300 M
0 24 V
0 24 V
0 23 V
0 22 V
0 22 V
0 22 V
0 21 V
0 20 V
0 20 V
0 20 V
0 19 V
0 19 V
0 19 V
0 18 V
0 18 V
0 17 V
0 17 V
0 17 V
0 17 V
0 16 V
0 16 V
0 16 V
0 15 V
0 16 V
0 15 V
0 14 V
0 15 V
0 14 V
0 14 V
0 14 V
0 14 V
0 14 V
0 13 V
0 13 V
0 13 V
0 13 V
0 12 V
0 13 V
0 12 V
0 12 V
0 12 V
0 12 V
0 12 V
0 11 V
0 12 V
0 11 V
0 11 V
0 11 V
0 11 V
0 11 V
0 10 V
0 11 V
0 10 V
0 11 V
0 10 V
0 10 V
0 10 V
0 10 V
0 9 V
0 10 V
0 10 V
0 9 V
0 9 V
0 10 V
0 9 V
0 9 V
0 9 V
0 9 V
0 9 V
0 9 V
0 8 V
0 9 V
0 8 V
0 9 V
0 8 V
0 8 V
0 9 V
0 8 V
0 8 V
0 8 V
0 8 V
0 8 V
0 8 V
0 7 V
0 8 V
0 8 V
0 7 V
0 8 V
0 7 V
0 8 V
0 7 V
0 7 V
0 8 V
0 7 V
0 7 V
0 7 V
0 7 V
0 7 V
0 7 V
1.000 UL
LTa
2855 300 M
0 85 V
0 77 V
0 72 V
0 67 V
0 62 V
0 59 V
0 55 V
0 52 V
0 49 V
0 47 V
0 45 V
0 42 V
0 41 V
0 39 V
0 38 V
0 36 V
0 35 V
0 34 V
0 32 V
0 31 V
0 31 V
0 29 V
0 29 V
0 28 V
0 26 V
0 27 V
0 25 V
0 25 V
0 24 V
0 24 V
0 23 V
0 22 V
0 22 V
0 22 V
0 20 V
0 21 V
0 20 V
0 20 V
0 19 V
0 19 V
0 18 V
0 19 V
0 17 V
0 18 V
0 17 V
0 17 V
0 16 V
0 17 V
0 16 V
0 15 V
0 16 V
0 15 V
0 15 V
0 15 V
0 15 V
0 14 V
0 14 V
0 14 V
0 14 V
0 13 V
0 13 V
0 13 V
0 13 V
0 13 V
0 13 V
0 12 V
0 13 V
0 12 V
0 12 V
0 11 V
0 12 V
0 12 V
0 11 V
0 11 V
0 11 V
0 11 V
0 11 V
0 11 V
0 11 V
0 10 V
0 11 V
0 10 V
0 10 V
0 10 V
0 10 V
0 10 V
0 10 V
0 9 V
0 10 V
0 9 V
0 10 V
0 9 V
0 9 V
0 9 V
0 9 V
0 9 V
0 9 V
0 9 V
0 8 V
1.000 UL
LTa
600 1504 M
10 0 V
9 0 V
10 0 V
9 0 V
9 0 V
9 0 V
9 0 V
9 0 V
9 0 V
8 0 V
9 0 V
8 0 V
8 0 V
8 0 V
8 0 V
8 0 V
8 0 V
8 0 V
7 0 V
8 0 V
7 0 V
7 0 V
8 0 V
7 0 V
7 0 V
7 0 V
7 0 V
7 0 V
6 0 V
7 0 V
7 0 V
6 0 V
7 0 V
6 0 V
6 0 V
7 0 V
6 0 V
6 0 V
6 0 V
6 0 V
6 0 V
6 0 V
6 0 V
6 0 V
6 0 V
5 0 V
6 0 V
5 0 V
6 0 V
5 0 V
6 0 V
5 0 V
6 0 V
5 0 V
5 0 V
5 0 V
6 0 V
5 0 V
5 0 V
5 0 V
5 0 V
5 0 V
5 0 V
5 0 V
5 0 V
4 0 V
5 0 V
5 0 V
5 0 V
4 0 V
5 0 V
4 0 V
5 0 V
5 0 V
4 0 V
4 0 V
5 0 V
4 0 V
5 0 V
4 0 V
4 0 V
5 0 V
4 0 V
4 0 V
4 0 V
4 0 V
5 0 V
4 0 V
4 0 V
4 0 V
4 0 V
4 0 V
4 0 V
4 0 V
4 0 V
4 0 V
3 0 V
4 0 V
4 0 V
1.000 UL
LTa
600 2505 M
199 0 V
150 0 V
121 0 V
101 0 V
87 0 V
76 0 V
67 0 V
61 0 V
56 0 V
51 0 V
47 0 V
43 0 V
41 0 V
38 0 V
36 0 V
34 0 V
33 0 V
30 0 V
30 0 V
27 0 V
27 0 V
26 0 V
24 0 V
24 0 V
23 0 V
22 0 V
21 0 V
20 0 V
20 0 V
19 0 V
19 0 V
18 0 V
17 0 V
17 0 V
17 0 V
16 0 V
16 0 V
15 0 V
15 0 V
15 0 V
14 0 V
14 0 V
14 0 V
13 0 V
13 0 V
13 0 V
13 0 V
12 0 V
12 0 V
12 0 V
12 0 V
11 0 V
11 0 V
11 0 V
11 0 V
11 0 V
10 0 V
10 0 V
10 0 V
10 0 V
10 0 V
10 0 V
9 0 V
9 0 V
10 0 V
9 0 V
9 0 V
8 0 V
9 0 V
9 0 V
8 0 V
8 0 V
9 0 V
8 0 V
8 0 V
8 0 V
7 0 V
8 0 V
8 0 V
7 0 V
8 0 V
7 0 V
7 0 V
7 0 V
7 0 V
7 0 V
7 0 V
7 0 V
7 0 V
7 0 V
6 0 V
7 0 V
6 0 V
7 0 V
6 0 V
6 0 V
7 0 V
6 0 V
6 0 V
0.500 UL
LTb
600 300 M
2840 0 V
0 2840 V
-2840 0 V
600 300 L
1.000 UP
stroke
grestore
end
showpage
}}%
\put(1278,1404){\makebox(0,0)[l]{Subcluster}}%
\put(2182,2638){\makebox(0,0)[l]{Main cluster}}%
\put(2020,50){\makebox(0,0){$M_{\mbox{gas}}\ [M_{\odot}]$}}%
\put(200,1720){%
\special{ps: gsave currentpoint currentpoint translate
270 rotate neg exch neg exch translate}%
\makebox(0,0)[b]{\shortstack{$M_{0}\ [M_{\odot}]$}}%
\special{ps: currentpoint grestore moveto}%
}%
\put(3440,200){\makebox(0,0){$10^{15}$}}%
\put(2020,200){\makebox(0,0){$10^{14}$}}%
\put(600,200){\makebox(0,0){$10^{13}$}}%
\put(550,2483){\makebox(0,0)[r]{$10^{16}$}}%
\put(550,300){\makebox(0,0)[r]{$10^{15}$}}%
\end{picture}%
\endgroup
 

%% file: figure/gnuplot/G_MainXrayLinear.tex
\begingroup%
  \makeatletter%
  \newcommand{\GNUPLOTspecial}{%
    \@sanitize\catcode`\%=14\relax\special}%
  \setlength{\unitlength}{0.1bp}%
\begin{picture}(3780,3780)(0,0)%
{\GNUPLOTspecial{"
/gnudict 256 dict def
gnudict begin
/Color true def
/Solid false def
/gnulinewidth 5.000 def
/userlinewidth gnulinewidth def
/vshift -33 def
/dl {10.0 mul} def
/hpt_ 31.5 def
/vpt_ 31.5 def
/hpt hpt_ def
/vpt vpt_ def
/Rounded false def
/M {moveto} bind def
/L {lineto} bind def
/R {rmoveto} bind def
/V {rlineto} bind def
/N {newpath moveto} bind def
/C {setrgbcolor} bind def
/f {rlineto fill} bind def
/vpt2 vpt 2 mul def
/hpt2 hpt 2 mul def
/Lshow { currentpoint stroke M
  0 vshift R show } def
/Rshow { currentpoint stroke M
  dup stringwidth pop neg vshift R show } def
/Cshow { currentpoint stroke M
  dup stringwidth pop -2 div vshift R show } def
/UP { dup vpt_ mul /vpt exch def hpt_ mul /hpt exch def
  /hpt2 hpt 2 mul def /vpt2 vpt 2 mul def } def
/DL { Color {setrgbcolor Solid {pop []} if 0 setdash }
 {pop pop pop 0 setgray Solid {pop []} if 0 setdash} ifelse } def
/BL { stroke userlinewidth 2 mul setlinewidth
      Rounded { 1 setlinejoin 1 setlinecap } if } def
/AL { stroke userlinewidth 2 div setlinewidth
      Rounded { 1 setlinejoin 1 setlinecap } if } def
/UL { dup gnulinewidth mul /userlinewidth exch def
      dup 1 lt {pop 1} if 10 mul /udl exch def } def
/PL { stroke userlinewidth setlinewidth
      Rounded { 1 setlinejoin 1 setlinecap } if } def
/fatlinewidth 7.500 def
/FL { stroke fatlinewidth setlinewidth Rounded { 1 setlinejoin 1 setlinecap } if } def/LTw { PL [] 1 setgray } def
/LTb { BL [] 0 0 0 DL } def
/LTa { FL [1 udl mul 2 udl mul] 0 setdash 0 0 0 setrgbcolor } def
/LT0 { FL [] 1 0 0 DL } def
/LT1 { PL [4 dl 2 dl] 0 1 0 DL } def
/LT2 { PL [2 dl 3 dl] 0 0 1 DL } def
/LT3 { PL [1 dl 1.5 dl] 1 0 1 DL } def
/LT4 { PL [5 dl 2 dl 1 dl 2 dl] 0 1 1 DL } def
/LT5 { PL [4 dl 3 dl 1 dl 3 dl] 1 1 0 DL } def
/LT6 { PL [2 dl 2 dl 2 dl 4 dl] 0 0 0 DL } def
/LT7 { PL [2 dl 2 dl 2 dl 2 dl 2 dl 4 dl] 1 0.3 0 DL } def
/LT8 { PL [2 dl 2 dl 2 dl 2 dl 2 dl 2 dl 2 dl 4 dl] 0.5 0.5 0.5 DL } def
/Pnt { stroke [] 0 setdash
   gsave 1 setlinecap M 0 0 V stroke grestore } def
/Dia { stroke [] 0 setdash 2 copy vpt add M
  hpt neg vpt neg V hpt vpt neg V
  hpt vpt V hpt neg vpt V closepath stroke
  Pnt } def
/Pls { stroke [] 0 setdash vpt sub M 0 vpt2 V
  currentpoint stroke M
  hpt neg vpt neg R hpt2 0 V stroke
  } def
/Box { stroke [] 0 setdash 2 copy exch hpt sub exch vpt add M
  0 vpt2 neg V hpt2 0 V 0 vpt2 V
  hpt2 neg 0 V closepath stroke
  Pnt } def
/Crs { stroke [] 0 setdash exch hpt sub exch vpt add M
  hpt2 vpt2 neg V currentpoint stroke M
  hpt2 neg 0 R hpt2 vpt2 V stroke } def
/TriU { stroke [] 0 setdash 2 copy vpt 1.12 mul add M
  hpt neg vpt -1.62 mul V
  hpt 2 mul 0 V
  hpt neg vpt 1.62 mul V closepath stroke
  Pnt  } def
/Star { 2 copy Pls Crs } def
/BoxF { stroke [] 0 setdash exch hpt sub exch vpt add M
  0 vpt2 neg V  hpt2 0 V  0 vpt2 V
  hpt2 neg 0 V  closepath fill } def
/TriUF { stroke [] 0 setdash vpt 1.12 mul add M
  hpt neg vpt -1.62 mul V
  hpt 2 mul 0 V
  hpt neg vpt 1.62 mul V closepath fill } def
/TriD { stroke [] 0 setdash 2 copy vpt 1.12 mul sub M
  hpt neg vpt 1.62 mul V
  hpt 2 mul 0 V
  hpt neg vpt -1.62 mul V closepath stroke
  Pnt  } def
/TriDF { stroke [] 0 setdash vpt 1.12 mul sub M
  hpt neg vpt 1.62 mul V
  hpt 2 mul 0 V
  hpt neg vpt -1.62 mul V closepath fill} def
/DiaF { stroke [] 0 setdash vpt add M
  hpt neg vpt neg V hpt vpt neg V
  hpt vpt V hpt neg vpt V closepath fill } def
/Pent { stroke [] 0 setdash 2 copy gsave
  translate 0 hpt M 4 {72 rotate 0 hpt L} repeat
  closepath stroke grestore Pnt } def
/PentF { stroke [] 0 setdash gsave
  translate 0 hpt M 4 {72 rotate 0 hpt L} repeat
  closepath fill grestore } def
/Circle { stroke [] 0 setdash 2 copy
  hpt 0 360 arc stroke Pnt } def
/CircleF { stroke [] 0 setdash hpt 0 360 arc fill } def
/C0 { BL [] 0 setdash 2 copy moveto vpt 90 450  arc } bind def
/C1 { BL [] 0 setdash 2 copy        moveto
       2 copy  vpt 0 90 arc closepath fill
               vpt 0 360 arc closepath } bind def
/C2 { BL [] 0 setdash 2 copy moveto
       2 copy  vpt 90 180 arc closepath fill
               vpt 0 360 arc closepath } bind def
/C3 { BL [] 0 setdash 2 copy moveto
       2 copy  vpt 0 180 arc closepath fill
               vpt 0 360 arc closepath } bind def
/C4 { BL [] 0 setdash 2 copy moveto
       2 copy  vpt 180 270 arc closepath fill
               vpt 0 360 arc closepath } bind def
/C5 { BL [] 0 setdash 2 copy moveto
       2 copy  vpt 0 90 arc
       2 copy moveto
       2 copy  vpt 180 270 arc closepath fill
               vpt 0 360 arc } bind def
/C6 { BL [] 0 setdash 2 copy moveto
      2 copy  vpt 90 270 arc closepath fill
              vpt 0 360 arc closepath } bind def
/C7 { BL [] 0 setdash 2 copy moveto
      2 copy  vpt 0 270 arc closepath fill
              vpt 0 360 arc closepath } bind def
/C8 { BL [] 0 setdash 2 copy moveto
      2 copy vpt 270 360 arc closepath fill
              vpt 0 360 arc closepath } bind def
/C9 { BL [] 0 setdash 2 copy moveto
      2 copy  vpt 270 450 arc closepath fill
              vpt 0 360 arc closepath } bind def
/C10 { BL [] 0 setdash 2 copy 2 copy moveto vpt 270 360 arc closepath fill
       2 copy moveto
       2 copy vpt 90 180 arc closepath fill
               vpt 0 360 arc closepath } bind def
/C11 { BL [] 0 setdash 2 copy moveto
       2 copy  vpt 0 180 arc closepath fill
       2 copy moveto
       2 copy  vpt 270 360 arc closepath fill
               vpt 0 360 arc closepath } bind def
/C12 { BL [] 0 setdash 2 copy moveto
       2 copy  vpt 180 360 arc closepath fill
               vpt 0 360 arc closepath } bind def
/C13 { BL [] 0 setdash  2 copy moveto
       2 copy  vpt 0 90 arc closepath fill
       2 copy moveto
       2 copy  vpt 180 360 arc closepath fill
               vpt 0 360 arc closepath } bind def
/C14 { BL [] 0 setdash 2 copy moveto
       2 copy  vpt 90 360 arc closepath fill
               vpt 0 360 arc } bind def
/C15 { BL [] 0 setdash 2 copy vpt 0 360 arc closepath fill
               vpt 0 360 arc closepath } bind def
/Rec   { newpath 4 2 roll moveto 1 index 0 rlineto 0 exch rlineto
       neg 0 rlineto closepath } bind def
/Square { dup Rec } bind def
/Bsquare { vpt sub exch vpt sub exch vpt2 Square } bind def
/S0 { BL [] 0 setdash 2 copy moveto 0 vpt rlineto BL Bsquare } bind def
/S1 { BL [] 0 setdash 2 copy vpt Square fill Bsquare } bind def
/S2 { BL [] 0 setdash 2 copy exch vpt sub exch vpt Square fill Bsquare } bind def
/S3 { BL [] 0 setdash 2 copy exch vpt sub exch vpt2 vpt Rec fill Bsquare } bind def
/S4 { BL [] 0 setdash 2 copy exch vpt sub exch vpt sub vpt Square fill Bsquare } bind def
/S5 { BL [] 0 setdash 2 copy 2 copy vpt Square fill
       exch vpt sub exch vpt sub vpt Square fill Bsquare } bind def
/S6 { BL [] 0 setdash 2 copy exch vpt sub exch vpt sub vpt vpt2 Rec fill Bsquare } bind def
/S7 { BL [] 0 setdash 2 copy exch vpt sub exch vpt sub vpt vpt2 Rec fill
       2 copy vpt Square fill
       Bsquare } bind def
/S8 { BL [] 0 setdash 2 copy vpt sub vpt Square fill Bsquare } bind def
/S9 { BL [] 0 setdash 2 copy vpt sub vpt vpt2 Rec fill Bsquare } bind def
/S10 { BL [] 0 setdash 2 copy vpt sub vpt Square fill 2 copy exch vpt sub exch vpt Square fill
       Bsquare } bind def
/S11 { BL [] 0 setdash 2 copy vpt sub vpt Square fill 2 copy exch vpt sub exch vpt2 vpt Rec fill
       Bsquare } bind def
/S12 { BL [] 0 setdash 2 copy exch vpt sub exch vpt sub vpt2 vpt Rec fill Bsquare } bind def
/S13 { BL [] 0 setdash 2 copy exch vpt sub exch vpt sub vpt2 vpt Rec fill
       2 copy vpt Square fill Bsquare } bind def
/S14 { BL [] 0 setdash 2 copy exch vpt sub exch vpt sub vpt2 vpt Rec fill
       2 copy exch vpt sub exch vpt Square fill Bsquare } bind def
/S15 { BL [] 0 setdash 2 copy Bsquare fill Bsquare } bind def
/D0 { gsave translate 45 rotate 0 0 S0 stroke grestore } bind def
/D1 { gsave translate 45 rotate 0 0 S1 stroke grestore } bind def
/D2 { gsave translate 45 rotate 0 0 S2 stroke grestore } bind def
/D3 { gsave translate 45 rotate 0 0 S3 stroke grestore } bind def
/D4 { gsave translate 45 rotate 0 0 S4 stroke grestore } bind def
/D5 { gsave translate 45 rotate 0 0 S5 stroke grestore } bind def
/D6 { gsave translate 45 rotate 0 0 S6 stroke grestore } bind def
/D7 { gsave translate 45 rotate 0 0 S7 stroke grestore } bind def
/D8 { gsave translate 45 rotate 0 0 S8 stroke grestore } bind def
/D9 { gsave translate 45 rotate 0 0 S9 stroke grestore } bind def
/D10 { gsave translate 45 rotate 0 0 S10 stroke grestore } bind def
/D11 { gsave translate 45 rotate 0 0 S11 stroke grestore } bind def
/D12 { gsave translate 45 rotate 0 0 S12 stroke grestore } bind def
/D13 { gsave translate 45 rotate 0 0 S13 stroke grestore } bind def
/D14 { gsave translate 45 rotate 0 0 S14 stroke grestore } bind def
/D15 { gsave translate 45 rotate 0 0 S15 stroke grestore } bind def
/DiaE { stroke [] 0 setdash vpt add M
  hpt neg vpt neg V hpt vpt neg V
  hpt vpt V hpt neg vpt V closepath stroke } def
/BoxE { stroke [] 0 setdash exch hpt sub exch vpt add M
  0 vpt2 neg V hpt2 0 V 0 vpt2 V
  hpt2 neg 0 V closepath stroke } def
/TriUE { stroke [] 0 setdash vpt 1.12 mul add M
  hpt neg vpt -1.62 mul V
  hpt 2 mul 0 V
  hpt neg vpt 1.62 mul V closepath stroke } def
/TriDE { stroke [] 0 setdash vpt 1.12 mul sub M
  hpt neg vpt 1.62 mul V
  hpt 2 mul 0 V
  hpt neg vpt -1.62 mul V closepath stroke } def
/PentE { stroke [] 0 setdash gsave
  translate 0 hpt M 4 {72 rotate 0 hpt L} repeat
  closepath stroke grestore } def
/CircE { stroke [] 0 setdash 
  hpt 0 360 arc stroke } def
/Opaque { gsave closepath 1 setgray fill grestore 0 setgray closepath } def
/DiaW { stroke [] 0 setdash vpt add M
  hpt neg vpt neg V hpt vpt neg V
  hpt vpt V hpt neg vpt V Opaque stroke } def
/BoxW { stroke [] 0 setdash exch hpt sub exch vpt add M
  0 vpt2 neg V hpt2 0 V 0 vpt2 V
  hpt2 neg 0 V Opaque stroke } def
/TriUW { stroke [] 0 setdash vpt 1.12 mul add M
  hpt neg vpt -1.62 mul V
  hpt 2 mul 0 V
  hpt neg vpt 1.62 mul V Opaque stroke } def
/TriDW { stroke [] 0 setdash vpt 1.12 mul sub M
  hpt neg vpt 1.62 mul V
  hpt 2 mul 0 V
  hpt neg vpt -1.62 mul V Opaque stroke } def
/PentW { stroke [] 0 setdash gsave
  translate 0 hpt M 4 {72 rotate 0 hpt L} repeat
  Opaque stroke grestore } def
/CircW { stroke [] 0 setdash 
  hpt 0 360 arc Opaque stroke } def
/BoxFill { gsave Rec 1 setgray fill grestore } def
/BoxColFill {
  gsave Rec
  /Fillden exch def
  currentrgbcolor
  /ColB exch def /ColG exch def /ColR exch def
  /ColR ColR Fillden mul Fillden sub 1 add def
  /ColG ColG Fillden mul Fillden sub 1 add def
  /ColB ColB Fillden mul Fillden sub 1 add def
  ColR ColG ColB setrgbcolor
  fill grestore } def
%
%
/PatternFill { gsave /PFa [ 9 2 roll ] def
    PFa 0 get PFa 2 get 2 div add PFa 1 get PFa 3 get 2 div add translate
    PFa 2 get -2 div PFa 3 get -2 div PFa 2 get PFa 3 get Rec
    gsave 1 setgray fill grestore clip
    currentlinewidth 0.5 mul setlinewidth
    /PFs PFa 2 get dup mul PFa 3 get dup mul add sqrt def
    0 0 M PFa 5 get rotate PFs -2 div dup translate
	0 1 PFs PFa 4 get div 1 add floor cvi
	{ PFa 4 get mul 0 M 0 PFs V } for
    0 PFa 6 get ne {
	0 1 PFs PFa 4 get div 1 add floor cvi
	{ PFa 4 get mul 0 2 1 roll M PFs 0 V } for
    } if
    stroke grestore } def
/Symbol-Oblique /Symbol findfont [1 0 .167 1 0 0] makefont
dup length dict begin {1 index /FID eq {pop pop} {def} ifelse} forall
currentdict end definefont pop
end
gnudict begin
gsave
0 0 translate
0.100 0.100 scale
0 setgray
newpath
0.500 UL
LTb
350 300 M
63 0 V
3317 0 R
-63 0 V
0.500 UL
LTb
350 397 M
31 0 V
3349 0 R
-31 0 V
350 493 M
31 0 V
3349 0 R
-31 0 V
350 590 M
31 0 V
3349 0 R
-31 0 V
350 686 M
31 0 V
3349 0 R
-31 0 V
350 783 M
63 0 V
3317 0 R
-63 0 V
0.500 UL
LTb
350 879 M
31 0 V
3349 0 R
-31 0 V
350 976 M
31 0 V
3349 0 R
-31 0 V
350 1073 M
31 0 V
3349 0 R
-31 0 V
350 1169 M
31 0 V
3349 0 R
-31 0 V
350 1266 M
63 0 V
3317 0 R
-63 0 V
0.500 UL
LTb
350 1362 M
31 0 V
3349 0 R
-31 0 V
350 1459 M
31 0 V
3349 0 R
-31 0 V
350 1555 M
31 0 V
3349 0 R
-31 0 V
350 1652 M
31 0 V
3349 0 R
-31 0 V
350 1749 M
63 0 V
3317 0 R
-63 0 V
0.500 UL
LTb
350 1845 M
31 0 V
3349 0 R
-31 0 V
350 1942 M
31 0 V
3349 0 R
-31 0 V
350 2038 M
31 0 V
3349 0 R
-31 0 V
350 2135 M
31 0 V
3349 0 R
-31 0 V
350 2231 M
63 0 V
3317 0 R
-63 0 V
0.500 UL
LTb
350 2328 M
31 0 V
3349 0 R
-31 0 V
350 2425 M
31 0 V
3349 0 R
-31 0 V
350 2521 M
31 0 V
3349 0 R
-31 0 V
350 2618 M
31 0 V
3349 0 R
-31 0 V
350 2714 M
63 0 V
3317 0 R
-63 0 V
0.500 UL
LTb
350 2811 M
31 0 V
3349 0 R
-31 0 V
350 2907 M
31 0 V
3349 0 R
-31 0 V
350 3004 M
31 0 V
3349 0 R
-31 0 V
350 3101 M
31 0 V
3349 0 R
-31 0 V
350 3197 M
63 0 V
3317 0 R
-63 0 V
0.500 UL
LTb
350 3294 M
31 0 V
3349 0 R
-31 0 V
350 3390 M
31 0 V
3349 0 R
-31 0 V
350 3487 M
31 0 V
3349 0 R
-31 0 V
350 3583 M
31 0 V
3349 0 R
-31 0 V
350 3680 M
63 0 V
3317 0 R
-63 0 V
0.500 UL
LTb
350 300 M
0 63 V
0 3317 R
0 -63 V
0.500 UL
LTb
479 300 M
0 31 V
0 3349 R
0 -31 V
608 300 M
0 31 V
0 3349 R
0 -31 V
737 300 M
0 31 V
0 3349 R
0 -31 V
866 300 M
0 31 V
0 3349 R
0 -31 V
995 300 M
0 63 V
0 3317 R
0 -63 V
0.500 UL
LTb
1124 300 M
0 31 V
0 3349 R
0 -31 V
1253 300 M
0 31 V
0 3349 R
0 -31 V
1382 300 M
0 31 V
0 3349 R
0 -31 V
1510 300 M
0 31 V
0 3349 R
0 -31 V
1639 300 M
0 63 V
0 3317 R
0 -63 V
0.500 UL
LTb
1768 300 M
0 31 V
0 3349 R
0 -31 V
1897 300 M
0 31 V
0 3349 R
0 -31 V
2026 300 M
0 31 V
0 3349 R
0 -31 V
2155 300 M
0 31 V
0 3349 R
0 -31 V
2284 300 M
0 63 V
0 3317 R
0 -63 V
0.500 UL
LTb
2413 300 M
0 31 V
0 3349 R
0 -31 V
2542 300 M
0 31 V
0 3349 R
0 -31 V
2671 300 M
0 31 V
0 3349 R
0 -31 V
2800 300 M
0 31 V
0 3349 R
0 -31 V
2929 300 M
0 63 V
0 3317 R
0 -63 V
0.500 UL
LTb
3058 300 M
0 31 V
0 3349 R
0 -31 V
3187 300 M
0 31 V
0 3349 R
0 -31 V
3316 300 M
0 31 V
0 3349 R
0 -31 V
3445 300 M
0 31 V
0 3349 R
0 -31 V
3573 300 M
0 63 V
0 3317 R
0 -63 V
0.500 UL
LTb
3702 300 M
0 31 V
0 3349 R
0 -31 V
0.500 UL
LTb
350 300 M
3380 0 V
0 3380 V
-3380 0 V
350 300 L
LTb
LTb
1.000 UP
1.000 UL
LT0
350 783 M
1 258 V
2 106 V
1 81 V
1 67 V
1 59 V
2 52 V
1 48 V
1 44 V
2 41 V
1 39 V
1 36 V
1 34 V
2 33 V
1 31 V
1 30 V
2 28 V
1 27 V
1 27 V
1 25 V
2 24 V
1 24 V
1 23 V
2 22 V
1 21 V
1 21 V
2 20 V
1 20 V
1 19 V
1 19 V
2 18 V
1 17 V
1 18 V
2 16 V
1 17 V
1 16 V
1 16 V
2 15 V
1 15 V
1 14 V
2 15 V
1 14 V
1 13 V
1 14 V
2 13 V
1 13 V
1 13 V
2 12 V
1 12 V
1 12 V
1 12 V
2 11 V
1 12 V
1 11 V
2 11 V
1 10 V
1 11 V
1 10 V
2 10 V
1 10 V
1 10 V
2 10 V
1 9 V
1 10 V
2 9 V
1 9 V
1 9 V
1 8 V
2 9 V
1 9 V
1 8 V
2 8 V
1 8 V
1 8 V
1 8 V
2 8 V
1 8 V
1 7 V
2 8 V
1 7 V
1 7 V
1 7 V
2 7 V
1 7 V
1 7 V
2 7 V
1 6 V
1 7 V
1 6 V
2 7 V
1 6 V
1 6 V
2 6 V
1 7 V
1 6 V
1 5 V
2 6 V
1 6 V
1 6 V
2 5 V
1 6 V
1 5 V
2 6 V
1 5 V
1 5 V
stroke
484 2787 M
1 6 V
2 5 V
1 5 V
1 5 V
2 5 V
1 5 V
1 5 V
1 5 V
2 4 V
1 5 V
1 5 V
2 4 V
1 5 V
1 4 V
1 5 V
2 4 V
1 5 V
1 4 V
2 4 V
1 4 V
1 4 V
1 5 V
2 4 V
1 4 V
1 4 V
2 4 V
1 4 V
1 3 V
1 4 V
2 4 V
1 4 V
1 4 V
2 3 V
1 4 V
1 3 V
2 4 V
1 4 V
1 3 V
1 4 V
2 3 V
1 3 V
1 4 V
2 3 V
1 3 V
1 4 V
1 3 V
2 3 V
1 3 V
1 3 V
2 3 V
1 4 V
1 3 V
1 3 V
2 3 V
1 3 V
1 3 V
2 3 V
1 2 V
1 3 V
1 3 V
2 3 V
1 3 V
1 2 V
2 3 V
1 3 V
1 3 V
1 2 V
2 3 V
1 3 V
1 2 V
2 3 V
1 2 V
1 3 V
2 2 V
1 3 V
1 2 V
1 3 V
2 2 V
1 2 V
1 3 V
2 2 V
1 2 V
1 3 V
1 2 V
2 2 V
1 3 V
1 2 V
2 2 V
1 2 V
1 2 V
1 3 V
2 2 V
1 2 V
1 2 V
2 2 V
1 2 V
1 2 V
1 2 V
2 2 V
1 2 V
1 2 V
2 2 V
1 2 V
1 2 V
stroke
618 3122 M
1 2 V
2 2 V
1 2 V
1 2 V
2 1 V
1 2 V
1 2 V
2 2 V
1 2 V
1 1 V
1 2 V
2 2 V
1 2 V
1 1 V
2 2 V
1 2 V
1 1 V
1 2 V
2 2 V
1 1 V
1 2 V
2 2 V
1 1 V
1 2 V
1 1 V
2 2 V
1 1 V
1 2 V
2 1 V
1 2 V
1 1 V
1 2 V
2 1 V
1 2 V
1 1 V
2 1 V
1 2 V
1 1 V
1 2 V
2 1 V
1 1 V
1 2 V
2 1 V
1 1 V
1 2 V
2 1 V
1 1 V
1 1 V
1 2 V
2 1 V
1 1 V
1 1 V
2 2 V
1 1 V
1 1 V
1 1 V
2 1 V
1 2 V
1 1 V
2 1 V
1 1 V
1 1 V
1 1 V
2 1 V
1 1 V
1 1 V
2 1 V
1 1 V
1 2 V
1 1 V
2 1 V
1 1 V
1 1 V
2 1 V
1 1 V
1 1 V
1 0 V
2 1 V
1 1 V
1 1 V
2 1 V
1 1 V
1 1 V
1 1 V
2 1 V
1 1 V
1 1 V
2 0 V
1 1 V
1 1 V
2 1 V
1 1 V
1 1 V
1 0 V
2 1 V
1 1 V
1 1 V
2 0 V
1 1 V
1 1 V
1 1 V
2 0 V
1 1 V
1 1 V
stroke
752 3252 M
2 1 V
1 0 V
1 1 V
1 1 V
2 0 V
1 1 V
1 1 V
2 0 V
1 1 V
1 0 V
1 1 V
2 1 V
1 0 V
1 1 V
2 0 V
1 1 V
1 1 V
1 0 V
2 1 V
1 0 V
1 1 V
2 0 V
1 1 V
1 0 V
2 1 V
1 0 V
1 1 V
1 0 V
2 1 V
1 0 V
1 1 V
2 0 V
1 1 V
1 0 V
1 1 V
2 0 V
1 0 V
1 1 V
2 0 V
1 1 V
1 0 V
1 0 V
2 1 V
1 0 V
1 0 V
2 1 V
1 0 V
1 1 V
1 0 V
2 0 V
1 0 V
1 1 V
2 0 V
1 0 V
1 1 V
1 0 V
2 0 V
1 1 V
1 0 V
2 0 V
1 0 V
1 1 V
2 0 V
1 0 V
1 0 V
1 0 V
2 1 V
1 0 V
1 0 V
2 0 V
1 0 V
1 1 V
1 0 V
2 0 V
1 0 V
1 0 V
2 0 V
1 1 V
1 0 V
1 0 V
2 0 V
1 0 V
1 0 V
2 0 V
1 0 V
1 1 V
1 0 V
2 0 V
1 0 V
1 0 V
2 0 V
1 0 V
1 0 V
1 0 V
2 0 V
1 0 V
1 0 V
2 0 V
1 0 V
1 0 V
2 0 V
1 0 V
1 0 V
1 0 V
stroke
886 3285 M
2 0 V
1 0 V
1 0 V
2 0 V
1 0 V
1 0 V
1 0 V
2 0 V
1 0 V
1 0 V
2 0 V
1 0 V
1 0 V
1 0 V
2 0 V
1 0 V
1 0 V
2 -1 V
1 0 V
1 0 V
1 0 V
2 0 V
1 0 V
1 0 V
2 0 V
1 -1 V
1 0 V
1 0 V
2 0 V
1 0 V
1 0 V
2 0 V
1 -1 V
1 0 V
2 0 V
1 0 V
1 0 V
1 -1 V
2 0 V
1 0 V
1 0 V
2 0 V
1 -1 V
1 0 V
1 0 V
2 0 V
1 0 V
1 -1 V
2 0 V
1 0 V
1 0 V
1 -1 V
2 0 V
1 0 V
1 0 V
2 -1 V
1 0 V
1 0 V
1 -1 V
2 0 V
1 0 V
1 0 V
2 -1 V
1 0 V
1 0 V
1 -1 V
2 0 V
1 0 V
1 -1 V
2 0 V
1 0 V
1 -1 V
2 0 V
1 0 V
1 -1 V
1 0 V
2 0 V
1 -1 V
1 0 V
2 0 V
1 -1 V
1 0 V
1 -1 V
2 0 V
1 0 V
1 -1 V
2 0 V
1 0 V
1 -1 V
1 0 V
2 -1 V
1 0 V
1 -1 V
2 0 V
1 0 V
1 -1 V
1 0 V
2 -1 V
1 0 V
1 0 V
2 -1 V
1 0 V
1 -1 V
1 0 V
stroke
1020 3260 M
2 -1 V
1 0 V
1 -1 V
2 0 V
1 -1 V
1 0 V
2 -1 V
1 0 V
1 0 V
1 -1 V
2 0 V
1 -1 V
1 0 V
2 -1 V
1 0 V
1 -1 V
1 0 V
2 -1 V
1 0 V
1 -1 V
2 0 V
1 -1 V
1 0 V
1 -1 V
2 -1 V
1 0 V
1 -1 V
2 0 V
1 -1 V
1 0 V
1 -1 V
2 0 V
1 -1 V
1 0 V
2 -1 V
1 0 V
1 -1 V
1 -1 V
2 0 V
1 -1 V
1 0 V
2 -1 V
1 0 V
1 -1 V
1 -1 V
2 0 V
1 -1 V
1 0 V
2 -1 V
1 -1 V
1 0 V
2 -1 V
1 0 V
1 -1 V
1 -1 V
2 0 V
1 -1 V
1 0 V
2 -1 V
1 -1 V
1 0 V
1 -1 V
2 0 V
1 -1 V
1 -1 V
2 0 V
1 -1 V
1 -1 V
1 0 V
2 -1 V
1 -1 V
1 0 V
2 -1 V
1 0 V
1 -1 V
1 -1 V
2 0 V
1 -1 V
1 -1 V
2 0 V
1 -1 V
1 -1 V
1 0 V
2 -1 V
1 -1 V
1 0 V
2 -1 V
1 -1 V
1 0 V
2 -1 V
1 -1 V
1 0 V
1 -1 V
2 -1 V
1 -1 V
1 0 V
2 -1 V
1 -1 V
1 0 V
1 -1 V
2 -1 V
1 0 V
1 -1 V
2 -1 V
stroke
1155 3198 M
1 -1 V
1 0 V
1 -1 V
2 -1 V
1 0 V
1 -1 V
2 -1 V
1 -1 V
1 0 V
1 -1 V
2 -1 V
1 0 V
1 -1 V
2 -1 V
1 -1 V
1 0 V
1 -1 V
2 -1 V
1 -1 V
1 0 V
2 -1 V
1 -1 V
1 -1 V
2 0 V
1 -1 V
1 -1 V
1 -1 V
2 0 V
1 -1 V
1 -1 V
2 -1 V
1 0 V
1 -1 V
1 -1 V
2 -1 V
1 0 V
1 -1 V
2 -1 V
1 -1 V
1 0 V
1 -1 V
2 -1 V
1 -1 V
1 0 V
2 -1 V
1 -1 V
1 -1 V
1 -1 V
2 0 V
1 -1 V
1 -1 V
2 -1 V
1 0 V
1 -1 V
1 -1 V
2 -1 V
1 -1 V
1 0 V
2 -1 V
1 -1 V
1 -1 V
2 -1 V
1 0 V
1 -1 V
1 -1 V
2 -1 V
1 -1 V
1 0 V
2 -1 V
1 -1 V
1 -1 V
1 -1 V
2 0 V
1 -1 V
1 -1 V
2 -1 V
1 -1 V
1 0 V
1 -1 V
2 -1 V
1 -1 V
1 -1 V
2 0 V
1 -1 V
1 -1 V
1 -1 V
2 -1 V
1 -1 V
1 0 V
2 -1 V
1 -1 V
1 -1 V
1 -1 V
2 0 V
1 -1 V
1 -1 V
2 -1 V
1 -1 V
1 -1 V
2 0 V
1 -1 V
1 -1 V
1 -1 V
2 -1 V
stroke
1289 3117 M
1 -1 V
1 0 V
2 -1 V
1 -1 V
1 -1 V
1 -1 V
2 -1 V
1 0 V
1 -1 V
2 -1 V
1 -1 V
1 -1 V
1 -1 V
2 -1 V
1 0 V
1 -1 V
2 -1 V
1 -1 V
1 -1 V
1 -1 V
2 0 V
1 -1 V
1 -1 V
2 -1 V
1 -1 V
1 -1 V
1 0 V
2 -1 V
1 -1 V
1 -1 V
2 -1 V
1 -1 V
1 -1 V
2 0 V
1 -1 V
1 -1 V
1 -1 V
2 -1 V
1 -1 V
1 -1 V
2 0 V
1 -1 V
1 -1 V
1 -1 V
2 -1 V
1 -1 V
1 -1 V
2 0 V
1 -1 V
1 -1 V
1 -1 V
2 -1 V
1 -1 V
1 -1 V
2 0 V
1 -1 V
1 -1 V
1 -1 V
2 -1 V
1 -1 V
1 -1 V
2 0 V
1 -1 V
1 -1 V
1 -1 V
2 -1 V
1 -1 V
1 -1 V
2 -1 V
1 0 V
1 -1 V
2 -1 V
1 -1 V
1 -1 V
1 -1 V
2 -1 V
1 0 V
1 -1 V
2 -1 V
1 -1 V
1 -1 V
1 -1 V
2 -1 V
1 -1 V
1 0 V
2 -1 V
1 -1 V
1 -1 V
1 -1 V
2 -1 V
1 -1 V
1 0 V
2 -1 V
1 -1 V
1 -1 V
1 -1 V
2 -1 V
1 -1 V
1 -1 V
2 0 V
1 -1 V
1 -1 V
1 -1 V
2 -1 V
stroke
1423 3028 M
1 -1 V
1 -1 V
2 -1 V
1 0 V
1 -1 V
2 -1 V
1 -1 V
1 -1 V
1 -1 V
2 -1 V
1 -1 V
1 0 V
2 -1 V
1 -1 V
1 -1 V
1 -1 V
2 -1 V
1 -1 V
1 -1 V
2 0 V
1 -1 V
1 -1 V
1 -1 V
2 -1 V
1 -1 V
1 -1 V
2 0 V
1 -1 V
1 -1 V
1 -1 V
2 -1 V
1 -1 V
1 -1 V
2 -1 V
1 0 V
1 -1 V
1 -1 V
2 -1 V
1 -1 V
1 -1 V
2 -1 V
1 -1 V
1 0 V
1 -1 V
2 -1 V
1 -1 V
1 -1 V
2 -1 V
1 -1 V
1 -1 V
2 0 V
1 -1 V
1 -1 V
1 -1 V
2 -1 V
1 -1 V
1 -1 V
2 0 V
1 -1 V
1 -1 V
1 -1 V
2 -1 V
1 -1 V
1 -1 V
2 -1 V
1 0 V
1 -1 V
1 -1 V
2 -1 V
1 -1 V
1 -1 V
2 -1 V
1 0 V
1 -1 V
1 -1 V
2 -1 V
1 -1 V
1 -1 V
2 -1 V
1 -1 V
1 0 V
1 -1 V
2 -1 V
1 -1 V
1 -1 V
2 -1 V
1 -1 V
1 0 V
2 -1 V
1 -1 V
1 -1 V
1 -1 V
2 -1 V
1 -1 V
1 0 V
2 -1 V
1 -1 V
1 -1 V
1 -1 V
2 -1 V
1 -1 V
1 0 V
2 -1 V
1 -1 V
stroke
1557 2938 M
1 -1 V
1 -1 V
2 -1 V
1 -1 V
1 0 V
2 -1 V
1 -1 V
1 -1 V
1 -1 V
2 -1 V
1 -1 V
1 0 V
2 -1 V
1 -1 V
1 -1 V
1 -1 V
2 -1 V
1 -1 V
1 0 V
2 -1 V
1 -1 V
1 -1 V
2 -1 V
1 -1 V
1 0 V
1 -1 V
2 -1 V
1 -1 V
1 -1 V
2 -1 V
1 -1 V
1 0 V
1 -1 V
2 -1 V
1 -1 V
1 -1 V
2 -1 V
1 -1 V
1 0 V
1 -1 V
2 -1 V
1 -1 V
1 -1 V
2 -1 V
1 0 V
1 -1 V
1 -1 V
2 -1 V
1 -1 V
1 -1 V
2 0 V
1 -1 V
1 -1 V
1 -1 V
2 -1 V
1 -1 V
1 0 V
2 -1 V
1 -1 V
1 -1 V
2 -1 V
1 -1 V
1 0 V
1 -1 V
2 -1 V
1 -1 V
1 -1 V
2 -1 V
1 0 V
1 -1 V
1 -1 V
2 -1 V
1 -1 V
1 -1 V
2 0 V
1 -1 V
1 -1 V
1 -1 V
2 -1 V
1 -1 V
1 0 V
2 -1 V
1 -1 V
1 -1 V
1 -1 V
2 -1 V
1 0 V
1 -1 V
2 -1 V
1 -1 V
1 -1 V
1 0 V
2 -1 V
1 -1 V
1 -1 V
2 -1 V
1 -1 V
1 0 V
2 -1 V
1 -1 V
1 -1 V
1 -1 V
2 0 V
1 -1 V
stroke
1691 2851 M
1 -1 V
2 -1 V
1 -1 V
1 -1 V
1 0 V
2 -1 V
1 -1 V
1 -1 V
2 -1 V
1 0 V
1 -1 V
1 -1 V
2 -1 V
1 -1 V
1 0 V
2 -1 V
1 -1 V
1 -1 V
1 -1 V
2 0 V
1 -1 V
1 -1 V
2 -1 V
1 -1 V
1 0 V
1 -1 V
2 -1 V
1 -1 V
1 -1 V
2 0 V
1 -1 V
1 -1 V
2 -1 V
1 -1 V
1 0 V
1 -1 V
2 -1 V
1 -1 V
1 -1 V
2 0 V
1 -1 V
1 -1 V
1 -1 V
2 -1 V
1 0 V
1 -1 V
2 -1 V
1 -1 V
1 -1 V
1 0 V
2 -1 V
1 -1 V
1 -1 V
2 0 V
1 -1 V
1 -1 V
1 -1 V
2 -1 V
1 0 V
1 -1 V
2 -1 V
1 -1 V
1 0 V
1 -1 V
2 -1 V
1 -1 V
1 -1 V
2 0 V
1 -1 V
1 -1 V
2 -1 V
1 0 V
1 -1 V
1 -1 V
2 -1 V
1 -1 V
1 0 V
2 -1 V
1 -1 V
1 -1 V
1 0 V
2 -1 V
1 -1 V
1 -1 V
2 0 V
1 -1 V
1 -1 V
1 -1 V
2 -1 V
1 0 V
1 -1 V
2 -1 V
1 -1 V
1 0 V
1 -1 V
2 -1 V
1 -1 V
1 0 V
2 -1 V
1 -1 V
1 -1 V
1 0 V
2 -1 V
1 -1 V
stroke
1825 2769 M
1 -1 V
2 0 V
1 -1 V
1 -1 V
1 -1 V
2 0 V
1 -1 V
1 -1 V
2 -1 V
1 0 V
1 -1 V
2 -1 V
1 -1 V
1 0 V
1 -1 V
2 -1 V
1 -1 V
1 0 V
2 -1 V
1 -1 V
1 -1 V
1 0 V
2 -1 V
1 -1 V
1 -1 V
2 0 V
1 -1 V
1 -1 V
1 -1 V
2 0 V
1 -1 V
1 -1 V
2 -1 V
1 0 V
1 -1 V
1 -1 V
2 -1 V
1 0 V
1 -1 V
2 -1 V
1 0 V
1 -1 V
1 -1 V
2 -1 V
1 0 V
1 -1 V
2 -1 V
1 -1 V
1 0 V
2 -1 V
1 -1 V
1 0 V
1 -1 V
2 -1 V
1 -1 V
1 0 V
2 -1 V
1 -1 V
1 -1 V
1 0 V
2 -1 V
1 -1 V
1 0 V
2 -1 V
1 -1 V
1 -1 V
1 0 V
2 -1 V
1 -1 V
1 0 V
2 -1 V
1 -1 V
1 -1 V
1 0 V
2 -1 V
1 -1 V
1 0 V
2 -1 V
1 -1 V
1 -1 V
1 0 V
2 -1 V
1 -1 V
1 0 V
2 -1 V
1 -1 V
1 0 V
2 -1 V
1 -1 V
1 -1 V
1 0 V
2 -1 V
1 -1 V
1 0 V
2 -1 V
1 -1 V
1 0 V
1 -1 V
2 -1 V
1 -1 V
1 0 V
2 -1 V
1 -1 V
1 0 V
stroke
1959 2694 M
1 -1 V
2 -1 V
1 0 V
1 -1 V
2 -1 V
1 -1 V
1 0 V
1 -1 V
2 -1 V
1 0 V
1 -1 V
2 -1 V
1 0 V
1 -1 V
1 -1 V
2 0 V
1 -1 V
1 -1 V
2 0 V
1 -1 V
1 -1 V
2 0 V
1 -1 V
1 -1 V
1 -1 V
2 0 V
1 -1 V
1 -1 V
2 0 V
1 -1 V
1 -1 V
1 0 V
2 -1 V
1 -1 V
1 0 V
2 -1 V
1 -1 V
1 0 V
1 -1 V
2 -1 V
1 0 V
1 -1 V
2 -1 V
1 0 V
1 -1 V
1 -1 V
2 0 V
1 -1 V
1 -1 V
2 0 V
1 -1 V
1 -1 V
1 0 V
2 -1 V
1 -1 V
1 0 V
2 -1 V
1 -1 V
1 0 V
2 -1 V
1 -1 V
1 0 V
1 -1 V
2 -1 V
1 0 V
1 -1 V
2 0 V
1 -1 V
1 -1 V
1 0 V
2 -1 V
1 -1 V
1 0 V
2 -1 V
1 -1 V
1 0 V
1 -1 V
2 -1 V
1 0 V
1 -1 V
2 -1 V
1 0 V
1 -1 V
1 0 V
2 -1 V
1 -1 V
1 0 V
2 -1 V
1 -1 V
1 0 V
1 -1 V
2 -1 V
1 0 V
1 -1 V
2 -1 V
1 0 V
1 -1 V
2 0 V
1 -1 V
1 -1 V
1 0 V
2 -1 V
1 -1 V
1 0 V
stroke
2093 2625 M
2 -1 V
1 -1 V
1 0 V
1 -1 V
2 0 V
1 -1 V
1 -1 V
2 0 V
1 -1 V
1 -1 V
1 0 V
2 -1 V
1 0 V
1 -1 V
2 -1 V
1 0 V
1 -1 V
1 -1 V
2 0 V
1 -1 V
1 0 V
2 -1 V
1 -1 V
1 0 V
1 -1 V
2 0 V
1 -1 V
1 -1 V
2 0 V
1 -1 V
1 -1 V
2 0 V
1 -1 V
1 0 V
1 -1 V
2 -1 V
1 0 V
1 -1 V
2 0 V
1 -1 V
1 -1 V
1 0 V
2 -1 V
1 0 V
1 -1 V
2 -1 V
1 0 V
1 -1 V
1 0 V
2 -1 V
1 -1 V
1 0 V
2 -1 V
1 -1 V
1 0 V
1 -1 V
2 0 V
1 -1 V
1 0 V
2 -1 V
1 -1 V
1 0 V
1 -1 V
2 0 V
1 -1 V
1 -1 V
2 0 V
1 -1 V
1 0 V
1 -1 V
2 -1 V
1 0 V
1 -1 V
2 0 V
1 -1 V
1 -1 V
2 0 V
1 -1 V
1 0 V
1 -1 V
2 -1 V
1 0 V
1 -1 V
2 0 V
1 -1 V
1 0 V
1 -1 V
2 -1 V
1 0 V
1 -1 V
2 0 V
1 -1 V
1 -1 V
1 0 V
2 -1 V
1 0 V
1 -1 V
2 0 V
1 -1 V
1 -1 V
1 0 V
2 -1 V
1 0 V
1 -1 V
stroke
2227 2562 M
2 0 V
1 -1 V
1 -1 V
1 0 V
2 -1 V
1 0 V
1 -1 V
2 0 V
1 -1 V
1 -1 V
2 0 V
1 -1 V
1 0 V
1 -1 V
2 0 V
1 -1 V
1 -1 V
2 0 V
1 -1 V
1 0 V
1 -1 V
2 0 V
1 -1 V
1 0 V
2 -1 V
1 -1 V
1 0 V
1 -1 V
2 0 V
1 -1 V
1 0 V
2 -1 V
1 0 V
1 -1 V
1 -1 V
2 0 V
1 -1 V
1 0 V
2 -1 V
1 0 V
1 -1 V
1 0 V
2 -1 V
1 -1 V
1 0 V
2 -1 V
1 0 V
1 -1 V
2 0 V
1 -1 V
1 0 V
1 -1 V
2 0 V
1 -1 V
1 -1 V
2 0 V
1 -1 V
1 0 V
1 -1 V
2 0 V
1 -1 V
1 0 V
2 -1 V
1 0 V
1 -1 V
1 -1 V
2 0 V
1 -1 V
1 0 V
2 -1 V
1 0 V
1 -1 V
1 0 V
2 -1 V
1 0 V
1 -1 V
2 0 V
1 -1 V
1 0 V
1 -1 V
2 -1 V
1 0 V
1 -1 V
2 0 V
1 -1 V
1 0 V
2 -1 V
1 0 V
1 -1 V
1 0 V
2 -1 V
1 0 V
1 -1 V
2 0 V
1 -1 V
1 0 V
1 -1 V
2 0 V
1 -1 V
1 -1 V
2 0 V
1 -1 V
1 0 V
1 -1 V
stroke
2361 2505 M
2 0 V
1 -1 V
1 0 V
2 -1 V
1 0 V
1 -1 V
1 0 V
2 -1 V
1 0 V
1 -1 V
2 0 V
1 -1 V
1 0 V
1 -1 V
2 0 V
1 -1 V
1 0 V
2 -1 V
1 0 V
1 -1 V
2 0 V
1 -1 V
1 0 V
1 -1 V
2 0 V
1 -1 V
1 0 V
2 -1 V
1 0 V
1 -1 V
1 -1 V
2 0 V
1 -1 V
1 0 V
2 -1 V
1 0 V
1 -1 V
1 0 V
2 -1 V
1 0 V
1 -1 V
2 0 V
1 -1 V
1 0 V
1 -1 V
2 0 V
1 -1 V
1 0 V
2 -1 V
1 0 V
1 -1 V
1 0 V
2 -1 V
1 0 V
1 0 V
2 -1 V
1 0 V
1 -1 V
2 0 V
1 -1 V
1 0 V
1 -1 V
2 0 V
1 -1 V
1 0 V
2 -1 V
1 0 V
1 -1 V
1 0 V
2 -1 V
1 0 V
1 -1 V
2 0 V
1 -1 V
1 0 V
1 -1 V
2 0 V
1 -1 V
1 0 V
2 -1 V
1 0 V
1 -1 V
1 0 V
2 -1 V
1 0 V
1 -1 V
2 0 V
1 -1 V
1 0 V
1 0 V
2 -1 V
1 0 V
1 -1 V
2 0 V
1 -1 V
1 0 V
2 -1 V
1 0 V
1 -1 V
1 0 V
2 -1 V
1 0 V
1 -1 V
2 0 V
stroke
2496 2454 M
1 -1 V
1 0 V
1 -1 V
2 0 V
1 0 V
1 -1 V
2 0 V
1 -1 V
1 0 V
1 -1 V
2 0 V
1 -1 V
1 0 V
2 -1 V
1 0 V
1 -1 V
1 0 V
2 -1 V
1 0 V
1 0 V
2 -1 V
1 0 V
1 -1 V
1 0 V
2 -1 V
1 0 V
1 -1 V
2 0 V
1 -1 V
1 0 V
1 -1 V
2 0 V
1 0 V
1 -1 V
2 0 V
1 -1 V
1 0 V
2 -1 V
1 0 V
1 -1 V
1 0 V
2 -1 V
1 0 V
1 0 V
2 -1 V
1 0 V
1 -1 V
1 0 V
2 -1 V
1 0 V
1 -1 V
2 0 V
1 -1 V
1 0 V
1 0 V
2 -1 V
1 0 V
1 -1 V
2 0 V
1 -1 V
1 0 V
1 -1 V
2 0 V
1 0 V
1 -1 V
2 0 V
1 -1 V
1 0 V
1 -1 V
2 0 V
1 -1 V
1 0 V
2 0 V
1 -1 V
1 0 V
2 -1 V
1 0 V
1 -1 V
1 0 V
2 -1 V
1 0 V
1 0 V
2 -1 V
1 0 V
1 -1 V
1 0 V
2 -1 V
1 0 V
1 0 V
2 -1 V
1 0 V
1 -1 V
1 0 V
2 -1 V
1 0 V
1 -1 V
2 0 V
1 0 V
1 -1 V
1 0 V
2 -1 V
1 0 V
1 -1 V
2 0 V
stroke
2630 2407 M
1 0 V
1 -1 V
1 0 V
2 -1 V
1 0 V
1 -1 V
2 0 V
1 0 V
1 -1 V
2 0 V
1 -1 V
1 0 V
1 -1 V
2 0 V
1 0 V
1 -1 V
2 0 V
1 -1 V
1 0 V
1 0 V
2 -1 V
1 0 V
1 -1 V
2 0 V
1 -1 V
1 0 V
1 0 V
2 -1 V
1 0 V
1 -1 V
2 0 V
1 0 V
1 -1 V
1 0 V
2 -1 V
1 0 V
1 -1 V
2 0 V
1 0 V
1 -1 V
1 0 V
2 -1 V
1 0 V
1 0 V
2 -1 V
1 0 V
1 -1 V
2 0 V
1 -1 V
1 0 V
1 0 V
2 -1 V
1 0 V
1 -1 V
2 0 V
1 0 V
1 -1 V
1 0 V
2 -1 V
1 0 V
1 0 V
2 -1 V
1 0 V
1 -1 V
1 0 V
2 0 V
1 -1 V
1 0 V
2 -1 V
1 0 V
1 0 V
1 -1 V
2 0 V
1 -1 V
1 0 V
2 0 V
1 -1 V
1 0 V
1 -1 V
2 0 V
1 0 V
1 -1 V
2 0 V
1 -1 V
1 0 V
2 0 V
1 -1 V
1 0 V
1 -1 V
2 0 V
1 0 V
1 -1 V
2 0 V
1 -1 V
1 0 V
1 0 V
2 -1 V
1 0 V
1 -1 V
2 0 V
1 0 V
1 -1 V
1 0 V
2 -1 V
stroke
2764 2364 M
1 0 V
1 0 V
2 -1 V
1 0 V
1 -1 V
1 0 V
2 0 V
1 -1 V
1 0 V
2 0 V
1 -1 V
1 0 V
1 -1 V
2 0 V
1 0 V
1 -1 V
2 0 V
1 -1 V
1 0 V
2 0 V
1 -1 V
1 0 V
1 -1 V
2 0 V
1 0 V
1 -1 V
2 0 V
1 0 V
1 -1 V
1 0 V
2 -1 V
1 0 V
1 0 V
2 -1 V
1 0 V
1 0 V
1 -1 V
2 0 V
1 -1 V
1 0 V
2 0 V
1 -1 V
1 0 V
1 -1 V
2 0 V
1 0 V
1 -1 V
2 0 V
1 0 V
1 -1 V
1 0 V
2 -1 V
1 0 V
1 0 V
2 -1 V
1 0 V
1 0 V
2 -1 V
1 0 V
1 0 V
1 -1 V
2 0 V
1 -1 V
1 0 V
2 0 V
1 -1 V
1 0 V
1 0 V
2 -1 V
1 0 V
1 -1 V
2 0 V
1 0 V
1 -1 V
1 0 V
2 0 V
1 -1 V
1 0 V
2 -1 V
1 0 V
1 0 V
1 -1 V
2 0 V
1 0 V
1 -1 V
2 0 V
1 0 V
1 -1 V
1 0 V
2 -1 V
1 0 V
1 0 V
2 -1 V
1 0 V
1 0 V
2 -1 V
1 0 V
1 0 V
1 -1 V
2 0 V
1 0 V
1 -1 V
2 0 V
1 -1 V
stroke
2898 2325 M
1 0 V
1 0 V
2 -1 V
1 0 V
1 0 V
2 -1 V
1 0 V
1 0 V
1 -1 V
2 0 V
1 0 V
1 -1 V
2 0 V
1 -1 V
1 0 V
1 0 V
2 -1 V
1 0 V
1 0 V
2 -1 V
1 0 V
1 0 V
1 -1 V
2 0 V
1 0 V
1 -1 V
2 0 V
1 0 V
1 -1 V
1 0 V
2 -1 V
1 0 V
1 0 V
2 -1 V
1 0 V
1 0 V
2 -1 V
1 0 V
1 0 V
1 -1 V
2 0 V
1 0 V
1 -1 V
2 0 V
1 0 V
1 -1 V
1 0 V
2 0 V
1 -1 V
1 0 V
2 0 V
1 -1 V
1 0 V
1 -1 V
2 0 V
1 0 V
1 -1 V
2 0 V
1 0 V
1 -1 V
1 0 V
2 0 V
1 -1 V
1 0 V
2 0 V
1 -1 V
1 0 V
1 0 V
2 -1 V
1 0 V
1 0 V
2 -1 V
1 0 V
1 0 V
2 -1 V
1 0 V
1 0 V
1 -1 V
2 0 V
1 0 V
1 -1 V
2 0 V
1 0 V
1 -1 V
1 0 V
2 0 V
1 -1 V
1 0 V
2 0 V
1 -1 V
1 0 V
1 0 V
2 -1 V
1 0 V
1 0 V
2 -1 V
1 0 V
1 0 V
1 -1 V
2 0 V
1 0 V
1 -1 V
2 0 V
1 0 V
stroke
3032 2290 M
1 -1 V
1 0 V
2 0 V
1 -1 V
1 0 V
2 0 V
1 -1 V
1 0 V
2 0 V
1 -1 V
1 0 V
1 0 V
2 -1 V
1 0 V
1 0 V
2 -1 V
1 0 V
1 0 V
1 -1 V
2 0 V
1 0 V
1 0 V
2 -1 V
1 0 V
1 0 V
1 -1 V
2 0 V
1 0 V
1 -1 V
2 0 V
1 0 V
1 -1 V
1 0 V
2 0 V
1 -1 V
1 0 V
2 0 V
1 -1 V
1 0 V
1 0 V
2 -1 V
1 0 V
1 0 V
2 -1 V
1 0 V
1 0 V
2 0 V
1 -1 V
1 0 V
1 0 V
2 -1 V
1 0 V
1 0 V
2 -1 V
1 0 V
1 0 V
1 -1 V
2 0 V
1 0 V
1 -1 V
2 0 V
1 0 V
1 -1 V
1 0 V
2 0 V
1 0 V
1 -1 V
2 0 V
1 0 V
1 -1 V
1 0 V
2 0 V
1 -1 V
1 0 V
2 0 V
1 -1 V
1 0 V
1 0 V
2 -1 V
1 0 V
1 0 V
2 0 V
1 -1 V
1 0 V
2 0 V
1 -1 V
1 0 V
1 0 V
2 -1 V
1 0 V
1 0 V
2 -1 V
1 0 V
1 0 V
1 0 V
2 -1 V
1 0 V
1 0 V
2 -1 V
1 0 V
1 0 V
1 -1 V
2 0 V
1 0 V
stroke
3166 2257 M
1 0 V
2 -1 V
1 0 V
1 0 V
1 -1 V
2 0 V
1 0 V
1 -1 V
2 0 V
1 0 V
1 -1 V
1 0 V
2 0 V
1 0 V
1 -1 V
2 0 V
1 0 V
1 -1 V
2 0 V
1 0 V
1 -1 V
1 0 V
2 0 V
1 0 V
1 -1 V
2 0 V
1 0 V
1 -1 V
1 0 V
2 0 V
1 0 V
1 -1 V
2 0 V
1 0 V
1 -1 V
1 0 V
2 0 V
1 -1 V
1 0 V
2 0 V
1 0 V
1 -1 V
1 0 V
2 0 V
1 -1 V
1 0 V
2 0 V
1 0 V
1 -1 V
1 0 V
2 0 V
1 -1 V
1 0 V
2 0 V
1 -1 V
1 0 V
2 0 V
1 0 V
1 -1 V
1 0 V
2 0 V
1 -1 V
1 0 V
2 0 V
1 0 V
1 -1 V
1 0 V
2 0 V
1 -1 V
1 0 V
2 0 V
1 0 V
1 -1 V
1 0 V
2 0 V
1 -1 V
1 0 V
2 0 V
1 0 V
1 -1 V
1 0 V
2 0 V
1 -1 V
1 0 V
2 0 V
1 0 V
1 -1 V
1 0 V
2 0 V
1 -1 V
1 0 V
2 0 V
1 0 V
1 -1 V
1 0 V
2 0 V
1 -1 V
1 0 V
2 0 V
1 0 V
1 -1 V
2 0 V
1 0 V
1 -1 V
stroke
3300 2226 M
1 0 V
2 0 V
1 0 V
1 -1 V
2 0 V
1 0 V
1 -1 V
1 0 V
2 0 V
1 0 V
1 -1 V
2 0 V
1 0 V
1 0 V
1 -1 V
2 0 V
1 0 V
1 -1 V
2 0 V
1 0 V
1 0 V
1 -1 V
2 0 V
1 0 V
1 -1 V
2 0 V
1 0 V
1 0 V
1 -1 V
2 0 V
1 0 V
1 0 V
2 -1 V
1 0 V
1 0 V
2 -1 V
1 0 V
1 0 V
1 0 V
2 -1 V
1 0 V
1 0 V
2 0 V
1 -1 V
1 0 V
1 0 V
2 -1 V
1 0 V
1 0 V
2 0 V
1 -1 V
1 0 V
1 0 V
2 0 V
1 -1 V
1 0 V
2 0 V
1 -1 V
1 0 V
1 0 V
2 0 V
1 -1 V
1 0 V
2 0 V
1 0 V
1 -1 V
1 0 V
2 0 V
1 0 V
1 -1 V
2 0 V
1 0 V
1 -1 V
2 0 V
1 0 V
1 0 V
1 -1 V
2 0 V
1 0 V
1 0 V
2 -1 V
1 0 V
1 0 V
1 0 V
2 -1 V
1 0 V
1 0 V
2 -1 V
1 0 V
1 0 V
1 0 V
2 -1 V
1 0 V
1 0 V
2 0 V
1 -1 V
1 0 V
1 0 V
2 0 V
1 -1 V
1 0 V
2 0 V
1 0 V
1 -1 V
stroke
3434 2198 M
1 0 V
2 0 V
1 0 V
1 -1 V
2 0 V
1 0 V
1 -1 V
2 0 V
1 0 V
1 0 V
1 -1 V
2 0 V
1 0 V
1 0 V
2 -1 V
1 0 V
1 0 V
1 0 V
2 -1 V
1 0 V
1 0 V
2 0 V
1 -1 V
1 0 V
1 0 V
2 0 V
1 -1 V
1 0 V
2 0 V
1 0 V
1 -1 V
1 0 V
2 0 V
1 0 V
1 -1 V
2 0 V
1 0 V
1 0 V
1 -1 V
2 0 V
1 0 V
1 0 V
2 -1 V
1 0 V
1 0 V
2 -1 V
1 0 V
1 0 V
1 0 V
2 -1 V
1 0 V
1 0 V
2 0 V
1 -1 V
1 0 V
1 0 V
2 0 V
1 -1 V
1 0 V
2 0 V
1 0 V
1 -1 V
1 0 V
2 0 V
1 0 V
1 -1 V
2 0 V
1 0 V
1 0 V
1 -1 V
2 0 V
1 0 V
1 0 V
2 -1 V
1 0 V
1 0 V
1 0 V
2 0 V
1 -1 V
1 0 V
2 0 V
1 0 V
1 -1 V
2 0 V
1 0 V
1 0 V
1 -1 V
2 0 V
1 0 V
1 0 V
2 -1 V
1 0 V
1 0 V
1 0 V
2 -1 V
1 0 V
1 0 V
2 0 V
1 -1 V
1 0 V
1 0 V
2 0 V
1 -1 V
1 0 V
stroke
3568 2172 M
2 0 V
1 0 V
1 -1 V
1 0 V
2 0 V
1 0 V
1 -1 V
2 0 V
1 0 V
1 0 V
1 -1 V
2 0 V
1 0 V
1 0 V
2 0 V
1 -1 V
1 0 V
2 0 V
1 0 V
1 -1 V
1 0 V
2 0 V
1 0 V
1 -1 V
2 0 V
1 0 V
1 0 V
1 -1 V
2 0 V
1 0 V
1 0 V
2 -1 V
1 0 V
1 0 V
1 0 V
2 0 V
1 -1 V
1 0 V
2 0 V
1 0 V
1 -1 V
1 0 V
2 0 V
1 0 V
1 -1 V
2 0 V
1 0 V
1 0 V
1 -1 V
2 0 V
1 0 V
1 0 V
2 -1 V
1 0 V
1 0 V
1 0 V
2 0 V
1 -1 V
1 0 V
2 0 V
1 0 V
1 -1 V
2 0 V
1 0 V
1 0 V
1 -1 V
2 0 V
1 0 V
1 0 V
2 0 V
1 -1 V
1 0 V
1 0 V
2 0 V
1 -1 V
1 0 V
2 0 V
1 0 V
1 -1 V
1 0 V
2 0 V
1 0 V
1 0 V
2 -1 V
1 0 V
1 0 V
1 0 V
2 -1 V
1 0 V
1 0 V
2 0 V
1 -1 V
1 0 V
1 0 V
2 0 V
1 0 V
1 -1 V
2 0 V
1 0 V
1 0 V
2 -1 V
1 0 V
1 0 V
1 0 V
stroke
3702 2148 M
2 0 V
1 -1 V
1 0 V
2 0 V
1 0 V
1 -1 V
1 0 V
2 0 V
1 0 V
1 -1 V
2 0 V
1 0 V
1 0 V
1 0 V
2 -1 V
1 0 V
1 0 V
2 0 V
1 -1 V
1 0 V
1 0 V
1.000 UL
LTa
885 300 M
0 7 V
0 7 V
0 6 V
0 7 V
0 7 V
0 7 V
0 6 V
0 7 V
0 7 V
0 7 V
0 7 V
0 6 V
0 7 V
0 7 V
0 7 V
0 6 V
0 7 V
0 7 V
0 7 V
0 6 V
0 7 V
0 7 V
0 7 V
0 7 V
0 6 V
0 7 V
0 7 V
0 7 V
0 6 V
0 7 V
0 7 V
0 7 V
0 7 V
0 6 V
0 7 V
0 7 V
0 7 V
0 6 V
0 7 V
0 7 V
0 7 V
0 6 V
0 7 V
0 7 V
0 7 V
0 7 V
0 6 V
0 7 V
0 7 V
0 7 V
0 6 V
0 7 V
0 7 V
0 7 V
0 7 V
0 6 V
0 7 V
0 7 V
0 7 V
0 6 V
0 7 V
0 7 V
0 7 V
0 7 V
0 6 V
0 7 V
0 7 V
0 7 V
0 6 V
0 7 V
0 7 V
0 7 V
0 6 V
0 7 V
0 7 V
0 7 V
0 7 V
0 6 V
0 7 V
0 7 V
0 7 V
0 6 V
0 7 V
0 7 V
0 7 V
0 7 V
0 6 V
0 7 V
0 7 V
0 7 V
0 6 V
0 7 V
0 7 V
0 7 V
0 6 V
0 7 V
0 7 V
0 7 V
0 7 V
0 6 V
0 7 V
0 7 V
0 7 V
0 6 V
stroke
885 1004 M
0 7 V
0 7 V
0 7 V
0 7 V
0 6 V
0 7 V
0 7 V
0 7 V
0 6 V
0 7 V
0 7 V
0 7 V
0 7 V
0 6 V
0 7 V
0 7 V
0 7 V
0 6 V
0 7 V
0 7 V
0 7 V
0 6 V
0 7 V
0 7 V
0 7 V
0 7 V
0 6 V
0 7 V
0 7 V
0 7 V
0 6 V
0 7 V
0 7 V
0 7 V
0 7 V
0 6 V
0 7 V
0 7 V
0 7 V
0 6 V
0 7 V
0 7 V
0 7 V
0 6 V
0 7 V
0 7 V
0 7 V
0 7 V
0 6 V
0 7 V
0 7 V
0 7 V
0 6 V
0 7 V
0 7 V
0 7 V
0 7 V
0 6 V
0 7 V
0 7 V
0 7 V
0 6 V
0 7 V
0 7 V
0 7 V
0 7 V
0 6 V
0 7 V
0 7 V
0 7 V
0 6 V
0 7 V
0 7 V
0 7 V
0 6 V
0 7 V
0 7 V
0 7 V
0 7 V
0 6 V
0 7 V
0 7 V
0 7 V
0 6 V
0 7 V
0 7 V
0 7 V
0 7 V
0 6 V
0 7 V
0 7 V
0 7 V
0 6 V
0 7 V
0 7 V
0 7 V
0 6 V
0 7 V
0 7 V
0 7 V
0 7 V
0 6 V
0 7 V
0 7 V
stroke
885 1709 M
0 7 V
0 6 V
0 7 V
0 7 V
0 7 V
0 7 V
0 6 V
0 7 V
0 7 V
0 7 V
0 6 V
0 7 V
0 7 V
0 7 V
0 7 V
0 6 V
0 7 V
0 7 V
0 7 V
0 6 V
0 7 V
0 7 V
0 7 V
0 6 V
0 7 V
0 7 V
0 7 V
0 7 V
0 6 V
0 7 V
0 7 V
0 7 V
0 6 V
0 7 V
0 7 V
0 7 V
0 7 V
0 6 V
0 7 V
0 7 V
0 7 V
0 6 V
0 7 V
0 7 V
0 7 V
0 6 V
0 7 V
0 7 V
0 7 V
0 7 V
0 6 V
0 7 V
0 7 V
0 7 V
0 6 V
0 7 V
0 7 V
0 7 V
0 7 V
0 6 V
0 7 V
0 7 V
0 7 V
0 6 V
0 7 V
0 7 V
0 7 V
0 6 V
0 7 V
0 7 V
0 7 V
0 7 V
0 6 V
0 7 V
0 7 V
0 7 V
0 6 V
0 7 V
0 7 V
0 7 V
0 7 V
0 6 V
0 7 V
0 7 V
0 7 V
0 6 V
0 7 V
0 7 V
0 7 V
0 7 V
0 6 V
0 7 V
0 7 V
0 7 V
0 6 V
0 7 V
0 7 V
0 7 V
0 6 V
0 7 V
0 7 V
0 7 V
0 7 V
0 6 V
stroke
885 2413 M
0 7 V
0 7 V
0 7 V
0 6 V
0 7 V
0 7 V
0 7 V
0 7 V
0 6 V
0 7 V
0 7 V
0 7 V
0 6 V
0 7 V
0 7 V
0 7 V
0 6 V
0 7 V
0 7 V
0 7 V
0 7 V
0 6 V
0 7 V
0 7 V
0 7 V
0 6 V
0 7 V
0 7 V
0 7 V
0 7 V
0 6 V
0 7 V
0 7 V
0 7 V
0 6 V
0 7 V
0 7 V
0 7 V
0 7 V
0 6 V
0 7 V
0 7 V
0 7 V
0 6 V
0 7 V
0 7 V
0 7 V
0 6 V
0 7 V
0 7 V
0 7 V
0 7 V
0 6 V
0 7 V
0 7 V
0 7 V
0 6 V
0 7 V
0 7 V
0 7 V
0 7 V
0 6 V
0 7 V
0 7 V
0 7 V
0 6 V
0 7 V
0 7 V
0 7 V
0 6 V
0 7 V
0 7 V
0 7 V
0 7 V
0 6 V
0 7 V
0 7 V
0 7 V
0 6 V
0 7 V
0 7 V
0 7 V
0 7 V
0 6 V
0 7 V
0 7 V
0 7 V
0 6 V
0 7 V
0 7 V
0 7 V
0 7 V
0 6 V
0 7 V
0 7 V
0 7 V
0 6 V
0 7 V
0 7 V
0 7 V
0 6 V
0 7 V
0 7 V
0 7 V
stroke
885 3118 M
0 7 V
0 6 V
0 7 V
0 7 V
0 7 V
0 6 V
0 7 V
0 7 V
0 7 V
0 7 V
0 6 V
0 7 V
0 7 V
0 7 V
0 6 V
0 7 V
0 7 V
0 7 V
0 6 V
0 7 V
0 7 V
0 7 V
0 7 V
0 6 V
0 7 V
0 7 V
0 7 V
0 6 V
0 7 V
0 7 V
0 7 V
0 7 V
0 6 V
0 7 V
0 7 V
0 7 V
0 6 V
0 7 V
0 7 V
0 7 V
0 7 V
0 6 V
0 7 V
0 7 V
0 7 V
0 6 V
0 7 V
0 7 V
0 7 V
0 6 V
0 7 V
0 7 V
0 7 V
0 7 V
0 6 V
0 7 V
0 7 V
0 7 V
0 6 V
0 7 V
0 7 V
0 7 V
0 7 V
0 6 V
0 7 V
0 7 V
0 7 V
0 6 V
0 7 V
0 7 V
0 7 V
0 6 V
0 7 V
0 7 V
0 7 V
0 7 V
0 6 V
0 7 V
0 7 V
0 7 V
0 6 V
0 7 V
0 7 V
1.000 UL
LTa
350 2143 M
7 0 V
7 0 V
6 0 V
7 0 V
7 0 V
7 0 V
6 0 V
7 0 V
7 0 V
7 0 V
7 0 V
6 0 V
7 0 V
7 0 V
7 0 V
6 0 V
7 0 V
7 0 V
7 0 V
6 0 V
7 0 V
7 0 V
7 0 V
7 0 V
6 0 V
7 0 V
7 0 V
7 0 V
6 0 V
7 0 V
7 0 V
7 0 V
7 0 V
6 0 V
7 0 V
7 0 V
7 0 V
6 0 V
7 0 V
7 0 V
7 0 V
6 0 V
7 0 V
7 0 V
7 0 V
7 0 V
6 0 V
7 0 V
7 0 V
7 0 V
6 0 V
7 0 V
7 0 V
7 0 V
7 0 V
6 0 V
7 0 V
7 0 V
7 0 V
6 0 V
7 0 V
7 0 V
7 0 V
7 0 V
6 0 V
7 0 V
7 0 V
7 0 V
6 0 V
7 0 V
7 0 V
7 0 V
6 0 V
7 0 V
7 0 V
7 0 V
7 0 V
6 0 V
7 0 V
7 0 V
7 0 V
6 0 V
7 0 V
7 0 V
7 0 V
7 0 V
6 0 V
7 0 V
7 0 V
7 0 V
6 0 V
7 0 V
7 0 V
7 0 V
6 0 V
7 0 V
7 0 V
7 0 V
7 0 V
6 0 V
7 0 V
7 0 V
7 0 V
6 0 V
stroke
1054 2143 M
7 0 V
7 0 V
7 0 V
7 0 V
6 0 V
7 0 V
7 0 V
7 0 V
6 0 V
7 0 V
7 0 V
7 0 V
7 0 V
6 0 V
7 0 V
7 0 V
7 0 V
6 0 V
7 0 V
7 0 V
7 0 V
6 0 V
7 0 V
7 0 V
7 0 V
7 0 V
6 0 V
7 0 V
7 0 V
7 0 V
6 0 V
7 0 V
7 0 V
7 0 V
7 0 V
6 0 V
7 0 V
7 0 V
7 0 V
6 0 V
7 0 V
7 0 V
7 0 V
6 0 V
7 0 V
7 0 V
7 0 V
7 0 V
6 0 V
7 0 V
7 0 V
7 0 V
6 0 V
7 0 V
7 0 V
7 0 V
7 0 V
6 0 V
7 0 V
7 0 V
7 0 V
6 0 V
7 0 V
7 0 V
7 0 V
7 0 V
6 0 V
7 0 V
7 0 V
7 0 V
6 0 V
7 0 V
7 0 V
7 0 V
6 0 V
7 0 V
7 0 V
7 0 V
7 0 V
6 0 V
7 0 V
7 0 V
7 0 V
6 0 V
7 0 V
7 0 V
7 0 V
7 0 V
6 0 V
7 0 V
7 0 V
7 0 V
6 0 V
7 0 V
7 0 V
7 0 V
6 0 V
7 0 V
7 0 V
7 0 V
7 0 V
6 0 V
7 0 V
7 0 V
stroke
1759 2143 M
7 0 V
6 0 V
7 0 V
7 0 V
7 0 V
7 0 V
6 0 V
7 0 V
7 0 V
7 0 V
6 0 V
7 0 V
7 0 V
7 0 V
7 0 V
6 0 V
7 0 V
7 0 V
7 0 V
6 0 V
7 0 V
7 0 V
7 0 V
6 0 V
7 0 V
7 0 V
7 0 V
7 0 V
6 0 V
7 0 V
7 0 V
7 0 V
6 0 V
7 0 V
7 0 V
7 0 V
7 0 V
6 0 V
7 0 V
7 0 V
7 0 V
6 0 V
7 0 V
7 0 V
7 0 V
6 0 V
7 0 V
7 0 V
7 0 V
7 0 V
6 0 V
7 0 V
7 0 V
7 0 V
6 0 V
7 0 V
7 0 V
7 0 V
7 0 V
6 0 V
7 0 V
7 0 V
7 0 V
6 0 V
7 0 V
7 0 V
7 0 V
6 0 V
7 0 V
7 0 V
7 0 V
7 0 V
6 0 V
7 0 V
7 0 V
7 0 V
6 0 V
7 0 V
7 0 V
7 0 V
7 0 V
6 0 V
7 0 V
7 0 V
7 0 V
6 0 V
7 0 V
7 0 V
7 0 V
7 0 V
6 0 V
7 0 V
7 0 V
7 0 V
6 0 V
7 0 V
7 0 V
7 0 V
6 0 V
7 0 V
7 0 V
7 0 V
7 0 V
6 0 V
stroke
2463 2143 M
7 0 V
7 0 V
7 0 V
6 0 V
7 0 V
7 0 V
7 0 V
7 0 V
6 0 V
7 0 V
7 0 V
7 0 V
6 0 V
7 0 V
7 0 V
7 0 V
6 0 V
7 0 V
7 0 V
7 0 V
7 0 V
6 0 V
7 0 V
7 0 V
7 0 V
6 0 V
7 0 V
7 0 V
7 0 V
7 0 V
6 0 V
7 0 V
7 0 V
7 0 V
6 0 V
7 0 V
7 0 V
7 0 V
7 0 V
6 0 V
7 0 V
7 0 V
7 0 V
6 0 V
7 0 V
7 0 V
7 0 V
6 0 V
7 0 V
7 0 V
7 0 V
7 0 V
6 0 V
7 0 V
7 0 V
7 0 V
6 0 V
7 0 V
7 0 V
7 0 V
7 0 V
6 0 V
7 0 V
7 0 V
7 0 V
6 0 V
7 0 V
7 0 V
7 0 V
6 0 V
7 0 V
7 0 V
7 0 V
7 0 V
6 0 V
7 0 V
7 0 V
7 0 V
6 0 V
7 0 V
7 0 V
7 0 V
7 0 V
6 0 V
7 0 V
7 0 V
7 0 V
6 0 V
7 0 V
7 0 V
7 0 V
7 0 V
6 0 V
7 0 V
7 0 V
7 0 V
6 0 V
7 0 V
7 0 V
7 0 V
6 0 V
7 0 V
7 0 V
7 0 V
stroke
3168 2143 M
7 0 V
6 0 V
7 0 V
7 0 V
7 0 V
6 0 V
7 0 V
7 0 V
7 0 V
7 0 V
6 0 V
7 0 V
7 0 V
7 0 V
6 0 V
7 0 V
7 0 V
7 0 V
6 0 V
7 0 V
7 0 V
7 0 V
7 0 V
6 0 V
7 0 V
7 0 V
7 0 V
6 0 V
7 0 V
7 0 V
7 0 V
7 0 V
6 0 V
7 0 V
7 0 V
7 0 V
6 0 V
7 0 V
7 0 V
7 0 V
7 0 V
6 0 V
7 0 V
7 0 V
7 0 V
6 0 V
7 0 V
7 0 V
7 0 V
6 0 V
7 0 V
7 0 V
7 0 V
7 0 V
6 0 V
7 0 V
7 0 V
7 0 V
6 0 V
7 0 V
7 0 V
7 0 V
7 0 V
6 0 V
7 0 V
7 0 V
7 0 V
6 0 V
7 0 V
7 0 V
7 0 V
6 0 V
7 0 V
7 0 V
7 0 V
7 0 V
6 0 V
7 0 V
7 0 V
7 0 V
6 0 V
7 0 V
7 0 V
0.500 UL
LTb
350 300 M
3380 0 V
0 3380 V
-3380 0 V
350 300 L
1.000 UP
stroke
grestore
end
showpage
}}%
\put(2040,50){\makebox(0,0){$r\ [\mbox{kpc}]$}}%
\put(100,1990){%
\special{ps: gsave currentpoint currentpoint translate
270 rotate neg exch neg exch translate}%
\makebox(0,0)[b]{\shortstack{$\mathcal{G}(r) \equiv G(r)/G_N$}}%
\special{ps: currentpoint grestore moveto}%
}%
\put(3573,200){\makebox(0,0){ 2500}}%
\put(2929,200){\makebox(0,0){ 2000}}%
\put(2284,200){\makebox(0,0){ 1500}}%
\put(1639,200){\makebox(0,0){ 1000}}%
\put(995,200){\makebox(0,0){ 500}}%
\put(350,200){\makebox(0,0){ 0}}%
\put(300,3680){\makebox(0,0)[r]{ 7}}%
\put(300,3197){\makebox(0,0)[r]{ 6}}%
\put(300,2714){\makebox(0,0)[r]{ 5}}%
\put(300,2231){\makebox(0,0)[r]{ 4}}%
\put(300,1749){\makebox(0,0)[r]{ 3}}%
\put(300,1266){\makebox(0,0)[r]{ 2}}%
\put(300,783){\makebox(0,0)[r]{ 1}}%
\put(300,300){\makebox(0,0)[r]{ 0}}%
\end{picture}%
\endgroup
 

%% file: figure/gnuplot/G_MainXrayLog.tex
\begingroup%
  \makeatletter%
  \newcommand{\GNUPLOTspecial}{%
    \@sanitize\catcode`\%=14\relax\special}%
  \setlength{\unitlength}{0.1bp}%
\begin{picture}(3780,3780)(0,0)%
{\GNUPLOTspecial{"
/gnudict 256 dict def
gnudict begin
/Color true def
/Solid false def
/gnulinewidth 5.000 def
/userlinewidth gnulinewidth def
/vshift -33 def
/dl {10.0 mul} def
/hpt_ 31.5 def
/vpt_ 31.5 def
/hpt hpt_ def
/vpt vpt_ def
/Rounded false def
/M {moveto} bind def
/L {lineto} bind def
/R {rmoveto} bind def
/V {rlineto} bind def
/N {newpath moveto} bind def
/C {setrgbcolor} bind def
/f {rlineto fill} bind def
/vpt2 vpt 2 mul def
/hpt2 hpt 2 mul def
/Lshow { currentpoint stroke M
  0 vshift R show } def
/Rshow { currentpoint stroke M
  dup stringwidth pop neg vshift R show } def
/Cshow { currentpoint stroke M
  dup stringwidth pop -2 div vshift R show } def
/UP { dup vpt_ mul /vpt exch def hpt_ mul /hpt exch def
  /hpt2 hpt 2 mul def /vpt2 vpt 2 mul def } def
/DL { Color {setrgbcolor Solid {pop []} if 0 setdash }
 {pop pop pop 0 setgray Solid {pop []} if 0 setdash} ifelse } def
/BL { stroke userlinewidth 2 mul setlinewidth
      Rounded { 1 setlinejoin 1 setlinecap } if } def
/AL { stroke userlinewidth 2 div setlinewidth
      Rounded { 1 setlinejoin 1 setlinecap } if } def
/UL { dup gnulinewidth mul /userlinewidth exch def
      dup 1 lt {pop 1} if 10 mul /udl exch def } def
/PL { stroke userlinewidth setlinewidth
      Rounded { 1 setlinejoin 1 setlinecap } if } def
/fatlinewidth 7.500 def
/FL { stroke fatlinewidth setlinewidth Rounded { 1 setlinejoin 1 setlinecap } if } def/LTw { PL [] 1 setgray } def
/LTb { BL [] 0 0 0 DL } def
/LTa { FL [1 udl mul 2 udl mul] 0 setdash 0 0 0 setrgbcolor } def
/LT0 { FL [] 1 0 0 DL } def
/LT1 { PL [4 dl 2 dl] 0 1 0 DL } def
/LT2 { PL [2 dl 3 dl] 0 0 1 DL } def
/LT3 { PL [1 dl 1.5 dl] 1 0 1 DL } def
/LT4 { PL [5 dl 2 dl 1 dl 2 dl] 0 1 1 DL } def
/LT5 { PL [4 dl 3 dl 1 dl 3 dl] 1 1 0 DL } def
/LT6 { PL [2 dl 2 dl 2 dl 4 dl] 0 0 0 DL } def
/LT7 { PL [2 dl 2 dl 2 dl 2 dl 2 dl 4 dl] 1 0.3 0 DL } def
/LT8 { PL [2 dl 2 dl 2 dl 2 dl 2 dl 2 dl 2 dl 4 dl] 0.5 0.5 0.5 DL } def
/Pnt { stroke [] 0 setdash
   gsave 1 setlinecap M 0 0 V stroke grestore } def
/Dia { stroke [] 0 setdash 2 copy vpt add M
  hpt neg vpt neg V hpt vpt neg V
  hpt vpt V hpt neg vpt V closepath stroke
  Pnt } def
/Pls { stroke [] 0 setdash vpt sub M 0 vpt2 V
  currentpoint stroke M
  hpt neg vpt neg R hpt2 0 V stroke
  } def
/Box { stroke [] 0 setdash 2 copy exch hpt sub exch vpt add M
  0 vpt2 neg V hpt2 0 V 0 vpt2 V
  hpt2 neg 0 V closepath stroke
  Pnt } def
/Crs { stroke [] 0 setdash exch hpt sub exch vpt add M
  hpt2 vpt2 neg V currentpoint stroke M
  hpt2 neg 0 R hpt2 vpt2 V stroke } def
/TriU { stroke [] 0 setdash 2 copy vpt 1.12 mul add M
  hpt neg vpt -1.62 mul V
  hpt 2 mul 0 V
  hpt neg vpt 1.62 mul V closepath stroke
  Pnt  } def
/Star { 2 copy Pls Crs } def
/BoxF { stroke [] 0 setdash exch hpt sub exch vpt add M
  0 vpt2 neg V  hpt2 0 V  0 vpt2 V
  hpt2 neg 0 V  closepath fill } def
/TriUF { stroke [] 0 setdash vpt 1.12 mul add M
  hpt neg vpt -1.62 mul V
  hpt 2 mul 0 V
  hpt neg vpt 1.62 mul V closepath fill } def
/TriD { stroke [] 0 setdash 2 copy vpt 1.12 mul sub M
  hpt neg vpt 1.62 mul V
  hpt 2 mul 0 V
  hpt neg vpt -1.62 mul V closepath stroke
  Pnt  } def
/TriDF { stroke [] 0 setdash vpt 1.12 mul sub M
  hpt neg vpt 1.62 mul V
  hpt 2 mul 0 V
  hpt neg vpt -1.62 mul V closepath fill} def
/DiaF { stroke [] 0 setdash vpt add M
  hpt neg vpt neg V hpt vpt neg V
  hpt vpt V hpt neg vpt V closepath fill } def
/Pent { stroke [] 0 setdash 2 copy gsave
  translate 0 hpt M 4 {72 rotate 0 hpt L} repeat
  closepath stroke grestore Pnt } def
/PentF { stroke [] 0 setdash gsave
  translate 0 hpt M 4 {72 rotate 0 hpt L} repeat
  closepath fill grestore } def
/Circle { stroke [] 0 setdash 2 copy
  hpt 0 360 arc stroke Pnt } def
/CircleF { stroke [] 0 setdash hpt 0 360 arc fill } def
/C0 { BL [] 0 setdash 2 copy moveto vpt 90 450  arc } bind def
/C1 { BL [] 0 setdash 2 copy        moveto
       2 copy  vpt 0 90 arc closepath fill
               vpt 0 360 arc closepath } bind def
/C2 { BL [] 0 setdash 2 copy moveto
       2 copy  vpt 90 180 arc closepath fill
               vpt 0 360 arc closepath } bind def
/C3 { BL [] 0 setdash 2 copy moveto
       2 copy  vpt 0 180 arc closepath fill
               vpt 0 360 arc closepath } bind def
/C4 { BL [] 0 setdash 2 copy moveto
       2 copy  vpt 180 270 arc closepath fill
               vpt 0 360 arc closepath } bind def
/C5 { BL [] 0 setdash 2 copy moveto
       2 copy  vpt 0 90 arc
       2 copy moveto
       2 copy  vpt 180 270 arc closepath fill
               vpt 0 360 arc } bind def
/C6 { BL [] 0 setdash 2 copy moveto
      2 copy  vpt 90 270 arc closepath fill
              vpt 0 360 arc closepath } bind def
/C7 { BL [] 0 setdash 2 copy moveto
      2 copy  vpt 0 270 arc closepath fill
              vpt 0 360 arc closepath } bind def
/C8 { BL [] 0 setdash 2 copy moveto
      2 copy vpt 270 360 arc closepath fill
              vpt 0 360 arc closepath } bind def
/C9 { BL [] 0 setdash 2 copy moveto
      2 copy  vpt 270 450 arc closepath fill
              vpt 0 360 arc closepath } bind def
/C10 { BL [] 0 setdash 2 copy 2 copy moveto vpt 270 360 arc closepath fill
       2 copy moveto
       2 copy vpt 90 180 arc closepath fill
               vpt 0 360 arc closepath } bind def
/C11 { BL [] 0 setdash 2 copy moveto
       2 copy  vpt 0 180 arc closepath fill
       2 copy moveto
       2 copy  vpt 270 360 arc closepath fill
               vpt 0 360 arc closepath } bind def
/C12 { BL [] 0 setdash 2 copy moveto
       2 copy  vpt 180 360 arc closepath fill
               vpt 0 360 arc closepath } bind def
/C13 { BL [] 0 setdash  2 copy moveto
       2 copy  vpt 0 90 arc closepath fill
       2 copy moveto
       2 copy  vpt 180 360 arc closepath fill
               vpt 0 360 arc closepath } bind def
/C14 { BL [] 0 setdash 2 copy moveto
       2 copy  vpt 90 360 arc closepath fill
               vpt 0 360 arc } bind def
/C15 { BL [] 0 setdash 2 copy vpt 0 360 arc closepath fill
               vpt 0 360 arc closepath } bind def
/Rec   { newpath 4 2 roll moveto 1 index 0 rlineto 0 exch rlineto
       neg 0 rlineto closepath } bind def
/Square { dup Rec } bind def
/Bsquare { vpt sub exch vpt sub exch vpt2 Square } bind def
/S0 { BL [] 0 setdash 2 copy moveto 0 vpt rlineto BL Bsquare } bind def
/S1 { BL [] 0 setdash 2 copy vpt Square fill Bsquare } bind def
/S2 { BL [] 0 setdash 2 copy exch vpt sub exch vpt Square fill Bsquare } bind def
/S3 { BL [] 0 setdash 2 copy exch vpt sub exch vpt2 vpt Rec fill Bsquare } bind def
/S4 { BL [] 0 setdash 2 copy exch vpt sub exch vpt sub vpt Square fill Bsquare } bind def
/S5 { BL [] 0 setdash 2 copy 2 copy vpt Square fill
       exch vpt sub exch vpt sub vpt Square fill Bsquare } bind def
/S6 { BL [] 0 setdash 2 copy exch vpt sub exch vpt sub vpt vpt2 Rec fill Bsquare } bind def
/S7 { BL [] 0 setdash 2 copy exch vpt sub exch vpt sub vpt vpt2 Rec fill
       2 copy vpt Square fill
       Bsquare } bind def
/S8 { BL [] 0 setdash 2 copy vpt sub vpt Square fill Bsquare } bind def
/S9 { BL [] 0 setdash 2 copy vpt sub vpt vpt2 Rec fill Bsquare } bind def
/S10 { BL [] 0 setdash 2 copy vpt sub vpt Square fill 2 copy exch vpt sub exch vpt Square fill
       Bsquare } bind def
/S11 { BL [] 0 setdash 2 copy vpt sub vpt Square fill 2 copy exch vpt sub exch vpt2 vpt Rec fill
       Bsquare } bind def
/S12 { BL [] 0 setdash 2 copy exch vpt sub exch vpt sub vpt2 vpt Rec fill Bsquare } bind def
/S13 { BL [] 0 setdash 2 copy exch vpt sub exch vpt sub vpt2 vpt Rec fill
       2 copy vpt Square fill Bsquare } bind def
/S14 { BL [] 0 setdash 2 copy exch vpt sub exch vpt sub vpt2 vpt Rec fill
       2 copy exch vpt sub exch vpt Square fill Bsquare } bind def
/S15 { BL [] 0 setdash 2 copy Bsquare fill Bsquare } bind def
/D0 { gsave translate 45 rotate 0 0 S0 stroke grestore } bind def
/D1 { gsave translate 45 rotate 0 0 S1 stroke grestore } bind def
/D2 { gsave translate 45 rotate 0 0 S2 stroke grestore } bind def
/D3 { gsave translate 45 rotate 0 0 S3 stroke grestore } bind def
/D4 { gsave translate 45 rotate 0 0 S4 stroke grestore } bind def
/D5 { gsave translate 45 rotate 0 0 S5 stroke grestore } bind def
/D6 { gsave translate 45 rotate 0 0 S6 stroke grestore } bind def
/D7 { gsave translate 45 rotate 0 0 S7 stroke grestore } bind def
/D8 { gsave translate 45 rotate 0 0 S8 stroke grestore } bind def
/D9 { gsave translate 45 rotate 0 0 S9 stroke grestore } bind def
/D10 { gsave translate 45 rotate 0 0 S10 stroke grestore } bind def
/D11 { gsave translate 45 rotate 0 0 S11 stroke grestore } bind def
/D12 { gsave translate 45 rotate 0 0 S12 stroke grestore } bind def
/D13 { gsave translate 45 rotate 0 0 S13 stroke grestore } bind def
/D14 { gsave translate 45 rotate 0 0 S14 stroke grestore } bind def
/D15 { gsave translate 45 rotate 0 0 S15 stroke grestore } bind def
/DiaE { stroke [] 0 setdash vpt add M
  hpt neg vpt neg V hpt vpt neg V
  hpt vpt V hpt neg vpt V closepath stroke } def
/BoxE { stroke [] 0 setdash exch hpt sub exch vpt add M
  0 vpt2 neg V hpt2 0 V 0 vpt2 V
  hpt2 neg 0 V closepath stroke } def
/TriUE { stroke [] 0 setdash vpt 1.12 mul add M
  hpt neg vpt -1.62 mul V
  hpt 2 mul 0 V
  hpt neg vpt 1.62 mul V closepath stroke } def
/TriDE { stroke [] 0 setdash vpt 1.12 mul sub M
  hpt neg vpt 1.62 mul V
  hpt 2 mul 0 V
  hpt neg vpt -1.62 mul V closepath stroke } def
/PentE { stroke [] 0 setdash gsave
  translate 0 hpt M 4 {72 rotate 0 hpt L} repeat
  closepath stroke grestore } def
/CircE { stroke [] 0 setdash 
  hpt 0 360 arc stroke } def
/Opaque { gsave closepath 1 setgray fill grestore 0 setgray closepath } def
/DiaW { stroke [] 0 setdash vpt add M
  hpt neg vpt neg V hpt vpt neg V
  hpt vpt V hpt neg vpt V Opaque stroke } def
/BoxW { stroke [] 0 setdash exch hpt sub exch vpt add M
  0 vpt2 neg V hpt2 0 V 0 vpt2 V
  hpt2 neg 0 V Opaque stroke } def
/TriUW { stroke [] 0 setdash vpt 1.12 mul add M
  hpt neg vpt -1.62 mul V
  hpt 2 mul 0 V
  hpt neg vpt 1.62 mul V Opaque stroke } def
/TriDW { stroke [] 0 setdash vpt 1.12 mul sub M
  hpt neg vpt 1.62 mul V
  hpt 2 mul 0 V
  hpt neg vpt -1.62 mul V Opaque stroke } def
/PentW { stroke [] 0 setdash gsave
  translate 0 hpt M 4 {72 rotate 0 hpt L} repeat
  Opaque stroke grestore } def
/CircW { stroke [] 0 setdash 
  hpt 0 360 arc Opaque stroke } def
/BoxFill { gsave Rec 1 setgray fill grestore } def
/BoxColFill {
  gsave Rec
  /Fillden exch def
  currentrgbcolor
  /ColB exch def /ColG exch def /ColR exch def
  /ColR ColR Fillden mul Fillden sub 1 add def
  /ColG ColG Fillden mul Fillden sub 1 add def
  /ColB ColB Fillden mul Fillden sub 1 add def
  ColR ColG ColB setrgbcolor
  fill grestore } def
%
%
/PatternFill { gsave /PFa [ 9 2 roll ] def
    PFa 0 get PFa 2 get 2 div add PFa 1 get PFa 3 get 2 div add translate
    PFa 2 get -2 div PFa 3 get -2 div PFa 2 get PFa 3 get Rec
    gsave 1 setgray fill grestore clip
    currentlinewidth 0.5 mul setlinewidth
    /PFs PFa 2 get dup mul PFa 3 get dup mul add sqrt def
    0 0 M PFa 5 get rotate PFs -2 div dup translate
	0 1 PFs PFa 4 get div 1 add floor cvi
	{ PFa 4 get mul 0 M 0 PFs V } for
    0 PFa 6 get ne {
	0 1 PFs PFa 4 get div 1 add floor cvi
	{ PFa 4 get mul 0 2 1 roll M PFs 0 V } for
    } if
    stroke grestore } def
/Symbol-Oblique /Symbol findfont [1 0 .167 1 0 0] makefont
dup length dict begin {1 index /FID eq {pop pop} {def} ifelse} forall
currentdict end definefont pop
end
gnudict begin
gsave
0 0 translate
0.100 0.100 scale
0 setgray
newpath
0.500 UL
LTb
350 300 M
63 0 V
3317 0 R
-63 0 V
0.500 UL
LTb
350 397 M
31 0 V
3349 0 R
-31 0 V
350 493 M
31 0 V
3349 0 R
-31 0 V
350 590 M
31 0 V
3349 0 R
-31 0 V
350 686 M
31 0 V
3349 0 R
-31 0 V
350 783 M
63 0 V
3317 0 R
-63 0 V
0.500 UL
LTb
350 879 M
31 0 V
3349 0 R
-31 0 V
350 976 M
31 0 V
3349 0 R
-31 0 V
350 1073 M
31 0 V
3349 0 R
-31 0 V
350 1169 M
31 0 V
3349 0 R
-31 0 V
350 1266 M
63 0 V
3317 0 R
-63 0 V
0.500 UL
LTb
350 1362 M
31 0 V
3349 0 R
-31 0 V
350 1459 M
31 0 V
3349 0 R
-31 0 V
350 1555 M
31 0 V
3349 0 R
-31 0 V
350 1652 M
31 0 V
3349 0 R
-31 0 V
350 1749 M
63 0 V
3317 0 R
-63 0 V
0.500 UL
LTb
350 1845 M
31 0 V
3349 0 R
-31 0 V
350 1942 M
31 0 V
3349 0 R
-31 0 V
350 2038 M
31 0 V
3349 0 R
-31 0 V
350 2135 M
31 0 V
3349 0 R
-31 0 V
350 2231 M
63 0 V
3317 0 R
-63 0 V
0.500 UL
LTb
350 2328 M
31 0 V
3349 0 R
-31 0 V
350 2425 M
31 0 V
3349 0 R
-31 0 V
350 2521 M
31 0 V
3349 0 R
-31 0 V
350 2618 M
31 0 V
3349 0 R
-31 0 V
350 2714 M
63 0 V
3317 0 R
-63 0 V
0.500 UL
LTb
350 2811 M
31 0 V
3349 0 R
-31 0 V
350 2907 M
31 0 V
3349 0 R
-31 0 V
350 3004 M
31 0 V
3349 0 R
-31 0 V
350 3101 M
31 0 V
3349 0 R
-31 0 V
350 3197 M
63 0 V
3317 0 R
-63 0 V
0.500 UL
LTb
350 3294 M
31 0 V
3349 0 R
-31 0 V
350 3390 M
31 0 V
3349 0 R
-31 0 V
350 3487 M
31 0 V
3349 0 R
-31 0 V
350 3583 M
31 0 V
3349 0 R
-31 0 V
350 3680 M
63 0 V
3317 0 R
-63 0 V
0.500 UL
LTb
350 300 M
0 63 V
0 3317 R
0 -63 V
0.500 UL
LTb
350 300 M
0 31 V
0 3349 R
0 -31 V
1748 300 M
0 63 V
0 3317 R
0 -63 V
0.500 UL
LTb
350 300 M
0 31 V
0 3349 R
0 -31 V
771 300 M
0 31 V
0 3349 R
0 -31 V
1017 300 M
0 31 V
0 3349 R
0 -31 V
1191 300 M
0 31 V
0 3349 R
0 -31 V
1327 300 M
0 31 V
0 3349 R
0 -31 V
1437 300 M
0 31 V
0 3349 R
0 -31 V
1531 300 M
0 31 V
0 3349 R
0 -31 V
1612 300 M
0 31 V
0 3349 R
0 -31 V
1684 300 M
0 31 V
0 3349 R
0 -31 V
1748 300 M
0 31 V
0 3349 R
0 -31 V
3145 300 M
0 63 V
0 3317 R
0 -63 V
0.500 UL
LTb
1748 300 M
0 31 V
0 3349 R
0 -31 V
2168 300 M
0 31 V
0 3349 R
0 -31 V
2414 300 M
0 31 V
0 3349 R
0 -31 V
2589 300 M
0 31 V
0 3349 R
0 -31 V
2724 300 M
0 31 V
0 3349 R
0 -31 V
2835 300 M
0 31 V
0 3349 R
0 -31 V
2929 300 M
0 31 V
0 3349 R
0 -31 V
3010 300 M
0 31 V
0 3349 R
0 -31 V
3081 300 M
0 31 V
0 3349 R
0 -31 V
3145 300 M
0 31 V
0 3349 R
0 -31 V
0.500 UL
LTb
350 300 M
3380 0 V
0 3380 V
-3380 0 V
350 300 L
LTb
LTb
1.000 UP
1.000 UL
LT0
350 1578 M
58 36 V
53 34 V
48 33 V
45 31 V
42 30 V
39 28 V
37 27 V
35 27 V
33 25 V
31 24 V
29 24 V
29 23 V
27 22 V
25 21 V
25 21 V
24 20 V
23 20 V
22 19 V
21 19 V
21 18 V
20 17 V
19 18 V
19 16 V
18 17 V
17 16 V
17 16 V
17 15 V
16 15 V
16 14 V
15 15 V
15 14 V
15 13 V
14 14 V
14 13 V
14 13 V
13 13 V
13 12 V
13 12 V
13 12 V
12 12 V
12 11 V
12 12 V
11 11 V
12 11 V
11 10 V
11 11 V
10 10 V
11 10 V
10 10 V
10 10 V
11 10 V
9 9 V
10 10 V
10 9 V
9 9 V
9 9 V
9 8 V
9 9 V
9 9 V
9 8 V
9 8 V
8 8 V
9 8 V
8 8 V
8 8 V
8 8 V
8 7 V
8 8 V
7 7 V
8 7 V
8 7 V
7 7 V
7 7 V
8 7 V
7 7 V
7 6 V
7 7 V
7 6 V
7 7 V
7 6 V
6 6 V
7 6 V
6 7 V
7 6 V
6 5 V
7 6 V
6 6 V
6 6 V
6 5 V
7 6 V
6 5 V
6 6 V
5 5 V
6 5 V
6 6 V
6 5 V
6 5 V
5 5 V
6 5 V
5 5 V
6 5 V
5 5 V
6 4 V
5 5 V
stroke
1827 2837 M
5 5 V
6 4 V
5 5 V
5 4 V
5 5 V
5 4 V
5 5 V
5 4 V
5 4 V
5 4 V
5 4 V
5 5 V
5 4 V
4 4 V
5 4 V
5 4 V
4 4 V
5 3 V
5 4 V
4 4 V
5 4 V
4 4 V
5 3 V
4 4 V
4 3 V
5 4 V
4 4 V
4 3 V
5 4 V
4 3 V
4 3 V
4 4 V
4 3 V
4 3 V
5 4 V
4 3 V
4 3 V
4 3 V
4 3 V
4 3 V
4 4 V
3 3 V
4 3 V
4 3 V
4 3 V
4 3 V
4 3 V
3 2 V
4 3 V
4 3 V
3 3 V
4 3 V
4 2 V
3 3 V
4 3 V
4 3 V
3 2 V
4 3 V
3 3 V
4 2 V
3 3 V
4 2 V
3 3 V
4 2 V
3 3 V
3 2 V
4 3 V
3 2 V
3 2 V
4 3 V
3 2 V
3 2 V
3 3 V
4 2 V
3 2 V
3 3 V
3 2 V
3 2 V
4 2 V
3 2 V
3 3 V
3 2 V
3 2 V
3 2 V
3 2 V
3 2 V
3 2 V
3 2 V
3 2 V
3 2 V
3 2 V
3 2 V
3 2 V
3 2 V
3 2 V
3 2 V
3 2 V
3 2 V
2 1 V
3 2 V
3 2 V
3 2 V
3 2 V
3 1 V
stroke
2221 3140 M
2 2 V
3 2 V
3 2 V
3 1 V
2 2 V
3 2 V
3 1 V
2 2 V
3 2 V
3 1 V
2 2 V
3 2 V
3 1 V
2 2 V
3 1 V
3 2 V
2 1 V
3 2 V
2 1 V
3 2 V
2 1 V
3 2 V
2 1 V
3 2 V
2 1 V
3 1 V
2 2 V
3 1 V
2 2 V
3 1 V
2 1 V
3 2 V
2 1 V
3 1 V
2 2 V
2 1 V
3 1 V
2 1 V
2 2 V
3 1 V
2 1 V
2 1 V
3 2 V
2 1 V
2 1 V
3 1 V
2 1 V
2 2 V
3 1 V
2 1 V
2 1 V
2 1 V
3 1 V
2 1 V
2 1 V
2 1 V
3 1 V
2 1 V
2 2 V
2 1 V
2 1 V
2 1 V
3 1 V
2 1 V
2 1 V
2 1 V
2 0 V
2 1 V
2 1 V
3 1 V
2 1 V
2 1 V
2 1 V
2 1 V
2 1 V
2 1 V
2 1 V
2 0 V
2 1 V
2 1 V
2 1 V
2 1 V
2 1 V
2 0 V
2 1 V
2 1 V
2 1 V
2 0 V
2 1 V
2 1 V
2 1 V
2 0 V
2 1 V
2 1 V
2 1 V
2 0 V
2 1 V
2 1 V
2 0 V
2 1 V
2 1 V
2 0 V
1 1 V
2 0 V
stroke
2457 3258 M
2 1 V
2 1 V
2 0 V
2 1 V
2 0 V
1 1 V
2 1 V
2 0 V
2 1 V
2 0 V
2 1 V
1 0 V
2 1 V
2 0 V
2 1 V
2 0 V
2 1 V
1 0 V
2 1 V
2 0 V
2 1 V
1 0 V
2 1 V
2 0 V
2 1 V
1 0 V
2 0 V
2 1 V
2 0 V
1 1 V
2 0 V
2 0 V
2 1 V
1 0 V
2 0 V
2 1 V
1 0 V
2 1 V
2 0 V
1 0 V
2 0 V
2 1 V
1 0 V
2 0 V
2 1 V
1 0 V
2 0 V
2 1 V
1 0 V
2 0 V
2 0 V
1 1 V
2 0 V
1 0 V
2 0 V
2 0 V
1 1 V
2 0 V
1 0 V
2 0 V
2 0 V
1 1 V
2 0 V
1 0 V
2 0 V
1 0 V
2 0 V
2 1 V
1 0 V
2 0 V
1 0 V
2 0 V
1 0 V
2 0 V
1 0 V
2 1 V
1 0 V
2 0 V
1 0 V
2 0 V
1 0 V
2 0 V
1 0 V
2 0 V
1 0 V
2 0 V
1 0 V
2 0 V
1 0 V
2 0 V
1 0 V
2 0 V
1 0 V
2 0 V
1 0 V
2 0 V
1 0 V
2 0 V
1 0 V
1 0 V
2 0 V
1 0 V
2 0 V
1 0 V
stroke
2627 3285 M
2 0 V
1 0 V
1 0 V
2 0 V
1 0 V
2 0 V
1 0 V
1 -1 V
2 0 V
1 0 V
2 0 V
1 0 V
1 0 V
2 0 V
1 0 V
2 -1 V
1 0 V
1 0 V
2 0 V
1 0 V
1 0 V
2 0 V
1 -1 V
1 0 V
2 0 V
1 0 V
1 0 V
2 -1 V
1 0 V
1 0 V
2 0 V
1 0 V
1 -1 V
2 0 V
1 0 V
1 0 V
2 0 V
1 -1 V
1 0 V
2 0 V
1 0 V
1 -1 V
2 0 V
1 0 V
1 0 V
1 -1 V
2 0 V
1 0 V
1 -1 V
2 0 V
1 0 V
1 0 V
1 -1 V
2 0 V
1 0 V
1 -1 V
1 0 V
2 0 V
1 -1 V
1 0 V
1 0 V
2 -1 V
1 0 V
1 0 V
1 -1 V
2 0 V
1 0 V
1 -1 V
1 0 V
2 0 V
1 -1 V
1 0 V
1 -1 V
1 0 V
2 0 V
1 -1 V
1 0 V
1 0 V
1 -1 V
2 0 V
1 -1 V
1 0 V
1 -1 V
1 0 V
2 0 V
1 -1 V
1 0 V
1 -1 V
1 0 V
1 0 V
2 -1 V
1 0 V
1 -1 V
1 0 V
1 -1 V
2 0 V
1 -1 V
1 0 V
1 -1 V
1 0 V
1 -1 V
1 0 V
2 0 V
1 -1 V
stroke
2760 3255 M
1 0 V
1 -1 V
1 0 V
1 -1 V
1 0 V
2 -1 V
1 0 V
1 -1 V
1 0 V
1 -1 V
1 0 V
1 -1 V
1 0 V
2 -1 V
1 -1 V
1 0 V
1 -1 V
1 0 V
1 -1 V
1 0 V
1 -1 V
1 0 V
2 -1 V
1 0 V
1 -1 V
1 0 V
1 -1 V
1 -1 V
1 0 V
1 -1 V
1 0 V
1 -1 V
1 0 V
1 -1 V
2 -1 V
1 0 V
1 -1 V
1 0 V
1 -1 V
1 -1 V
1 0 V
1 -1 V
1 0 V
1 -1 V
1 -1 V
1 0 V
1 -1 V
1 0 V
1 -1 V
1 -1 V
2 0 V
1 -1 V
1 0 V
1 -1 V
1 -1 V
1 0 V
1 -1 V
1 -1 V
1 0 V
1 -1 V
1 -1 V
1 0 V
1 -1 V
1 0 V
1 -1 V
1 -1 V
1 0 V
1 -1 V
1 -1 V
1 0 V
1 -1 V
1 -1 V
1 0 V
1 -1 V
1 -1 V
1 0 V
1 -1 V
1 -1 V
1 0 V
1 -1 V
1 -1 V
1 0 V
1 -1 V
1 -1 V
1 -1 V
1 0 V
1 -1 V
1 -1 V
1 0 V
1 -1 V
1 -1 V
1 0 V
1 -1 V
1 -1 V
1 -1 V
1 0 V
1 -1 V
1 -1 V
1 0 V
1 -1 V
1 -1 V
1 -1 V
1 0 V
0 -1 V
stroke
2868 3191 M
1 -1 V
1 0 V
1 -1 V
1 -1 V
1 -1 V
1 0 V
1 -1 V
1 -1 V
1 -1 V
1 0 V
1 -1 V
1 -1 V
1 -1 V
1 0 V
1 -1 V
1 -1 V
1 -1 V
1 -1 V
1 -1 V
1 -1 V
1 0 V
1 -1 V
1 -1 V
1 -1 V
1 0 V
1 -1 V
1 -1 V
1 -1 V
1 0 V
0 -1 V
1 -1 V
1 -1 V
1 0 V
1 -1 V
1 -1 V
1 -1 V
1 -1 V
1 0 V
1 -1 V
1 -1 V
0 -1 V
1 0 V
1 -1 V
1 -1 V
1 -1 V
1 -1 V
1 0 V
1 -1 V
1 -1 V
0 -1 V
1 -1 V
1 0 V
1 -1 V
1 -1 V
1 -1 V
1 -1 V
1 0 V
0 -1 V
1 -1 V
1 -1 V
1 -1 V
1 0 V
1 -1 V
1 -1 V
1 -1 V
0 -1 V
1 0 V
1 -1 V
1 -1 V
1 -1 V
1 -1 V
1 0 V
0 -1 V
1 -1 V
1 -1 V
1 -1 V
1 -1 V
1 0 V
1 -1 V
0 -1 V
1 -1 V
1 -1 V
1 0 V
1 -1 V
1 -1 V
1 -1 V
0 -1 V
1 -1 V
1 0 V
1 -1 V
1 -1 V
1 -1 V
0 -1 V
1 -1 V
1 0 V
1 -1 V
1 -1 V
1 -1 V
0 -1 V
1 -1 V
1 0 V
1 -1 V
1 -1 V
1 -1 V
stroke
2962 3108 M
0 -1 V
1 -1 V
1 -1 V
1 0 V
1 -1 V
0 -1 V
1 -1 V
1 -1 V
1 -1 V
1 0 V
0 -1 V
1 -1 V
1 -1 V
1 -1 V
1 -1 V
1 0 V
0 -1 V
1 -1 V
1 -1 V
1 -1 V
1 -1 V
0 -1 V
1 0 V
1 -1 V
1 -1 V
0 -1 V
1 -1 V
1 -1 V
1 -1 V
1 0 V
0 -1 V
1 -1 V
1 -1 V
1 -1 V
1 -1 V
0 -1 V
1 0 V
1 -1 V
1 -1 V
0 -1 V
1 -1 V
1 -1 V
1 -1 V
1 0 V
0 -1 V
1 -1 V
1 -1 V
1 -1 V
0 -1 V
1 -1 V
1 0 V
1 -1 V
1 -1 V
0 -1 V
1 -1 V
1 -1 V
1 -1 V
0 -1 V
1 0 V
1 -1 V
1 -1 V
0 -1 V
1 -1 V
1 -1 V
1 -1 V
1 -1 V
1 -1 V
1 -1 V
0 -1 V
1 -1 V
1 -1 V
1 -1 V
1 -1 V
1 -1 V
1 -1 V
0 -1 V
1 -1 V
1 -1 V
1 0 V
0 -1 V
1 -1 V
1 -1 V
1 -1 V
0 -1 V
1 -1 V
1 -1 V
1 0 V
0 -1 V
1 -1 V
1 -1 V
0 -1 V
1 -1 V
1 -1 V
1 -1 V
1 -1 V
1 -1 V
1 -1 V
0 -1 V
1 -1 V
1 -1 V
0 -1 V
1 0 V
1 -1 V
1 -1 V
stroke
3044 3016 M
0 -1 V
1 -1 V
1 -1 V
0 -1 V
1 -1 V
1 0 V
1 -1 V
0 -1 V
1 -1 V
1 -1 V
0 -1 V
1 -1 V
1 0 V
1 -1 V
0 -1 V
1 -1 V
1 -1 V
0 -1 V
1 -1 V
1 -1 V
1 -1 V
1 -1 V
1 -1 V
0 -1 V
1 -1 V
1 -1 V
0 -1 V
1 0 V
1 -1 V
0 -1 V
1 -1 V
1 -1 V
0 -1 V
1 -1 V
1 -1 V
1 0 V
0 -1 V
1 -1 V
1 -1 V
0 -1 V
1 -1 V
1 -1 V
1 -1 V
1 -1 V
0 -1 V
1 -1 V
1 -1 V
0 -1 V
1 -1 V
1 0 V
0 -1 V
1 -1 V
1 -1 V
0 -1 V
1 -1 V
1 -1 V
1 -1 V
1 -1 V
1 -1 V
0 -1 V
1 -1 V
1 -1 V
0 -1 V
1 0 V
1 -1 V
0 -1 V
1 -1 V
0 -1 V
1 -1 V
1 -1 V
1 -1 V
1 -1 V
0 -1 V
1 -1 V
1 -1 V
0 -1 V
1 0 V
1 -1 V
0 -1 V
1 -1 V
1 -1 V
0 -1 V
1 -1 V
1 0 V
0 -1 V
1 -1 V
1 -1 V
0 -1 V
1 -1 V
1 -1 V
1 -1 V
0 -1 V
1 -1 V
1 -1 V
0 -1 V
1 -1 V
1 0 V
0 -1 V
1 -1 V
1 -1 V
0 -1 V
1 -1 V
0 -1 V
1 0 V
stroke
3117 2922 M
1 -1 V
0 -1 V
1 -1 V
1 -1 V
0 -1 V
1 0 V
1 -1 V
0 -1 V
1 -1 V
0 -1 V
1 -1 V
1 -1 V
1 -1 V
1 -1 V
0 -1 V
1 -1 V
0 -1 V
1 -1 V
1 0 V
0 -1 V
1 -1 V
1 -1 V
0 -1 V
1 -1 V
1 -1 V
1 -1 V
0 -1 V
1 -1 V
1 -1 V
1 -1 V
0 -1 V
1 -1 V
1 -1 V
0 -1 V
1 0 V
0 -1 V
1 -1 V
1 -1 V
0 -1 V
1 -1 V
1 -1 V
1 -1 V
0 -1 V
1 -1 V
1 -1 V
1 -1 V
0 -1 V
1 -1 V
1 -1 V
0 -1 V
1 0 V
0 -1 V
1 -1 V
1 -1 V
0 -1 V
1 -1 V
1 -1 V
1 -1 V
0 -1 V
1 -1 V
0 -1 V
1 0 V
0 -1 V
1 -1 V
1 -1 V
0 -1 V
1 0 V
0 -1 V
1 -1 V
1 -1 V
0 -1 V
1 -1 V
1 -1 V
1 -1 V
0 -1 V
1 -1 V
1 -1 V
0 -1 V
1 -1 V
1 -1 V
0 -1 V
1 0 V
0 -1 V
1 -1 V
1 -1 V
0 -1 V
1 0 V
0 -1 V
1 -1 V
0 -1 V
1 -1 V
1 0 V
0 -1 V
1 -1 V
0 -1 V
1 -1 V
1 -1 V
1 -1 V
0 -1 V
1 -1 V
1 -1 V
0 -1 V
1 -1 V
1 -1 V
stroke
3186 2827 M
1 -1 V
0 -1 V
1 -1 V
0 -1 V
1 0 V
1 -1 V
0 -1 V
1 -1 V
0 -1 V
1 0 V
0 -1 V
1 -1 V
0 -1 V
1 -1 V
1 0 V
0 -1 V
1 -1 V
0 -1 V
1 -1 V
1 -1 V
0 -1 V
1 -1 V
1 0 V
0 -1 V
1 -1 V
0 -1 V
1 -1 V
1 -1 V
0 -1 V
1 -1 V
1 0 V
0 -1 V
1 -1 V
0 -1 V
1 -1 V
1 -1 V
0 -1 V
1 -1 V
1 0 V
0 -1 V
1 -1 V
0 -1 V
1 -1 V
1 -1 V
0 -1 V
1 -1 V
1 -1 V
0 -1 V
1 -1 V
1 0 V
0 -1 V
1 -1 V
0 -1 V
1 -1 V
1 -1 V
0 -1 V
1 -1 V
1 -1 V
0 -1 V
1 -1 V
1 0 V
0 -1 V
1 -1 V
0 -1 V
1 0 V
0 -1 V
1 -1 V
0 -1 V
1 0 V
0 -1 V
1 -1 V
0 -1 V
1 0 V
0 -1 V
1 -1 V
0 -1 V
1 0 V
1 -1 V
0 -1 V
1 -1 V
1 -1 V
0 -1 V
1 -1 V
1 -1 V
0 -1 V
1 -1 V
1 -1 V
0 -1 V
1 -1 V
1 -1 V
0 -1 V
1 -1 V
1 -1 V
0 -1 V
1 -1 V
1 0 V
0 -1 V
1 -1 V
0 -1 V
1 0 V
0 -1 V
1 -1 V
1 -1 V
0 -1 V
stroke
3249 2737 M
1 -1 V
1 -1 V
0 -1 V
1 -1 V
1 -1 V
0 -1 V
1 0 V
0 -1 V
1 -1 V
0 -1 V
1 0 V
0 -1 V
1 -1 V
0 -1 V
1 0 V
0 -1 V
1 -1 V
1 -1 V
0 -1 V
1 -1 V
1 -1 V
0 -1 V
1 0 V
0 -1 V
1 -1 V
0 -1 V
1 0 V
0 -1 V
1 -1 V
1 -1 V
0 -1 V
1 -1 V
1 -1 V
0 -1 V
1 0 V
0 -1 V
1 -1 V
1 -1 V
0 -1 V
1 -1 V
1 -1 V
0 -1 V
1 0 V
0 -1 V
1 -1 V
1 -1 V
0 -1 V
1 -1 V
1 -1 V
0 -1 V
1 0 V
0 -1 V
1 -1 V
0 -1 V
1 -1 V
0 -1 V
1 0 V
0 -1 V
1 -1 V
1 -1 V
0 -1 V
1 0 V
0 -1 V
1 -1 V
1 -1 V
0 -1 V
1 0 V
0 -1 V
1 -1 V
1 -1 V
0 -1 V
1 -1 V
1 -1 V
0 -1 V
1 -1 V
0 -1 V
1 0 V
0 -1 V
1 -1 V
1 -1 V
0 -1 V
1 0 V
0 -1 V
1 -1 V
1 -1 V
0 -1 V
1 0 V
0 -1 V
1 -1 V
0 -1 V
1 -1 V
1 -1 V
0 -1 V
1 0 V
0 -1 V
1 -1 V
1 -1 V
0 -1 V
1 0 V
0 -1 V
1 -1 V
0 -1 V
1 -1 V
1 -1 V
stroke
3311 2649 M
1 -1 V
0 -1 V
1 0 V
0 -1 V
1 -1 V
1 -1 V
0 -1 V
1 -1 V
0 -1 V
1 0 V
0 -1 V
1 -1 V
1 -1 V
1 -1 V
0 -1 V
1 -1 V
0 -1 V
1 0 V
0 -1 V
1 -1 V
1 -1 V
0 -1 V
1 0 V
0 -1 V
1 -1 V
0 -1 V
1 0 V
0 -1 V
1 -1 V
1 -1 V
0 -1 V
1 -1 V
1 -1 V
0 -1 V
1 0 V
0 -1 V
1 -1 V
0 -1 V
1 0 V
0 -1 V
1 -1 V
1 -1 V
0 -1 V
1 0 V
0 -1 V
1 -1 V
0 -1 V
1 0 V
0 -1 V
1 0 V
0 -1 V
1 -1 V
0 -1 V
1 -1 V
1 -1 V
1 -1 V
0 -1 V
1 0 V
0 -1 V
1 -1 V
0 -1 V
1 0 V
0 -1 V
1 0 V
0 -1 V
0 -1 V
1 0 V
0 -1 V
1 0 V
0 -1 V
1 -1 V
0 -1 V
1 -1 V
1 -1 V
1 -1 V
0 -1 V
1 -1 V
1 -1 V
1 -1 V
0 -1 V
1 -1 V
1 -1 V
0 -1 V
1 0 V
0 -1 V
1 -1 V
0 -1 V
1 0 V
0 -1 V
1 0 V
0 -1 V
0 -1 V
1 0 V
0 -1 V
1 0 V
0 -1 V
1 0 V
0 -1 V
0 -1 V
1 0 V
0 -1 V
1 0 V
0 -1 V
1 -1 V
stroke
3369 2568 M
0 -1 V
1 0 V
0 -1 V
1 0 V
0 -1 V
0 -1 V
1 0 V
0 -1 V
1 0 V
0 -1 V
1 0 V
0 -1 V
0 -1 V
1 0 V
0 -1 V
1 0 V
0 -1 V
1 -1 V
0 -1 V
1 0 V
0 -1 V
1 -1 V
1 -1 V
0 -1 V
1 0 V
0 -1 V
1 -1 V
1 -1 V
0 -1 V
1 -1 V
1 -1 V
0 -1 V
1 0 V
0 -1 V
1 0 V
0 -1 V
1 -1 V
0 -1 V
1 0 V
0 -1 V
1 0 V
0 -1 V
1 -1 V
0 -1 V
1 0 V
0 -1 V
1 -1 V
1 -1 V
0 -1 V
1 0 V
0 -1 V
1 -1 V
0 -1 V
1 0 V
0 -1 V
1 0 V
0 -1 V
1 -1 V
1 -1 V
0 -1 V
1 -1 V
1 -1 V
0 -1 V
1 0 V
0 -1 V
1 0 V
0 -1 V
1 -1 V
1 -1 V
0 -1 V
1 -1 V
1 -1 V
0 -1 V
1 0 V
0 -1 V
1 0 V
0 -1 V
1 -1 V
0 -1 V
1 0 V
0 -1 V
1 0 V
0 -1 V
0 -1 V
1 0 V
0 -1 V
1 0 V
0 -1 V
1 -1 V
1 -1 V
0 -1 V
1 0 V
0 -1 V
1 -1 V
1 -1 V
0 -1 V
1 0 V
0 -1 V
1 0 V
0 -1 V
1 -1 V
0 -1 V
1 0 V
0 -1 V
stroke
3424 2493 M
1 0 V
0 -1 V
1 -1 V
1 -1 V
0 -1 V
1 -1 V
0 -1 V
1 0 V
0 -1 V
1 0 V
0 -1 V
1 -1 V
0 -1 V
1 0 V
0 -1 V
1 0 V
0 -1 V
1 -1 V
1 -1 V
0 -1 V
1 0 V
0 -1 V
1 0 V
0 -1 V
1 0 V
0 -1 V
1 -1 V
0 -1 V
1 0 V
0 -1 V
1 0 V
0 -1 V
1 -1 V
0 -1 V
1 0 V
0 -1 V
1 -1 V
1 -1 V
0 -1 V
1 0 V
0 -1 V
1 -1 V
1 -1 V
0 -1 V
1 0 V
0 -1 V
1 0 V
0 -1 V
1 -1 V
1 -1 V
0 -1 V
1 0 V
0 -1 V
1 -1 V
1 -1 V
0 -1 V
1 0 V
0 -1 V
1 0 V
0 -1 V
1 -1 V
1 -1 V
0 -1 V
1 0 V
0 -1 V
1 -1 V
1 -1 V
0 -1 V
1 0 V
0 -1 V
1 0 V
0 -1 V
1 -1 V
0 -1 V
1 0 V
0 -1 V
1 0 V
0 -1 V
1 -1 V
1 -1 V
0 -1 V
1 0 V
0 -1 V
1 0 V
0 -1 V
1 -1 V
0 -1 V
1 0 V
0 -1 V
1 0 V
0 -1 V
1 0 V
0 -1 V
1 -1 V
0 -1 V
1 0 V
0 -1 V
1 0 V
0 -1 V
1 -1 V
1 -1 V
1 -1 V
0 -1 V
1 0 V
stroke
3482 2419 M
0 -1 V
1 -1 V
1 -1 V
0 -1 V
1 0 V
0 -1 V
1 0 V
0 -1 V
1 -1 V
1 -1 V
0 -1 V
1 0 V
0 -1 V
1 0 V
0 -1 V
1 -1 V
1 -1 V
0 -1 V
1 0 V
0 -1 V
1 0 V
0 -1 V
1 -1 V
0 -1 V
1 0 V
0 -1 V
1 0 V
0 -1 V
1 0 V
0 -1 V
1 -1 V
0 -1 V
1 0 V
0 -1 V
1 0 V
0 -1 V
1 0 V
0 -1 V
1 0 V
0 -1 V
1 -1 V
1 -1 V
0 -1 V
1 0 V
0 -1 V
1 0 V
0 -1 V
1 -1 V
1 -1 V
0 -1 V
1 0 V
0 -1 V
1 0 V
0 -1 V
1 0 V
0 -1 V
1 -1 V
1 -1 V
0 -1 V
1 0 V
0 -1 V
1 0 V
0 -1 V
1 -1 V
0 -1 V
1 0 V
0 -1 V
1 0 V
0 -1 V
1 0 V
0 -1 V
1 -1 V
1 -1 V
0 -1 V
1 0 V
0 -1 V
1 0 V
0 -1 V
1 0 V
0 -1 V
1 -1 V
1 -1 V
0 -1 V
1 0 V
0 -1 V
1 0 V
0 -1 V
1 0 V
0 -1 V
1 -1 V
0 -1 V
1 0 V
0 -1 V
1 0 V
0 -1 V
1 0 V
0 -1 V
1 0 V
0 -1 V
1 -1 V
1 -1 V
1 -1 V
0 -1 V
1 0 V
stroke
3539 2349 M
0 -1 V
1 -1 V
1 -1 V
1 -1 V
0 -1 V
1 0 V
0 -1 V
1 0 V
0 -1 V
1 0 V
0 -1 V
1 -1 V
1 -1 V
1 -1 V
0 -1 V
1 0 V
0 -1 V
1 0 V
0 -1 V
1 -1 V
1 -1 V
1 -1 V
1 -1 V
0 -1 V
1 0 V
0 -1 V
1 0 V
0 -1 V
1 0 V
0 -1 V
1 0 V
0 -1 V
0 -1 V
1 0 V
0 -1 V
1 0 V
0 -1 V
1 0 V
0 -1 V
1 0 V
0 -1 V
1 -1 V
1 -1 V
1 -1 V
0 -1 V
1 0 V
0 -1 V
1 0 V
0 -1 V
1 0 V
0 -1 V
1 0 V
0 -1 V
1 -1 V
1 -1 V
1 -1 V
0 -1 V
1 0 V
0 -1 V
1 0 V
0 -1 V
1 0 V
0 -1 V
1 -1 V
1 -1 V
1 -1 V
0 -1 V
1 0 V
0 -1 V
1 0 V
0 -1 V
1 0 V
0 -1 V
1 0 V
0 -1 V
1 0 V
0 -1 V
1 -1 V
1 -1 V
0 -1 V
1 0 V
0 -1 V
1 0 V
0 -1 V
1 0 V
0 -1 V
1 0 V
0 -1 V
1 0 V
0 -1 V
1 -1 V
1 -1 V
0 -1 V
1 0 V
0 -1 V
1 0 V
0 -1 V
1 0 V
0 -1 V
1 0 V
0 -1 V
1 0 V
0 -1 V
1 0 V
stroke
3598 2281 M
0 -1 V
1 0 V
0 -1 V
1 -1 V
1 -1 V
0 -1 V
1 0 V
0 -1 V
1 0 V
0 -1 V
1 0 V
0 -1 V
1 0 V
0 -1 V
1 0 V
0 -1 V
1 0 V
0 -1 V
1 -1 V
1 0 V
0 -1 V
1 -1 V
1 -1 V
0 -1 V
1 0 V
0 -1 V
1 0 V
0 -1 V
1 0 V
0 -1 V
1 0 V
0 -1 V
1 0 V
0 -1 V
1 0 V
0 -1 V
1 0 V
0 -1 V
1 -1 V
1 -1 V
1 -1 V
0 -1 V
1 0 V
0 -1 V
1 0 V
0 -1 V
1 0 V
0 -1 V
1 0 V
0 -1 V
1 0 V
0 -1 V
1 0 V
0 -1 V
1 0 V
0 -1 V
1 0 V
0 -1 V
1 -1 V
1 -1 V
1 -1 V
1 -1 V
0 -1 V
1 0 V
0 -1 V
1 0 V
0 -1 V
1 0 V
0 -1 V
1 0 V
0 -1 V
1 0 V
0 -1 V
1 0 V
0 -1 V
1 0 V
0 -1 V
1 0 V
0 -1 V
1 0 V
0 -1 V
1 0 V
0 -1 V
1 -1 V
1 0 V
0 -1 V
1 -1 V
1 -1 V
0 -1 V
1 0 V
0 -1 V
1 0 V
0 -1 V
1 0 V
0 -1 V
1 0 V
0 -1 V
1 0 V
0 -1 V
1 0 V
0 -1 V
1 0 V
0 -1 V
1 0 V
stroke
3655 2219 M
0 -1 V
1 0 V
0 -1 V
1 0 V
0 -1 V
1 0 V
0 -1 V
1 0 V
0 -1 V
1 0 V
0 -1 V
1 0 V
0 -1 V
1 0 V
0 -1 V
1 -1 V
1 -1 V
1 -1 V
1 -1 V
0 -1 V
1 0 V
1 -1 V
1 -1 V
1 -1 V
0 -1 V
1 0 V
0 -1 V
1 0 V
0 -1 V
1 0 V
0 -1 V
1 0 V
0 -1 V
1 0 V
0 -1 V
1 0 V
0 -1 V
1 0 V
0 -1 V
1 0 V
0 -1 V
1 0 V
0 -1 V
1 0 V
0 -1 V
1 0 V
0 -1 V
1 0 V
0 -1 V
1 0 V
0 -1 V
1 0 V
0 -1 V
1 0 V
0 -1 V
1 0 V
0 -1 V
1 0 V
0 -1 V
1 0 V
0 -1 V
1 0 V
0 -1 V
1 0 V
0 -1 V
1 0 V
0 -1 V
1 0 V
0 -1 V
1 0 V
0 -1 V
1 0 V
0 -1 V
1 0 V
0 -1 V
1 0 V
0 -1 V
1 0 V
0 -1 V
1 0 V
0 -1 V
1 0 V
0 -1 V
1 0 V
0 -1 V
1 0 V
0 -1 V
1 0 V
0 -1 V
1 0 V
0 -1 V
1 0 V
0 -1 V
1 0 V
0 -1 V
1 0 V
0 -1 V
1 0 V
0 -1 V
1 0 V
0 -1 V
1 0 V
0 -1 V
1 0 V
stroke
3710 2163 M
0 -1 V
1 0 V
0 -1 V
1 0 V
0 -1 V
1 0 V
0 -1 V
1 0 V
0 -1 V
1 0 V
0 -1 V
1 0 V
0 -1 V
1 0 V
0 -1 V
1 0 V
0 -1 V
1 0 V
0 -1 V
1 0 V
0 -1 V
1 0 V
0 -1 V
1 0 V
0 -1 V
1 0 V
0 -1 V
1 0 V
0 -1 V
1 0 V
0 -1 V
1 0 V
0 -1 V
1 0 V
0 -1 V
1 0 V
1 -1 V
0 -1 V
1 0 V
1.000 UL
LTa
2611 300 M
0 7 V
0 7 V
0 6 V
0 7 V
0 7 V
0 7 V
0 6 V
0 7 V
0 7 V
0 7 V
0 7 V
0 6 V
0 7 V
0 7 V
0 7 V
0 6 V
0 7 V
0 7 V
0 7 V
0 6 V
0 7 V
0 7 V
0 7 V
0 7 V
0 6 V
0 7 V
0 7 V
0 7 V
0 6 V
0 7 V
0 7 V
0 7 V
0 7 V
0 6 V
0 7 V
0 7 V
0 7 V
0 6 V
0 7 V
0 7 V
0 7 V
0 6 V
0 7 V
0 7 V
0 7 V
0 7 V
0 6 V
0 7 V
0 7 V
0 7 V
0 6 V
0 7 V
0 7 V
0 7 V
0 7 V
0 6 V
0 7 V
0 7 V
0 7 V
0 6 V
0 7 V
0 7 V
0 7 V
0 7 V
0 6 V
0 7 V
0 7 V
0 7 V
0 6 V
0 7 V
0 7 V
0 7 V
0 6 V
0 7 V
0 7 V
0 7 V
0 7 V
0 6 V
0 7 V
0 7 V
0 7 V
0 6 V
0 7 V
0 7 V
0 7 V
0 7 V
0 6 V
0 7 V
0 7 V
0 7 V
0 6 V
0 7 V
0 7 V
0 7 V
0 6 V
0 7 V
0 7 V
0 7 V
0 7 V
0 6 V
0 7 V
0 7 V
0 7 V
0 6 V
stroke
2611 1004 M
0 7 V
0 7 V
0 7 V
0 7 V
0 6 V
0 7 V
0 7 V
0 7 V
0 6 V
0 7 V
0 7 V
0 7 V
0 7 V
0 6 V
0 7 V
0 7 V
0 7 V
0 6 V
0 7 V
0 7 V
0 7 V
0 6 V
0 7 V
0 7 V
0 7 V
0 7 V
0 6 V
0 7 V
0 7 V
0 7 V
0 6 V
0 7 V
0 7 V
0 7 V
0 7 V
0 6 V
0 7 V
0 7 V
0 7 V
0 6 V
0 7 V
0 7 V
0 7 V
0 6 V
0 7 V
0 7 V
0 7 V
0 7 V
0 6 V
0 7 V
0 7 V
0 7 V
0 6 V
0 7 V
0 7 V
0 7 V
0 7 V
0 6 V
0 7 V
0 7 V
0 7 V
0 6 V
0 7 V
0 7 V
0 7 V
0 7 V
0 6 V
0 7 V
0 7 V
0 7 V
0 6 V
0 7 V
0 7 V
0 7 V
0 6 V
0 7 V
0 7 V
0 7 V
0 7 V
0 6 V
0 7 V
0 7 V
0 7 V
0 6 V
0 7 V
0 7 V
0 7 V
0 7 V
0 6 V
0 7 V
0 7 V
0 7 V
0 6 V
0 7 V
0 7 V
0 7 V
0 6 V
0 7 V
0 7 V
0 7 V
0 7 V
0 6 V
0 7 V
0 7 V
stroke
2611 1709 M
0 7 V
0 6 V
0 7 V
0 7 V
0 7 V
0 7 V
0 6 V
0 7 V
0 7 V
0 7 V
0 6 V
0 7 V
0 7 V
0 7 V
0 7 V
0 6 V
0 7 V
0 7 V
0 7 V
0 6 V
0 7 V
0 7 V
0 7 V
0 6 V
0 7 V
0 7 V
0 7 V
0 7 V
0 6 V
0 7 V
0 7 V
0 7 V
0 6 V
0 7 V
0 7 V
0 7 V
0 7 V
0 6 V
0 7 V
0 7 V
0 7 V
0 6 V
0 7 V
0 7 V
0 7 V
0 6 V
0 7 V
0 7 V
0 7 V
0 7 V
0 6 V
0 7 V
0 7 V
0 7 V
0 6 V
0 7 V
0 7 V
0 7 V
0 7 V
0 6 V
0 7 V
0 7 V
0 7 V
0 6 V
0 7 V
0 7 V
0 7 V
0 6 V
0 7 V
0 7 V
0 7 V
0 7 V
0 6 V
0 7 V
0 7 V
0 7 V
0 6 V
0 7 V
0 7 V
0 7 V
0 7 V
0 6 V
0 7 V
0 7 V
0 7 V
0 6 V
0 7 V
0 7 V
0 7 V
0 7 V
0 6 V
0 7 V
0 7 V
0 7 V
0 6 V
0 7 V
0 7 V
0 7 V
0 6 V
0 7 V
0 7 V
0 7 V
0 7 V
0 6 V
stroke
2611 2413 M
0 7 V
0 7 V
0 7 V
0 6 V
0 7 V
0 7 V
0 7 V
0 7 V
0 6 V
0 7 V
0 7 V
0 7 V
0 6 V
0 7 V
0 7 V
0 7 V
0 6 V
0 7 V
0 7 V
0 7 V
0 7 V
0 6 V
0 7 V
0 7 V
0 7 V
0 6 V
0 7 V
0 7 V
0 7 V
0 7 V
0 6 V
0 7 V
0 7 V
0 7 V
0 6 V
0 7 V
0 7 V
0 7 V
0 7 V
0 6 V
0 7 V
0 7 V
0 7 V
0 6 V
0 7 V
0 7 V
0 7 V
0 6 V
0 7 V
0 7 V
0 7 V
0 7 V
0 6 V
0 7 V
0 7 V
0 7 V
0 6 V
0 7 V
0 7 V
0 7 V
0 7 V
0 6 V
0 7 V
0 7 V
0 7 V
0 6 V
0 7 V
0 7 V
0 7 V
0 6 V
0 7 V
0 7 V
0 7 V
0 7 V
0 6 V
0 7 V
0 7 V
0 7 V
0 6 V
0 7 V
0 7 V
0 7 V
0 7 V
0 6 V
0 7 V
0 7 V
0 7 V
0 6 V
0 7 V
0 7 V
0 7 V
0 7 V
0 6 V
0 7 V
0 7 V
0 7 V
0 6 V
0 7 V
0 7 V
0 7 V
0 6 V
0 7 V
0 7 V
0 7 V
stroke
2611 3118 M
0 7 V
0 6 V
0 7 V
0 7 V
0 7 V
0 6 V
0 7 V
0 7 V
0 7 V
0 7 V
0 6 V
0 7 V
0 7 V
0 7 V
0 6 V
0 7 V
0 7 V
0 7 V
0 6 V
0 7 V
0 7 V
0 7 V
0 7 V
0 6 V
0 7 V
0 7 V
0 7 V
0 6 V
0 7 V
0 7 V
0 7 V
0 7 V
0 6 V
0 7 V
0 7 V
0 7 V
0 6 V
0 7 V
0 7 V
0 7 V
0 7 V
0 6 V
0 7 V
0 7 V
0 7 V
0 6 V
0 7 V
0 7 V
0 7 V
0 6 V
0 7 V
0 7 V
0 7 V
0 7 V
0 6 V
0 7 V
0 7 V
0 7 V
0 6 V
0 7 V
0 7 V
0 7 V
0 7 V
0 6 V
0 7 V
0 7 V
0 7 V
0 6 V
0 7 V
0 7 V
0 7 V
0 6 V
0 7 V
0 7 V
0 7 V
0 7 V
0 6 V
0 7 V
0 7 V
0 7 V
0 6 V
0 7 V
0 7 V
1.000 UL
LTa
350 2143 M
255 0 V
180 0 V
138 0 V
112 0 V
95 0 V
82 0 V
73 0 V
64 0 V
58 0 V
54 0 V
49 0 V
45 0 V
42 0 V
39 0 V
37 0 V
35 0 V
33 0 V
32 0 V
29 0 V
29 0 V
27 0 V
26 0 V
24 0 V
24 0 V
23 0 V
23 0 V
21 0 V
21 0 V
19 0 V
20 0 V
19 0 V
18 0 V
17 0 V
17 0 V
17 0 V
16 0 V
16 0 V
16 0 V
15 0 V
14 0 V
15 0 V
14 0 V
13 0 V
14 0 V
13 0 V
12 0 V
13 0 V
12 0 V
12 0 V
12 0 V
12 0 V
11 0 V
11 0 V
11 0 V
11 0 V
11 0 V
10 0 V
10 0 V
10 0 V
10 0 V
10 0 V
9 0 V
10 0 V
9 0 V
9 0 V
9 0 V
9 0 V
9 0 V
8 0 V
9 0 V
8 0 V
9 0 V
8 0 V
8 0 V
8 0 V
8 0 V
7 0 V
8 0 V
7 0 V
8 0 V
7 0 V
8 0 V
7 0 V
7 0 V
7 0 V
7 0 V
7 0 V
7 0 V
6 0 V
7 0 V
6 0 V
7 0 V
6 0 V
7 0 V
6 0 V
6 0 V
6 0 V
7 0 V
6 0 V
6 0 V
5 0 V
6 0 V
6 0 V
6 0 V
stroke
2787 2143 M
6 0 V
5 0 V
6 0 V
5 0 V
6 0 V
5 0 V
6 0 V
5 0 V
5 0 V
6 0 V
5 0 V
5 0 V
5 0 V
5 0 V
5 0 V
5 0 V
5 0 V
5 0 V
5 0 V
5 0 V
5 0 V
4 0 V
5 0 V
5 0 V
5 0 V
4 0 V
5 0 V
4 0 V
5 0 V
4 0 V
5 0 V
4 0 V
5 0 V
4 0 V
4 0 V
5 0 V
4 0 V
4 0 V
4 0 V
4 0 V
5 0 V
4 0 V
4 0 V
4 0 V
4 0 V
4 0 V
4 0 V
4 0 V
4 0 V
4 0 V
3 0 V
4 0 V
4 0 V
4 0 V
4 0 V
4 0 V
3 0 V
4 0 V
4 0 V
3 0 V
4 0 V
4 0 V
3 0 V
4 0 V
3 0 V
4 0 V
3 0 V
4 0 V
3 0 V
4 0 V
3 0 V
4 0 V
3 0 V
4 0 V
3 0 V
3 0 V
4 0 V
3 0 V
3 0 V
3 0 V
4 0 V
3 0 V
3 0 V
3 0 V
4 0 V
3 0 V
3 0 V
3 0 V
3 0 V
3 0 V
3 0 V
3 0 V
3 0 V
3 0 V
4 0 V
3 0 V
3 0 V
3 0 V
2 0 V
3 0 V
3 0 V
3 0 V
3 0 V
3 0 V
stroke
3202 2143 M
3 0 V
3 0 V
3 0 V
3 0 V
2 0 V
3 0 V
3 0 V
3 0 V
3 0 V
2 0 V
3 0 V
3 0 V
3 0 V
2 0 V
3 0 V
3 0 V
2 0 V
3 0 V
3 0 V
2 0 V
3 0 V
3 0 V
2 0 V
3 0 V
2 0 V
3 0 V
3 0 V
2 0 V
3 0 V
2 0 V
3 0 V
2 0 V
3 0 V
2 0 V
3 0 V
2 0 V
3 0 V
2 0 V
3 0 V
2 0 V
2 0 V
3 0 V
2 0 V
3 0 V
2 0 V
2 0 V
3 0 V
2 0 V
2 0 V
3 0 V
2 0 V
2 0 V
3 0 V
2 0 V
2 0 V
3 0 V
2 0 V
2 0 V
2 0 V
3 0 V
2 0 V
2 0 V
2 0 V
3 0 V
2 0 V
2 0 V
2 0 V
2 0 V
3 0 V
2 0 V
2 0 V
2 0 V
2 0 V
2 0 V
3 0 V
2 0 V
2 0 V
2 0 V
2 0 V
2 0 V
2 0 V
2 0 V
2 0 V
2 0 V
2 0 V
3 0 V
2 0 V
2 0 V
2 0 V
2 0 V
2 0 V
2 0 V
2 0 V
2 0 V
2 0 V
2 0 V
2 0 V
2 0 V
2 0 V
2 0 V
2 0 V
1 0 V
2 0 V
2 0 V
stroke
3446 2143 M
2 0 V
2 0 V
2 0 V
2 0 V
2 0 V
2 0 V
2 0 V
2 0 V
2 0 V
1 0 V
2 0 V
2 0 V
2 0 V
2 0 V
2 0 V
2 0 V
1 0 V
2 0 V
2 0 V
2 0 V
2 0 V
1 0 V
2 0 V
2 0 V
2 0 V
2 0 V
1 0 V
2 0 V
2 0 V
2 0 V
2 0 V
1 0 V
2 0 V
2 0 V
2 0 V
1 0 V
2 0 V
2 0 V
1 0 V
2 0 V
2 0 V
2 0 V
1 0 V
2 0 V
2 0 V
1 0 V
2 0 V
2 0 V
1 0 V
2 0 V
2 0 V
1 0 V
2 0 V
2 0 V
1 0 V
2 0 V
2 0 V
1 0 V
2 0 V
2 0 V
1 0 V
2 0 V
1 0 V
2 0 V
2 0 V
1 0 V
2 0 V
1 0 V
2 0 V
2 0 V
1 0 V
2 0 V
1 0 V
2 0 V
1 0 V
2 0 V
2 0 V
1 0 V
2 0 V
1 0 V
2 0 V
1 0 V
2 0 V
1 0 V
2 0 V
1 0 V
2 0 V
1 0 V
2 0 V
1 0 V
2 0 V
1 0 V
2 0 V
1 0 V
2 0 V
1 0 V
2 0 V
1 0 V
2 0 V
1 0 V
2 0 V
1 0 V
2 0 V
1 0 V
stroke
3620 2143 M
1 0 V
2 0 V
1 0 V
2 0 V
1 0 V
2 0 V
1 0 V
2 0 V
1 0 V
1 0 V
2 0 V
1 0 V
2 0 V
1 0 V
1 0 V
2 0 V
1 0 V
2 0 V
1 0 V
1 0 V
2 0 V
1 0 V
2 0 V
1 0 V
1 0 V
2 0 V
1 0 V
1 0 V
2 0 V
1 0 V
1 0 V
2 0 V
1 0 V
2 0 V
1 0 V
1 0 V
2 0 V
1 0 V
1 0 V
2 0 V
1 0 V
1 0 V
1 0 V
2 0 V
1 0 V
1 0 V
2 0 V
1 0 V
1 0 V
2 0 V
1 0 V
1 0 V
2 0 V
1 0 V
1 0 V
1 0 V
2 0 V
1 0 V
1 0 V
1 0 V
2 0 V
1 0 V
1 0 V
2 0 V
1 0 V
1 0 V
1 0 V
2 0 V
1 0 V
1 0 V
1 0 V
2 0 V
1 0 V
1 0 V
1 0 V
1 0 V
2 0 V
1 0 V
1 0 V
1 0 V
2 0 V
1 0 V
1 0 V
0.500 UL
LTb
350 300 M
3380 0 V
0 3380 V
-3380 0 V
350 300 L
1.000 UP
stroke
grestore
end
showpage
}}%
\put(2040,50){\makebox(0,0){$r\ [\mbox{kpc}]$}}%
\put(100,1990){%
\special{ps: gsave currentpoint currentpoint translate
270 rotate neg exch neg exch translate}%
\makebox(0,0)[b]{\shortstack{$\mathcal{G}(r) \equiv G(r)/G_N$}}%
\special{ps: currentpoint grestore moveto}%
}%
\put(3145,200){\makebox(0,0){ 1000}}%
\put(1748,200){\makebox(0,0){ 100}}%
\put(350,200){\makebox(0,0){ 10}}%
\put(300,3680){\makebox(0,0)[r]{ 7}}%
\put(300,3197){\makebox(0,0)[r]{ 6}}%
\put(300,2714){\makebox(0,0)[r]{ 5}}%
\put(300,2231){\makebox(0,0)[r]{ 4}}%
\put(300,1749){\makebox(0,0)[r]{ 3}}%
\put(300,1266){\makebox(0,0)[r]{ 2}}%
\put(300,783){\makebox(0,0)[r]{ 1}}%
\put(300,300){\makebox(0,0)[r]{ 0}}%
\end{picture}%
\endgroup
 

%% file: figure/gnuplot/MainXray.tex
\begingroup%
  \makeatletter%
  \newcommand{\GNUPLOTspecial}{%
    \@sanitize\catcode`\%=14\relax\special}%
  \setlength{\unitlength}{0.1bp}%
\begin{picture}(4050,3780)(100,0)%
{\GNUPLOTspecial{"
/gnudict 256 dict def
gnudict begin
/Color true def
/Solid false def
/gnulinewidth 5.000 def
/userlinewidth gnulinewidth def
/vshift -33 def
/dl {10.0 mul} def
/hpt_ 31.5 def
/vpt_ 31.5 def
/hpt hpt_ def
/vpt vpt_ def
/Rounded false def
/M {moveto} bind def
/L {lineto} bind def
/R {rmoveto} bind def
/V {rlineto} bind def
/N {newpath moveto} bind def
/C {setrgbcolor} bind def
/f {rlineto fill} bind def
/vpt2 vpt 2 mul def
/hpt2 hpt 2 mul def
/Lshow { currentpoint stroke M
  0 vshift R show } def
/Rshow { currentpoint stroke M
  dup stringwidth pop neg vshift R show } def
/Cshow { currentpoint stroke M
  dup stringwidth pop -2 div vshift R show } def
/UP { dup vpt_ mul /vpt exch def hpt_ mul /hpt exch def
  /hpt2 hpt 2 mul def /vpt2 vpt 2 mul def } def
/DL { Color {setrgbcolor Solid {pop []} if 0 setdash }
 {pop pop pop 0 setgray Solid {pop []} if 0 setdash} ifelse } def
/BL { stroke userlinewidth 2 mul setlinewidth
      Rounded { 1 setlinejoin 1 setlinecap } if } def
/AL { stroke userlinewidth 2 div setlinewidth
      Rounded { 1 setlinejoin 1 setlinecap } if } def
/UL { dup gnulinewidth mul /userlinewidth exch def
      dup 1 lt {pop 1} if 10 mul /udl exch def } def
/PL { stroke userlinewidth setlinewidth
      Rounded { 1 setlinejoin 1 setlinecap } if } def
/LTw { PL [] 1 setgray } def
/LTb { BL [] 0 0 0 DL } def
/LTa { AL [1 udl mul 2 udl mul] 0 setdash 0 0 0 setrgbcolor } def
/fatlinewidth 7.500 def
/gatlinewidth 10.000 def
/FL { stroke fatlinewidth setlinewidth Rounded { 1 setlinejoin 1 setlinecap } if } def
/GL { stroke gatlinewidth setlinewidth Rounded { 1 setlinejoin 1 setlinecap } if } def
/LT0 { FL [] 1 0 0 DL } def
/LT1 { GL [2 dl 3 dl] 0 0 1 DL } def
/LT2 { GL [5 dl 2 dl 1 dl 2 dl] 0 0 0 DL } def
/LT3 { GL [4 dl 2 dl] 1 0 1 DL } def/LT4 { PL [5 dl 2 dl 1 dl 2 dl] 0 1 1 DL } def
/LT5 { PL [4 dl 3 dl 1 dl 3 dl] 1 1 0 DL } def
/LT6 { PL [2 dl 2 dl 2 dl 4 dl] 0 0 0 DL } def
/LT7 { PL [2 dl 2 dl 2 dl 2 dl 2 dl 4 dl] 1 0.3 0 DL } def
/LT8 { PL [2 dl 2 dl 2 dl 2 dl 2 dl 2 dl 2 dl 4 dl] 0.5 0.5 0.5 DL } def
/Pnt { stroke [] 0 setdash
   gsave 1 setlinecap M 0 0 V stroke grestore } def
/Dia { stroke [] 0 setdash 2 copy vpt add M
  hpt neg vpt neg V hpt vpt neg V
  hpt vpt V hpt neg vpt V closepath stroke
  Pnt } def
/Pls { stroke [] 0 setdash vpt sub M 0 vpt2 V
  currentpoint stroke M
  hpt neg vpt neg R hpt2 0 V stroke
  } def
/Box { stroke [] 0 setdash 2 copy exch hpt sub exch vpt add M
  0 vpt2 neg V hpt2 0 V 0 vpt2 V
  hpt2 neg 0 V closepath stroke
  Pnt } def
/Crs { stroke [] 0 setdash exch hpt sub exch vpt add M
  hpt2 vpt2 neg V currentpoint stroke M
  hpt2 neg 0 R hpt2 vpt2 V stroke } def
/TriU { stroke [] 0 setdash 2 copy vpt 1.12 mul add M
  hpt neg vpt -1.62 mul V
  hpt 2 mul 0 V
  hpt neg vpt 1.62 mul V closepath stroke
  Pnt  } def
/Star { 2 copy Pls Crs } def
/BoxF { stroke [] 0 setdash exch hpt sub exch vpt add M
  0 vpt2 neg V  hpt2 0 V  0 vpt2 V
  hpt2 neg 0 V  closepath fill } def
/TriUF { stroke [] 0 setdash vpt 1.12 mul add M
  hpt neg vpt -1.62 mul V
  hpt 2 mul 0 V
  hpt neg vpt 1.62 mul V closepath fill } def
/TriD { stroke [] 0 setdash 2 copy vpt 1.12 mul sub M
  hpt neg vpt 1.62 mul V
  hpt 2 mul 0 V
  hpt neg vpt -1.62 mul V closepath stroke
  Pnt  } def
/TriDF { stroke [] 0 setdash vpt 1.12 mul sub M
  hpt neg vpt 1.62 mul V
  hpt 2 mul 0 V
  hpt neg vpt -1.62 mul V closepath fill} def
/DiaF { stroke [] 0 setdash vpt add M
  hpt neg vpt neg V hpt vpt neg V
  hpt vpt V hpt neg vpt V closepath fill } def
/Pent { stroke [] 0 setdash 2 copy gsave
  translate 0 hpt M 4 {72 rotate 0 hpt L} repeat
  closepath stroke grestore Pnt } def
/PentF { stroke [] 0 setdash gsave
  translate 0 hpt M 4 {72 rotate 0 hpt L} repeat
  closepath fill grestore } def
/Circle { stroke [] 0 setdash 2 copy
  hpt 0 360 arc stroke Pnt } def
/CircleF { stroke [] 0 setdash hpt 0 360 arc fill } def
/C0 { BL [] 0 setdash 2 copy moveto vpt 90 450  arc } bind def
/C1 { BL [] 0 setdash 2 copy        moveto
       2 copy  vpt 0 90 arc closepath fill
               vpt 0 360 arc closepath } bind def
/C2 { BL [] 0 setdash 2 copy moveto
       2 copy  vpt 90 180 arc closepath fill
               vpt 0 360 arc closepath } bind def
/C3 { BL [] 0 setdash 2 copy moveto
       2 copy  vpt 0 180 arc closepath fill
               vpt 0 360 arc closepath } bind def
/C4 { BL [] 0 setdash 2 copy moveto
       2 copy  vpt 180 270 arc closepath fill
               vpt 0 360 arc closepath } bind def
/C5 { BL [] 0 setdash 2 copy moveto
       2 copy  vpt 0 90 arc
       2 copy moveto
       2 copy  vpt 180 270 arc closepath fill
               vpt 0 360 arc } bind def
/C6 { BL [] 0 setdash 2 copy moveto
      2 copy  vpt 90 270 arc closepath fill
              vpt 0 360 arc closepath } bind def
/C7 { BL [] 0 setdash 2 copy moveto
      2 copy  vpt 0 270 arc closepath fill
              vpt 0 360 arc closepath } bind def
/C8 { BL [] 0 setdash 2 copy moveto
      2 copy vpt 270 360 arc closepath fill
              vpt 0 360 arc closepath } bind def
/C9 { BL [] 0 setdash 2 copy moveto
      2 copy  vpt 270 450 arc closepath fill
              vpt 0 360 arc closepath } bind def
/C10 { BL [] 0 setdash 2 copy 2 copy moveto vpt 270 360 arc closepath fill
       2 copy moveto
       2 copy vpt 90 180 arc closepath fill
               vpt 0 360 arc closepath } bind def
/C11 { BL [] 0 setdash 2 copy moveto
       2 copy  vpt 0 180 arc closepath fill
       2 copy moveto
       2 copy  vpt 270 360 arc closepath fill
               vpt 0 360 arc closepath } bind def
/C12 { BL [] 0 setdash 2 copy moveto
       2 copy  vpt 180 360 arc closepath fill
               vpt 0 360 arc closepath } bind def
/C13 { BL [] 0 setdash  2 copy moveto
       2 copy  vpt 0 90 arc closepath fill
       2 copy moveto
       2 copy  vpt 180 360 arc closepath fill
               vpt 0 360 arc closepath } bind def
/C14 { BL [] 0 setdash 2 copy moveto
       2 copy  vpt 90 360 arc closepath fill
               vpt 0 360 arc } bind def
/C15 { BL [] 0 setdash 2 copy vpt 0 360 arc closepath fill
               vpt 0 360 arc closepath } bind def
/Rec   { newpath 4 2 roll moveto 1 index 0 rlineto 0 exch rlineto
       neg 0 rlineto closepath } bind def
/Square { dup Rec } bind def
/Bsquare { vpt sub exch vpt sub exch vpt2 Square } bind def
/S0 { BL [] 0 setdash 2 copy moveto 0 vpt rlineto BL Bsquare } bind def
/S1 { BL [] 0 setdash 2 copy vpt Square fill Bsquare } bind def
/S2 { BL [] 0 setdash 2 copy exch vpt sub exch vpt Square fill Bsquare } bind def
/S3 { BL [] 0 setdash 2 copy exch vpt sub exch vpt2 vpt Rec fill Bsquare } bind def
/S4 { BL [] 0 setdash 2 copy exch vpt sub exch vpt sub vpt Square fill Bsquare } bind def
/S5 { BL [] 0 setdash 2 copy 2 copy vpt Square fill
       exch vpt sub exch vpt sub vpt Square fill Bsquare } bind def
/S6 { BL [] 0 setdash 2 copy exch vpt sub exch vpt sub vpt vpt2 Rec fill Bsquare } bind def
/S7 { BL [] 0 setdash 2 copy exch vpt sub exch vpt sub vpt vpt2 Rec fill
       2 copy vpt Square fill
       Bsquare } bind def
/S8 { BL [] 0 setdash 2 copy vpt sub vpt Square fill Bsquare } bind def
/S9 { BL [] 0 setdash 2 copy vpt sub vpt vpt2 Rec fill Bsquare } bind def
/S10 { BL [] 0 setdash 2 copy vpt sub vpt Square fill 2 copy exch vpt sub exch vpt Square fill
       Bsquare } bind def
/S11 { BL [] 0 setdash 2 copy vpt sub vpt Square fill 2 copy exch vpt sub exch vpt2 vpt Rec fill
       Bsquare } bind def
/S12 { BL [] 0 setdash 2 copy exch vpt sub exch vpt sub vpt2 vpt Rec fill Bsquare } bind def
/S13 { BL [] 0 setdash 2 copy exch vpt sub exch vpt sub vpt2 vpt Rec fill
       2 copy vpt Square fill Bsquare } bind def
/S14 { BL [] 0 setdash 2 copy exch vpt sub exch vpt sub vpt2 vpt Rec fill
       2 copy exch vpt sub exch vpt Square fill Bsquare } bind def
/S15 { BL [] 0 setdash 2 copy Bsquare fill Bsquare } bind def
/D0 { gsave translate 45 rotate 0 0 S0 stroke grestore } bind def
/D1 { gsave translate 45 rotate 0 0 S1 stroke grestore } bind def
/D2 { gsave translate 45 rotate 0 0 S2 stroke grestore } bind def
/D3 { gsave translate 45 rotate 0 0 S3 stroke grestore } bind def
/D4 { gsave translate 45 rotate 0 0 S4 stroke grestore } bind def
/D5 { gsave translate 45 rotate 0 0 S5 stroke grestore } bind def
/D6 { gsave translate 45 rotate 0 0 S6 stroke grestore } bind def
/D7 { gsave translate 45 rotate 0 0 S7 stroke grestore } bind def
/D8 { gsave translate 45 rotate 0 0 S8 stroke grestore } bind def
/D9 { gsave translate 45 rotate 0 0 S9 stroke grestore } bind def
/D10 { gsave translate 45 rotate 0 0 S10 stroke grestore } bind def
/D11 { gsave translate 45 rotate 0 0 S11 stroke grestore } bind def
/D12 { gsave translate 45 rotate 0 0 S12 stroke grestore } bind def
/D13 { gsave translate 45 rotate 0 0 S13 stroke grestore } bind def
/D14 { gsave translate 45 rotate 0 0 S14 stroke grestore } bind def
/D15 { gsave translate 45 rotate 0 0 S15 stroke grestore } bind def
/DiaE { stroke [] 0 setdash vpt add M
  hpt neg vpt neg V hpt vpt neg V
  hpt vpt V hpt neg vpt V closepath stroke } def
/BoxE { stroke [] 0 setdash exch hpt sub exch vpt add M
  0 vpt2 neg V hpt2 0 V 0 vpt2 V
  hpt2 neg 0 V closepath stroke } def
/TriUE { stroke [] 0 setdash vpt 1.12 mul add M
  hpt neg vpt -1.62 mul V
  hpt 2 mul 0 V
  hpt neg vpt 1.62 mul V closepath stroke } def
/TriDE { stroke [] 0 setdash vpt 1.12 mul sub M
  hpt neg vpt 1.62 mul V
  hpt 2 mul 0 V
  hpt neg vpt -1.62 mul V closepath stroke } def
/PentE { stroke [] 0 setdash gsave
  translate 0 hpt M 4 {72 rotate 0 hpt L} repeat
  closepath stroke grestore } def
/CircE { stroke [] 0 setdash 
  hpt 0 360 arc stroke } def
/Opaque { gsave closepath 1 setgray fill grestore 0 setgray closepath } def
/DiaW { stroke [] 0 setdash vpt add M
  hpt neg vpt neg V hpt vpt neg V
  hpt vpt V hpt neg vpt V Opaque stroke } def
/BoxW { stroke [] 0 setdash exch hpt sub exch vpt add M
  0 vpt2 neg V hpt2 0 V 0 vpt2 V
  hpt2 neg 0 V Opaque stroke } def
/TriUW { stroke [] 0 setdash vpt 1.12 mul add M
  hpt neg vpt -1.62 mul V
  hpt 2 mul 0 V
  hpt neg vpt 1.62 mul V Opaque stroke } def
/TriDW { stroke [] 0 setdash vpt 1.12 mul sub M
  hpt neg vpt 1.62 mul V
  hpt 2 mul 0 V
  hpt neg vpt -1.62 mul V Opaque stroke } def
/PentW { stroke [] 0 setdash gsave
  translate 0 hpt M 4 {72 rotate 0 hpt L} repeat
  Opaque stroke grestore } def
/CircW { stroke [] 0 setdash 
  hpt 0 360 arc Opaque stroke } def
/BoxFill { gsave Rec 1 setgray fill grestore } def
/BoxColFill {
  gsave Rec
  /Fillden exch def
  currentrgbcolor
  /ColB exch def /ColG exch def /ColR exch def
  /ColR ColR Fillden mul Fillden sub 1 add def
  /ColG ColG Fillden mul Fillden sub 1 add def
  /ColB ColB Fillden mul Fillden sub 1 add def
  ColR ColG ColB setrgbcolor
  fill grestore } def
%
%
/PatternFill { gsave /PFa [ 9 2 roll ] def
    PFa 0 get PFa 2 get 2 div add PFa 1 get PFa 3 get 2 div add translate
    PFa 2 get -2 div PFa 3 get -2 div PFa 2 get PFa 3 get Rec
    gsave 1 setgray fill grestore clip
    currentlinewidth 0.5 mul setlinewidth
    /PFs PFa 2 get dup mul PFa 3 get dup mul add sqrt def
    0 0 M PFa 5 get rotate PFs -2 div dup translate
	0 1 PFs PFa 4 get div 1 add floor cvi
	{ PFa 4 get mul 0 M 0 PFs V } for
    0 PFa 6 get ne {
	0 1 PFs PFa 4 get div 1 add floor cvi
	{ PFa 4 get mul 0 2 1 roll M PFs 0 V } for
    } if
    stroke grestore } def
/Symbol-Oblique /Symbol findfont [1 0 .167 1 0 0] makefont
dup length dict begin {1 index /FID eq {pop pop} {def} ifelse} forall
currentdict end definefont pop
end
gnudict begin
gsave
0 0 translate
0.100 0.100 scale
0 setgray
newpath
0.500 UL
LTb
600 3680 M
63 0 V
3317 0 R
-63 0 V
0.500 UL
LTb
600 3103 M
63 0 V
3317 0 R
-63 0 V
0.500 UL
LTb
600 2525 M
63 0 V
3317 0 R
-63 0 V
0.500 UL
LTb
600 1948 M
63 0 V
3317 0 R
-63 0 V
0.500 UL
LTb
600 1371 M
63 0 V
3317 0 R
-63 0 V
0.500 UL
LTb
600 793 M
63 0 V
3317 0 R
-63 0 V
0.500 UL
LTb
600 300 M
0 63 V
0 3317 R
0 -63 V
0.500 UL
LTb
600 300 M
0 31 V
0 3349 R
0 -31 V
2290 300 M
0 63 V
0 3317 R
0 -63 V
0.500 UL
LTb
600 300 M
0 31 V
0 3349 R
0 -31 V
1109 300 M
0 31 V
0 3349 R
0 -31 V
1406 300 M
0 31 V
0 3349 R
0 -31 V
1617 300 M
0 31 V
0 3349 R
0 -31 V
1781 300 M
0 31 V
0 3349 R
0 -31 V
1915 300 M
0 31 V
0 3349 R
0 -31 V
2028 300 M
0 31 V
0 3349 R
0 -31 V
2126 300 M
0 31 V
0 3349 R
0 -31 V
2213 300 M
0 31 V
0 3349 R
0 -31 V
2290 300 M
0 31 V
0 3349 R
0 -31 V
3980 300 M
0 63 V
0 3317 R
0 -63 V
0.500 UL
LTb
2290 300 M
0 31 V
0 3349 R
0 -31 V
2799 300 M
0 31 V
0 3349 R
0 -31 V
3096 300 M
0 31 V
0 3349 R
0 -31 V
3307 300 M
0 31 V
0 3349 R
0 -31 V
3471 300 M
0 31 V
0 3349 R
0 -31 V
3605 300 M
0 31 V
0 3349 R
0 -31 V
3718 300 M
0 31 V
0 3349 R
0 -31 V
3816 300 M
0 31 V
0 3349 R
0 -31 V
3903 300 M
0 31 V
0 3349 R
0 -31 V
3980 300 M
0 31 V
0 3349 R
0 -31 V
0.500 UL
LTb
600 300 M
3380 0 V
0 3380 V
-3380 0 V
600 300 L
LTb
LTb
1.000 UP
1.000 UL
LT0
600 1327 M
70 0 V
64 0 V
59 148 V
54 0 V
51 0 V
47 0 V
44 0 V
42 92 V
40 0 V
38 120 V
36 0 V
34 0 V
32 0 V
32 0 V
30 43 V
28 37 V
28 61 V
27 0 V
25 0 V
25 0 V
24 49 V
24 0 V
22 0 V
22 0 V
21 21 V
21 38 V
20 16 V
20 0 V
19 31 V
18 0 V
19 0 V
17 0 V
18 40 V
16 23 V
17 0 V
16 22 V
16 0 V
15 0 V
15 10 V
15 18 V
15 0 V
14 26 V
14 0 V
14 16 V
13 15 V
13 0 V
13 0 V
13 14 V
13 0 V
12 7 V
12 19 V
12 12 V
12 11 V
11 0 V
12 11 V
11 0 V
11 10 V
11 0 V
11 25 V
10 0 V
11 9 V
10 0 V
10 13 V
10 8 V
10 0 V
10 0 V
9 12 V
10 7 V
9 15 V
9 0 V
9 13 V
9 0 V
9 0 V
9 7 V
9 3 V
8 16 V
9 6 V
8 5 V
8 0 V
9 6 V
8 5 V
8 6 V
8 5 V
8 3 V
7 5 V
8 10 V
8 2 V
7 9 V
8 0 V
7 9 V
7 0 V
8 0 V
7 11 V
7 8 V
7 4 V
7 4 V
7 4 V
6 4 V
7 2 V
7 4 V
7 5 V
6 7 V
7 4 V
6 3 V
stroke
2386 2496 M
7 7 V
6 6 V
6 0 V
6 0 V
7 7 V
6 4 V
6 5 V
6 3 V
6 9 V
6 0 V
6 3 V
6 2 V
5 6 V
6 4 V
6 3 V
6 5 V
5 5 V
6 0 V
5 4 V
6 4 V
5 5 V
6 0 V
5 4 V
5 7 V
6 4 V
5 2 V
5 3 V
5 4 V
6 2 V
5 0 V
5 4 V
5 8 V
5 2 V
5 2 V
5 4 V
5 2 V
4 2 V
5 4 V
5 2 V
5 6 V
5 2 V
4 2 V
5 4 V
5 3 V
4 2 V
5 2 V
5 3 V
4 2 V
5 6 V
4 1 V
5 5 V
4 3 V
4 0 V
5 3 V
4 2 V
4 3 V
5 4 V
4 3 V
4 4 V
5 1 V
4 2 V
4 5 V
4 0 V
4 2 V
4 1 V
4 6 V
4 3 V
5 2 V
4 2 V
4 3 V
4 2 V
3 0 V
4 1 V
4 7 V
4 2 V
4 3 V
4 2 V
4 2 V
4 2 V
3 1 V
4 4 V
4 0 V
4 5 V
3 1 V
4 3 V
4 4 V
3 1 V
4 1 V
4 3 V
3 1 V
4 1 V
3 4 V
4 4 V
4 2 V
3 1 V
4 2 V
3 2 V
4 2 V
3 0 V
3 6 V
4 1 V
3 3 V
4 0 V
3 4 V
stroke
2862 2798 M
3 1 V
4 1 V
3 1 V
3 4 V
4 2 V
3 3 V
3 1 V
3 3 V
4 1 V
3 2 V
3 0 V
3 4 V
4 3 V
3 2 V
3 2 V
3 0 V
3 2 V
3 3 V
3 1 V
3 2 V
3 2 V
4 1 V
3 3 V
3 2 V
3 2 V
3 1 V
3 2 V
3 1 V
3 1 V
3 4 V
3 2 V
3 2 V
2 2 V
3 1 V
3 1 V
3 2 V
3 1 V
3 3 V
3 2 V
3 1 V
2 1 V
3 3 V
3 1 V
3 2 V
3 1 V
3 1 V
2 3 V
3 2 V
3 1 V
3 3 V
2 0 V
3 2 V
3 0 V
2 3 V
3 1 V
3 2 V
2 2 V
3 2 V
3 1 V
2 2 V
3 1 V
3 2 V
2 0 V
3 3 V
3 1 V
2 3 V
3 1 V
2 1 V
3 2 V
2 1 V
3 1 V
2 2 V
3 2 V
2 2 V
3 0 V
3 2 V
2 2 V
2 1 V
3 1 V
2 2 V
3 2 V
2 1 V
3 1 V
2 2 V
3 2 V
2 1 V
2 0 V
3 2 V
2 1 V
3 2 V
2 2 V
2 1 V
3 2 V
2 0 V
2 2 V
3 1 V
2 1 V
2 2 V
3 2 V
2 1 V
2 1 V
3 1 V
2 2 V
2 1 V
stroke
3148 2968 M
3 1 V
2 1 V
2 2 V
2 2 V
3 2 V
2 0 V
2 1 V
2 2 V
3 0 V
2 2 V
2 1 V
2 2 V
2 1 V
3 1 V
2 1 V
2 2 V
2 1 V
2 1 V
2 1 V
3 1 V
2 1 V
2 3 V
2 0 V
2 2 V
2 0 V
2 2 V
2 1 V
2 1 V
3 2 V
2 1 V
2 1 V
2 1 V
2 1 V
2 1 V
2 1 V
2 1 V
2 2 V
2 2 V
2 0 V
2 1 V
2 2 V
2 0 V
2 2 V
2 0 V
2 2 V
2 1 V
2 1 V
2 1 V
2 2 V
2 1 V
2 0 V
2 1 V
2 2 V
2 1 V
2 1 V
2 1 V
2 2 V
2 0 V
2 1 V
2 1 V
2 1 V
2 1 V
1 1 V
2 1 V
2 2 V
2 1 V
2 1 V
2 1 V
2 0 V
2 1 V
2 1 V
1 2 V
2 1 V
2 1 V
2 1 V
2 1 V
2 0 V
1 2 V
2 1 V
2 1 V
2 1 V
2 0 V
2 2 V
1 1 V
2 0 V
2 1 V
2 1 V
2 1 V
1 2 V
2 1 V
2 1 V
2 0 V
2 1 V
1 1 V
2 1 V
2 1 V
2 1 V
1 1 V
2 1 V
2 1 V
2 0 V
1 2 V
2 1 V
2 0 V
stroke
3354 3082 M
1 1 V
2 1 V
2 1 V
2 1 V
1 1 V
2 1 V
2 0 V
1 1 V
2 1 V
2 2 V
1 0 V
2 1 V
2 1 V
1 0 V
2 2 V
2 0 V
1 1 V
2 1 V
2 1 V
1 0 V
2 2 V
2 1 V
1 0 V
2 0 V
2 1 V
1 1 V
2 1 V
1 2 V
2 0 V
2 1 V
1 0 V
2 1 V
1 1 V
2 1 V
2 0 V
1 2 V
2 0 V
1 1 V
2 1 V
2 1 V
1 0 V
2 1 V
1 0 V
2 2 V
1 1 V
2 0 V
2 1 V
1 1 V
2 0 V
1 1 V
2 1 V
1 1 V
2 1 V
1 0 V
2 1 V
1 1 V
2 0 V
1 1 V
2 1 V
1 0 V
2 1 V
1 1 V
2 1 V
1 1 V
2 0 V
1 1 V
2 1 V
1 0 V
2 1 V
1 1 V
2 1 V
1 0 V
2 0 V
1 1 V
2 1 V
1 1 V
2 0 V
1 1 V
2 1 V
1 0 V
1 1 V
2 1 V
1 1 V
2 0 V
1 1 V
2 1 V
1 0 V
2 1 V
1 0 V
1 1 V
2 1 V
1 0 V
2 1 V
1 1 V
1 1 V
2 0 V
1 1 V
2 0 V
1 1 V
1 0 V
2 1 V
1 1 V
2 0 V
1 1 V
stroke
3514 3160 M
1 1 V
2 1 V
1 0 V
2 0 V
1 1 V
1 1 V
2 0 V
1 1 V
1 1 V
2 0 V
1 1 V
1 0 V
2 1 V
1 1 V
2 0 V
1 1 V
1 1 V
2 0 V
1 1 V
1 0 V
2 1 V
1 0 V
1 1 V
2 0 V
1 1 V
1 1 V
1 1 V
2 0 V
1 0 V
1 1 V
2 0 V
1 1 V
1 0 V
2 1 V
1 1 V
1 0 V
2 1 V
1 0 V
1 1 V
1 0 V
2 1 V
1 0 V
1 1 V
2 1 V
1 0 V
1 1 V
1 0 V
2 1 V
1 0 V
1 1 V
1 1 V
2 0 V
1 1 V
1 0 V
1 1 V
2 0 V
1 0 V
1 1 V
1 1 V
2 0 V
1 1 V
1 0 V
1 1 V
2 0 V
1 1 V
1 0 V
1 1 V
2 0 V
1 1 V
1 0 V
1 1 V
2 0 V
1 1 V
1 0 V
1 1 V
1 0 V
2 1 V
1 0 V
1 1 V
1 0 V
1 1 V
2 0 V
1 1 V
1 0 V
1 1 V
1 0 V
2 1 V
1 0 V
1 1 V
1 0 V
1 0 V
2 1 V
1 0 V
1 1 V
1 0 V
1 1 V
1 0 V
2 1 V
1 0 V
1 1 V
1 0 V
1 1 V
1 0 V
2 1 V
stroke
3646 3216 M
1 0 V
1 0 V
1 1 V
1 0 V
1 0 V
1 1 V
2 1 V
1 0 V
1 1 V
1 0 V
1 0 V
1 1 V
1 0 V
2 1 V
1 0 V
1 1 V
1 0 V
1 0 V
1 1 V
1 0 V
1 1 V
2 0 V
1 1 V
1 0 V
1 1 V
1 0 V
1 1 V
1 0 V
1 0 V
1 1 V
2 0 V
1 0 V
1 1 V
1 0 V
1 1 V
1 0 V
1 1 V
1 0 V
1 0 V
1 1 V
2 0 V
1 1 V
1 0 V
1 1 V
1 0 V
1 0 V
1 1 V
1 0 V
1 1 V
1 0 V
1 0 V
1 1 V
1 0 V
2 1 V
1 0 V
1 0 V
1 1 V
1 0 V
1 1 V
1 0 V
1 1 V
1 0 V
1 0 V
1 1 V
1 0 V
1 0 V
1 1 V
1 0 V
1 1 V
1 0 V
1 0 V
1 1 V
2 0 V
1 1 V
1 0 V
1 0 V
1 1 V
1 0 V
1 0 V
1 1 V
1 0 V
1 0 V
1 1 V
1 0 V
1 1 V
1 0 V
1 1 V
1 0 V
1 0 V
1 0 V
1 1 V
1 0 V
1 1 V
1 0 V
1 1 V
1 0 V
1 0 V
1 0 V
1 1 V
1 0 V
1 1 V
1 0 V
1 1 V
1 0 V
stroke
3757 3258 M
1 0 V
1 0 V
1 1 V
1 0 V
1 1 V
1 0 V
1 0 V
1 1 V
1 0 V
1 0 V
1 1 V
1 0 V
1 0 V
1 1 V
1 0 V
1 1 V
1 0 V
1 0 V
1 0 V
1 1 V
1 0 V
1 1 V
1 0 V
1 0 V
1 1 V
1 0 V
1 0 V
1 1 V
1 0 V
1 0 V
1 1 V
1 0 V
1 0 V
1 0 V
1 1 V
1 0 V
1 0 V
1 1 V
1 0 V
1 0 V
1 1 V
1 0 V
1 0 V
1 1 V
1 0 V
1 0 V
1 0 V
1 1 V
1 0 V
1 0 V
1 0 V
1 1 V
1 0 V
1 0 V
1 1 V
1 0 V
1 0 V
1 1 V
1 0 V
1 0 V
1 0 V
1 1 V
1 0 V
1 0 V
1 0 V
1 1 V
1 0 V
1 0 V
1 1 V
1 0 V
1 0 V
1 0 V
1 1 V
1 0 V
1 0 V
1 0 V
1 1 V
1 0 V
1 0 V
1 0 V
1 1 V
1 0 V
1 0 V
1 0 V
0 1 V
1 0 V
1 0 V
1 0 V
1 0 V
1 1 V
1 0 V
1 0 V
1 0 V
1 0 V
1 1 V
1 0 V
1 0 V
1 0 V
1 0 V
1 0 V
0 1 V
1 0 V
1 0 V
1 0 V
stroke
3859 3287 M
1 0 V
1 0 V
1 1 V
1 0 V
1 0 V
1 0 V
1 0 V
1 1 V
1 0 V
1 0 V
1 0 V
1 0 V
1 0 V
1 1 V
1 0 V
1 0 V
1 0 V
1 0 V
1 0 V
1 1 V
1 0 V
1 0 V
1 0 V
1 0 V
1 0 V
1 0 V
0 1 V
1 0 V
1 0 V
1 0 V
1 0 V
1 0 V
1 0 V
1 1 V
1 0 V
1 0 V
1 0 V
1 0 V
1 0 V
1 0 V
1 0 V
0 1 V
1 0 V
1 0 V
1 0 V
1 0 V
1 0 V
1 0 V
1 0 V
1 1 V
1 0 V
1 0 V
1 0 V
1 0 V
1 0 V
1 0 V
1 0 V
1 0 V
1 1 V
1 0 V
1 0 V
1 0 V
1 0 V
1 0 V
1 0 V
1 0 V
1 0 V
1 1 V
1 0 V
1 0 V
1 0 V
1 0 V
1 0 V
1 0 V
1 0 V
1 0 V
1 0 V
1 1 V
1 0 V
1 0 V
1 0 V
1 0 V
1 0 V
1 0 V
1 0 V
1 0 V
1 0 V
1 0 V
0 1 V
1 0 V
1 0 V
1 0 V
1 0 V
1 0 V
1 0 V
1 0 V
1 0 V
1 0 V
1 0 V
1 0 V
1 1 V
1 0 V
1 0 V
1 0 V
stroke
3960 3300 M
1 0 V
1 0 V
1 0 V
1 0 V
1 0 V
1 0 V
1 0 V
1 0 V
1 0 V
1 0 V
1 0 V
1 1 V
1 0 V
1 0 V
1 0 V
1 0 V
1 0 V
1 0 V
1 0 V
1 0 V
1.000 UL
LT1
600 1291 M
7 5 V
7 4 V
6 5 V
7 4 V
7 5 V
7 5 V
6 4 V
7 5 V
7 5 V
7 4 V
7 5 V
6 4 V
7 5 V
7 5 V
7 4 V
6 5 V
7 5 V
7 4 V
7 5 V
6 4 V
7 5 V
7 5 V
7 4 V
7 5 V
6 5 V
7 4 V
7 5 V
7 4 V
6 5 V
7 5 V
7 4 V
7 5 V
7 5 V
6 4 V
7 5 V
7 4 V
7 5 V
6 5 V
7 4 V
7 5 V
7 5 V
6 4 V
7 5 V
7 4 V
7 5 V
7 5 V
6 4 V
7 5 V
7 4 V
7 5 V
6 5 V
7 4 V
7 5 V
7 5 V
7 4 V
6 5 V
7 4 V
7 5 V
7 5 V
6 4 V
7 5 V
7 5 V
7 4 V
7 5 V
6 4 V
7 5 V
7 5 V
7 4 V
6 5 V
7 5 V
7 4 V
7 5 V
6 4 V
7 5 V
7 5 V
7 4 V
7 5 V
6 5 V
7 4 V
7 5 V
7 4 V
6 5 V
7 5 V
7 4 V
7 5 V
7 4 V
6 5 V
7 5 V
7 4 V
7 5 V
6 5 V
7 4 V
7 5 V
7 4 V
6 5 V
7 5 V
7 4 V
7 5 V
7 4 V
6 5 V
7 5 V
7 4 V
7 5 V
6 5 V
stroke
1304 1772 M
7 4 V
7 5 V
7 4 V
7 5 V
6 5 V
7 4 V
7 5 V
7 4 V
6 5 V
7 5 V
7 4 V
7 5 V
7 5 V
6 4 V
7 5 V
7 4 V
7 5 V
6 5 V
7 4 V
7 5 V
7 4 V
6 5 V
7 5 V
7 4 V
7 5 V
7 4 V
6 5 V
7 5 V
7 4 V
7 5 V
6 4 V
7 5 V
7 5 V
7 4 V
7 5 V
6 4 V
7 5 V
7 5 V
7 4 V
6 5 V
7 4 V
7 5 V
7 5 V
6 4 V
7 5 V
7 4 V
7 5 V
7 5 V
6 4 V
7 5 V
7 4 V
7 5 V
6 5 V
7 4 V
7 5 V
7 4 V
7 5 V
6 5 V
7 4 V
7 5 V
7 4 V
6 5 V
7 4 V
7 5 V
7 5 V
7 4 V
6 5 V
7 4 V
7 5 V
7 5 V
6 4 V
7 5 V
7 4 V
7 5 V
6 4 V
7 5 V
7 5 V
7 4 V
7 5 V
6 4 V
7 5 V
7 4 V
7 5 V
6 4 V
7 5 V
7 5 V
7 4 V
7 5 V
6 4 V
7 5 V
7 4 V
7 5 V
6 4 V
7 5 V
7 5 V
7 4 V
6 5 V
7 4 V
7 5 V
7 4 V
7 5 V
6 4 V
7 5 V
7 4 V
stroke
2009 2248 M
7 5 V
6 5 V
7 4 V
7 5 V
7 4 V
7 5 V
6 4 V
7 5 V
7 4 V
7 5 V
6 4 V
7 5 V
7 4 V
7 5 V
7 4 V
6 5 V
7 4 V
7 5 V
7 4 V
6 5 V
7 4 V
7 5 V
7 4 V
6 5 V
7 4 V
7 5 V
7 4 V
7 5 V
6 4 V
7 5 V
7 4 V
7 5 V
6 4 V
7 4 V
7 5 V
7 4 V
7 5 V
6 4 V
7 5 V
7 4 V
7 5 V
6 4 V
7 4 V
7 5 V
7 4 V
6 5 V
7 4 V
7 5 V
7 4 V
7 4 V
6 5 V
7 4 V
7 5 V
7 4 V
6 4 V
7 5 V
7 4 V
7 5 V
7 4 V
6 4 V
7 5 V
7 4 V
7 4 V
6 5 V
7 4 V
7 4 V
7 5 V
6 4 V
7 4 V
7 5 V
7 4 V
7 4 V
6 5 V
7 4 V
7 4 V
7 5 V
6 4 V
7 4 V
7 4 V
7 5 V
7 4 V
6 4 V
7 5 V
7 4 V
7 4 V
6 4 V
7 4 V
7 5 V
7 4 V
7 4 V
6 4 V
7 5 V
7 4 V
7 4 V
6 4 V
7 4 V
7 4 V
7 5 V
6 4 V
7 4 V
7 4 V
7 4 V
7 4 V
6 4 V
stroke
2713 2703 M
7 5 V
7 4 V
7 4 V
6 4 V
7 4 V
7 4 V
7 4 V
7 4 V
6 4 V
7 4 V
7 4 V
7 4 V
6 4 V
7 4 V
7 4 V
7 4 V
6 4 V
7 4 V
7 4 V
7 4 V
7 4 V
6 4 V
7 4 V
7 4 V
7 4 V
6 4 V
7 4 V
7 4 V
7 4 V
7 3 V
6 4 V
7 4 V
7 4 V
7 4 V
6 4 V
7 4 V
7 3 V
7 4 V
7 4 V
6 4 V
7 3 V
7 4 V
7 4 V
6 4 V
7 3 V
7 4 V
7 4 V
6 4 V
7 3 V
7 4 V
7 4 V
7 3 V
6 4 V
7 4 V
7 3 V
7 4 V
6 3 V
7 4 V
7 4 V
7 3 V
7 4 V
6 3 V
7 4 V
7 3 V
7 4 V
6 3 V
7 4 V
7 3 V
7 4 V
6 3 V
7 4 V
7 3 V
7 4 V
7 3 V
6 3 V
7 4 V
7 3 V
7 4 V
6 3 V
7 3 V
7 4 V
7 3 V
7 3 V
6 4 V
7 3 V
7 3 V
7 4 V
6 3 V
7 3 V
7 3 V
7 4 V
7 3 V
6 3 V
7 3 V
7 3 V
7 4 V
6 3 V
7 3 V
7 3 V
7 3 V
6 3 V
7 3 V
7 4 V
7 3 V
stroke
3418 3082 M
7 3 V
6 3 V
7 3 V
7 3 V
7 3 V
6 3 V
7 3 V
7 3 V
7 3 V
7 3 V
6 3 V
7 3 V
7 3 V
7 3 V
6 3 V
7 3 V
7 3 V
7 3 V
6 3 V
7 2 V
7 3 V
7 3 V
7 3 V
6 3 V
7 3 V
7 3 V
7 2 V
6 3 V
7 3 V
7 3 V
7 3 V
7 2 V
6 3 V
7 3 V
7 3 V
7 2 V
6 3 V
7 3 V
7 2 V
7 3 V
7 3 V
6 2 V
7 3 V
7 3 V
7 2 V
6 3 V
7 3 V
7 2 V
7 3 V
6 3 V
7 2 V
7 3 V
7 2 V
7 3 V
6 2 V
7 3 V
7 3 V
7 2 V
6 3 V
7 2 V
7 3 V
7 2 V
7 3 V
6 2 V
7 3 V
7 2 V
7 3 V
6 2 V
7 3 V
7 2 V
7 2 V
6 3 V
7 2 V
7 3 V
7 2 V
7 2 V
6 3 V
7 2 V
7 3 V
7 2 V
6 2 V
7 3 V
7 2 V
1.000 UL
LT2
600 1587 M
70 0 V
64 0 V
59 148 V
54 0 V
51 0 V
47 0 V
44 0 V
42 92 V
40 0 V
38 120 V
36 0 V
34 0 V
32 0 V
32 0 V
30 44 V
28 37 V
28 61 V
27 0 V
25 0 V
25 0 V
24 49 V
24 0 V
22 0 V
22 0 V
21 22 V
21 38 V
20 17 V
20 0 V
19 31 V
18 0 V
19 0 V
17 0 V
18 41 V
16 23 V
17 0 V
16 22 V
16 0 V
15 0 V
15 10 V
15 19 V
15 0 V
14 27 V
14 0 V
14 16 V
13 15 V
13 0 V
13 0 V
13 14 V
13 0 V
12 7 V
12 19 V
12 13 V
12 11 V
11 0 V
12 11 V
11 0 V
11 11 V
11 0 V
11 25 V
10 0 V
11 9 V
10 0 V
10 14 V
10 8 V
10 0 V
10 0 V
9 12 V
10 8 V
9 15 V
9 0 V
9 15 V
9 0 V
9 0 V
9 6 V
9 4 V
8 16 V
9 6 V
8 6 V
8 0 V
9 6 V
8 6 V
8 5 V
8 6 V
8 2 V
7 6 V
8 10 V
8 2 V
7 10 V
8 0 V
7 10 V
7 0 V
8 0 V
7 11 V
7 9 V
7 4 V
7 4 V
7 4 V
6 4 V
7 2 V
7 4 V
7 6 V
6 7 V
7 4 V
6 3 V
stroke
2386 2774 M
7 7 V
6 7 V
6 0 V
6 0 V
7 7 V
6 5 V
6 4 V
6 4 V
6 9 V
6 0 V
6 3 V
6 3 V
5 5 V
6 4 V
6 3 V
6 5 V
5 6 V
6 0 V
5 4 V
6 5 V
5 5 V
6 0 V
5 3 V
5 8 V
6 4 V
5 3 V
5 2 V
5 4 V
6 3 V
5 0 V
5 4 V
5 8 V
5 3 V
5 2 V
5 4 V
5 2 V
4 2 V
5 4 V
5 2 V
5 6 V
5 2 V
4 2 V
5 5 V
5 3 V
4 2 V
5 2 V
5 3 V
4 2 V
5 6 V
4 2 V
5 5 V
4 3 V
4 0 V
5 3 V
4 3 V
4 3 V
5 3 V
4 3 V
4 5 V
5 1 V
4 2 V
4 6 V
4 0 V
4 1 V
4 2 V
4 6 V
4 3 V
5 2 V
4 2 V
4 3 V
4 2 V
3 0 V
4 2 V
4 7 V
4 2 V
4 3 V
4 2 V
4 2 V
4 2 V
3 1 V
4 5 V
4 0 V
4 5 V
3 1 V
4 3 V
4 4 V
3 1 V
4 1 V
4 3 V
3 1 V
4 2 V
3 4 V
4 4 V
4 2 V
3 1 V
4 2 V
3 2 V
4 2 V
3 0 V
3 7 V
4 0 V
3 3 V
4 0 V
3 4 V
stroke
2862 3089 M
3 2 V
4 1 V
3 1 V
3 4 V
4 2 V
3 3 V
3 1 V
3 3 V
4 1 V
3 2 V
3 0 V
3 4 V
4 3 V
3 2 V
3 2 V
3 0 V
3 3 V
3 2 V
3 1 V
3 3 V
3 1 V
4 1 V
3 3 V
3 2 V
3 3 V
3 0 V
3 2 V
3 2 V
3 0 V
3 4 V
3 3 V
3 1 V
2 2 V
3 1 V
3 1 V
3 2 V
3 1 V
3 3 V
3 2 V
3 1 V
2 1 V
3 3 V
3 1 V
3 2 V
3 1 V
3 1 V
2 3 V
3 1 V
3 2 V
3 3 V
2 0 V
3 1 V
3 1 V
2 2 V
3 1 V
3 3 V
2 1 V
3 2 V
3 1 V
2 2 V
3 1 V
3 2 V
2 0 V
3 3 V
3 1 V
2 3 V
3 0 V
2 1 V
3 3 V
2 0 V
3 1 V
2 2 V
3 2 V
2 2 V
3 0 V
3 2 V
2 1 V
2 1 V
3 2 V
2 1 V
3 2 V
2 1 V
3 1 V
2 2 V
3 2 V
2 0 V
2 1 V
3 2 V
2 0 V
3 2 V
2 2 V
2 1 V
3 2 V
2 0 V
2 2 V
3 1 V
2 1 V
2 2 V
3 1 V
2 1 V
2 1 V
3 1 V
2 1 V
2 2 V
stroke
3148 3255 M
3 1 V
2 0 V
2 3 V
2 1 V
3 2 V
2 0 V
2 1 V
2 1 V
3 0 V
2 2 V
2 2 V
2 1 V
2 1 V
3 1 V
2 1 V
2 2 V
2 0 V
2 2 V
2 0 V
3 2 V
2 0 V
2 3 V
2 0 V
2 1 V
2 1 V
2 1 V
2 1 V
2 1 V
3 2 V
2 1 V
2 1 V
2 1 V
2 0 V
2 1 V
2 1 V
2 1 V
2 2 V
2 2 V
2 0 V
2 1 V
2 1 V
2 1 V
2 1 V
2 0 V
2 2 V
2 1 V
2 1 V
2 1 V
2 1 V
2 1 V
2 1 V
2 0 V
2 2 V
2 1 V
2 1 V
2 1 V
2 1 V
2 0 V
2 1 V
2 2 V
2 0 V
2 1 V
1 1 V
2 1 V
2 2 V
2 0 V
2 1 V
2 1 V
2 0 V
2 1 V
2 1 V
1 2 V
2 1 V
2 0 V
2 1 V
2 1 V
2 1 V
1 1 V
2 1 V
2 1 V
2 1 V
2 0 V
2 2 V
1 1 V
2 0 V
2 1 V
2 1 V
2 1 V
1 1 V
2 1 V
2 1 V
2 0 V
2 1 V
1 1 V
2 1 V
2 0 V
2 2 V
1 1 V
2 0 V
2 1 V
2 1 V
1 1 V
2 1 V
2 0 V
stroke
3354 3357 M
1 1 V
2 1 V
2 1 V
2 1 V
1 1 V
2 1 V
2 0 V
1 0 V
2 2 V
2 1 V
1 1 V
2 0 V
2 1 V
1 1 V
2 1 V
2 0 V
1 1 V
2 1 V
2 1 V
1 0 V
2 2 V
2 0 V
1 1 V
2 0 V
2 1 V
1 1 V
2 1 V
1 1 V
2 1 V
2 0 V
1 1 V
2 0 V
1 1 V
2 1 V
2 1 V
1 1 V
2 0 V
1 1 V
2 1 V
2 1 V
1 0 V
2 1 V
1 0 V
2 2 V
1 1 V
2 0 V
2 1 V
1 1 V
2 0 V
1 1 V
2 0 V
1 1 V
2 1 V
1 1 V
2 1 V
1 0 V
2 1 V
1 1 V
2 0 V
1 1 V
2 1 V
1 0 V
2 2 V
1 0 V
2 1 V
1 0 V
2 1 V
1 0 V
2 1 V
1 1 V
2 1 V
1 1 V
2 0 V
1 1 V
2 1 V
1 0 V
2 1 V
1 1 V
2 1 V
1 0 V
1 1 V
2 1 V
1 0 V
2 0 V
1 1 V
2 1 V
1 1 V
2 0 V
1 1 V
1 1 V
2 0 V
1 1 V
2 0 V
1 1 V
1 1 V
2 1 V
1 0 V
2 1 V
1 0 V
1 1 V
2 1 V
1 0 V
2 1 V
1 0 V
stroke
3514 3432 M
1 1 V
2 1 V
1 1 V
2 0 V
1 1 V
1 0 V
2 1 V
1 1 V
1 0 V
2 1 V
1 0 V
1 1 V
2 1 V
1 0 V
2 1 V
1 0 V
1 1 V
2 1 V
1 0 V
1 1 V
2 1 V
1 0 V
1 0 V
2 1 V
1 1 V
1 1 V
1 0 V
2 1 V
1 0 V
1 1 V
2 0 V
1 1 V
1 0 V
2 1 V
1 1 V
1 0 V
2 1 V
1 0 V
1 1 V
1 0 V
2 1 V
1 1 V
1 0 V
2 1 V
1 0 V
1 1 V
1 0 V
2 1 V
1 0 V
1 1 V
1 1 V
2 0 V
1 1 V
1 0 V
1 1 V
2 0 V
1 1 V
1 0 V
1 1 V
2 0 V
1 1 V
1 1 V
1 0 V
2 1 V
1 0 V
1 0 V
1 1 V
2 1 V
1 0 V
1 1 V
1 0 V
2 1 V
1 0 V
1 1 V
1 0 V
1 1 V
2 0 V
1 1 V
1 0 V
1 1 V
1 0 V
2 1 V
1 0 V
1 0 V
1 1 V
1 1 V
2 0 V
1 1 V
1 0 V
1 1 V
1 0 V
2 1 V
1 0 V
1 1 V
1 0 V
1 1 V
1 0 V
2 1 V
1 0 V
1 1 V
1 0 V
1 1 V
1 0 V
2 1 V
stroke
3646 3490 M
1 0 V
1 1 V
1 0 V
1 0 V
1 1 V
1 0 V
2 1 V
1 0 V
1 1 V
1 0 V
1 1 V
1 0 V
1 1 V
2 0 V
1 1 V
1 0 V
1 1 V
1 0 V
1 1 V
1 0 V
1 0 V
2 1 V
1 0 V
1 1 V
1 0 V
1 1 V
1 0 V
1 1 V
1 0 V
1 1 V
2 0 V
1 1 V
1 0 V
1 1 V
1 0 V
1 0 V
1 1 V
1 0 V
1 1 V
1 0 V
2 1 V
1 0 V
1 1 V
1 0 V
1 0 V
1 1 V
1 0 V
1 1 V
1 0 V
1 1 V
1 0 V
1 1 V
1 0 V
2 0 V
1 1 V
1 0 V
1 1 V
1 0 V
1 0 V
1 1 V
1 0 V
1 1 V
1 0 V
1 1 V
1 0 V
1 1 V
1 0 V
1 0 V
1 1 V
1 0 V
1 0 V
1 1 V
2 0 V
1 1 V
1 0 V
1 1 V
1 0 V
1 0 V
1 1 V
1 0 V
1 0 V
1 1 V
1 0 V
1 1 V
1 0 V
1 0 V
1 1 V
1 0 V
1 0 V
1 1 V
1 0 V
1 1 V
1 0 V
1 0 V
1 1 V
1 0 V
1 1 V
1 0 V
1 0 V
1 1 V
1 0 V
1 0 V
1 1 V
1 0 V
stroke
3757 3534 M
1 0 V
1 1 V
1 0 V
1 1 V
1 0 V
1 0 V
1 1 V
1 0 V
1 0 V
1 1 V
1 0 V
1 0 V
1 1 V
1 0 V
1 0 V
1 1 V
1 0 V
1 1 V
1 0 V
1 0 V
1 0 V
1 1 V
1 0 V
1 0 V
0 1 V
1 0 V
1 0 V
1 1 V
1 0 V
1 0 V
1 1 V
1 0 V
1 0 V
1 1 V
1 0 V
1 0 V
1 0 V
1 1 V
1 0 V
1 0 V
1 1 V
1 0 V
1 0 V
1 0 V
0 1 V
1 0 V
1 0 V
1 1 V
1 0 V
1 0 V
1 1 V
1 0 V
1 0 V
1 1 V
1 0 V
1 0 V
1 0 V
1 1 V
1 0 V
1 1 V
1 0 V
1 0 V
1 0 V
1 1 V
1 0 V
1 0 V
1 0 V
1 1 V
1 0 V
1 1 V
1 0 V
1 0 V
1 0 V
1 1 V
1 0 V
1 0 V
1 0 V
1 1 V
1 0 V
1 1 V
1 0 V
1 0 V
1 0 V
1 0 V
1 1 V
1 0 V
1 0 V
1 1 V
1 0 V
1 0 V
1 0 V
1 1 V
1 0 V
1 0 V
1 0 V
1 1 V
1 0 V
1 0 V
1 0 V
1 0 V
1 1 V
1 0 V
1 0 V
1 0 V
stroke
3859 3565 M
1 1 V
1 0 V
1 0 V
1 0 V
1 0 V
1 1 V
1 0 V
1 0 V
1 0 V
1 1 V
1 0 V
1 0 V
1 0 V
1 0 V
1 0 V
1 1 V
1 0 V
1 0 V
1 0 V
1 0 V
1 0 V
1 1 V
1 0 V
1 0 V
1 0 V
1 0 V
1 0 V
1 1 V
1 0 V
1 0 V
1 0 V
1 0 V
1 0 V
1 0 V
1 1 V
1 0 V
1 0 V
1 0 V
1 0 V
1 0 V
1 0 V
1 0 V
1 1 V
1 0 V
1 0 V
1 0 V
1 0 V
1 0 V
1 0 V
1 0 V
1 0 V
1 1 V
1 0 V
1 0 V
1 0 V
1 0 V
1 0 V
1 0 V
1 0 V
1 0 V
1 0 V
1 1 V
1 0 V
1 0 V
1 0 V
1 0 V
1 0 V
1 0 V
1 0 V
1 0 V
1 0 V
1 0 V
1 0 V
1 1 V
1 0 V
1 0 V
1 0 V
1 0 V
1 0 V
1 0 V
1 0 V
1 0 V
1 0 V
1 0 V
1 0 V
1 0 V
1 0 V
1 0 V
1 0 V
1 0 V
0 1 V
1 0 V
1 0 V
1 0 V
1 0 V
1 0 V
1 0 V
1 0 V
1 0 V
1 0 V
1 0 V
1 0 V
1 0 V
1 0 V
stroke
3962 3577 M
1 0 V
1 0 V
1 0 V
1 0 V
1 0 V
1 1 V
1 0 V
1 0 V
1 0 V
1 0 V
1 0 V
1 0 V
1 0 V
1 0 V
1 0 V
1 0 V
1 0 V
1 0 V
1.000 UL
LT3
1583 300 M
16 125 V
18 0 V
19 0 V
17 0 V
18 134 V
16 65 V
17 0 V
16 55 V
16 0 V
15 0 V
15 34 V
15 45 V
15 0 V
14 43 V
14 0 V
14 29 V
13 37 V
13 0 V
13 0 V
13 29 V
13 0 V
12 8 V
12 41 V
12 26 V
12 21 V
11 0 V
12 25 V
11 0 V
11 26 V
11 0 V
11 47 V
10 0 V
11 16 V
10 0 V
10 29 V
10 24 V
10 0 V
10 0 V
9 21 V
10 15 V
9 34 V
9 0 V
9 34 V
9 0 V
9 0 V
9 15 V
9 9 V
8 30 V
9 10 V
8 15 V
8 0 V
9 12 V
8 14 V
8 11 V
8 13 V
8 3 V
7 9 V
8 20 V
8 6 V
7 20 V
8 0 V
7 20 V
7 0 V
8 0 V
7 19 V
7 15 V
7 8 V
7 7 V
7 8 V
6 6 V
7 4 V
7 8 V
7 8 V
6 12 V
7 5 V
6 5 V
7 12 V
6 12 V
6 0 V
6 0 V
7 11 V
6 5 V
6 6 V
6 7 V
6 13 V
6 0 V
6 4 V
6 5 V
5 10 V
6 6 V
6 1 V
6 7 V
5 7 V
6 0 V
5 5 V
6 9 V
5 8 V
6 0 V
5 3 V
5 10 V
6 6 V
5 5 V
5 2 V
5 8 V
stroke
2547 1737 M
6 3 V
5 0 V
5 7 V
5 11 V
5 4 V
5 2 V
5 6 V
5 3 V
4 4 V
5 7 V
5 4 V
5 10 V
5 2 V
4 2 V
5 9 V
5 6 V
4 3 V
5 3 V
5 6 V
4 3 V
5 9 V
4 2 V
5 9 V
4 6 V
4 0 V
5 5 V
4 5 V
4 5 V
5 5 V
4 6 V
4 7 V
5 3 V
4 2 V
4 10 V
4 0 V
4 2 V
4 3 V
4 10 V
4 6 V
5 2 V
4 4 V
4 5 V
4 4 V
3 0 V
4 2 V
4 12 V
4 4 V
4 4 V
4 5 V
4 3 V
4 3 V
3 2 V
4 8 V
4 0 V
4 9 V
3 1 V
4 4 V
4 7 V
3 2 V
4 2 V
4 5 V
3 2 V
4 2 V
3 6 V
4 6 V
4 3 V
3 2 V
4 3 V
3 5 V
4 3 V
3 0 V
3 9 V
4 2 V
3 4 V
4 0 V
3 7 V
3 3 V
4 1 V
3 1 V
3 6 V
4 3 V
3 5 V
3 2 V
3 4 V
4 2 V
3 2 V
3 1 V
3 6 V
4 4 V
3 3 V
3 4 V
3 0 V
3 3 V
3 3 V
3 2 V
3 4 V
3 3 V
4 1 V
3 4 V
3 3 V
3 3 V
3 1 V
3 3 V
3 2 V
stroke
2951 2148 M
3 1 V
3 5 V
3 4 V
3 2 V
2 2 V
3 2 V
3 1 V
3 3 V
3 2 V
3 4 V
3 2 V
3 2 V
2 1 V
3 4 V
3 2 V
3 2 V
3 1 V
3 1 V
2 5 V
3 2 V
3 2 V
3 4 V
2 0 V
3 2 V
3 1 V
2 3 V
3 2 V
3 3 V
2 2 V
3 3 V
3 1 V
2 2 V
3 2 V
3 2 V
2 1 V
3 3 V
3 2 V
2 4 V
3 0 V
2 2 V
3 3 V
2 0 V
3 1 V
2 3 V
3 2 V
2 3 V
3 0 V
3 2 V
2 2 V
2 1 V
3 2 V
2 2 V
3 2 V
2 2 V
3 1 V
2 2 V
3 3 V
2 0 V
2 1 V
3 2 V
2 1 V
3 2 V
2 2 V
2 2 V
3 1 V
2 1 V
2 1 V
3 2 V
2 1 V
2 2 V
3 2 V
2 1 V
2 1 V
3 2 V
2 1 V
2 2 V
3 0 V
2 1 V
2 3 V
2 1 V
3 2 V
2 0 V
2 1 V
2 2 V
3 0 V
2 2 V
2 2 V
2 1 V
2 1 V
3 2 V
2 1 V
2 1 V
2 1 V
2 2 V
2 0 V
3 2 V
2 1 V
2 2 V
2 1 V
2 1 V
2 1 V
2 1 V
2 1 V
2 1 V
stroke
3209 2329 M
3 2 V
2 2 V
2 1 V
2 1 V
2 1 V
2 1 V
2 1 V
2 1 V
2 2 V
2 2 V
2 0 V
2 1 V
2 2 V
2 1 V
2 1 V
2 0 V
2 2 V
2 2 V
2 1 V
2 1 V
2 1 V
2 1 V
2 1 V
2 1 V
2 2 V
2 1 V
2 2 V
2 1 V
2 1 V
2 1 V
2 1 V
2 1 V
2 0 V
2 2 V
1 2 V
2 1 V
2 2 V
2 0 V
2 2 V
2 1 V
2 0 V
2 1 V
2 2 V
1 2 V
2 1 V
2 1 V
2 1 V
2 1 V
2 1 V
1 2 V
2 1 V
2 1 V
2 2 V
2 0 V
2 2 V
1 1 V
2 1 V
2 1 V
2 1 V
2 2 V
1 1 V
2 1 V
2 2 V
2 1 V
2 0 V
1 1 V
2 2 V
2 0 V
2 2 V
1 2 V
2 1 V
2 1 V
2 1 V
1 1 V
2 1 V
2 0 V
1 2 V
2 2 V
2 1 V
2 1 V
1 1 V
2 1 V
2 1 V
1 0 V
2 2 V
2 2 V
1 1 V
2 1 V
2 0 V
1 1 V
2 2 V
2 0 V
1 1 V
2 2 V
2 1 V
1 1 V
2 1 V
2 1 V
1 1 V
2 0 V
2 2 V
1 1 V
2 1 V
1 2 V
stroke
3400 2451 M
2 1 V
2 0 V
1 1 V
2 1 V
1 1 V
2 1 V
2 1 V
1 1 V
2 1 V
1 1 V
2 1 V
2 1 V
1 1 V
2 1 V
1 0 V
2 2 V
1 2 V
2 0 V
2 1 V
1 2 V
2 0 V
1 1 V
2 1 V
1 1 V
2 2 V
1 0 V
2 2 V
1 1 V
2 1 V
1 1 V
2 0 V
1 1 V
2 2 V
1 1 V
2 1 V
1 1 V
2 1 V
1 1 V
2 1 V
1 1 V
2 1 V
1 1 V
2 2 V
1 0 V
2 1 V
1 1 V
2 2 V
1 0 V
2 1 V
1 2 V
2 1 V
1 1 V
1 1 V
2 1 V
1 1 V
2 0 V
1 1 V
2 2 V
1 1 V
2 1 V
1 0 V
1 2 V
2 1 V
1 0 V
2 1 V
1 1 V
1 2 V
2 1 V
1 1 V
2 1 V
1 0 V
1 1 V
2 1 V
1 1 V
2 1 V
1 1 V
1 1 V
2 1 V
1 1 V
2 1 V
1 1 V
1 1 V
2 1 V
1 1 V
1 1 V
2 1 V
1 1 V
1 0 V
2 1 V
1 1 V
2 1 V
1 1 V
1 1 V
2 1 V
1 1 V
1 1 V
2 1 V
1 0 V
1 1 V
2 1 V
1 1 V
1 1 V
1 1 V
2 1 V
stroke
3552 2553 M
1 1 V
1 0 V
2 1 V
1 1 V
1 1 V
2 1 V
1 1 V
1 1 V
2 1 V
1 0 V
1 1 V
1 1 V
2 0 V
1 1 V
1 2 V
2 0 V
1 1 V
1 1 V
1 0 V
2 1 V
1 1 V
1 1 V
1 1 V
2 0 V
1 1 V
1 1 V
1 1 V
2 0 V
1 1 V
1 1 V
1 1 V
2 0 V
1 1 V
1 1 V
1 1 V
2 1 V
1 0 V
1 1 V
1 0 V
2 1 V
1 1 V
1 1 V
1 0 V
2 1 V
1 0 V
1 1 V
1 1 V
1 0 V
2 2 V
1 0 V
1 1 V
1 0 V
1 1 V
2 1 V
1 0 V
1 0 V
1 2 V
1 1 V
2 0 V
1 1 V
1 1 V
1 0 V
1 1 V
2 0 V
1 1 V
1 1 V
1 1 V
1 0 V
1 1 V
2 1 V
1 0 V
1 1 V
1 1 V
1 1 V
1 0 V
2 1 V
1 1 V
1 0 V
1 1 V
1 1 V
1 0 V
1 1 V
2 1 V
1 1 V
1 0 V
1 1 V
1 1 V
1 0 V
1 1 V
2 1 V
1 0 V
1 1 V
1 1 V
1 1 V
1 1 V
1 0 V
1 1 V
2 0 V
1 1 V
1 1 V
1 1 V
1 0 V
1 1 V
1 1 V
stroke
3677 2629 M
1 0 V
1 1 V
2 1 V
1 1 V
1 0 V
1 1 V
1 1 V
1 1 V
1 0 V
1 1 V
1 1 V
1 0 V
2 1 V
1 1 V
1 0 V
1 1 V
1 0 V
1 1 V
1 1 V
1 0 V
1 2 V
1 0 V
1 1 V
1 0 V
1 1 V
2 1 V
1 1 V
1 0 V
1 1 V
1 0 V
1 1 V
1 0 V
1 1 V
1 1 V
1 0 V
1 1 V
1 1 V
1 0 V
1 1 V
1 1 V
1 0 V
1 1 V
1 0 V
1 1 V
2 1 V
1 0 V
1 0 V
1 1 V
1 1 V
1 0 V
1 1 V
1 0 V
1 1 V
1 1 V
1 0 V
1 1 V
1 0 V
1 1 V
1 0 V
1 1 V
1 0 V
1 1 V
1 0 V
1 1 V
1 0 V
1 0 V
1 1 V
1 1 V
1 0 V
1 1 V
1 0 V
1 1 V
1 0 V
1 0 V
1 1 V
1 1 V
1 0 V
1 1 V
1 0 V
1 1 V
1 0 V
1 0 V
1 1 V
1 1 V
1 0 V
1 1 V
1 0 V
1 0 V
1 1 V
1 0 V
1 1 V
1 0 V
1 0 V
1 1 V
1 0 V
1 1 V
1 0 V
1 1 V
1 0 V
1 1 V
1 1 V
1 0 V
1 1 V
1 0 V
stroke
3785 2688 M
1 1 V
1 0 V
1 1 V
1 0 V
1 1 V
1 0 V
1 0 V
1 1 V
1 0 V
1 1 V
1 0 V
1 1 V
1 1 V
1 0 V
1 0 V
0 1 V
1 1 V
1 0 V
1 0 V
1 1 V
1 1 V
1 0 V
1 0 V
1 1 V
1 1 V
1 0 V
1 1 V
1 0 V
1 1 V
1 1 V
1 0 V
1 1 V
1 0 V
1 1 V
1 0 V
1 1 V
1 0 V
1 0 V
1 1 V
1 0 V
0 1 V
1 1 V
1 0 V
1 1 V
1 0 V
1 0 V
1 1 V
1 0 V
1 1 V
1 0 V
0 1 V
1 0 V
1 1 V
1 0 V
1 0 V
1 1 V
1 0 V
1 1 V
1 0 V
0 1 V
1 0 V
1 1 V
1 0 V
1 1 V
1 0 V
1 1 V
1 0 V
1 0 V
0 1 V
1 0 V
1 1 V
1 0 V
1 1 V
1 0 V
1 0 V
1 0 V
0 1 V
1 0 V
1 1 V
1 0 V
1 0 V
1 1 V
1 0 V
1 1 V
1 0 V
1 0 V
1 1 V
1 0 V
1 0 V
0 1 V
1 0 V
1 0 V
1 1 V
1 0 V
1 0 V
1 0 V
0 1 V
1 0 V
1 0 V
1 1 V
1 0 V
1 0 V
1 0 V
1 1 V
stroke
3881 2734 M
1 0 V
1 0 V
1 0 V
1 1 V
1 0 V
1 0 V
1 0 V
1 1 V
1 0 V
1 0 V
1 0 V
1 0 V
1 1 V
1 0 V
1 0 V
1 0 V
1 0 V
1 1 V
1 0 V
1 0 V
1 0 V
1 0 V
1 1 V
1 0 V
1 0 V
1 0 V
1 0 V
1 0 V
1 0 V
1 0 V
1 0 V
0 1 V
1 0 V
1 0 V
1 0 V
1 0 V
1 0 V
1 0 V
1 0 V
1 0 V
1 0 V
1 0 V
1 0 V
1 0 V
1 1 V
1 0 V
1 0 V
1 0 V
1 0 V
1 0 V
1 0 V
1 0 V
1 0 V
1 0 V
1 0 V
1 0 V
1 0 V
1 0 V
1 0 V
1 0 V
1 0 V
1 0 V
1 0 V
1 0 V
1 0 V
1 0 V
1 0 V
1 0 V
1 0 V
1 0 V
1 0 V
1 0 V
1 0 V
1 0 V
1 0 V
1 0 V
1 0 V
1 0 V
1 0 V
1 0 V
1 0 V
1 0 V
1 0 V
1 0 V
1 0 V
1 0 V
1 0 V
1 0 V
1 0 V
1 0 V
1 0 V
1 1 V
1 0 V
0 -1 V
1 1 V
1 0 V
1 0 V
1 0 V
1 0 V
1 0 V
1 0 V
0.500 UL
LTb
600 300 M
3380 0 V
0 3380 V
-3380 0 V
600 300 L
1.000 UP
stroke
grestore
end
showpage
}}%
\put(2290,50){\makebox(0,0){$R\ [\mbox{kpc}]$}}%
\put(200,1990){%
\special{ps: gsave currentpoint currentpoint translate
270 rotate neg exch neg exch translate}%
\makebox(0,0)[b]{\shortstack{$M(R)\ [M_{\odot}]$}}%
\special{ps: currentpoint grestore moveto}%
}%
\put(3980,200){\makebox(0,0){ 1000}}%
\put(2290,200){\makebox(0,0){ 100}}%
\put(600,200){\makebox(0,0){ 10}}%
\put(550,793){\makebox(0,0)[r]{$10^{10}$}}%
\put(550,1371){\makebox(0,0)[r]{$10^{11}$}}%
\put(550,1948){\makebox(0,0)[r]{$10^{12}$}}%
\put(550,2525){\makebox(0,0)[r]{$10^{13}$}}%
\put(550,3103){\makebox(0,0)[r]{$10^{14}$}}%
\put(550,3680){\makebox(0,0)[r]{$10^{15}$}}%
\end{picture}%
\endgroup
 

%% file: figure/gnuplot/Kappa_MainXray.tex
\begingroup%
  \makeatletter%
  \newcommand{\GNUPLOTspecial}{%
    \@sanitize\catcode`\%=14\relax\special}%
  \setlength{\unitlength}{0.1bp}%
\begin{picture}(3780,3780)(0,0)%
{\GNUPLOTspecial{"
/gnudict 256 dict def
gnudict begin
/Color true def
/Solid false def
/gnulinewidth 5.000 def
/userlinewidth gnulinewidth def
/vshift -33 def
/dl {10.0 mul} def
/hpt_ 31.5 def
/vpt_ 31.5 def
/hpt hpt_ def
/vpt vpt_ def
/Rounded false def
/M {moveto} bind def
/L {lineto} bind def
/R {rmoveto} bind def
/V {rlineto} bind def
/N {newpath moveto} bind def
/C {setrgbcolor} bind def
/f {rlineto fill} bind def
/vpt2 vpt 2 mul def
/hpt2 hpt 2 mul def
/Lshow { currentpoint stroke M
  0 vshift R show } def
/Rshow { currentpoint stroke M
  dup stringwidth pop neg vshift R show } def
/Cshow { currentpoint stroke M
  dup stringwidth pop -2 div vshift R show } def
/UP { dup vpt_ mul /vpt exch def hpt_ mul /hpt exch def
  /hpt2 hpt 2 mul def /vpt2 vpt 2 mul def } def
/DL { Color {setrgbcolor Solid {pop []} if 0 setdash }
 {pop pop pop 0 setgray Solid {pop []} if 0 setdash} ifelse } def
/BL { stroke userlinewidth 2 mul setlinewidth
      Rounded { 1 setlinejoin 1 setlinecap } if } def
/AL { stroke userlinewidth 2 div setlinewidth
      Rounded { 1 setlinejoin 1 setlinecap } if } def
/UL { dup gnulinewidth mul /userlinewidth exch def
      dup 1 lt {pop 1} if 10 mul /udl exch def } def
/PL { stroke userlinewidth setlinewidth
      Rounded { 1 setlinejoin 1 setlinecap } if } def
/LTw { PL [] 1 setgray } def
/LTb { BL [] 0 0 0 DL } def
/LTa { AL [1 udl mul 2 udl mul] 0 setdash 0 0 0 setrgbcolor } def
/fatlinewidth 7.500 def
/gatlinewidth 10.000 def
/FL { stroke fatlinewidth setlinewidth Rounded { 1 setlinejoin 1 setlinecap } if } def
/GL { stroke gatlinewidth setlinewidth Rounded { 1 setlinejoin 1 setlinecap } if } def
/LT0 { FL [] 0 0 0 DL } def
/LT1 { GL [4 dl 2 dl] 1 0 0 DL } def
/LT2 { GL [2 dl 3 dl] 0 0.5 0 DL } def
/LT3 { PL [1 dl 1.5 dl] 1 0 1 DL } def
/LT4 { PL [5 dl 2 dl 1 dl 2 dl] 0 1 1 DL } def
/LT5 { PL [4 dl 3 dl 1 dl 3 dl] 1 1 0 DL } def
/LT6 { PL [2 dl 2 dl 2 dl 4 dl] 0 0 0 DL } def
/LT7 { PL [2 dl 2 dl 2 dl 2 dl 2 dl 4 dl] 1 0.3 0 DL } def
/LT8 { PL [2 dl 2 dl 2 dl 2 dl 2 dl 2 dl 2 dl 4 dl] 0.5 0.5 0.5 DL } def
/Pnt { stroke [] 0 setdash
   gsave 1 setlinecap M 0 0 V stroke grestore } def
/Dia { stroke [] 0 setdash 2 copy vpt add M
  hpt neg vpt neg V hpt vpt neg V
  hpt vpt V hpt neg vpt V closepath stroke
  Pnt } def
/Pls { stroke [] 0 setdash vpt sub M 0 vpt2 V
  currentpoint stroke M
  hpt neg vpt neg R hpt2 0 V stroke
  } def
/Box { stroke [] 0 setdash 2 copy exch hpt sub exch vpt add M
  0 vpt2 neg V hpt2 0 V 0 vpt2 V
  hpt2 neg 0 V closepath stroke
  Pnt } def
/Crs { stroke [] 0 setdash exch hpt sub exch vpt add M
  hpt2 vpt2 neg V currentpoint stroke M
  hpt2 neg 0 R hpt2 vpt2 V stroke } def
/TriU { stroke [] 0 setdash 2 copy vpt 1.12 mul add M
  hpt neg vpt -1.62 mul V
  hpt 2 mul 0 V
  hpt neg vpt 1.62 mul V closepath stroke
  Pnt  } def
/Star { 2 copy Pls Crs } def
/BoxF { stroke [] 0 setdash exch hpt sub exch vpt add M
  0 vpt2 neg V  hpt2 0 V  0 vpt2 V
  hpt2 neg 0 V  closepath fill } def
/TriUF { stroke [] 0 setdash vpt 1.12 mul add M
  hpt neg vpt -1.62 mul V
  hpt 2 mul 0 V
  hpt neg vpt 1.62 mul V closepath fill } def
/TriD { stroke [] 0 setdash 2 copy vpt 1.12 mul sub M
  hpt neg vpt 1.62 mul V
  hpt 2 mul 0 V
  hpt neg vpt -1.62 mul V closepath stroke
  Pnt  } def
/TriDF { stroke [] 0 setdash vpt 1.12 mul sub M
  hpt neg vpt 1.62 mul V
  hpt 2 mul 0 V
  hpt neg vpt -1.62 mul V closepath fill} def
/DiaF { stroke [] 0 setdash vpt add M
  hpt neg vpt neg V hpt vpt neg V
  hpt vpt V hpt neg vpt V closepath fill } def
/Pent { stroke [] 0 setdash 2 copy gsave
  translate 0 hpt M 4 {72 rotate 0 hpt L} repeat
  closepath stroke grestore Pnt } def
/PentF { stroke [] 0 setdash gsave
  translate 0 hpt M 4 {72 rotate 0 hpt L} repeat
  closepath fill grestore } def
/Circle { stroke [] 0 setdash 2 copy
  hpt 0 360 arc stroke Pnt } def
/CircleF { stroke [] 0 setdash hpt 0 360 arc fill } def
/C0 { BL [] 0 setdash 2 copy moveto vpt 90 450  arc } bind def
/C1 { BL [] 0 setdash 2 copy        moveto
       2 copy  vpt 0 90 arc closepath fill
               vpt 0 360 arc closepath } bind def
/C2 { BL [] 0 setdash 2 copy moveto
       2 copy  vpt 90 180 arc closepath fill
               vpt 0 360 arc closepath } bind def
/C3 { BL [] 0 setdash 2 copy moveto
       2 copy  vpt 0 180 arc closepath fill
               vpt 0 360 arc closepath } bind def
/C4 { BL [] 0 setdash 2 copy moveto
       2 copy  vpt 180 270 arc closepath fill
               vpt 0 360 arc closepath } bind def
/C5 { BL [] 0 setdash 2 copy moveto
       2 copy  vpt 0 90 arc
       2 copy moveto
       2 copy  vpt 180 270 arc closepath fill
               vpt 0 360 arc } bind def
/C6 { BL [] 0 setdash 2 copy moveto
      2 copy  vpt 90 270 arc closepath fill
              vpt 0 360 arc closepath } bind def
/C7 { BL [] 0 setdash 2 copy moveto
      2 copy  vpt 0 270 arc closepath fill
              vpt 0 360 arc closepath } bind def
/C8 { BL [] 0 setdash 2 copy moveto
      2 copy vpt 270 360 arc closepath fill
              vpt 0 360 arc closepath } bind def
/C9 { BL [] 0 setdash 2 copy moveto
      2 copy  vpt 270 450 arc closepath fill
              vpt 0 360 arc closepath } bind def
/C10 { BL [] 0 setdash 2 copy 2 copy moveto vpt 270 360 arc closepath fill
       2 copy moveto
       2 copy vpt 90 180 arc closepath fill
               vpt 0 360 arc closepath } bind def
/C11 { BL [] 0 setdash 2 copy moveto
       2 copy  vpt 0 180 arc closepath fill
       2 copy moveto
       2 copy  vpt 270 360 arc closepath fill
               vpt 0 360 arc closepath } bind def
/C12 { BL [] 0 setdash 2 copy moveto
       2 copy  vpt 180 360 arc closepath fill
               vpt 0 360 arc closepath } bind def
/C13 { BL [] 0 setdash  2 copy moveto
       2 copy  vpt 0 90 arc closepath fill
       2 copy moveto
       2 copy  vpt 180 360 arc closepath fill
               vpt 0 360 arc closepath } bind def
/C14 { BL [] 0 setdash 2 copy moveto
       2 copy  vpt 90 360 arc closepath fill
               vpt 0 360 arc } bind def
/C15 { BL [] 0 setdash 2 copy vpt 0 360 arc closepath fill
               vpt 0 360 arc closepath } bind def
/Rec   { newpath 4 2 roll moveto 1 index 0 rlineto 0 exch rlineto
       neg 0 rlineto closepath } bind def
/Square { dup Rec } bind def
/Bsquare { vpt sub exch vpt sub exch vpt2 Square } bind def
/S0 { BL [] 0 setdash 2 copy moveto 0 vpt rlineto BL Bsquare } bind def
/S1 { BL [] 0 setdash 2 copy vpt Square fill Bsquare } bind def
/S2 { BL [] 0 setdash 2 copy exch vpt sub exch vpt Square fill Bsquare } bind def
/S3 { BL [] 0 setdash 2 copy exch vpt sub exch vpt2 vpt Rec fill Bsquare } bind def
/S4 { BL [] 0 setdash 2 copy exch vpt sub exch vpt sub vpt Square fill Bsquare } bind def
/S5 { BL [] 0 setdash 2 copy 2 copy vpt Square fill
       exch vpt sub exch vpt sub vpt Square fill Bsquare } bind def
/S6 { BL [] 0 setdash 2 copy exch vpt sub exch vpt sub vpt vpt2 Rec fill Bsquare } bind def
/S7 { BL [] 0 setdash 2 copy exch vpt sub exch vpt sub vpt vpt2 Rec fill
       2 copy vpt Square fill
       Bsquare } bind def
/S8 { BL [] 0 setdash 2 copy vpt sub vpt Square fill Bsquare } bind def
/S9 { BL [] 0 setdash 2 copy vpt sub vpt vpt2 Rec fill Bsquare } bind def
/S10 { BL [] 0 setdash 2 copy vpt sub vpt Square fill 2 copy exch vpt sub exch vpt Square fill
       Bsquare } bind def
/S11 { BL [] 0 setdash 2 copy vpt sub vpt Square fill 2 copy exch vpt sub exch vpt2 vpt Rec fill
       Bsquare } bind def
/S12 { BL [] 0 setdash 2 copy exch vpt sub exch vpt sub vpt2 vpt Rec fill Bsquare } bind def
/S13 { BL [] 0 setdash 2 copy exch vpt sub exch vpt sub vpt2 vpt Rec fill
       2 copy vpt Square fill Bsquare } bind def
/S14 { BL [] 0 setdash 2 copy exch vpt sub exch vpt sub vpt2 vpt Rec fill
       2 copy exch vpt sub exch vpt Square fill Bsquare } bind def
/S15 { BL [] 0 setdash 2 copy Bsquare fill Bsquare } bind def
/D0 { gsave translate 45 rotate 0 0 S0 stroke grestore } bind def
/D1 { gsave translate 45 rotate 0 0 S1 stroke grestore } bind def
/D2 { gsave translate 45 rotate 0 0 S2 stroke grestore } bind def
/D3 { gsave translate 45 rotate 0 0 S3 stroke grestore } bind def
/D4 { gsave translate 45 rotate 0 0 S4 stroke grestore } bind def
/D5 { gsave translate 45 rotate 0 0 S5 stroke grestore } bind def
/D6 { gsave translate 45 rotate 0 0 S6 stroke grestore } bind def
/D7 { gsave translate 45 rotate 0 0 S7 stroke grestore } bind def
/D8 { gsave translate 45 rotate 0 0 S8 stroke grestore } bind def
/D9 { gsave translate 45 rotate 0 0 S9 stroke grestore } bind def
/D10 { gsave translate 45 rotate 0 0 S10 stroke grestore } bind def
/D11 { gsave translate 45 rotate 0 0 S11 stroke grestore } bind def
/D12 { gsave translate 45 rotate 0 0 S12 stroke grestore } bind def
/D13 { gsave translate 45 rotate 0 0 S13 stroke grestore } bind def
/D14 { gsave translate 45 rotate 0 0 S14 stroke grestore } bind def
/D15 { gsave translate 45 rotate 0 0 S15 stroke grestore } bind def
/DiaE { stroke [] 0 setdash vpt add M
  hpt neg vpt neg V hpt vpt neg V
  hpt vpt V hpt neg vpt V closepath stroke } def
/BoxE { stroke [] 0 setdash exch hpt sub exch vpt add M
  0 vpt2 neg V hpt2 0 V 0 vpt2 V
  hpt2 neg 0 V closepath stroke } def
/TriUE { stroke [] 0 setdash vpt 1.12 mul add M
  hpt neg vpt -1.62 mul V
  hpt 2 mul 0 V
  hpt neg vpt 1.62 mul V closepath stroke } def
/TriDE { stroke [] 0 setdash vpt 1.12 mul sub M
  hpt neg vpt 1.62 mul V
  hpt 2 mul 0 V
  hpt neg vpt -1.62 mul V closepath stroke } def
/PentE { stroke [] 0 setdash gsave
  translate 0 hpt M 4 {72 rotate 0 hpt L} repeat
  closepath stroke grestore } def
/CircE { stroke [] 0 setdash 
  hpt 0 360 arc stroke } def
/Opaque { gsave closepath 1 setgray fill grestore 0 setgray closepath } def
/DiaW { stroke [] 0 setdash vpt add M
  hpt neg vpt neg V hpt vpt neg V
  hpt vpt V hpt neg vpt V Opaque stroke } def
/BoxW { stroke [] 0 setdash exch hpt sub exch vpt add M
  0 vpt2 neg V hpt2 0 V 0 vpt2 V
  hpt2 neg 0 V Opaque stroke } def
/TriUW { stroke [] 0 setdash vpt 1.12 mul add M
  hpt neg vpt -1.62 mul V
  hpt 2 mul 0 V
  hpt neg vpt 1.62 mul V Opaque stroke } def
/TriDW { stroke [] 0 setdash vpt 1.12 mul sub M
  hpt neg vpt 1.62 mul V
  hpt 2 mul 0 V
  hpt neg vpt -1.62 mul V Opaque stroke } def
/PentW { stroke [] 0 setdash gsave
  translate 0 hpt M 4 {72 rotate 0 hpt L} repeat
  Opaque stroke grestore } def
/CircW { stroke [] 0 setdash 
  hpt 0 360 arc Opaque stroke } def
/BoxFill { gsave Rec 1 setgray fill grestore } def
/BoxColFill {
  gsave Rec
  /Fillden exch def
  currentrgbcolor
  /ColB exch def /ColG exch def /ColR exch def
  /ColR ColR Fillden mul Fillden sub 1 add def
  /ColG ColG Fillden mul Fillden sub 1 add def
  /ColB ColB Fillden mul Fillden sub 1 add def
  ColR ColG ColB setrgbcolor
  fill grestore } def
%
%
/PatternFill { gsave /PFa [ 9 2 roll ] def
    PFa 0 get PFa 2 get 2 div add PFa 1 get PFa 3 get 2 div add translate
    PFa 2 get -2 div PFa 3 get -2 div PFa 2 get PFa 3 get Rec
    gsave 1 setgray fill grestore clip
    currentlinewidth 0.5 mul setlinewidth
    /PFs PFa 2 get dup mul PFa 3 get dup mul add sqrt def
    0 0 M PFa 5 get rotate PFs -2 div dup translate
	0 1 PFs PFa 4 get div 1 add floor cvi
	{ PFa 4 get mul 0 M 0 PFs V } for
    0 PFa 6 get ne {
	0 1 PFs PFa 4 get div 1 add floor cvi
	{ PFa 4 get mul 0 2 1 roll M PFs 0 V } for
    } if
    stroke grestore } def
/Symbol-Oblique /Symbol findfont [1 0 .167 1 0 0] makefont
dup length dict begin {1 index /FID eq {pop pop} {def} ifelse} forall
currentdict end definefont pop
end
gnudict begin
gsave
0 0 translate
0.100 0.100 scale
0 setgray
newpath
0.500 UL
LTb
450 300 M
63 0 V
3317 0 R
-63 0 V
0.500 UL
LTb
450 469 M
31 0 V
3349 0 R
-31 0 V
450 638 M
31 0 V
3349 0 R
-31 0 V
450 807 M
31 0 V
3349 0 R
-31 0 V
450 976 M
63 0 V
3317 0 R
-63 0 V
0.500 UL
LTb
450 1145 M
31 0 V
3349 0 R
-31 0 V
450 1314 M
31 0 V
3349 0 R
-31 0 V
450 1483 M
31 0 V
3349 0 R
-31 0 V
450 1652 M
63 0 V
3317 0 R
-63 0 V
0.500 UL
LTb
450 1821 M
31 0 V
3349 0 R
-31 0 V
450 1990 M
31 0 V
3349 0 R
-31 0 V
450 2159 M
31 0 V
3349 0 R
-31 0 V
450 2328 M
63 0 V
3317 0 R
-63 0 V
0.500 UL
LTb
450 2497 M
31 0 V
3349 0 R
-31 0 V
450 2666 M
31 0 V
3349 0 R
-31 0 V
450 2835 M
31 0 V
3349 0 R
-31 0 V
450 3004 M
63 0 V
3317 0 R
-63 0 V
0.500 UL
LTb
450 3173 M
31 0 V
3349 0 R
-31 0 V
450 3342 M
31 0 V
3349 0 R
-31 0 V
450 3511 M
31 0 V
3349 0 R
-31 0 V
450 3680 M
63 0 V
3317 0 R
-63 0 V
0.500 UL
LTb
450 300 M
0 63 V
0 3317 R
0 -63 V
0.500 UL
LTb
619 300 M
0 31 V
0 3349 R
0 -31 V
788 300 M
0 31 V
0 3349 R
0 -31 V
957 300 M
0 31 V
0 3349 R
0 -31 V
1126 300 M
0 31 V
0 3349 R
0 -31 V
1295 300 M
0 63 V
0 3317 R
0 -63 V
0.500 UL
LTb
1464 300 M
0 31 V
0 3349 R
0 -31 V
1633 300 M
0 31 V
0 3349 R
0 -31 V
1802 300 M
0 31 V
0 3349 R
0 -31 V
1971 300 M
0 31 V
0 3349 R
0 -31 V
2140 300 M
0 63 V
0 3317 R
0 -63 V
0.500 UL
LTb
2309 300 M
0 31 V
0 3349 R
0 -31 V
2478 300 M
0 31 V
0 3349 R
0 -31 V
2647 300 M
0 31 V
0 3349 R
0 -31 V
2816 300 M
0 31 V
0 3349 R
0 -31 V
2985 300 M
0 63 V
0 3317 R
0 -63 V
0.500 UL
LTb
3154 300 M
0 31 V
0 3349 R
0 -31 V
3323 300 M
0 31 V
0 3349 R
0 -31 V
3492 300 M
0 31 V
0 3349 R
0 -31 V
3661 300 M
0 31 V
0 3349 R
0 -31 V
3830 300 M
0 63 V
0 3317 R
0 -63 V
0.500 UL
LTb
0.500 UL
LTb
450 300 M
3380 0 V
0 3380 V
-3380 0 V
450 300 L
LTb
LTb
1.000 UP
1.000 UL
LT0
450 517 M
7 1 V
7 2 V
6 2 V
7 1 V
7 2 V
7 2 V
6 2 V
7 2 V
7 2 V
7 1 V
7 2 V
6 2 V
7 2 V
7 2 V
7 2 V
6 2 V
7 2 V
7 2 V
7 2 V
6 2 V
7 3 V
7 2 V
7 2 V
7 2 V
6 2 V
7 3 V
7 2 V
7 2 V
6 3 V
7 2 V
7 2 V
7 3 V
7 2 V
6 3 V
7 2 V
7 3 V
7 2 V
6 3 V
7 3 V
7 2 V
7 3 V
6 3 V
7 3 V
7 3 V
7 2 V
7 3 V
6 3 V
7 3 V
7 3 V
7 3 V
6 3 V
7 3 V
7 4 V
7 3 V
7 3 V
6 3 V
7 4 V
7 3 V
7 3 V
6 4 V
7 3 V
7 4 V
7 3 V
7 4 V
6 4 V
7 3 V
7 4 V
7 4 V
6 4 V
7 4 V
7 4 V
7 4 V
6 4 V
7 4 V
7 5 V
7 4 V
7 4 V
6 5 V
7 4 V
7 5 V
7 4 V
6 5 V
7 5 V
7 5 V
7 4 V
7 5 V
6 5 V
7 6 V
7 5 V
7 5 V
6 5 V
7 6 V
7 5 V
7 6 V
6 5 V
7 6 V
7 6 V
7 6 V
7 6 V
6 6 V
7 6 V
7 6 V
7 7 V
6 6 V
stroke
1154 875 M
7 7 V
7 6 V
7 7 V
7 7 V
6 7 V
7 7 V
7 7 V
7 7 V
6 8 V
7 7 V
7 8 V
7 8 V
7 8 V
6 8 V
7 8 V
7 8 V
7 9 V
6 8 V
7 9 V
7 9 V
7 8 V
6 10 V
7 9 V
7 9 V
7 10 V
7 9 V
6 10 V
7 10 V
7 10 V
7 11 V
6 10 V
7 11 V
7 10 V
7 11 V
7 12 V
6 11 V
7 11 V
7 12 V
7 12 V
6 12 V
7 12 V
7 13 V
7 13 V
6 12 V
7 14 V
7 13 V
7 13 V
7 14 V
6 14 V
7 14 V
7 15 V
7 14 V
6 15 V
7 15 V
7 15 V
7 16 V
7 16 V
6 15 V
7 17 V
7 16 V
7 17 V
6 17 V
7 17 V
7 17 V
7 18 V
7 18 V
6 18 V
7 19 V
7 19 V
7 19 V
6 19 V
7 19 V
7 20 V
7 20 V
6 20 V
7 21 V
7 20 V
7 21 V
7 22 V
6 21 V
7 22 V
7 22 V
7 22 V
6 22 V
7 23 V
7 22 V
7 23 V
7 23 V
6 24 V
7 23 V
7 24 V
7 23 V
6 24 V
7 24 V
7 24 V
7 24 V
6 25 V
7 24 V
7 24 V
7 24 V
7 25 V
6 24 V
7 24 V
7 24 V
stroke
1859 2467 M
7 24 V
6 24 V
7 23 V
7 23 V
7 24 V
7 22 V
6 23 V
7 22 V
7 21 V
7 22 V
6 20 V
7 20 V
7 20 V
7 19 V
7 18 V
6 18 V
7 17 V
7 15 V
7 16 V
6 14 V
7 13 V
7 12 V
7 11 V
6 10 V
7 9 V
7 7 V
7 7 V
7 5 V
6 3 V
7 3 V
7 1 V
7 -1 V
6 -2 V
7 -3 V
7 -5 V
7 -6 V
7 -7 V
6 -8 V
7 -9 V
7 -7 V
7 -3 V
6 0 V
7 3 V
7 7 V
7 9 V
6 8 V
7 7 V
7 6 V
7 5 V
7 3 V
6 2 V
7 1 V
7 -1 V
7 -3 V
6 -3 V
7 -5 V
7 -7 V
7 -7 V
7 -9 V
6 -10 V
7 -11 V
7 -12 V
7 -13 V
6 -14 V
7 -16 V
7 -15 V
7 -17 V
6 -18 V
7 -18 V
7 -19 V
7 -20 V
7 -20 V
6 -20 V
7 -22 V
7 -21 V
7 -22 V
6 -23 V
7 -22 V
7 -24 V
7 -23 V
7 -23 V
6 -24 V
7 -24 V
7 -24 V
7 -24 V
6 -24 V
7 -25 V
7 -24 V
7 -24 V
7 -24 V
6 -25 V
7 -24 V
7 -24 V
7 -24 V
6 -24 V
7 -23 V
7 -24 V
7 -23 V
6 -24 V
7 -23 V
7 -23 V
7 -22 V
7 -23 V
6 -22 V
stroke
2563 1970 M
7 -22 V
7 -22 V
7 -22 V
6 -21 V
7 -22 V
7 -21 V
7 -20 V
7 -21 V
6 -20 V
7 -20 V
7 -20 V
7 -19 V
6 -19 V
7 -19 V
7 -19 V
7 -19 V
6 -18 V
7 -18 V
7 -18 V
7 -17 V
7 -17 V
6 -17 V
7 -17 V
7 -16 V
7 -17 V
6 -15 V
7 -16 V
7 -16 V
7 -15 V
7 -15 V
6 -15 V
7 -14 V
7 -15 V
7 -14 V
6 -14 V
7 -14 V
7 -13 V
7 -13 V
7 -14 V
6 -12 V
7 -13 V
7 -13 V
7 -12 V
6 -12 V
7 -12 V
7 -12 V
7 -11 V
6 -11 V
7 -12 V
7 -11 V
7 -10 V
7 -11 V
6 -10 V
7 -11 V
7 -10 V
7 -10 V
6 -10 V
7 -9 V
7 -10 V
7 -9 V
7 -9 V
6 -10 V
7 -8 V
7 -9 V
7 -9 V
6 -8 V
7 -9 V
7 -8 V
7 -8 V
6 -8 V
7 -8 V
7 -8 V
7 -8 V
7 -7 V
6 -8 V
7 -7 V
7 -7 V
7 -7 V
6 -7 V
7 -7 V
7 -7 V
7 -6 V
7 -7 V
6 -6 V
7 -7 V
7 -6 V
7 -6 V
6 -6 V
7 -6 V
7 -6 V
7 -6 V
7 -6 V
6 -5 V
7 -6 V
7 -5 V
7 -6 V
6 -5 V
7 -5 V
7 -5 V
7 -6 V
6 -5 V
7 -5 V
7 -4 V
7 -5 V
stroke
3268 758 M
7 -5 V
6 -5 V
7 -4 V
7 -5 V
7 -4 V
6 -5 V
7 -4 V
7 -4 V
7 -5 V
7 -4 V
6 -4 V
7 -4 V
7 -4 V
7 -4 V
6 -4 V
7 -4 V
7 -4 V
7 -3 V
6 -4 V
7 -4 V
7 -3 V
7 -4 V
7 -3 V
6 -4 V
7 -3 V
7 -3 V
7 -4 V
6 -3 V
7 -3 V
7 -3 V
7 -4 V
7 -3 V
6 -3 V
7 -3 V
7 -3 V
7 -3 V
6 -3 V
7 -3 V
7 -2 V
7 -3 V
7 -3 V
6 -3 V
7 -3 V
7 -2 V
7 -3 V
6 -3 V
7 -2 V
7 -3 V
7 -2 V
6 -3 V
7 -2 V
7 -3 V
7 -2 V
7 -2 V
6 -3 V
7 -2 V
7 -2 V
7 -3 V
6 -2 V
7 -2 V
7 -2 V
7 -2 V
7 -3 V
6 -2 V
7 -2 V
7 -2 V
7 -2 V
6 -2 V
7 -2 V
7 -2 V
7 -2 V
6 -2 V
7 -2 V
7 -1 V
7 -2 V
7 -2 V
6 -2 V
7 -2 V
7 -2 V
7 -1 V
6 -2 V
7 -2 V
7 -1 V
1.000 UL
LT1
450 552 M
3 7 V
3 8 V
2 7 V
3 6 V
3 6 V
3 5 V
2 3 V
3 3 V
3 1 V
3 0 V
2 -1 V
3 -2 V
3 -4 V
3 -6 V
2 -7 V
3 -8 V
3 -10 V
3 -11 V
2 -13 V
3 -13 V
3 -14 V
3 -15 V
2 -16 V
3 -16 V
3 -17 V
3 -17 V
2 -17 V
3 -18 V
3 -17 V
3 -16 V
2 -16 V
3 -14 V
3 -13 V
3 -11 V
1 -6 V
99 0 R
1 2 V
3 5 V
3 2 V
3 -3 V
2 -6 V
21 0 R
1 6 V
3 14 V
3 15 V
3 15 V
2 14 V
3 14 V
3 13 V
3 12 V
2 12 V
3 10 V
3 8 V
3 8 V
2 6 V
3 5 V
3 3 V
3 3 V
2 1 V
3 1 V
3 0 V
3 0 V
2 -1 V
3 0 V
3 0 V
3 1 V
2 1 V
3 2 V
3 3 V
3 4 V
2 4 V
3 5 V
3 6 V
3 6 V
2 6 V
3 7 V
3 6 V
3 7 V
2 6 V
3 7 V
3 6 V
3 6 V
2 6 V
3 5 V
3 5 V
3 5 V
2 4 V
3 4 V
3 4 V
3 3 V
2 3 V
3 3 V
3 3 V
3 2 V
2 2 V
3 3 V
3 2 V
3 2 V
2 2 V
3 2 V
3 1 V
3 1 V
2 1 V
3 0 V
stroke
846 605 M
3 0 V
3 -1 V
2 -1 V
3 -3 V
3 -3 V
3 -4 V
2 -4 V
3 -5 V
3 -6 V
3 -7 V
2 -7 V
3 -8 V
3 -8 V
3 -9 V
2 -9 V
3 -8 V
3 -9 V
3 -8 V
2 -7 V
3 -7 V
3 -5 V
3 -5 V
2 -3 V
3 -2 V
3 0 V
3 1 V
2 2 V
3 3 V
3 4 V
3 4 V
2 5 V
3 4 V
3 5 V
3 5 V
3 5 V
2 5 V
3 5 V
3 5 V
3 4 V
2 4 V
3 4 V
3 4 V
3 3 V
2 3 V
3 2 V
3 2 V
3 1 V
2 1 V
3 0 V
3 0 V
3 -1 V
2 -2 V
3 -3 V
3 -3 V
3 -4 V
2 -5 V
3 -4 V
3 -6 V
3 -5 V
2 -5 V
3 -5 V
3 -5 V
3 -4 V
2 -4 V
3 -3 V
3 -4 V
3 -3 V
2 -2 V
3 -3 V
3 -2 V
3 -3 V
2 -2 V
3 -3 V
3 -2 V
3 -3 V
2 -2 V
3 -2 V
3 -2 V
3 -2 V
2 0 V
3 1 V
3 1 V
3 3 V
2 5 V
3 6 V
3 6 V
3 8 V
2 8 V
3 8 V
3 8 V
3 8 V
2 8 V
3 7 V
3 7 V
3 6 V
2 6 V
3 6 V
3 7 V
3 7 V
2 7 V
3 8 V
3 9 V
3 10 V
2 11 V
stroke
1132 624 M
3 12 V
3 13 V
3 14 V
2 14 V
3 15 V
3 14 V
3 14 V
2 13 V
3 12 V
3 12 V
3 10 V
2 10 V
3 9 V
3 9 V
3 8 V
2 8 V
3 9 V
3 9 V
3 9 V
2 9 V
3 10 V
3 10 V
3 9 V
2 10 V
3 9 V
3 9 V
3 9 V
2 9 V
3 7 V
3 8 V
3 6 V
2 6 V
3 6 V
3 6 V
3 5 V
2 6 V
3 5 V
3 6 V
3 7 V
2 7 V
3 8 V
3 8 V
3 9 V
2 9 V
3 10 V
3 10 V
3 11 V
2 12 V
3 12 V
3 13 V
3 13 V
2 13 V
3 13 V
3 13 V
3 13 V
2 13 V
3 12 V
3 12 V
3 11 V
2 10 V
3 10 V
3 9 V
3 8 V
2 7 V
3 6 V
3 6 V
3 4 V
2 4 V
3 3 V
3 2 V
3 3 V
2 5 V
3 6 V
3 10 V
3 13 V
2 19 V
3 24 V
3 30 V
3 35 V
2 34 V
3 28 V
3 21 V
3 15 V
2 9 V
3 5 V
3 2 V
3 0 V
2 -1 V
3 -1 V
3 1 V
3 2 V
2 4 V
3 6 V
3 7 V
3 8 V
2 10 V
3 10 V
3 11 V
3 12 V
2 12 V
3 12 V
3 13 V
3 13 V
2 12 V
stroke
1418 1671 M
3 13 V
3 12 V
3 12 V
3 11 V
2 11 V
3 9 V
3 9 V
3 8 V
2 7 V
3 6 V
3 6 V
3 5 V
2 4 V
3 5 V
3 4 V
3 5 V
2 4 V
3 5 V
3 4 V
3 5 V
2 5 V
3 5 V
3 5 V
3 5 V
2 6 V
3 6 V
3 6 V
3 7 V
2 6 V
3 7 V
3 7 V
3 8 V
2 8 V
3 8 V
3 8 V
3 9 V
2 9 V
3 10 V
3 10 V
3 11 V
2 11 V
3 13 V
3 14 V
3 15 V
2 17 V
3 18 V
3 20 V
3 20 V
2 21 V
3 21 V
3 21 V
3 19 V
2 19 V
3 17 V
3 14 V
3 13 V
2 11 V
3 10 V
3 9 V
3 8 V
2 8 V
3 7 V
3 8 V
3 8 V
2 9 V
3 10 V
3 11 V
3 13 V
2 13 V
3 15 V
3 16 V
3 16 V
2 17 V
3 18 V
3 17 V
3 18 V
2 18 V
3 17 V
3 17 V
3 14 V
2 14 V
3 11 V
3 9 V
3 5 V
2 3 V
3 2 V
3 2 V
3 3 V
2 7 V
3 8 V
3 9 V
3 11 V
2 11 V
3 12 V
3 11 V
3 11 V
2 10 V
3 9 V
3 8 V
3 7 V
2 7 V
3 7 V
3 6 V
3 6 V
stroke
1705 2732 M
2 7 V
3 6 V
3 7 V
3 8 V
2 7 V
3 7 V
3 7 V
3 8 V
2 7 V
3 6 V
3 7 V
3 6 V
2 6 V
3 5 V
3 5 V
3 4 V
2 4 V
3 3 V
3 3 V
3 3 V
2 2 V
3 1 V
3 2 V
3 1 V
2 0 V
3 1 V
3 1 V
3 0 V
2 1 V
3 1 V
3 2 V
3 2 V
2 2 V
3 3 V
3 3 V
3 4 V
2 4 V
3 4 V
3 4 V
3 4 V
2 5 V
3 5 V
3 4 V
49 9 R
1 -1 V
1 -1 V
0 -1 V
1 0 V
0 -1 V
1 -1 V
1 -1 V
1 -1 V
0 -1 V
1 0 V
0 -1 V
1 -1 V
1 -1 V
0 -1 V
1 0 V
0 -1 V
1 -1 V
1 -1 V
1 -1 V
0 -1 V
1 -1 V
1 -1 V
0 -1 V
1 -1 V
1 -1 V
0 -1 V
1 -1 V
0 -1 V
1 -1 V
0 -1 V
1 -1 V
1 -1 V
0 -1 V
1 -1 V
0 -1 V
1 -1 V
0 -1 V
1 -1 V
0 -1 V
1 -2 V
0 -1 V
1 -1 V
0 -1 V
1 -1 V
0 -1 V
1 -1 V
0 -2 V
1 -1 V
0 -1 V
1 -1 V
0 -2 V
1 -1 V
0 -2 V
1 -1 V
0 -1 V
1 -2 V
0 -1 V
1 -2 V
0 -1 V
1 -2 V
stroke
1907 2849 M
0 -1 V
1 -2 V
0 -2 V
1 -1 V
0 -2 V
1 -2 V
0 -1 V
1 -2 V
0 -2 V
1 -1 V
0 -2 V
1 -2 V
0 -2 V
1 -1 V
0 -2 V
1 -2 V
0 -2 V
1 -2 V
0 -2 V
1 -2 V
0 -1 V
1 -2 V
0 -2 V
1 -2 V
0 -2 V
1 -2 V
0 -2 V
1 -2 V
0 -2 V
1 -1 V
0 -2 V
1 -2 V
0 -2 V
1 -2 V
0 -2 V
1 -2 V
0 -2 V
1 -2 V
0 -1 V
1 -2 V
1 -2 V
0 -2 V
1 -2 V
0 -2 V
1 -1 V
0 -2 V
1 -2 V
0 -2 V
1 -2 V
0 -1 V
1 -2 V
0 -2 V
1 -1 V
0 -2 V
1 -2 V
0 -2 V
1 -1 V
0 -2 V
1 -2 V
0 -1 V
1 -2 V
0 -1 V
1 -2 V
0 -2 V
1 -1 V
0 -2 V
1 -1 V
0 -2 V
1 -1 V
0 -2 V
1 -2 V
0 -1 V
1 -2 V
0 -1 V
1 -2 V
0 -1 V
1 -2 V
0 -1 V
1 -2 V
0 -1 V
1 -1 V
0 -2 V
1 -1 V
0 -2 V
1 -1 V
0 -2 V
1 -1 V
0 -2 V
1 -1 V
0 -1 V
1 -2 V
0 -1 V
1 -2 V
0 -1 V
1 -2 V
0 -1 V
1 -1 V
0 -2 V
1 -1 V
0 -2 V
1 -1 V
0 -1 V
1 -2 V
0 -1 V
stroke
1959 2677 M
1 -2 V
0 -1 V
1 -1 V
0 -2 V
1 -1 V
0 -1 V
1 -2 V
0 -1 V
1 -2 V
0 -1 V
1 -1 V
0 -2 V
1 -1 V
0 -2 V
1 -1 V
0 -1 V
1 -2 V
0 -1 V
1 -1 V
0 -2 V
1 -1 V
0 -2 V
1 -1 V
1 -1 V
0 -2 V
1 -1 V
0 -2 V
1 -1 V
0 -1 V
1 -2 V
0 -1 V
1 -2 V
0 -1 V
1 -2 V
0 -1 V
1 -1 V
0 -2 V
1 -1 V
0 -2 V
1 -1 V
0 -2 V
1 -1 V
0 -2 V
1 -1 V
0 -1 V
1 -2 V
0 -1 V
1 -2 V
0 -1 V
1 -2 V
0 -1 V
1 -2 V
0 -1 V
1 -2 V
0 -1 V
1 -2 V
0 -1 V
1 -2 V
0 -2 V
1 -1 V
0 -2 V
1 -1 V
0 -2 V
1 -1 V
0 -2 V
1 -2 V
0 -1 V
1 -2 V
0 -2 V
1 -1 V
0 -2 V
1 -1 V
0 -2 V
1 -2 V
0 -2 V
1 -1 V
0 -2 V
1 -2 V
0 -1 V
1 -2 V
0 -2 V
1 -2 V
0 -1 V
1 -2 V
0 -2 V
1 -2 V
0 -2 V
1 -1 V
0 -2 V
1 -2 V
0 -2 V
1 -2 V
0 -2 V
1 -2 V
0 -2 V
1 -2 V
0 -1 V
1 -2 V
0 -2 V
1 -2 V
0 -2 V
1 -2 V
0 -2 V
1 -2 V
stroke
2012 2513 M
0 -3 V
1 -2 V
0 -2 V
1 -2 V
1 -2 V
0 -2 V
1 -2 V
0 -2 V
1 -2 V
0 -2 V
1 -2 V
0 -2 V
1 -2 V
0 -2 V
1 -2 V
0 -2 V
1 -2 V
0 -2 V
1 -2 V
0 -2 V
1 -1 V
0 -2 V
1 -2 V
0 -1 V
1 -2 V
0 -1 V
1 -2 V
0 -1 V
1 -1 V
0 -1 V
1 -1 V
0 -2 V
1 0 V
0 -1 V
1 -1 V
0 -1 V
1 0 V
0 -1 V
1 0 V
0 -1 V
1 0 V
1 0 V
1 1 V
1 0 V
0 1 V
1 1 V
0 1 V
1 1 V
1 1 V
0 2 V
1 1 V
0 1 V
1 1 V
0 1 V
1 1 V
1 1 V
0 1 V
1 1 V
1 0 V
1 0 V
1 0 V
1 -1 V
0 -1 V
1 0 V
0 -1 V
1 -1 V
0 -2 V
1 -1 V
0 -1 V
1 -2 V
0 -1 V
1 -2 V
0 -2 V
1 -1 V
0 -2 V
1 -2 V
0 -2 V
1 -2 V
0 -3 V
1 -2 V
0 -2 V
1 -3 V
1 -2 V
0 -2 V
1 -3 V
0 -2 V
1 -3 V
0 -3 V
1 -2 V
0 -3 V
1 -3 V
0 -2 V
1 -3 V
0 -3 V
1 -2 V
0 -3 V
1 -3 V
0 -3 V
1 -2 V
0 -3 V
1 -3 V
0 -2 V
1 -3 V
0 -2 V
stroke
2069 2375 M
1 -3 V
0 -2 V
1 -3 V
0 -2 V
1 -2 V
0 -3 V
1 -2 V
0 -2 V
1 -3 V
0 -2 V
1 -2 V
0 -2 V
1 -2 V
0 -2 V
1 -2 V
0 -2 V
1 -2 V
0 -2 V
1 -2 V
0 -1 V
1 -2 V
0 -2 V
1 -2 V
0 -1 V
1 -2 V
0 -1 V
1 -2 V
0 -2 V
1 -1 V
0 -2 V
1 -1 V
0 -1 V
1 -2 V
0 -1 V
1 -1 V
0 -2 V
1 -1 V
0 -1 V
1 -1 V
0 -2 V
1 -1 V
0 -1 V
1 -1 V
0 -1 V
1 -1 V
0 -1 V
1 -1 V
0 -1 V
1 -1 V
0 -1 V
1 -1 V
1 -1 V
0 -1 V
1 -1 V
0 -1 V
1 0 V
0 -1 V
1 -1 V
1 -1 V
0 -1 V
1 0 V
0 -1 V
1 0 V
1 -1 V
0 -1 V
1 0 V
0 -1 V
1 0 V
1 -1 V
1 -1 V
1 0 V
0 -1 V
1 0 V
0 -1 V
1 0 V
1 -1 V
1 0 V
1 -1 V
1 0 V
1 -1 V
1 0 V
1 0 V
0 -1 V
1 0 V
1 0 V
1 0 V
0 -1 V
1 0 V
1 0 V
1 0 V
1 -1 V
53 -166 R
4 -2 V
3 -3 V
3 -3 V
4 -3 V
3 -3 V
3 -4 V
4 -5 V
3 -6 V
3 -7 V
3 -7 V
4 -9 V
3 -10 V
stroke
2217 2041 M
3 -10 V
4 -11 V
3 -10 V
3 -11 V
4 -10 V
3 -10 V
3 -8 V
4 -8 V
3 -7 V
3 -7 V
4 -5 V
3 -4 V
3 -3 V
3 -3 V
4 -1 V
3 -1 V
3 -2 V
4 -2 V
3 -2 V
3 -4 V
4 -5 V
3 -8 V
3 -9 V
4 -12 V
3 -14 V
3 -16 V
4 -19 V
3 -22 V
3 -24 V
3 -27 V
4 -28 V
3 -29 V
3 -27 V
4 -23 V
3 -16 V
3 -8 V
4 4 V
3 16 V
3 26 V
4 26 V
3 18 V
3 7 V
3 1 V
4 -6 V
3 -10 V
3 -12 V
4 -13 V
3 -11 V
3 -9 V
4 -7 V
3 -5 V
3 -2 V
4 -1 V
3 1 V
3 3 V
4 4 V
3 6 V
3 6 V
3 7 V
4 7 V
3 7 V
3 6 V
4 5 V
3 3 V
3 1 V
4 -1 V
3 -4 V
3 -5 V
4 -7 V
3 -9 V
3 -11 V
4 -12 V
3 -13 V
3 -13 V
3 -15 V
4 -14 V
3 -13 V
3 -13 V
4 -11 V
3 -9 V
3 -7 V
4 -5 V
3 -3 V
3 -3 V
4 -1 V
3 -1 V
3 -1 V
4 -1 V
3 -1 V
3 -2 V
3 -3 V
4 -3 V
3 -4 V
3 -3 V
4 -2 V
3 -1 V
3 0 V
4 3 V
3 4 V
3 5 V
4 6 V
3 6 V
3 7 V
4 5 V
stroke
2562 1558 M
3 4 V
3 3 V
3 2 V
4 1 V
3 0 V
3 -1 V
4 -1 V
3 -1 V
3 -1 V
4 -1 V
3 0 V
3 -1 V
4 0 V
3 0 V
3 1 V
4 -1 V
3 0 V
3 -1 V
3 -2 V
4 -2 V
3 -2 V
3 -3 V
4 -3 V
3 -2 V
3 -3 V
4 -2 V
3 -1 V
3 -1 V
4 0 V
3 1 V
3 3 V
4 4 V
3 6 V
3 9 V
3 10 V
4 13 V
3 16 V
3 16 V
4 16 V
3 16 V
3 14 V
4 12 V
3 8 V
3 4 V
4 0 V
3 -2 V
3 -4 V
4 -6 V
3 -7 V
3 -7 V
3 -7 V
4 -6 V
3 -5 V
3 -2 V
4 -1 V
3 0 V
3 2 V
4 3 V
3 4 V
3 5 V
4 6 V
3 6 V
3 7 V
3 6 V
4 7 V
3 7 V
3 7 V
4 7 V
3 6 V
3 7 V
4 6 V
3 5 V
3 6 V
4 5 V
3 4 V
3 5 V
4 4 V
3 3 V
3 3 V
3 3 V
4 3 V
3 4 V
3 3 V
4 5 V
3 5 V
3 6 V
4 7 V
3 8 V
3 8 V
4 7 V
3 7 V
3 7 V
4 5 V
3 4 V
3 4 V
3 2 V
4 3 V
3 2 V
3 2 V
4 3 V
3 3 V
3 4 V
4 4 V
3 6 V
stroke
2906 1877 M
3 7 V
4 8 V
3 8 V
3 9 V
4 10 V
3 11 V
3 11 V
3 11 V
4 12 V
3 11 V
3 10 V
4 10 V
3 9 V
3 8 V
4 7 V
3 6 V
3 5 V
4 5 V
3 6 V
3 5 V
4 7 V
3 7 V
3 9 V
3 11 V
4 12 V
3 14 V
3 17 V
4 18 V
3 20 V
3 22 V
4 25 V
3 26 V
3 25 V
4 20 V
3 14 V
3 3 V
4 -8 V
3 -24 V
3 -40 V
3 -44 V
4 -35 V
3 -16 V
3 -1 V
4 12 V
3 20 V
3 24 V
4 24 V
3 19 V
3 13 V
4 4 V
3 -1 V
3 -7 V
4 -10 V
3 -14 V
3 -15 V
3 -15 V
4 -15 V
3 -14 V
3 -15 V
4 -13 V
3 -13 V
3 -12 V
4 -12 V
3 -10 V
3 -10 V
4 -8 V
3 -7 V
3 -7 V
3 -5 V
4 -6 V
3 -5 V
3 -5 V
4 -5 V
3 -6 V
3 -6 V
4 -7 V
3 -8 V
3 -10 V
4 -10 V
3 -11 V
3 -14 V
4 -14 V
3 -15 V
3 -16 V
3 -17 V
4 -16 V
3 -15 V
3 -14 V
4 -13 V
3 -12 V
3 -9 V
4 -7 V
3 -6 V
3 -4 V
4 -4 V
3 -4 V
3 -4 V
4 -5 V
3 -5 V
3 -7 V
3 -8 V
4 -8 V
3 -9 V
3 -9 V
stroke
3250 1727 M
4 -9 V
3 -9 V
3 -10 V
4 -9 V
3 -10 V
3 -9 V
4 -10 V
3 -9 V
3 -9 V
4 -10 V
3 -9 V
3 -9 V
3 -9 V
4 -9 V
3 -10 V
3 -10 V
4 -10 V
3 -10 V
3 -11 V
4 -12 V
3 -13 V
3 -15 V
4 -17 V
3 -19 V
3 -22 V
4 -25 V
3 -29 V
3 -30 V
3 -27 V
4 -23 V
3 -20 V
3 -17 V
4 -15 V
3 -13 V
3 -12 V
4 -11 V
3 -11 V
3 -11 V
4 -12 V
3 -13 V
3 -13 V
4 -14 V
3 -16 V
3 -17 V
3 -18 V
4 -19 V
3 -19 V
3 -18 V
4 -18 V
3 -16 V
3 -14 V
4 -11 V
3 -8 V
3 -4 V
4 0 V
3 2 V
3 5 V
4 7 V
3 8 V
3 8 V
3 8 V
4 7 V
3 6 V
3 4 V
4 4 V
3 2 V
3 1 V
4 -1 V
3 -3 V
3 -4 V
4 -6 V
3 -7 V
3 -7 V
3 -8 V
4 -7 V
3 -7 V
3 -7 V
4 -6 V
3 -4 V
3 -4 V
4 -2 V
3 -2 V
3 -2 V
4 -2 V
3 -2 V
3 -3 V
4 -3 V
3 -5 V
3 -6 V
3 -7 V
4 -9 V
3 -11 V
3 -12 V
4 -14 V
3 -16 V
3 -16 V
4 -16 V
3 -16 V
3 -15 V
4 -12 V
3 -11 V
3 -8 V
4 -3 V
3 0 V
stroke
3595 772 M
3 3 V
3 7 V
4 8 V
3 8 V
3 9 V
4 8 V
3 5 V
3 3 V
4 2 V
3 -1 V
3 -2 V
4 -3 V
3 -4 V
3 -4 V
4 -5 V
3 -4 V
3 -5 V
3 -5 V
4 -4 V
3 -5 V
3 -4 V
4 -4 V
3 -4 V
3 -3 V
4 -2 V
3 -3 V
3 -2 V
4 -2 V
3 -1 V
3 -2 V
4 -2 V
3 -3 V
3 -2 V
3 -4 V
4 -4 V
3 -5 V
3 -6 V
4 -8 V
3 -8 V
3 -10 V
4 -9 V
3 -6 V
3 -3 V
4 0 V
3 3 V
3 4 V
4 5 V
3 6 V
3 7 V
3 6 V
4 5 V
3 5 V
3 2 V
4 1 V
3 -1 V
3 -4 V
4 -7 V
3 -10 V
3 -12 V
4 -15 V
3 -15 V
3 -16 V
4 -16 V
3 -17 V
3 -15 V
3 -14 V
4 -13 V
3 -10 V
3 -9 V
4 -6 V
3 -4 V
1.000 UL
LT2
868 402 M
14 2 V
14 2 V
15 3 V
14 2 V
15 2 V
14 3 V
15 2 V
15 3 V
14 2 V
15 3 V
14 2 V
14 3 V
15 3 V
14 3 V
14 2 V
15 3 V
14 3 V
14 3 V
15 3 V
15 4 V
15 3 V
14 3 V
14 3 V
15 4 V
14 3 V
15 3 V
14 4 V
14 4 V
15 3 V
14 4 V
14 4 V
16 4 V
14 4 V
15 4 V
14 4 V
14 5 V
15 4 V
14 5 V
14 4 V
15 5 V
14 5 V
15 4 V
14 5 V
14 6 V
16 5 V
14 5 V
15 5 V
14 6 V
14 5 V
15 6 V
14 6 V
14 5 V
15 6 V
14 6 V
14 6 V
15 6 V
15 6 V
15 6 V
14 6 V
14 7 V
15 6 V
14 7 V
15 6 V
14 6 V
14 7 V
15 6 V
14 7 V
14 6 V
15 6 V
15 6 V
15 7 V
14 5 V
14 7 V
15 6 V
14 6 V
14 6 V
15 6 V
14 7 V
14 6 V
15 6 V
14 5 V
16 6 V
14 6 V
14 4 V
15 6 V
14 5 V
15 4 V
14 1 V
14 -1 V
15 -1 V
14 -2 V
15 -3 V
14 -3 V
14 -3 V
16 -5 V
14 -4 V
15 -4 V
14 -6 V
14 -5 V
15 -6 V
14 -5 V
14 -7 V
15 -7 V
14 -5 V
stroke
2371 739 M
14 -5 V
15 -6 V
15 -2 V
15 -6 V
14 -3 V
14 -5 V
15 -4 V
14 1 V
14 0 V
15 2 V
14 2 V
15 4 V
14 3 V
14 3 V
15 5 V
15 8 V
15 5 V
14 6 V
14 2 V
15 0 V
14 4 V
14 7 V
15 7 V
14 2 V
14 -5 V
15 -16 V
14 -10 V
16 8 V
14 -34 V
14 -70 V
15 -8 V
14 -7 V
15 -7 V
14 -7 V
14 -7 V
15 -8 V
14 -7 V
14 -7 V
15 -7 V
14 -8 V
16 -5 V
14 -7 V
14 -8 V
15 -7 V
14 -7 V
14 -8 V
15 -9 V
14 -7 V
15 -8 V
14 -7 V
14 -11 V
15 -10 V
15 -7 V
15 -13 V
14 -19 V
14 -2 V
15 -1 V
14 -2 V
14 -1 V
15 -1 V
14 -2 V
14 -1 V
15 -1 V
14 -1 V
15 -2 V
15 -1 V
14 -1 V
15 -1 V
14 -1 V
15 -2 V
14 -1 V
14 -1 V
15 -1 V
14 -1 V
14 -1 V
15 -1 V
14 -1 V
16 -1 V
14 -1 V
14 -1 V
0.500 UL
LTb
450 300 M
3380 0 V
0 3380 V
-3380 0 V
450 300 L
1.000 UP
stroke
grestore
end
showpage
}}%
\put(2140,50){\makebox(0,0){$R\ [\mbox{kpc}]$}}%
\put(100,1990){%
\special{ps: gsave currentpoint currentpoint translate
270 rotate neg exch neg exch translate}%
\makebox(0,0)[b]{\shortstack{$\kappa(R)$}}%
\special{ps: currentpoint grestore moveto}%
}%
\put(3830,200){\makebox(0,0){ 1000}}%
\put(2985,200){\makebox(0,0){ 500}}%
\put(2140,200){\makebox(0,0){ 0}}%
\put(1295,200){\makebox(0,0){-500}}%
\put(450,200){\makebox(0,0){-1000}}%
\put(400,3680){\makebox(0,0)[r]{ 0.5}}%
\put(400,3004){\makebox(0,0)[r]{ 0.4}}%
\put(400,2328){\makebox(0,0)[r]{ 0.3}}%
\put(400,1652){\makebox(0,0)[r]{ 0.2}}%
\put(400,976){\makebox(0,0)[r]{ 0.1}}%
\put(400,300){\makebox(0,0)[r]{ 0}}%
\end{picture}%
\endgroup
 

%% file: figure/gnuplot/Kappa_MainXray_MainGalaxy.tex
\begingroup%
  \makeatletter%
  \newcommand{\GNUPLOTspecial}{%
    \@sanitize\catcode`\%=14\relax\special}%
  \setlength{\unitlength}{0.1bp}%
\begin{picture}(3780,3780)(0,0)%
{\GNUPLOTspecial{"
/gnudict 256 dict def
gnudict begin
/Color true def
/Solid false def
/gnulinewidth 5.000 def
/userlinewidth gnulinewidth def
/vshift -33 def
/dl {10.0 mul} def
/hpt_ 31.5 def
/vpt_ 31.5 def
/hpt hpt_ def
/vpt vpt_ def
/Rounded false def
/M {moveto} bind def
/L {lineto} bind def
/R {rmoveto} bind def
/V {rlineto} bind def
/N {newpath moveto} bind def
/C {setrgbcolor} bind def
/f {rlineto fill} bind def
/vpt2 vpt 2 mul def
/hpt2 hpt 2 mul def
/Lshow { currentpoint stroke M
  0 vshift R show } def
/Rshow { currentpoint stroke M
  dup stringwidth pop neg vshift R show } def
/Cshow { currentpoint stroke M
  dup stringwidth pop -2 div vshift R show } def
/UP { dup vpt_ mul /vpt exch def hpt_ mul /hpt exch def
  /hpt2 hpt 2 mul def /vpt2 vpt 2 mul def } def
/DL { Color {setrgbcolor Solid {pop []} if 0 setdash }
 {pop pop pop 0 setgray Solid {pop []} if 0 setdash} ifelse } def
/BL { stroke userlinewidth 2 mul setlinewidth
      Rounded { 1 setlinejoin 1 setlinecap } if } def
/AL { stroke userlinewidth 2 div setlinewidth
      Rounded { 1 setlinejoin 1 setlinecap } if } def
/UL { dup gnulinewidth mul /userlinewidth exch def
      dup 1 lt {pop 1} if 10 mul /udl exch def } def
/PL { stroke userlinewidth setlinewidth
      Rounded { 1 setlinejoin 1 setlinecap } if } def
/LTw { PL [] 1 setgray } def
/LTb { BL [] 0 0 0 DL } def
/LTa { AL [1 udl mul 2 udl mul] 0 setdash 0 0 0 setrgbcolor } def
/fatlinewidth 7.500 def
/gatlinewidth 10.000 def
/FL { stroke fatlinewidth setlinewidth Rounded { 1 setlinejoin 1 setlinecap } if } def
/GL { stroke gatlinewidth setlinewidth Rounded { 1 setlinejoin 1 setlinecap } if } def
/LT0 { FL [] 0 0 0 DL } def
/LT1 { GL [4 dl 2 dl] 1 0 0 DL } def
/LT2 { GL [2 dl 3 dl] 0 0.5 0 DL } def
/LT3 { PL [1 dl 1.5 dl] 1 0 1 DL } def
/LT4 { PL [5 dl 2 dl 1 dl 2 dl] 0 1 1 DL } def
/LT5 { PL [4 dl 3 dl 1 dl 3 dl] 1 1 0 DL } def
/LT6 { PL [2 dl 2 dl 2 dl 4 dl] 0 0 0 DL } def
/LT7 { PL [2 dl 2 dl 2 dl 2 dl 2 dl 4 dl] 1 0.3 0 DL } def
/LT8 { PL [2 dl 2 dl 2 dl 2 dl 2 dl 2 dl 2 dl 4 dl] 0.5 0.5 0.5 DL } def
/Pnt { stroke [] 0 setdash
   gsave 1 setlinecap M 0 0 V stroke grestore } def
/Dia { stroke [] 0 setdash 2 copy vpt add M
  hpt neg vpt neg V hpt vpt neg V
  hpt vpt V hpt neg vpt V closepath stroke
  Pnt } def
/Pls { stroke [] 0 setdash vpt sub M 0 vpt2 V
  currentpoint stroke M
  hpt neg vpt neg R hpt2 0 V stroke
  } def
/Box { stroke [] 0 setdash 2 copy exch hpt sub exch vpt add M
  0 vpt2 neg V hpt2 0 V 0 vpt2 V
  hpt2 neg 0 V closepath stroke
  Pnt } def
/Crs { stroke [] 0 setdash exch hpt sub exch vpt add M
  hpt2 vpt2 neg V currentpoint stroke M
  hpt2 neg 0 R hpt2 vpt2 V stroke } def
/TriU { stroke [] 0 setdash 2 copy vpt 1.12 mul add M
  hpt neg vpt -1.62 mul V
  hpt 2 mul 0 V
  hpt neg vpt 1.62 mul V closepath stroke
  Pnt  } def
/Star { 2 copy Pls Crs } def
/BoxF { stroke [] 0 setdash exch hpt sub exch vpt add M
  0 vpt2 neg V  hpt2 0 V  0 vpt2 V
  hpt2 neg 0 V  closepath fill } def
/TriUF { stroke [] 0 setdash vpt 1.12 mul add M
  hpt neg vpt -1.62 mul V
  hpt 2 mul 0 V
  hpt neg vpt 1.62 mul V closepath fill } def
/TriD { stroke [] 0 setdash 2 copy vpt 1.12 mul sub M
  hpt neg vpt 1.62 mul V
  hpt 2 mul 0 V
  hpt neg vpt -1.62 mul V closepath stroke
  Pnt  } def
/TriDF { stroke [] 0 setdash vpt 1.12 mul sub M
  hpt neg vpt 1.62 mul V
  hpt 2 mul 0 V
  hpt neg vpt -1.62 mul V closepath fill} def
/DiaF { stroke [] 0 setdash vpt add M
  hpt neg vpt neg V hpt vpt neg V
  hpt vpt V hpt neg vpt V closepath fill } def
/Pent { stroke [] 0 setdash 2 copy gsave
  translate 0 hpt M 4 {72 rotate 0 hpt L} repeat
  closepath stroke grestore Pnt } def
/PentF { stroke [] 0 setdash gsave
  translate 0 hpt M 4 {72 rotate 0 hpt L} repeat
  closepath fill grestore } def
/Circle { stroke [] 0 setdash 2 copy
  hpt 0 360 arc stroke Pnt } def
/CircleF { stroke [] 0 setdash hpt 0 360 arc fill } def
/C0 { BL [] 0 setdash 2 copy moveto vpt 90 450  arc } bind def
/C1 { BL [] 0 setdash 2 copy        moveto
       2 copy  vpt 0 90 arc closepath fill
               vpt 0 360 arc closepath } bind def
/C2 { BL [] 0 setdash 2 copy moveto
       2 copy  vpt 90 180 arc closepath fill
               vpt 0 360 arc closepath } bind def
/C3 { BL [] 0 setdash 2 copy moveto
       2 copy  vpt 0 180 arc closepath fill
               vpt 0 360 arc closepath } bind def
/C4 { BL [] 0 setdash 2 copy moveto
       2 copy  vpt 180 270 arc closepath fill
               vpt 0 360 arc closepath } bind def
/C5 { BL [] 0 setdash 2 copy moveto
       2 copy  vpt 0 90 arc
       2 copy moveto
       2 copy  vpt 180 270 arc closepath fill
               vpt 0 360 arc } bind def
/C6 { BL [] 0 setdash 2 copy moveto
      2 copy  vpt 90 270 arc closepath fill
              vpt 0 360 arc closepath } bind def
/C7 { BL [] 0 setdash 2 copy moveto
      2 copy  vpt 0 270 arc closepath fill
              vpt 0 360 arc closepath } bind def
/C8 { BL [] 0 setdash 2 copy moveto
      2 copy vpt 270 360 arc closepath fill
              vpt 0 360 arc closepath } bind def
/C9 { BL [] 0 setdash 2 copy moveto
      2 copy  vpt 270 450 arc closepath fill
              vpt 0 360 arc closepath } bind def
/C10 { BL [] 0 setdash 2 copy 2 copy moveto vpt 270 360 arc closepath fill
       2 copy moveto
       2 copy vpt 90 180 arc closepath fill
               vpt 0 360 arc closepath } bind def
/C11 { BL [] 0 setdash 2 copy moveto
       2 copy  vpt 0 180 arc closepath fill
       2 copy moveto
       2 copy  vpt 270 360 arc closepath fill
               vpt 0 360 arc closepath } bind def
/C12 { BL [] 0 setdash 2 copy moveto
       2 copy  vpt 180 360 arc closepath fill
               vpt 0 360 arc closepath } bind def
/C13 { BL [] 0 setdash  2 copy moveto
       2 copy  vpt 0 90 arc closepath fill
       2 copy moveto
       2 copy  vpt 180 360 arc closepath fill
               vpt 0 360 arc closepath } bind def
/C14 { BL [] 0 setdash 2 copy moveto
       2 copy  vpt 90 360 arc closepath fill
               vpt 0 360 arc } bind def
/C15 { BL [] 0 setdash 2 copy vpt 0 360 arc closepath fill
               vpt 0 360 arc closepath } bind def
/Rec   { newpath 4 2 roll moveto 1 index 0 rlineto 0 exch rlineto
       neg 0 rlineto closepath } bind def
/Square { dup Rec } bind def
/Bsquare { vpt sub exch vpt sub exch vpt2 Square } bind def
/S0 { BL [] 0 setdash 2 copy moveto 0 vpt rlineto BL Bsquare } bind def
/S1 { BL [] 0 setdash 2 copy vpt Square fill Bsquare } bind def
/S2 { BL [] 0 setdash 2 copy exch vpt sub exch vpt Square fill Bsquare } bind def
/S3 { BL [] 0 setdash 2 copy exch vpt sub exch vpt2 vpt Rec fill Bsquare } bind def
/S4 { BL [] 0 setdash 2 copy exch vpt sub exch vpt sub vpt Square fill Bsquare } bind def
/S5 { BL [] 0 setdash 2 copy 2 copy vpt Square fill
       exch vpt sub exch vpt sub vpt Square fill Bsquare } bind def
/S6 { BL [] 0 setdash 2 copy exch vpt sub exch vpt sub vpt vpt2 Rec fill Bsquare } bind def
/S7 { BL [] 0 setdash 2 copy exch vpt sub exch vpt sub vpt vpt2 Rec fill
       2 copy vpt Square fill
       Bsquare } bind def
/S8 { BL [] 0 setdash 2 copy vpt sub vpt Square fill Bsquare } bind def
/S9 { BL [] 0 setdash 2 copy vpt sub vpt vpt2 Rec fill Bsquare } bind def
/S10 { BL [] 0 setdash 2 copy vpt sub vpt Square fill 2 copy exch vpt sub exch vpt Square fill
       Bsquare } bind def
/S11 { BL [] 0 setdash 2 copy vpt sub vpt Square fill 2 copy exch vpt sub exch vpt2 vpt Rec fill
       Bsquare } bind def
/S12 { BL [] 0 setdash 2 copy exch vpt sub exch vpt sub vpt2 vpt Rec fill Bsquare } bind def
/S13 { BL [] 0 setdash 2 copy exch vpt sub exch vpt sub vpt2 vpt Rec fill
       2 copy vpt Square fill Bsquare } bind def
/S14 { BL [] 0 setdash 2 copy exch vpt sub exch vpt sub vpt2 vpt Rec fill
       2 copy exch vpt sub exch vpt Square fill Bsquare } bind def
/S15 { BL [] 0 setdash 2 copy Bsquare fill Bsquare } bind def
/D0 { gsave translate 45 rotate 0 0 S0 stroke grestore } bind def
/D1 { gsave translate 45 rotate 0 0 S1 stroke grestore } bind def
/D2 { gsave translate 45 rotate 0 0 S2 stroke grestore } bind def
/D3 { gsave translate 45 rotate 0 0 S3 stroke grestore } bind def
/D4 { gsave translate 45 rotate 0 0 S4 stroke grestore } bind def
/D5 { gsave translate 45 rotate 0 0 S5 stroke grestore } bind def
/D6 { gsave translate 45 rotate 0 0 S6 stroke grestore } bind def
/D7 { gsave translate 45 rotate 0 0 S7 stroke grestore } bind def
/D8 { gsave translate 45 rotate 0 0 S8 stroke grestore } bind def
/D9 { gsave translate 45 rotate 0 0 S9 stroke grestore } bind def
/D10 { gsave translate 45 rotate 0 0 S10 stroke grestore } bind def
/D11 { gsave translate 45 rotate 0 0 S11 stroke grestore } bind def
/D12 { gsave translate 45 rotate 0 0 S12 stroke grestore } bind def
/D13 { gsave translate 45 rotate 0 0 S13 stroke grestore } bind def
/D14 { gsave translate 45 rotate 0 0 S14 stroke grestore } bind def
/D15 { gsave translate 45 rotate 0 0 S15 stroke grestore } bind def
/DiaE { stroke [] 0 setdash vpt add M
  hpt neg vpt neg V hpt vpt neg V
  hpt vpt V hpt neg vpt V closepath stroke } def
/BoxE { stroke [] 0 setdash exch hpt sub exch vpt add M
  0 vpt2 neg V hpt2 0 V 0 vpt2 V
  hpt2 neg 0 V closepath stroke } def
/TriUE { stroke [] 0 setdash vpt 1.12 mul add M
  hpt neg vpt -1.62 mul V
  hpt 2 mul 0 V
  hpt neg vpt 1.62 mul V closepath stroke } def
/TriDE { stroke [] 0 setdash vpt 1.12 mul sub M
  hpt neg vpt 1.62 mul V
  hpt 2 mul 0 V
  hpt neg vpt -1.62 mul V closepath stroke } def
/PentE { stroke [] 0 setdash gsave
  translate 0 hpt M 4 {72 rotate 0 hpt L} repeat
  closepath stroke grestore } def
/CircE { stroke [] 0 setdash 
  hpt 0 360 arc stroke } def
/Opaque { gsave closepath 1 setgray fill grestore 0 setgray closepath } def
/DiaW { stroke [] 0 setdash vpt add M
  hpt neg vpt neg V hpt vpt neg V
  hpt vpt V hpt neg vpt V Opaque stroke } def
/BoxW { stroke [] 0 setdash exch hpt sub exch vpt add M
  0 vpt2 neg V hpt2 0 V 0 vpt2 V
  hpt2 neg 0 V Opaque stroke } def
/TriUW { stroke [] 0 setdash vpt 1.12 mul add M
  hpt neg vpt -1.62 mul V
  hpt 2 mul 0 V
  hpt neg vpt 1.62 mul V Opaque stroke } def
/TriDW { stroke [] 0 setdash vpt 1.12 mul sub M
  hpt neg vpt 1.62 mul V
  hpt 2 mul 0 V
  hpt neg vpt -1.62 mul V Opaque stroke } def
/PentW { stroke [] 0 setdash gsave
  translate 0 hpt M 4 {72 rotate 0 hpt L} repeat
  Opaque stroke grestore } def
/CircW { stroke [] 0 setdash 
  hpt 0 360 arc Opaque stroke } def
/BoxFill { gsave Rec 1 setgray fill grestore } def
/BoxColFill {
  gsave Rec
  /Fillden exch def
  currentrgbcolor
  /ColB exch def /ColG exch def /ColR exch def
  /ColR ColR Fillden mul Fillden sub 1 add def
  /ColG ColG Fillden mul Fillden sub 1 add def
  /ColB ColB Fillden mul Fillden sub 1 add def
  ColR ColG ColB setrgbcolor
  fill grestore } def
%
%
/PatternFill { gsave /PFa [ 9 2 roll ] def
    PFa 0 get PFa 2 get 2 div add PFa 1 get PFa 3 get 2 div add translate
    PFa 2 get -2 div PFa 3 get -2 div PFa 2 get PFa 3 get Rec
    gsave 1 setgray fill grestore clip
    currentlinewidth 0.5 mul setlinewidth
    /PFs PFa 2 get dup mul PFa 3 get dup mul add sqrt def
    0 0 M PFa 5 get rotate PFs -2 div dup translate
	0 1 PFs PFa 4 get div 1 add floor cvi
	{ PFa 4 get mul 0 M 0 PFs V } for
    0 PFa 6 get ne {
	0 1 PFs PFa 4 get div 1 add floor cvi
	{ PFa 4 get mul 0 2 1 roll M PFs 0 V } for
    } if
    stroke grestore } def
/Symbol-Oblique /Symbol findfont [1 0 .167 1 0 0] makefont
dup length dict begin {1 index /FID eq {pop pop} {def} ifelse} forall
currentdict end definefont pop
end
gnudict begin
gsave
0 0 translate
0.100 0.100 scale
0 setgray
newpath
0.500 UL
LTb
450 300 M
63 0 V
3317 0 R
-63 0 V
0.500 UL
LTb
450 469 M
31 0 V
3349 0 R
-31 0 V
450 638 M
31 0 V
3349 0 R
-31 0 V
450 807 M
31 0 V
3349 0 R
-31 0 V
450 976 M
63 0 V
3317 0 R
-63 0 V
0.500 UL
LTb
450 1145 M
31 0 V
3349 0 R
-31 0 V
450 1314 M
31 0 V
3349 0 R
-31 0 V
450 1483 M
31 0 V
3349 0 R
-31 0 V
450 1652 M
63 0 V
3317 0 R
-63 0 V
0.500 UL
LTb
450 1821 M
31 0 V
3349 0 R
-31 0 V
450 1990 M
31 0 V
3349 0 R
-31 0 V
450 2159 M
31 0 V
3349 0 R
-31 0 V
450 2328 M
63 0 V
3317 0 R
-63 0 V
0.500 UL
LTb
450 2497 M
31 0 V
3349 0 R
-31 0 V
450 2666 M
31 0 V
3349 0 R
-31 0 V
450 2835 M
31 0 V
3349 0 R
-31 0 V
450 3004 M
63 0 V
3317 0 R
-63 0 V
0.500 UL
LTb
450 3173 M
31 0 V
3349 0 R
-31 0 V
450 3342 M
31 0 V
3349 0 R
-31 0 V
450 3511 M
31 0 V
3349 0 R
-31 0 V
450 3680 M
63 0 V
3317 0 R
-63 0 V
0.500 UL
LTb
450 300 M
0 63 V
0 3317 R
0 -63 V
0.500 UL
LTb
619 300 M
0 31 V
0 3349 R
0 -31 V
788 300 M
0 31 V
0 3349 R
0 -31 V
957 300 M
0 31 V
0 3349 R
0 -31 V
1126 300 M
0 31 V
0 3349 R
0 -31 V
1295 300 M
0 63 V
0 3317 R
0 -63 V
0.500 UL
LTb
1464 300 M
0 31 V
0 3349 R
0 -31 V
1633 300 M
0 31 V
0 3349 R
0 -31 V
1802 300 M
0 31 V
0 3349 R
0 -31 V
1971 300 M
0 31 V
0 3349 R
0 -31 V
2140 300 M
0 63 V
0 3317 R
0 -63 V
0.500 UL
LTb
2309 300 M
0 31 V
0 3349 R
0 -31 V
2478 300 M
0 31 V
0 3349 R
0 -31 V
2647 300 M
0 31 V
0 3349 R
0 -31 V
2816 300 M
0 31 V
0 3349 R
0 -31 V
2985 300 M
0 63 V
0 3317 R
0 -63 V
0.500 UL
LTb
3154 300 M
0 31 V
0 3349 R
0 -31 V
3323 300 M
0 31 V
0 3349 R
0 -31 V
3492 300 M
0 31 V
0 3349 R
0 -31 V
3661 300 M
0 31 V
0 3349 R
0 -31 V
3830 300 M
0 63 V
0 3317 R
0 -63 V
0.500 UL
LTb
0.500 UL
LTb
450 300 M
3380 0 V
0 3380 V
-3380 0 V
450 300 L
LTb
LTb
1.000 UP
1.000 UL
LT0
775 631 M
6 3 V
6 2 V
5 3 V
6 2 V
6 3 V
5 3 V
6 2 V
6 3 V
6 3 V
5 2 V
6 3 V
6 3 V
5 3 V
6 3 V
6 2 V
6 3 V
5 3 V
6 3 V
6 3 V
5 3 V
6 3 V
6 3 V
6 3 V
5 3 V
6 4 V
6 3 V
5 3 V
6 3 V
6 4 V
6 3 V
5 3 V
6 4 V
6 3 V
5 4 V
6 3 V
6 4 V
6 3 V
5 4 V
6 3 V
6 4 V
5 4 V
6 4 V
6 3 V
6 4 V
5 4 V
6 4 V
6 4 V
6 4 V
5 4 V
6 4 V
6 4 V
5 5 V
6 4 V
6 4 V
6 5 V
5 4 V
6 4 V
6 5 V
5 4 V
6 5 V
6 5 V
6 4 V
5 5 V
6 5 V
6 5 V
5 4 V
6 5 V
6 5 V
6 5 V
5 5 V
6 6 V
6 5 V
5 5 V
6 5 V
6 6 V
6 5 V
5 6 V
6 5 V
6 6 V
5 6 V
6 5 V
6 6 V
6 6 V
5 6 V
6 6 V
6 6 V
5 6 V
6 6 V
6 7 V
6 6 V
5 6 V
6 7 V
6 6 V
6 7 V
5 7 V
6 6 V
6 7 V
5 7 V
6 7 V
6 7 V
6 7 V
5 7 V
6 8 V
6 7 V
stroke
1370 1095 M
5 7 V
6 8 V
6 7 V
6 8 V
5 8 V
6 8 V
6 8 V
5 8 V
6 8 V
6 8 V
6 8 V
5 8 V
6 9 V
6 8 V
5 9 V
6 9 V
6 8 V
6 9 V
5 9 V
6 9 V
6 9 V
5 9 V
6 10 V
6 9 V
6 10 V
5 9 V
6 10 V
6 10 V
5 9 V
6 10 V
6 10 V
6 10 V
5 11 V
6 10 V
6 10 V
6 11 V
5 10 V
6 11 V
6 11 V
5 10 V
6 11 V
6 11 V
6 11 V
5 11 V
6 11 V
6 12 V
5 11 V
6 12 V
6 11 V
6 12 V
5 11 V
6 12 V
6 12 V
5 12 V
6 11 V
6 12 V
6 12 V
5 12 V
6 13 V
6 12 V
5 12 V
6 12 V
6 12 V
6 13 V
5 12 V
6 12 V
6 13 V
5 12 V
6 13 V
6 12 V
6 12 V
5 13 V
6 12 V
6 12 V
5 13 V
6 12 V
6 12 V
6 12 V
5 13 V
6 12 V
6 12 V
6 12 V
5 11 V
6 12 V
6 12 V
5 11 V
6 12 V
6 11 V
6 11 V
5 11 V
6 11 V
6 10 V
5 11 V
6 10 V
6 10 V
6 10 V
5 10 V
6 9 V
6 9 V
5 9 V
6 8 V
6 9 V
6 8 V
5 8 V
stroke
1964 2183 M
6 7 V
6 7 V
5 7 V
6 6 V
6 6 V
6 6 V
5 6 V
6 5 V
6 5 V
5 4 V
6 3 V
6 4 V
6 3 V
5 3 V
6 2 V
6 2 V
5 1 V
6 0 V
6 1 V
6 0 V
5 -1 V
6 0 V
6 -2 V
6 -2 V
5 -2 V
6 -3 V
6 -3 V
5 -4 V
6 -4 V
6 -4 V
6 -5 V
5 -5 V
6 -6 V
6 -6 V
5 -7 V
6 -7 V
6 -7 V
6 -9 V
5 -8 V
6 -8 V
6 -8 V
5 -10 V
6 -9 V
6 -11 V
6 -10 V
5 -11 V
6 -11 V
6 -11 V
5 -11 V
6 -11 V
6 -12 V
6 -12 V
5 -14 V
6 -13 V
6 -11 V
5 -10 V
6 -14 V
6 -14 V
6 -14 V
5 -13 V
6 -12 V
6 -13 V
5 -12 V
6 -12 V
6 -14 V
6 -14 V
5 -15 V
6 -12 V
6 -6 V
6 -9 V
5 -14 V
6 -12 V
6 -11 V
5 -8 V
6 -7 V
6 -7 V
6 -6 V
5 -8 V
6 -8 V
6 -8 V
5 -7 V
6 -7 V
6 -6 V
6 -7 V
5 -7 V
6 -7 V
6 -7 V
5 -8 V
6 -7 V
6 -6 V
6 -8 V
5 -7 V
6 -8 V
6 -8 V
5 -8 V
6 -7 V
6 -8 V
6 -7 V
5 -8 V
6 -8 V
6 -8 V
5 -8 V
6 -7 V
6 -8 V
stroke
2559 1553 M
6 -8 V
5 -8 V
6 -8 V
6 -8 V
5 -8 V
6 -8 V
6 -8 V
6 -8 V
5 -8 V
6 -8 V
6 -8 V
5 -8 V
6 -8 V
6 -8 V
6 -8 V
5 -8 V
6 -8 V
6 -7 V
6 -8 V
5 -8 V
6 -8 V
6 -7 V
5 -8 V
6 -8 V
6 -8 V
6 -7 V
5 -8 V
6 -7 V
6 -8 V
5 -8 V
6 -7 V
6 -8 V
6 -7 V
5 -7 V
6 -8 V
6 -7 V
5 -7 V
6 -7 V
6 -8 V
6 -7 V
5 -7 V
6 -7 V
6 -7 V
5 -7 V
6 -7 V
6 -7 V
6 -7 V
5 -6 V
6 -7 V
6 -7 V
5 -6 V
6 -7 V
6 -7 V
6 -6 V
5 -7 V
6 -6 V
6 -7 V
5 -6 V
6 -6 V
6 -7 V
6 -6 V
5 -6 V
6 -6 V
6 -6 V
6 -6 V
5 -6 V
6 -6 V
6 -6 V
5 -6 V
6 -6 V
6 -6 V
6 -6 V
5 -5 V
6 -6 V
6 -6 V
5 -5 V
6 -6 V
6 -5 V
6 -6 V
5 -5 V
6 -6 V
6 -5 V
5 -5 V
6 -6 V
6 -5 V
6 -5 V
5 -5 V
6 -5 V
6 -5 V
5 -5 V
6 -5 V
6 -5 V
6 -5 V
5 -5 V
6 -5 V
6 -5 V
5 -4 V
6 -5 V
6 -5 V
6 -4 V
5 -5 V
6 -4 V
6 -5 V
5 -4 V
stroke
3153 876 M
6 -5 V
6 -4 V
6 -5 V
5 -4 V
6 -4 V
6 -4 V
6 -5 V
5 -4 V
6 -4 V
6 -4 V
5 -4 V
6 -4 V
6 -4 V
6 -4 V
5 -4 V
6 -4 V
6 -4 V
5 -4 V
6 -4 V
6 -3 V
6 -4 V
5 -4 V
6 -4 V
6 -3 V
5 -4 V
6 -3 V
6 -4 V
6 -4 V
5 -3 V
6 -4 V
6 -3 V
5 -3 V
6 -4 V
6 -3 V
6 -3 V
5 -4 V
6 -3 V
6 -3 V
5 -3 V
6 -4 V
6 -3 V
6 -3 V
5 -3 V
6 -3 V
6 -3 V
5 -3 V
6 -3 V
6 -3 V
6 -3 V
5 -3 V
6 -3 V
6 -3 V
6 -3 V
5 -2 V
6 -3 V
6 -3 V
5 -3 V
6 -2 V
6 -3 V
6 -3 V
5 -2 V
6 -3 V
6 -3 V
5 -2 V
6 -3 V
6 -2 V
6 -3 V
5 -2 V
6 -3 V
6 -2 V
5 -3 V
6 -2 V
6 -3 V
6 -2 V
5 -2 V
6 -3 V
6 -2 V
5 -2 V
6 -2 V
6 -3 V
6 -2 V
5 -2 V
6 -2 V
1.000 UL
LT1
450 552 M
3 7 V
3 8 V
2 7 V
3 6 V
3 6 V
3 5 V
2 3 V
3 3 V
3 1 V
3 0 V
2 -1 V
3 -2 V
3 -4 V
3 -6 V
2 -7 V
3 -8 V
3 -10 V
3 -11 V
2 -13 V
3 -13 V
3 -14 V
3 -15 V
2 -16 V
3 -16 V
3 -17 V
3 -17 V
2 -17 V
3 -18 V
3 -17 V
3 -16 V
2 -16 V
3 -14 V
3 -13 V
3 -11 V
1 -6 V
99 0 R
1 2 V
3 5 V
3 2 V
3 -3 V
2 -6 V
21 0 R
1 6 V
3 14 V
3 15 V
3 15 V
2 14 V
3 14 V
3 13 V
3 12 V
2 12 V
3 10 V
3 8 V
3 8 V
2 6 V
3 5 V
3 3 V
3 3 V
2 1 V
3 1 V
3 0 V
3 0 V
2 -1 V
3 0 V
3 0 V
3 1 V
2 1 V
3 2 V
3 3 V
3 4 V
2 4 V
3 5 V
3 6 V
3 6 V
2 6 V
3 7 V
3 6 V
3 7 V
2 6 V
3 7 V
3 6 V
3 6 V
2 6 V
3 5 V
3 5 V
3 5 V
2 4 V
3 4 V
3 4 V
3 3 V
2 3 V
3 3 V
3 3 V
3 2 V
2 2 V
3 3 V
3 2 V
3 2 V
2 2 V
3 2 V
3 1 V
3 1 V
2 1 V
3 0 V
stroke
846 605 M
3 0 V
3 -1 V
2 -1 V
3 -3 V
3 -3 V
3 -4 V
2 -4 V
3 -5 V
3 -6 V
3 -7 V
2 -7 V
3 -8 V
3 -8 V
3 -9 V
2 -9 V
3 -8 V
3 -9 V
3 -8 V
2 -7 V
3 -7 V
3 -5 V
3 -5 V
2 -3 V
3 -2 V
3 0 V
3 1 V
2 2 V
3 3 V
3 4 V
3 4 V
2 5 V
3 4 V
3 5 V
3 5 V
3 5 V
2 5 V
3 5 V
3 5 V
3 4 V
2 4 V
3 4 V
3 4 V
3 3 V
2 3 V
3 2 V
3 2 V
3 1 V
2 1 V
3 0 V
3 0 V
3 -1 V
2 -2 V
3 -3 V
3 -3 V
3 -4 V
2 -5 V
3 -4 V
3 -6 V
3 -5 V
2 -5 V
3 -5 V
3 -5 V
3 -4 V
2 -4 V
3 -3 V
3 -4 V
3 -3 V
2 -2 V
3 -3 V
3 -2 V
3 -3 V
2 -2 V
3 -3 V
3 -2 V
3 -3 V
2 -2 V
3 -2 V
3 -2 V
3 -2 V
2 0 V
3 1 V
3 1 V
3 3 V
2 5 V
3 6 V
3 6 V
3 8 V
2 8 V
3 8 V
3 8 V
3 8 V
2 8 V
3 7 V
3 7 V
3 6 V
2 6 V
3 6 V
3 7 V
3 7 V
2 7 V
3 8 V
3 9 V
3 10 V
2 11 V
stroke
1132 624 M
3 12 V
3 13 V
3 14 V
2 14 V
3 15 V
3 14 V
3 14 V
2 13 V
3 12 V
3 12 V
3 10 V
2 10 V
3 9 V
3 9 V
3 8 V
2 8 V
3 9 V
3 9 V
3 9 V
2 9 V
3 10 V
3 10 V
3 9 V
2 10 V
3 9 V
3 9 V
3 9 V
2 9 V
3 7 V
3 8 V
3 6 V
2 6 V
3 6 V
3 6 V
3 5 V
2 6 V
3 5 V
3 6 V
3 7 V
2 7 V
3 8 V
3 8 V
3 9 V
2 9 V
3 10 V
3 10 V
3 11 V
2 12 V
3 12 V
3 13 V
3 13 V
2 13 V
3 13 V
3 13 V
3 13 V
2 13 V
3 12 V
3 12 V
3 11 V
2 10 V
3 10 V
3 9 V
3 8 V
2 7 V
3 6 V
3 6 V
3 4 V
2 4 V
3 3 V
3 2 V
3 3 V
2 5 V
3 6 V
3 10 V
3 13 V
2 19 V
3 24 V
3 30 V
3 35 V
2 34 V
3 28 V
3 21 V
3 15 V
2 9 V
3 5 V
3 2 V
3 0 V
2 -1 V
3 -1 V
3 1 V
3 2 V
2 4 V
3 6 V
3 7 V
3 8 V
2 10 V
3 10 V
3 11 V
3 12 V
2 12 V
3 12 V
3 13 V
3 13 V
2 12 V
stroke
1418 1671 M
3 13 V
3 12 V
3 12 V
3 11 V
2 11 V
3 9 V
3 9 V
3 8 V
2 7 V
3 6 V
3 6 V
3 5 V
2 4 V
3 5 V
3 4 V
3 5 V
2 4 V
3 5 V
3 4 V
3 5 V
2 5 V
3 5 V
3 5 V
3 5 V
2 6 V
3 6 V
3 6 V
3 7 V
2 6 V
3 7 V
3 7 V
3 8 V
2 8 V
3 8 V
3 8 V
3 9 V
2 9 V
3 10 V
3 10 V
3 11 V
2 11 V
3 13 V
3 14 V
3 15 V
2 17 V
3 18 V
3 20 V
3 20 V
2 21 V
3 21 V
3 21 V
3 19 V
2 19 V
3 17 V
3 14 V
3 13 V
2 11 V
3 10 V
3 9 V
3 8 V
2 8 V
3 7 V
3 8 V
3 8 V
2 9 V
3 10 V
3 11 V
3 13 V
2 13 V
3 15 V
3 16 V
3 16 V
2 17 V
3 18 V
3 17 V
3 18 V
2 18 V
3 17 V
3 17 V
3 14 V
2 14 V
3 11 V
3 9 V
3 5 V
2 3 V
3 2 V
3 2 V
3 3 V
2 7 V
3 8 V
3 9 V
3 11 V
2 11 V
3 12 V
3 11 V
3 11 V
2 10 V
3 9 V
3 8 V
3 7 V
2 7 V
3 7 V
3 6 V
3 6 V
stroke
1705 2732 M
2 7 V
3 6 V
3 7 V
3 8 V
2 7 V
3 7 V
3 7 V
3 8 V
2 7 V
3 6 V
3 7 V
3 6 V
2 6 V
3 5 V
3 5 V
3 4 V
2 4 V
3 3 V
3 3 V
3 3 V
2 2 V
3 1 V
3 2 V
3 1 V
2 0 V
3 1 V
3 1 V
3 0 V
2 1 V
3 1 V
3 2 V
3 2 V
2 2 V
3 3 V
3 3 V
3 4 V
2 4 V
3 4 V
3 4 V
3 4 V
2 5 V
3 5 V
3 4 V
49 9 R
1 -1 V
1 -1 V
0 -1 V
1 0 V
0 -1 V
1 -1 V
1 -1 V
1 -1 V
0 -1 V
1 0 V
0 -1 V
1 -1 V
1 -1 V
0 -1 V
1 0 V
0 -1 V
1 -1 V
1 -1 V
1 -1 V
0 -1 V
1 -1 V
1 -1 V
0 -1 V
1 -1 V
1 -1 V
0 -1 V
1 -1 V
0 -1 V
1 -1 V
0 -1 V
1 -1 V
1 -1 V
0 -1 V
1 -1 V
0 -1 V
1 -1 V
0 -1 V
1 -1 V
0 -1 V
1 -2 V
0 -1 V
1 -1 V
0 -1 V
1 -1 V
0 -1 V
1 -1 V
0 -2 V
1 -1 V
0 -1 V
1 -1 V
0 -2 V
1 -1 V
0 -2 V
1 -1 V
0 -1 V
1 -2 V
0 -1 V
1 -2 V
0 -1 V
1 -2 V
stroke
1907 2849 M
0 -1 V
1 -2 V
0 -2 V
1 -1 V
0 -2 V
1 -2 V
0 -1 V
1 -2 V
0 -2 V
1 -1 V
0 -2 V
1 -2 V
0 -2 V
1 -1 V
0 -2 V
1 -2 V
0 -2 V
1 -2 V
0 -2 V
1 -2 V
0 -1 V
1 -2 V
0 -2 V
1 -2 V
0 -2 V
1 -2 V
0 -2 V
1 -2 V
0 -2 V
1 -1 V
0 -2 V
1 -2 V
0 -2 V
1 -2 V
0 -2 V
1 -2 V
0 -2 V
1 -2 V
0 -1 V
1 -2 V
1 -2 V
0 -2 V
1 -2 V
0 -2 V
1 -1 V
0 -2 V
1 -2 V
0 -2 V
1 -2 V
0 -1 V
1 -2 V
0 -2 V
1 -1 V
0 -2 V
1 -2 V
0 -2 V
1 -1 V
0 -2 V
1 -2 V
0 -1 V
1 -2 V
0 -1 V
1 -2 V
0 -2 V
1 -1 V
0 -2 V
1 -1 V
0 -2 V
1 -1 V
0 -2 V
1 -2 V
0 -1 V
1 -2 V
0 -1 V
1 -2 V
0 -1 V
1 -2 V
0 -1 V
1 -2 V
0 -1 V
1 -1 V
0 -2 V
1 -1 V
0 -2 V
1 -1 V
0 -2 V
1 -1 V
0 -2 V
1 -1 V
0 -1 V
1 -2 V
0 -1 V
1 -2 V
0 -1 V
1 -2 V
0 -1 V
1 -1 V
0 -2 V
1 -1 V
0 -2 V
1 -1 V
0 -1 V
1 -2 V
0 -1 V
stroke
1959 2677 M
1 -2 V
0 -1 V
1 -1 V
0 -2 V
1 -1 V
0 -1 V
1 -2 V
0 -1 V
1 -2 V
0 -1 V
1 -1 V
0 -2 V
1 -1 V
0 -2 V
1 -1 V
0 -1 V
1 -2 V
0 -1 V
1 -1 V
0 -2 V
1 -1 V
0 -2 V
1 -1 V
1 -1 V
0 -2 V
1 -1 V
0 -2 V
1 -1 V
0 -1 V
1 -2 V
0 -1 V
1 -2 V
0 -1 V
1 -2 V
0 -1 V
1 -1 V
0 -2 V
1 -1 V
0 -2 V
1 -1 V
0 -2 V
1 -1 V
0 -2 V
1 -1 V
0 -1 V
1 -2 V
0 -1 V
1 -2 V
0 -1 V
1 -2 V
0 -1 V
1 -2 V
0 -1 V
1 -2 V
0 -1 V
1 -2 V
0 -1 V
1 -2 V
0 -2 V
1 -1 V
0 -2 V
1 -1 V
0 -2 V
1 -1 V
0 -2 V
1 -2 V
0 -1 V
1 -2 V
0 -2 V
1 -1 V
0 -2 V
1 -1 V
0 -2 V
1 -2 V
0 -2 V
1 -1 V
0 -2 V
1 -2 V
0 -1 V
1 -2 V
0 -2 V
1 -2 V
0 -1 V
1 -2 V
0 -2 V
1 -2 V
0 -2 V
1 -1 V
0 -2 V
1 -2 V
0 -2 V
1 -2 V
0 -2 V
1 -2 V
0 -2 V
1 -2 V
0 -1 V
1 -2 V
0 -2 V
1 -2 V
0 -2 V
1 -2 V
0 -2 V
1 -2 V
stroke
2012 2513 M
0 -3 V
1 -2 V
0 -2 V
1 -2 V
1 -2 V
0 -2 V
1 -2 V
0 -2 V
1 -2 V
0 -2 V
1 -2 V
0 -2 V
1 -2 V
0 -2 V
1 -2 V
0 -2 V
1 -2 V
0 -2 V
1 -2 V
0 -2 V
1 -1 V
0 -2 V
1 -2 V
0 -1 V
1 -2 V
0 -1 V
1 -2 V
0 -1 V
1 -1 V
0 -1 V
1 -1 V
0 -2 V
1 0 V
0 -1 V
1 -1 V
0 -1 V
1 0 V
0 -1 V
1 0 V
0 -1 V
1 0 V
1 0 V
1 1 V
1 0 V
0 1 V
1 1 V
0 1 V
1 1 V
1 1 V
0 2 V
1 1 V
0 1 V
1 1 V
0 1 V
1 1 V
1 1 V
0 1 V
1 1 V
1 0 V
1 0 V
1 0 V
1 -1 V
0 -1 V
1 0 V
0 -1 V
1 -1 V
0 -2 V
1 -1 V
0 -1 V
1 -2 V
0 -1 V
1 -2 V
0 -2 V
1 -1 V
0 -2 V
1 -2 V
0 -2 V
1 -2 V
0 -3 V
1 -2 V
0 -2 V
1 -3 V
1 -2 V
0 -2 V
1 -3 V
0 -2 V
1 -3 V
0 -3 V
1 -2 V
0 -3 V
1 -3 V
0 -2 V
1 -3 V
0 -3 V
1 -2 V
0 -3 V
1 -3 V
0 -3 V
1 -2 V
0 -3 V
1 -3 V
0 -2 V
1 -3 V
0 -2 V
stroke
2069 2375 M
1 -3 V
0 -2 V
1 -3 V
0 -2 V
1 -2 V
0 -3 V
1 -2 V
0 -2 V
1 -3 V
0 -2 V
1 -2 V
0 -2 V
1 -2 V
0 -2 V
1 -2 V
0 -2 V
1 -2 V
0 -2 V
1 -2 V
0 -1 V
1 -2 V
0 -2 V
1 -2 V
0 -1 V
1 -2 V
0 -1 V
1 -2 V
0 -2 V
1 -1 V
0 -2 V
1 -1 V
0 -1 V
1 -2 V
0 -1 V
1 -1 V
0 -2 V
1 -1 V
0 -1 V
1 -1 V
0 -2 V
1 -1 V
0 -1 V
1 -1 V
0 -1 V
1 -1 V
0 -1 V
1 -1 V
0 -1 V
1 -1 V
0 -1 V
1 -1 V
1 -1 V
0 -1 V
1 -1 V
0 -1 V
1 0 V
0 -1 V
1 -1 V
1 -1 V
0 -1 V
1 0 V
0 -1 V
1 0 V
1 -1 V
0 -1 V
1 0 V
0 -1 V
1 0 V
1 -1 V
1 -1 V
1 0 V
0 -1 V
1 0 V
0 -1 V
1 0 V
1 -1 V
1 0 V
1 -1 V
1 0 V
1 -1 V
1 0 V
1 0 V
0 -1 V
1 0 V
1 0 V
1 0 V
0 -1 V
1 0 V
1 0 V
1 0 V
1 -1 V
53 -166 R
4 -2 V
3 -3 V
3 -3 V
4 -3 V
3 -3 V
3 -4 V
4 -5 V
3 -6 V
3 -7 V
3 -7 V
4 -9 V
3 -10 V
stroke
2217 2041 M
3 -10 V
4 -11 V
3 -10 V
3 -11 V
4 -10 V
3 -10 V
3 -8 V
4 -8 V
3 -7 V
3 -7 V
4 -5 V
3 -4 V
3 -3 V
3 -3 V
4 -1 V
3 -1 V
3 -2 V
4 -2 V
3 -2 V
3 -4 V
4 -5 V
3 -8 V
3 -9 V
4 -12 V
3 -14 V
3 -16 V
4 -19 V
3 -22 V
3 -24 V
3 -27 V
4 -28 V
3 -29 V
3 -27 V
4 -23 V
3 -16 V
3 -8 V
4 4 V
3 16 V
3 26 V
4 26 V
3 18 V
3 7 V
3 1 V
4 -6 V
3 -10 V
3 -12 V
4 -13 V
3 -11 V
3 -9 V
4 -7 V
3 -5 V
3 -2 V
4 -1 V
3 1 V
3 3 V
4 4 V
3 6 V
3 6 V
3 7 V
4 7 V
3 7 V
3 6 V
4 5 V
3 3 V
3 1 V
4 -1 V
3 -4 V
3 -5 V
4 -7 V
3 -9 V
3 -11 V
4 -12 V
3 -13 V
3 -13 V
3 -15 V
4 -14 V
3 -13 V
3 -13 V
4 -11 V
3 -9 V
3 -7 V
4 -5 V
3 -3 V
3 -3 V
4 -1 V
3 -1 V
3 -1 V
4 -1 V
3 -1 V
3 -2 V
3 -3 V
4 -3 V
3 -4 V
3 -3 V
4 -2 V
3 -1 V
3 0 V
4 3 V
3 4 V
3 5 V
4 6 V
3 6 V
3 7 V
4 5 V
stroke
2562 1558 M
3 4 V
3 3 V
3 2 V
4 1 V
3 0 V
3 -1 V
4 -1 V
3 -1 V
3 -1 V
4 -1 V
3 0 V
3 -1 V
4 0 V
3 0 V
3 1 V
4 -1 V
3 0 V
3 -1 V
3 -2 V
4 -2 V
3 -2 V
3 -3 V
4 -3 V
3 -2 V
3 -3 V
4 -2 V
3 -1 V
3 -1 V
4 0 V
3 1 V
3 3 V
4 4 V
3 6 V
3 9 V
3 10 V
4 13 V
3 16 V
3 16 V
4 16 V
3 16 V
3 14 V
4 12 V
3 8 V
3 4 V
4 0 V
3 -2 V
3 -4 V
4 -6 V
3 -7 V
3 -7 V
3 -7 V
4 -6 V
3 -5 V
3 -2 V
4 -1 V
3 0 V
3 2 V
4 3 V
3 4 V
3 5 V
4 6 V
3 6 V
3 7 V
3 6 V
4 7 V
3 7 V
3 7 V
4 7 V
3 6 V
3 7 V
4 6 V
3 5 V
3 6 V
4 5 V
3 4 V
3 5 V
4 4 V
3 3 V
3 3 V
3 3 V
4 3 V
3 4 V
3 3 V
4 5 V
3 5 V
3 6 V
4 7 V
3 8 V
3 8 V
4 7 V
3 7 V
3 7 V
4 5 V
3 4 V
3 4 V
3 2 V
4 3 V
3 2 V
3 2 V
4 3 V
3 3 V
3 4 V
4 4 V
3 6 V
stroke
2906 1877 M
3 7 V
4 8 V
3 8 V
3 9 V
4 10 V
3 11 V
3 11 V
3 11 V
4 12 V
3 11 V
3 10 V
4 10 V
3 9 V
3 8 V
4 7 V
3 6 V
3 5 V
4 5 V
3 6 V
3 5 V
4 7 V
3 7 V
3 9 V
3 11 V
4 12 V
3 14 V
3 17 V
4 18 V
3 20 V
3 22 V
4 25 V
3 26 V
3 25 V
4 20 V
3 14 V
3 3 V
4 -8 V
3 -24 V
3 -40 V
3 -44 V
4 -35 V
3 -16 V
3 -1 V
4 12 V
3 20 V
3 24 V
4 24 V
3 19 V
3 13 V
4 4 V
3 -1 V
3 -7 V
4 -10 V
3 -14 V
3 -15 V
3 -15 V
4 -15 V
3 -14 V
3 -15 V
4 -13 V
3 -13 V
3 -12 V
4 -12 V
3 -10 V
3 -10 V
4 -8 V
3 -7 V
3 -7 V
3 -5 V
4 -6 V
3 -5 V
3 -5 V
4 -5 V
3 -6 V
3 -6 V
4 -7 V
3 -8 V
3 -10 V
4 -10 V
3 -11 V
3 -14 V
4 -14 V
3 -15 V
3 -16 V
3 -17 V
4 -16 V
3 -15 V
3 -14 V
4 -13 V
3 -12 V
3 -9 V
4 -7 V
3 -6 V
3 -4 V
4 -4 V
3 -4 V
3 -4 V
4 -5 V
3 -5 V
3 -7 V
3 -8 V
4 -8 V
3 -9 V
3 -9 V
stroke
3250 1727 M
4 -9 V
3 -9 V
3 -10 V
4 -9 V
3 -10 V
3 -9 V
4 -10 V
3 -9 V
3 -9 V
4 -10 V
3 -9 V
3 -9 V
3 -9 V
4 -9 V
3 -10 V
3 -10 V
4 -10 V
3 -10 V
3 -11 V
4 -12 V
3 -13 V
3 -15 V
4 -17 V
3 -19 V
3 -22 V
4 -25 V
3 -29 V
3 -30 V
3 -27 V
4 -23 V
3 -20 V
3 -17 V
4 -15 V
3 -13 V
3 -12 V
4 -11 V
3 -11 V
3 -11 V
4 -12 V
3 -13 V
3 -13 V
4 -14 V
3 -16 V
3 -17 V
3 -18 V
4 -19 V
3 -19 V
3 -18 V
4 -18 V
3 -16 V
3 -14 V
4 -11 V
3 -8 V
3 -4 V
4 0 V
3 2 V
3 5 V
4 7 V
3 8 V
3 8 V
3 8 V
4 7 V
3 6 V
3 4 V
4 4 V
3 2 V
3 1 V
4 -1 V
3 -3 V
3 -4 V
4 -6 V
3 -7 V
3 -7 V
3 -8 V
4 -7 V
3 -7 V
3 -7 V
4 -6 V
3 -4 V
3 -4 V
4 -2 V
3 -2 V
3 -2 V
4 -2 V
3 -2 V
3 -3 V
4 -3 V
3 -5 V
3 -6 V
3 -7 V
4 -9 V
3 -11 V
3 -12 V
4 -14 V
3 -16 V
3 -16 V
4 -16 V
3 -16 V
3 -15 V
4 -12 V
3 -11 V
3 -8 V
4 -3 V
3 0 V
stroke
3595 772 M
3 3 V
3 7 V
4 8 V
3 8 V
3 9 V
4 8 V
3 5 V
3 3 V
4 2 V
3 -1 V
3 -2 V
4 -3 V
3 -4 V
3 -4 V
4 -5 V
3 -4 V
3 -5 V
3 -5 V
4 -4 V
3 -5 V
3 -4 V
4 -4 V
3 -4 V
3 -3 V
4 -2 V
3 -3 V
3 -2 V
4 -2 V
3 -1 V
3 -2 V
4 -2 V
3 -3 V
3 -2 V
3 -4 V
4 -4 V
3 -5 V
3 -6 V
4 -8 V
3 -8 V
3 -10 V
4 -9 V
3 -6 V
3 -3 V
4 0 V
3 3 V
3 4 V
4 5 V
3 6 V
3 7 V
3 6 V
4 5 V
3 5 V
3 2 V
4 1 V
3 -1 V
3 -4 V
4 -7 V
3 -10 V
3 -12 V
4 -15 V
3 -15 V
3 -16 V
4 -16 V
3 -17 V
3 -15 V
3 -14 V
4 -13 V
3 -10 V
3 -9 V
4 -6 V
3 -4 V
1.000 UL
LT2
868 402 M
14 2 V
14 2 V
15 3 V
14 2 V
15 2 V
14 3 V
15 2 V
15 3 V
14 2 V
15 3 V
14 2 V
14 3 V
15 3 V
14 3 V
14 2 V
15 3 V
14 3 V
14 3 V
15 3 V
15 4 V
15 3 V
14 3 V
14 3 V
15 4 V
14 3 V
15 3 V
14 4 V
14 4 V
15 3 V
14 4 V
14 4 V
16 4 V
14 4 V
15 4 V
14 4 V
14 5 V
15 4 V
14 5 V
14 4 V
15 5 V
14 5 V
15 4 V
14 5 V
14 6 V
16 5 V
14 5 V
15 5 V
14 6 V
14 5 V
15 6 V
14 6 V
14 5 V
15 6 V
14 6 V
14 6 V
15 6 V
15 6 V
15 6 V
14 6 V
14 7 V
15 6 V
14 7 V
15 6 V
14 6 V
14 7 V
15 6 V
14 7 V
14 6 V
15 6 V
15 6 V
15 7 V
14 5 V
14 7 V
15 6 V
14 6 V
14 6 V
15 6 V
14 7 V
14 6 V
15 6 V
14 5 V
16 6 V
14 6 V
14 4 V
15 6 V
14 5 V
15 4 V
14 1 V
14 -1 V
15 -1 V
14 -2 V
15 -3 V
14 -3 V
14 -3 V
16 -5 V
14 -4 V
15 -4 V
14 -6 V
14 -5 V
15 -6 V
14 -5 V
14 -7 V
15 -7 V
14 -5 V
stroke
2371 739 M
14 -5 V
15 -6 V
15 -2 V
15 -6 V
14 -3 V
14 -5 V
15 -4 V
14 1 V
14 0 V
15 2 V
14 2 V
15 4 V
14 3 V
14 3 V
15 5 V
15 8 V
15 5 V
14 6 V
14 2 V
15 0 V
14 4 V
14 7 V
15 7 V
14 2 V
14 -5 V
15 -16 V
14 -10 V
16 8 V
14 -34 V
14 -70 V
15 -8 V
14 -7 V
15 -7 V
14 -7 V
14 -7 V
15 -8 V
14 -7 V
14 -7 V
15 -7 V
14 -8 V
16 -5 V
14 -7 V
14 -8 V
15 -7 V
14 -7 V
14 -8 V
15 -9 V
14 -7 V
15 -8 V
14 -7 V
14 -11 V
15 -10 V
15 -7 V
15 -13 V
14 -19 V
14 -2 V
15 -1 V
14 -2 V
14 -1 V
15 -1 V
14 -2 V
14 -1 V
15 -1 V
14 -1 V
15 -2 V
15 -1 V
14 -1 V
15 -1 V
14 -1 V
15 -2 V
14 -1 V
14 -1 V
15 -1 V
14 -1 V
14 -1 V
15 -1 V
14 -1 V
16 -1 V
14 -1 V
14 -1 V
0.500 UL
LTb
450 300 M
3380 0 V
0 3380 V
-3380 0 V
450 300 L
1.000 UP
stroke
grestore
end
showpage
}}%
\put(2140,50){\makebox(0,0){$R\ [\mbox{kpc}]$}}%
\put(100,1990){%
\special{ps: gsave currentpoint currentpoint translate
270 rotate neg exch neg exch translate}%
\makebox(0,0)[b]{\shortstack{$\kappa(R)$}}%
\special{ps: currentpoint grestore moveto}%
}%
\put(3830,200){\makebox(0,0){ 1000}}%
\put(2985,200){\makebox(0,0){ 500}}%
\put(2140,200){\makebox(0,0){ 0}}%
\put(1295,200){\makebox(0,0){-500}}%
\put(450,200){\makebox(0,0){-1000}}%
\put(400,3680){\makebox(0,0)[r]{ 0.5}}%
\put(400,3004){\makebox(0,0)[r]{ 0.4}}%
\put(400,2328){\makebox(0,0)[r]{ 0.3}}%
\put(400,1652){\makebox(0,0)[r]{ 0.2}}%
\put(400,976){\makebox(0,0)[r]{ 0.1}}%
\put(400,300){\makebox(0,0)[r]{ 0}}%
\end{picture}%
\endgroup
 

%% file: figure/gnuplot/SigmaGalax.tex
\begingroup%
  \makeatletter%
  \newcommand{\GNUPLOTspecial}{%
    \@sanitize\catcode`\%=14\relax\special}%
  \setlength{\unitlength}{0.1bp}%
\begin{picture}(3780,3780)(0,0)%
{\GNUPLOTspecial{"
/gnudict 256 dict def
gnudict begin
/Color true def
/Solid false def
/gnulinewidth 5.000 def
/userlinewidth gnulinewidth def
/vshift -33 def
/dl {10.0 mul} def
/hpt_ 31.5 def
/vpt_ 31.5 def
/hpt hpt_ def
/vpt vpt_ def
/Rounded false def
/M {moveto} bind def
/L {lineto} bind def
/R {rmoveto} bind def
/V {rlineto} bind def
/N {newpath moveto} bind def
/C {setrgbcolor} bind def
/f {rlineto fill} bind def
/vpt2 vpt 2 mul def
/hpt2 hpt 2 mul def
/Lshow { currentpoint stroke M
  0 vshift R show } def
/Rshow { currentpoint stroke M
  dup stringwidth pop neg vshift R show } def
/Cshow { currentpoint stroke M
  dup stringwidth pop -2 div vshift R show } def
/UP { dup vpt_ mul /vpt exch def hpt_ mul /hpt exch def
  /hpt2 hpt 2 mul def /vpt2 vpt 2 mul def } def
/DL { Color {setrgbcolor Solid {pop []} if 0 setdash }
 {pop pop pop 0 setgray Solid {pop []} if 0 setdash} ifelse } def
/BL { stroke userlinewidth 2 mul setlinewidth
      Rounded { 1 setlinejoin 1 setlinecap } if } def
/AL { stroke userlinewidth 2 div setlinewidth
      Rounded { 1 setlinejoin 1 setlinecap } if } def
/UL { dup gnulinewidth mul /userlinewidth exch def
      dup 1 lt {pop 1} if 10 mul /udl exch def } def
/PL { stroke userlinewidth setlinewidth
      Rounded { 1 setlinejoin 1 setlinecap } if } def
/LTw { PL [] 1 setgray } def
/LTb { BL [] 0 0 0 DL } def
/LTa { AL [1 udl mul 2 udl mul] 0 setdash 0 0 0 setrgbcolor } def
/fatlinewidth 7.500 def
/gatlinewidth 10.000 def
/FL { stroke fatlinewidth setlinewidth Rounded { 1 setlinejoin 1 setlinecap } if } def
/GL { stroke gatlinewidth setlinewidth Rounded { 1 setlinejoin 1 setlinecap } if } def
/LT0 { GL [2 dl 3 dl] 0 0.5 0 DL } def
/LT1 { GL [4 dl 2 dl] 1 0 1 DL } def
/LT2 { FL [] 0.8 0.4 0 DL } def
/LT3 { PL [1 dl 1.5 dl] 1 0 1 DL } def
/LT4 { PL [5 dl 2 dl 1 dl 2 dl] 0 1 1 DL } def
/LT5 { PL [4 dl 3 dl 1 dl 3 dl] 1 1 0 DL } def
/LT6 { PL [2 dl 2 dl 2 dl 4 dl] 0 0 0 DL } def
/LT7 { PL [2 dl 2 dl 2 dl 2 dl 2 dl 4 dl] 1 0.3 0 DL } def
/LT8 { PL [2 dl 2 dl 2 dl 2 dl 2 dl 2 dl 2 dl 4 dl] 0.5 0.5 0.5 DL } def
/Pnt { stroke [] 0 setdash
   gsave 1 setlinecap M 0 0 V stroke grestore } def
/Dia { stroke [] 0 setdash 2 copy vpt add M
  hpt neg vpt neg V hpt vpt neg V
  hpt vpt V hpt neg vpt V closepath stroke
  Pnt } def
/Pls { stroke [] 0 setdash vpt sub M 0 vpt2 V
  currentpoint stroke M
  hpt neg vpt neg R hpt2 0 V stroke
  } def
/Box { stroke [] 0 setdash 2 copy exch hpt sub exch vpt add M
  0 vpt2 neg V hpt2 0 V 0 vpt2 V
  hpt2 neg 0 V closepath stroke
  Pnt } def
/Crs { stroke [] 0 setdash exch hpt sub exch vpt add M
  hpt2 vpt2 neg V currentpoint stroke M
  hpt2 neg 0 R hpt2 vpt2 V stroke } def
/TriU { stroke [] 0 setdash 2 copy vpt 1.12 mul add M
  hpt neg vpt -1.62 mul V
  hpt 2 mul 0 V
  hpt neg vpt 1.62 mul V closepath stroke
  Pnt  } def
/Star { 2 copy Pls Crs } def
/BoxF { stroke [] 0 setdash exch hpt sub exch vpt add M
  0 vpt2 neg V  hpt2 0 V  0 vpt2 V
  hpt2 neg 0 V  closepath fill } def
/TriUF { stroke [] 0 setdash vpt 1.12 mul add M
  hpt neg vpt -1.62 mul V
  hpt 2 mul 0 V
  hpt neg vpt 1.62 mul V closepath fill } def
/TriD { stroke [] 0 setdash 2 copy vpt 1.12 mul sub M
  hpt neg vpt 1.62 mul V
  hpt 2 mul 0 V
  hpt neg vpt -1.62 mul V closepath stroke
  Pnt  } def
/TriDF { stroke [] 0 setdash vpt 1.12 mul sub M
  hpt neg vpt 1.62 mul V
  hpt 2 mul 0 V
  hpt neg vpt -1.62 mul V closepath fill} def
/DiaF { stroke [] 0 setdash vpt add M
  hpt neg vpt neg V hpt vpt neg V
  hpt vpt V hpt neg vpt V closepath fill } def
/Pent { stroke [] 0 setdash 2 copy gsave
  translate 0 hpt M 4 {72 rotate 0 hpt L} repeat
  closepath stroke grestore Pnt } def
/PentF { stroke [] 0 setdash gsave
  translate 0 hpt M 4 {72 rotate 0 hpt L} repeat
  closepath fill grestore } def
/Circle { stroke [] 0 setdash 2 copy
  hpt 0 360 arc stroke Pnt } def
/CircleF { stroke [] 0 setdash hpt 0 360 arc fill } def
/C0 { BL [] 0 setdash 2 copy moveto vpt 90 450  arc } bind def
/C1 { BL [] 0 setdash 2 copy        moveto
       2 copy  vpt 0 90 arc closepath fill
               vpt 0 360 arc closepath } bind def
/C2 { BL [] 0 setdash 2 copy moveto
       2 copy  vpt 90 180 arc closepath fill
               vpt 0 360 arc closepath } bind def
/C3 { BL [] 0 setdash 2 copy moveto
       2 copy  vpt 0 180 arc closepath fill
               vpt 0 360 arc closepath } bind def
/C4 { BL [] 0 setdash 2 copy moveto
       2 copy  vpt 180 270 arc closepath fill
               vpt 0 360 arc closepath } bind def
/C5 { BL [] 0 setdash 2 copy moveto
       2 copy  vpt 0 90 arc
       2 copy moveto
       2 copy  vpt 180 270 arc closepath fill
               vpt 0 360 arc } bind def
/C6 { BL [] 0 setdash 2 copy moveto
      2 copy  vpt 90 270 arc closepath fill
              vpt 0 360 arc closepath } bind def
/C7 { BL [] 0 setdash 2 copy moveto
      2 copy  vpt 0 270 arc closepath fill
              vpt 0 360 arc closepath } bind def
/C8 { BL [] 0 setdash 2 copy moveto
      2 copy vpt 270 360 arc closepath fill
              vpt 0 360 arc closepath } bind def
/C9 { BL [] 0 setdash 2 copy moveto
      2 copy  vpt 270 450 arc closepath fill
              vpt 0 360 arc closepath } bind def
/C10 { BL [] 0 setdash 2 copy 2 copy moveto vpt 270 360 arc closepath fill
       2 copy moveto
       2 copy vpt 90 180 arc closepath fill
               vpt 0 360 arc closepath } bind def
/C11 { BL [] 0 setdash 2 copy moveto
       2 copy  vpt 0 180 arc closepath fill
       2 copy moveto
       2 copy  vpt 270 360 arc closepath fill
               vpt 0 360 arc closepath } bind def
/C12 { BL [] 0 setdash 2 copy moveto
       2 copy  vpt 180 360 arc closepath fill
               vpt 0 360 arc closepath } bind def
/C13 { BL [] 0 setdash  2 copy moveto
       2 copy  vpt 0 90 arc closepath fill
       2 copy moveto
       2 copy  vpt 180 360 arc closepath fill
               vpt 0 360 arc closepath } bind def
/C14 { BL [] 0 setdash 2 copy moveto
       2 copy  vpt 90 360 arc closepath fill
               vpt 0 360 arc } bind def
/C15 { BL [] 0 setdash 2 copy vpt 0 360 arc closepath fill
               vpt 0 360 arc closepath } bind def
/Rec   { newpath 4 2 roll moveto 1 index 0 rlineto 0 exch rlineto
       neg 0 rlineto closepath } bind def
/Square { dup Rec } bind def
/Bsquare { vpt sub exch vpt sub exch vpt2 Square } bind def
/S0 { BL [] 0 setdash 2 copy moveto 0 vpt rlineto BL Bsquare } bind def
/S1 { BL [] 0 setdash 2 copy vpt Square fill Bsquare } bind def
/S2 { BL [] 0 setdash 2 copy exch vpt sub exch vpt Square fill Bsquare } bind def
/S3 { BL [] 0 setdash 2 copy exch vpt sub exch vpt2 vpt Rec fill Bsquare } bind def
/S4 { BL [] 0 setdash 2 copy exch vpt sub exch vpt sub vpt Square fill Bsquare } bind def
/S5 { BL [] 0 setdash 2 copy 2 copy vpt Square fill
       exch vpt sub exch vpt sub vpt Square fill Bsquare } bind def
/S6 { BL [] 0 setdash 2 copy exch vpt sub exch vpt sub vpt vpt2 Rec fill Bsquare } bind def
/S7 { BL [] 0 setdash 2 copy exch vpt sub exch vpt sub vpt vpt2 Rec fill
       2 copy vpt Square fill
       Bsquare } bind def
/S8 { BL [] 0 setdash 2 copy vpt sub vpt Square fill Bsquare } bind def
/S9 { BL [] 0 setdash 2 copy vpt sub vpt vpt2 Rec fill Bsquare } bind def
/S10 { BL [] 0 setdash 2 copy vpt sub vpt Square fill 2 copy exch vpt sub exch vpt Square fill
       Bsquare } bind def
/S11 { BL [] 0 setdash 2 copy vpt sub vpt Square fill 2 copy exch vpt sub exch vpt2 vpt Rec fill
       Bsquare } bind def
/S12 { BL [] 0 setdash 2 copy exch vpt sub exch vpt sub vpt2 vpt Rec fill Bsquare } bind def
/S13 { BL [] 0 setdash 2 copy exch vpt sub exch vpt sub vpt2 vpt Rec fill
       2 copy vpt Square fill Bsquare } bind def
/S14 { BL [] 0 setdash 2 copy exch vpt sub exch vpt sub vpt2 vpt Rec fill
       2 copy exch vpt sub exch vpt Square fill Bsquare } bind def
/S15 { BL [] 0 setdash 2 copy Bsquare fill Bsquare } bind def
/D0 { gsave translate 45 rotate 0 0 S0 stroke grestore } bind def
/D1 { gsave translate 45 rotate 0 0 S1 stroke grestore } bind def
/D2 { gsave translate 45 rotate 0 0 S2 stroke grestore } bind def
/D3 { gsave translate 45 rotate 0 0 S3 stroke grestore } bind def
/D4 { gsave translate 45 rotate 0 0 S4 stroke grestore } bind def
/D5 { gsave translate 45 rotate 0 0 S5 stroke grestore } bind def
/D6 { gsave translate 45 rotate 0 0 S6 stroke grestore } bind def
/D7 { gsave translate 45 rotate 0 0 S7 stroke grestore } bind def
/D8 { gsave translate 45 rotate 0 0 S8 stroke grestore } bind def
/D9 { gsave translate 45 rotate 0 0 S9 stroke grestore } bind def
/D10 { gsave translate 45 rotate 0 0 S10 stroke grestore } bind def
/D11 { gsave translate 45 rotate 0 0 S11 stroke grestore } bind def
/D12 { gsave translate 45 rotate 0 0 S12 stroke grestore } bind def
/D13 { gsave translate 45 rotate 0 0 S13 stroke grestore } bind def
/D14 { gsave translate 45 rotate 0 0 S14 stroke grestore } bind def
/D15 { gsave translate 45 rotate 0 0 S15 stroke grestore } bind def
/DiaE { stroke [] 0 setdash vpt add M
  hpt neg vpt neg V hpt vpt neg V
  hpt vpt V hpt neg vpt V closepath stroke } def
/BoxE { stroke [] 0 setdash exch hpt sub exch vpt add M
  0 vpt2 neg V hpt2 0 V 0 vpt2 V
  hpt2 neg 0 V closepath stroke } def
/TriUE { stroke [] 0 setdash vpt 1.12 mul add M
  hpt neg vpt -1.62 mul V
  hpt 2 mul 0 V
  hpt neg vpt 1.62 mul V closepath stroke } def
/TriDE { stroke [] 0 setdash vpt 1.12 mul sub M
  hpt neg vpt 1.62 mul V
  hpt 2 mul 0 V
  hpt neg vpt -1.62 mul V closepath stroke } def
/PentE { stroke [] 0 setdash gsave
  translate 0 hpt M 4 {72 rotate 0 hpt L} repeat
  closepath stroke grestore } def
/CircE { stroke [] 0 setdash 
  hpt 0 360 arc stroke } def
/Opaque { gsave closepath 1 setgray fill grestore 0 setgray closepath } def
/DiaW { stroke [] 0 setdash vpt add M
  hpt neg vpt neg V hpt vpt neg V
  hpt vpt V hpt neg vpt V Opaque stroke } def
/BoxW { stroke [] 0 setdash exch hpt sub exch vpt add M
  0 vpt2 neg V hpt2 0 V 0 vpt2 V
  hpt2 neg 0 V Opaque stroke } def
/TriUW { stroke [] 0 setdash vpt 1.12 mul add M
  hpt neg vpt -1.62 mul V
  hpt 2 mul 0 V
  hpt neg vpt 1.62 mul V Opaque stroke } def
/TriDW { stroke [] 0 setdash vpt 1.12 mul sub M
  hpt neg vpt 1.62 mul V
  hpt 2 mul 0 V
  hpt neg vpt -1.62 mul V Opaque stroke } def
/PentW { stroke [] 0 setdash gsave
  translate 0 hpt M 4 {72 rotate 0 hpt L} repeat
  Opaque stroke grestore } def
/CircW { stroke [] 0 setdash 
  hpt 0 360 arc Opaque stroke } def
/BoxFill { gsave Rec 1 setgray fill grestore } def
/BoxColFill {
  gsave Rec
  /Fillden exch def
  currentrgbcolor
  /ColB exch def /ColG exch def /ColR exch def
  /ColR ColR Fillden mul Fillden sub 1 add def
  /ColG ColG Fillden mul Fillden sub 1 add def
  /ColB ColB Fillden mul Fillden sub 1 add def
  ColR ColG ColB setrgbcolor
  fill grestore } def
%
%
/PatternFill { gsave /PFa [ 9 2 roll ] def
    PFa 0 get PFa 2 get 2 div add PFa 1 get PFa 3 get 2 div add translate
    PFa 2 get -2 div PFa 3 get -2 div PFa 2 get PFa 3 get Rec
    gsave 1 setgray fill grestore clip
    currentlinewidth 0.5 mul setlinewidth
    /PFs PFa 2 get dup mul PFa 3 get dup mul add sqrt def
    0 0 M PFa 5 get rotate PFs -2 div dup translate
	0 1 PFs PFa 4 get div 1 add floor cvi
	{ PFa 4 get mul 0 M 0 PFs V } for
    0 PFa 6 get ne {
	0 1 PFs PFa 4 get div 1 add floor cvi
	{ PFa 4 get mul 0 2 1 roll M PFs 0 V } for
    } if
    stroke grestore } def
/Symbol-Oblique /Symbol findfont [1 0 .167 1 0 0] makefont
dup length dict begin {1 index /FID eq {pop pop} {def} ifelse} forall
currentdict end definefont pop
end
gnudict begin
gsave
0 0 translate
0.100 0.100 scale
0 setgray
newpath
0.500 UL
LTb
500 300 M
63 0 V
3317 0 R
-63 0 V
0.500 UL
LTb
500 435 M
31 0 V
3349 0 R
-31 0 V
500 570 M
31 0 V
3349 0 R
-31 0 V
500 706 M
31 0 V
3349 0 R
-31 0 V
500 841 M
31 0 V
3349 0 R
-31 0 V
500 976 M
63 0 V
3317 0 R
-63 0 V
0.500 UL
LTb
500 1111 M
31 0 V
3349 0 R
-31 0 V
500 1246 M
31 0 V
3349 0 R
-31 0 V
500 1382 M
31 0 V
3349 0 R
-31 0 V
500 1517 M
31 0 V
3349 0 R
-31 0 V
500 1652 M
63 0 V
3317 0 R
-63 0 V
0.500 UL
LTb
500 1787 M
31 0 V
3349 0 R
-31 0 V
500 1922 M
31 0 V
3349 0 R
-31 0 V
500 2058 M
31 0 V
3349 0 R
-31 0 V
500 2193 M
31 0 V
3349 0 R
-31 0 V
500 2328 M
63 0 V
3317 0 R
-63 0 V
0.500 UL
LTb
500 2463 M
31 0 V
3349 0 R
-31 0 V
500 2598 M
31 0 V
3349 0 R
-31 0 V
500 2734 M
31 0 V
3349 0 R
-31 0 V
500 2869 M
31 0 V
3349 0 R
-31 0 V
500 3004 M
63 0 V
3317 0 R
-63 0 V
0.500 UL
LTb
500 3139 M
31 0 V
3349 0 R
-31 0 V
500 3274 M
31 0 V
3349 0 R
-31 0 V
500 3410 M
31 0 V
3349 0 R
-31 0 V
500 3545 M
31 0 V
3349 0 R
-31 0 V
500 3680 M
63 0 V
3317 0 R
-63 0 V
0.500 UL
LTb
500 300 M
0 63 V
0 3317 R
0 -63 V
0.500 UL
LTb
669 300 M
0 31 V
0 3349 R
0 -31 V
838 300 M
0 31 V
0 3349 R
0 -31 V
1007 300 M
0 31 V
0 3349 R
0 -31 V
1176 300 M
0 31 V
0 3349 R
0 -31 V
1345 300 M
0 63 V
0 3317 R
0 -63 V
0.500 UL
LTb
1514 300 M
0 31 V
0 3349 R
0 -31 V
1683 300 M
0 31 V
0 3349 R
0 -31 V
1852 300 M
0 31 V
0 3349 R
0 -31 V
2021 300 M
0 31 V
0 3349 R
0 -31 V
2190 300 M
0 63 V
0 3317 R
0 -63 V
0.500 UL
LTb
2359 300 M
0 31 V
0 3349 R
0 -31 V
2528 300 M
0 31 V
0 3349 R
0 -31 V
2697 300 M
0 31 V
0 3349 R
0 -31 V
2866 300 M
0 31 V
0 3349 R
0 -31 V
3035 300 M
0 63 V
0 3317 R
0 -63 V
0.500 UL
LTb
3204 300 M
0 31 V
0 3349 R
0 -31 V
3373 300 M
0 31 V
0 3349 R
0 -31 V
3542 300 M
0 31 V
0 3349 R
0 -31 V
3711 300 M
0 31 V
0 3349 R
0 -31 V
3880 300 M
0 63 V
0 3317 R
0 -63 V
0.500 UL
LTb
0.500 UL
LTb
500 300 M
3380 0 V
0 3380 V
-3380 0 V
500 300 L
LTb
LTb
1.000 UP
1.000 UL
LT0
918 811 M
14 10 V
14 11 V
15 12 V
14 11 V
15 12 V
14 12 V
15 12 V
15 12 V
14 13 V
15 13 V
14 13 V
14 14 V
15 13 V
14 14 V
14 14 V
15 15 V
14 14 V
14 15 V
15 16 V
15 16 V
15 15 V
14 17 V
14 16 V
15 17 V
14 17 V
15 17 V
14 18 V
14 18 V
15 19 V
14 18 V
14 20 V
16 20 V
14 20 V
15 21 V
14 21 V
14 21 V
15 23 V
14 22 V
14 23 V
15 24 V
14 24 V
15 23 V
14 25 V
14 26 V
16 25 V
14 27 V
15 27 V
14 27 V
14 28 V
15 28 V
14 28 V
14 29 V
15 29 V
14 30 V
14 30 V
15 30 V
15 30 V
15 31 V
14 30 V
14 32 V
15 31 V
14 33 V
15 32 V
14 32 V
14 32 V
15 33 V
14 32 V
14 32 V
15 31 V
15 29 V
15 32 V
14 29 V
14 32 V
15 30 V
14 32 V
14 30 V
15 31 V
14 33 V
14 29 V
15 32 V
14 25 V
16 30 V
14 27 V
14 24 V
15 28 V
14 27 V
15 18 V
14 3 V
14 -2 V
15 -7 V
14 -11 V
15 -13 V
14 -14 V
14 -18 V
16 -21 V
14 -20 V
15 -24 V
14 -26 V
14 -27 V
15 -28 V
14 -27 V
14 -33 V
15 -38 V
14 -25 V
stroke
2421 2494 M
14 -24 V
15 -30 V
15 -11 V
15 -27 V
14 -15 V
14 -28 V
15 -21 V
14 6 V
14 0 V
15 11 V
14 11 V
15 18 V
14 15 V
14 16 V
15 27 V
15 39 V
15 25 V
14 30 V
14 11 V
15 -2 V
14 20 V
14 34 V
15 38 V
14 8 V
14 -26 V
15 -81 V
14 -50 V
16 40 V
14 -168 V
14 -352 V
15 -38 V
14 -33 V
15 -37 V
14 -36 V
14 -36 V
15 -37 V
14 -36 V
14 -36 V
15 -36 V
14 -36 V
16 -28 V
14 -36 V
14 -36 V
15 -37 V
14 -36 V
14 -37 V
15 -45 V
14 -38 V
15 -38 V
14 -38 V
14 -51 V
15 -51 V
15 -37 V
15 -65 V
14 -96 V
14 -7 V
15 -7 V
14 -7 V
14 -7 V
15 -6 V
14 -7 V
14 -6 V
15 -7 V
14 -6 V
15 -6 V
15 -7 V
14 -6 V
15 -5 V
14 -6 V
15 -6 V
14 -6 V
14 -5 V
15 -6 V
14 -5 V
14 -6 V
15 -5 V
14 -5 V
16 -6 V
14 -5 V
14 -5 V
1.000 UL
LT1
1242 300 M
8 67 V
15 33 V
14 -12 V
15 -24 V
14 -12 V
14 -2 V
15 -13 V
14 6 V
14 -14 V
16 22 V
14 50 V
15 -15 V
14 77 V
14 -16 V
15 19 V
14 -16 V
14 33 V
15 -18 V
14 54 V
15 73 V
14 -19 V
14 72 V
16 -47 V
14 59 V
15 -20 V
14 58 V
14 -22 V
15 91 V
14 86 V
14 -23 V
15 106 V
14 -24 V
14 116 V
15 -25 V
15 96 V
15 -26 V
14 65 V
14 -26 V
15 57 V
14 31 V
15 -26 V
14 -16 V
14 -25 V
15 -37 V
14 -26 V
14 -29 V
15 -24 V
15 0 V
15 -41 V
14 -21 V
14 -57 V
15 -19 V
14 -75 V
14 -17 V
15 -86 V
14 -14 V
14 -69 V
15 -75 V
14 -7 V
16 -57 V
14 -3 V
14 -79 V
15 2 V
14 -61 V
15 7 V
13 -42 V
404 0 R
2 7 V
14 17 V
14 67 V
15 17 V
15 -18 V
15 17 V
14 24 V
14 17 V
15 -8 V
14 17 V
14 17 V
15 40 V
14 16 V
14 69 V
15 15 V
14 59 V
16 17 V
14 49 V
14 73 V
15 15 V
14 38 V
15 15 V
14 24 V
14 14 V
15 51 V
14 13 V
14 81 V
15 76 V
14 12 V
16 158 V
14 13 V
14 46 V
15 11 V
14 12 V
14 11 V
15 36 V
14 11 V
stroke
3115 1449 M
15 23 V
14 -65 V
14 10 V
15 -79 V
15 11 V
15 -42 V
14 9 V
14 -76 V
15 9 V
14 -135 V
14 -129 V
15 8 V
14 -144 V
14 8 V
15 -136 V
14 8 V
15 -165 V
15 50 V
14 -219 V
15 -77 V
14 7 V
8 -25 V
1.000 UL
LT2
918 618 M
14 -15 V
14 4 V
15 51 V
14 4 V
15 30 V
14 -27 V
15 3 V
15 -69 V
14 5 V
15 -37 V
14 5 V
14 69 V
15 7 V
14 71 V
14 10 V
15 7 V
14 101 V
14 4 V
15 118 V
15 -54 V
15 149 V
14 7 V
14 128 V
15 47 V
14 7 V
15 -7 V
14 7 V
14 14 V
15 6 V
14 24 V
14 7 V
16 40 V
14 71 V
15 7 V
14 98 V
14 7 V
15 40 V
14 7 V
14 54 V
15 7 V
14 78 V
15 98 V
14 3 V
14 98 V
16 -20 V
14 84 V
15 7 V
14 85 V
14 6 V
15 119 V
14 115 V
14 6 V
15 136 V
14 6 V
14 146 V
15 3 V
15 125 V
15 7 V
14 94 V
14 7 V
15 88 V
14 64 V
15 7 V
14 17 V
14 3 V
15 -3 V
14 7 V
14 3 V
15 7 V
15 30 V
15 -10 V
14 7 V
14 -24 V
15 10 V
14 -44 V
14 14 V
15 -54 V
14 20 V
14 -40 V
15 -44 V
14 20 V
16 -30 V
14 27 V
14 -58 V
15 31 V
14 -34 V
15 23 V
14 -40 V
14 -74 V
15 6 V
14 -67 V
15 3 V
14 -61 V
14 0 V
16 -23 V
14 3 V
15 -41 V
14 -50 V
14 -4 V
15 -84 V
14 -4 V
14 -81 V
15 -17 V
14 -37 V
stroke
2421 2294 M
14 -3 V
15 -27 V
15 34 V
15 -11 V
14 31 V
14 -14 V
15 -23 V
14 23 V
14 -37 V
15 27 V
14 24 V
15 74 V
14 31 V
14 84 V
15 44 V
15 20 V
15 41 V
14 54 V
14 27 V
15 -7 V
14 37 V
14 48 V
15 81 V
14 24 V
14 44 V
15 -68 V
14 10 V
16 58 V
14 -122 V
14 -277 V
15 -24 V
14 7 V
15 -24 V
14 -13 V
14 -21 V
15 14 V
14 -24 V
14 47 V
15 38 V
14 -24 V
16 132 V
14 -24 V
14 10 V
15 -27 V
14 -23 V
14 -24 V
15 -10 V
14 -27 V
15 -17 V
14 -102 V
14 -40 V
15 -132 V
15 -24 V
15 -108 V
14 -88 V
14 -81 V
15 0 V
14 -142 V
14 -135 V
15 3 V
14 -152 V
14 0 V
15 -142 V
14 4 V
15 -173 V
15 44 V
14 -226 V
15 -81 V
14 0 V
15 -54 V
14 3 V
14 -88 V
15 0 V
14 -78 V
14 0 V
15 -43 V
14 -52 V
16 0 V
14 10 V
14 0 V
0.500 UL
LTb
500 300 M
3380 0 V
0 3380 V
-3380 0 V
500 300 L
1.000 UP
stroke
grestore
end
showpage
}}%
\put(2190,50){\makebox(0,0){$R\ [\mbox{kpc}]$}}%
\put(100,1990){%
\special{ps: gsave currentpoint currentpoint translate
270 rotate neg exch neg exch translate}%
\makebox(0,0)[b]{\shortstack{${\Sigma(R)}/{\Sigma_{c}}$}}%
\special{ps: currentpoint grestore moveto}%
}%
\put(3880,200){\makebox(0,0){ 1000}}%
\put(3035,200){\makebox(0,0){ 500}}%
\put(2190,200){\makebox(0,0){ 0}}%
\put(1345,200){\makebox(0,0){-500}}%
\put(500,200){\makebox(0,0){-1000}}%
\put(450,3680){\makebox(0,0)[r]{ 0.1}}%
\put(450,3004){\makebox(0,0)[r]{ 0.08}}%
\put(450,2328){\makebox(0,0)[r]{ 0.06}}%
\put(450,1652){\makebox(0,0)[r]{ 0.04}}%
\put(450,976){\makebox(0,0)[r]{ 0.02}}%
\put(450,300){\makebox(0,0)[r]{ 0}}%
\end{picture}%
\endgroup
 

%% file: figure/gnuplot/SigmaDM.tex
\begingroup%
  \makeatletter%
  \newcommand{\GNUPLOTspecial}{%
    \@sanitize\catcode`\%=14\relax\special}%
  \setlength{\unitlength}{0.1bp}%
\begin{picture}(3780,3780)(0,0)%
{\GNUPLOTspecial{"
/gnudict 256 dict def
gnudict begin
/Color true def
/Solid false def
/gnulinewidth 5.000 def
/userlinewidth gnulinewidth def
/vshift -33 def
/dl {10.0 mul} def
/hpt_ 31.5 def
/vpt_ 31.5 def
/hpt hpt_ def
/vpt vpt_ def
/Rounded false def
/M {moveto} bind def
/L {lineto} bind def
/R {rmoveto} bind def
/V {rlineto} bind def
/N {newpath moveto} bind def
/C {setrgbcolor} bind def
/f {rlineto fill} bind def
/vpt2 vpt 2 mul def
/hpt2 hpt 2 mul def
/Lshow { currentpoint stroke M
  0 vshift R show } def
/Rshow { currentpoint stroke M
  dup stringwidth pop neg vshift R show } def
/Cshow { currentpoint stroke M
  dup stringwidth pop -2 div vshift R show } def
/UP { dup vpt_ mul /vpt exch def hpt_ mul /hpt exch def
  /hpt2 hpt 2 mul def /vpt2 vpt 2 mul def } def
/DL { Color {setrgbcolor Solid {pop []} if 0 setdash }
 {pop pop pop 0 setgray Solid {pop []} if 0 setdash} ifelse } def
/BL { stroke userlinewidth 2 mul setlinewidth
      Rounded { 1 setlinejoin 1 setlinecap } if } def
/AL { stroke userlinewidth 2 div setlinewidth
      Rounded { 1 setlinejoin 1 setlinecap } if } def
/UL { dup gnulinewidth mul /userlinewidth exch def
      dup 1 lt {pop 1} if 10 mul /udl exch def } def
/PL { stroke userlinewidth setlinewidth
      Rounded { 1 setlinejoin 1 setlinecap } if } def
/LTw { PL [] 1 setgray } def
/LTb { BL [] 0 0 0 DL } def
/LTa { AL [1 udl mul 2 udl mul] 0 setdash 0 0 0 setrgbcolor } def
/gatlinewidth 10.000 def
/GL { stroke gatlinewidth setlinewidth Rounded { 1 setlinejoin 1 setlinecap } if } def
/LT0 { GL [2 dl 3 dl] 0 0.5 0 DL } def
/LT1 { GL [5 dl 2 dl 1 dl 2 dl] 0 0 0 DL } def/LT2 { PL [2 dl 3 dl] 0 0 1 DL } def
/LT3 { PL [1 dl 1.5 dl] 1 0 1 DL } def
/LT4 { PL [5 dl 2 dl 1 dl 2 dl] 0 1 1 DL } def
/LT5 { PL [4 dl 3 dl 1 dl 3 dl] 1 1 0 DL } def
/LT6 { PL [2 dl 2 dl 2 dl 4 dl] 0 0 0 DL } def
/LT7 { PL [2 dl 2 dl 2 dl 2 dl 2 dl 4 dl] 1 0.3 0 DL } def
/LT8 { PL [2 dl 2 dl 2 dl 2 dl 2 dl 2 dl 2 dl 4 dl] 0.5 0.5 0.5 DL } def
/Pnt { stroke [] 0 setdash
   gsave 1 setlinecap M 0 0 V stroke grestore } def
/Dia { stroke [] 0 setdash 2 copy vpt add M
  hpt neg vpt neg V hpt vpt neg V
  hpt vpt V hpt neg vpt V closepath stroke
  Pnt } def
/Pls { stroke [] 0 setdash vpt sub M 0 vpt2 V
  currentpoint stroke M
  hpt neg vpt neg R hpt2 0 V stroke
  } def
/Box { stroke [] 0 setdash 2 copy exch hpt sub exch vpt add M
  0 vpt2 neg V hpt2 0 V 0 vpt2 V
  hpt2 neg 0 V closepath stroke
  Pnt } def
/Crs { stroke [] 0 setdash exch hpt sub exch vpt add M
  hpt2 vpt2 neg V currentpoint stroke M
  hpt2 neg 0 R hpt2 vpt2 V stroke } def
/TriU { stroke [] 0 setdash 2 copy vpt 1.12 mul add M
  hpt neg vpt -1.62 mul V
  hpt 2 mul 0 V
  hpt neg vpt 1.62 mul V closepath stroke
  Pnt  } def
/Star { 2 copy Pls Crs } def
/BoxF { stroke [] 0 setdash exch hpt sub exch vpt add M
  0 vpt2 neg V  hpt2 0 V  0 vpt2 V
  hpt2 neg 0 V  closepath fill } def
/TriUF { stroke [] 0 setdash vpt 1.12 mul add M
  hpt neg vpt -1.62 mul V
  hpt 2 mul 0 V
  hpt neg vpt 1.62 mul V closepath fill } def
/TriD { stroke [] 0 setdash 2 copy vpt 1.12 mul sub M
  hpt neg vpt 1.62 mul V
  hpt 2 mul 0 V
  hpt neg vpt -1.62 mul V closepath stroke
  Pnt  } def
/TriDF { stroke [] 0 setdash vpt 1.12 mul sub M
  hpt neg vpt 1.62 mul V
  hpt 2 mul 0 V
  hpt neg vpt -1.62 mul V closepath fill} def
/DiaF { stroke [] 0 setdash vpt add M
  hpt neg vpt neg V hpt vpt neg V
  hpt vpt V hpt neg vpt V closepath fill } def
/Pent { stroke [] 0 setdash 2 copy gsave
  translate 0 hpt M 4 {72 rotate 0 hpt L} repeat
  closepath stroke grestore Pnt } def
/PentF { stroke [] 0 setdash gsave
  translate 0 hpt M 4 {72 rotate 0 hpt L} repeat
  closepath fill grestore } def
/Circle { stroke [] 0 setdash 2 copy
  hpt 0 360 arc stroke Pnt } def
/CircleF { stroke [] 0 setdash hpt 0 360 arc fill } def
/C0 { BL [] 0 setdash 2 copy moveto vpt 90 450  arc } bind def
/C1 { BL [] 0 setdash 2 copy        moveto
       2 copy  vpt 0 90 arc closepath fill
               vpt 0 360 arc closepath } bind def
/C2 { BL [] 0 setdash 2 copy moveto
       2 copy  vpt 90 180 arc closepath fill
               vpt 0 360 arc closepath } bind def
/C3 { BL [] 0 setdash 2 copy moveto
       2 copy  vpt 0 180 arc closepath fill
               vpt 0 360 arc closepath } bind def
/C4 { BL [] 0 setdash 2 copy moveto
       2 copy  vpt 180 270 arc closepath fill
               vpt 0 360 arc closepath } bind def
/C5 { BL [] 0 setdash 2 copy moveto
       2 copy  vpt 0 90 arc
       2 copy moveto
       2 copy  vpt 180 270 arc closepath fill
               vpt 0 360 arc } bind def
/C6 { BL [] 0 setdash 2 copy moveto
      2 copy  vpt 90 270 arc closepath fill
              vpt 0 360 arc closepath } bind def
/C7 { BL [] 0 setdash 2 copy moveto
      2 copy  vpt 0 270 arc closepath fill
              vpt 0 360 arc closepath } bind def
/C8 { BL [] 0 setdash 2 copy moveto
      2 copy vpt 270 360 arc closepath fill
              vpt 0 360 arc closepath } bind def
/C9 { BL [] 0 setdash 2 copy moveto
      2 copy  vpt 270 450 arc closepath fill
              vpt 0 360 arc closepath } bind def
/C10 { BL [] 0 setdash 2 copy 2 copy moveto vpt 270 360 arc closepath fill
       2 copy moveto
       2 copy vpt 90 180 arc closepath fill
               vpt 0 360 arc closepath } bind def
/C11 { BL [] 0 setdash 2 copy moveto
       2 copy  vpt 0 180 arc closepath fill
       2 copy moveto
       2 copy  vpt 270 360 arc closepath fill
               vpt 0 360 arc closepath } bind def
/C12 { BL [] 0 setdash 2 copy moveto
       2 copy  vpt 180 360 arc closepath fill
               vpt 0 360 arc closepath } bind def
/C13 { BL [] 0 setdash  2 copy moveto
       2 copy  vpt 0 90 arc closepath fill
       2 copy moveto
       2 copy  vpt 180 360 arc closepath fill
               vpt 0 360 arc closepath } bind def
/C14 { BL [] 0 setdash 2 copy moveto
       2 copy  vpt 90 360 arc closepath fill
               vpt 0 360 arc } bind def
/C15 { BL [] 0 setdash 2 copy vpt 0 360 arc closepath fill
               vpt 0 360 arc closepath } bind def
/Rec   { newpath 4 2 roll moveto 1 index 0 rlineto 0 exch rlineto
       neg 0 rlineto closepath } bind def
/Square { dup Rec } bind def
/Bsquare { vpt sub exch vpt sub exch vpt2 Square } bind def
/S0 { BL [] 0 setdash 2 copy moveto 0 vpt rlineto BL Bsquare } bind def
/S1 { BL [] 0 setdash 2 copy vpt Square fill Bsquare } bind def
/S2 { BL [] 0 setdash 2 copy exch vpt sub exch vpt Square fill Bsquare } bind def
/S3 { BL [] 0 setdash 2 copy exch vpt sub exch vpt2 vpt Rec fill Bsquare } bind def
/S4 { BL [] 0 setdash 2 copy exch vpt sub exch vpt sub vpt Square fill Bsquare } bind def
/S5 { BL [] 0 setdash 2 copy 2 copy vpt Square fill
       exch vpt sub exch vpt sub vpt Square fill Bsquare } bind def
/S6 { BL [] 0 setdash 2 copy exch vpt sub exch vpt sub vpt vpt2 Rec fill Bsquare } bind def
/S7 { BL [] 0 setdash 2 copy exch vpt sub exch vpt sub vpt vpt2 Rec fill
       2 copy vpt Square fill
       Bsquare } bind def
/S8 { BL [] 0 setdash 2 copy vpt sub vpt Square fill Bsquare } bind def
/S9 { BL [] 0 setdash 2 copy vpt sub vpt vpt2 Rec fill Bsquare } bind def
/S10 { BL [] 0 setdash 2 copy vpt sub vpt Square fill 2 copy exch vpt sub exch vpt Square fill
       Bsquare } bind def
/S11 { BL [] 0 setdash 2 copy vpt sub vpt Square fill 2 copy exch vpt sub exch vpt2 vpt Rec fill
       Bsquare } bind def
/S12 { BL [] 0 setdash 2 copy exch vpt sub exch vpt sub vpt2 vpt Rec fill Bsquare } bind def
/S13 { BL [] 0 setdash 2 copy exch vpt sub exch vpt sub vpt2 vpt Rec fill
       2 copy vpt Square fill Bsquare } bind def
/S14 { BL [] 0 setdash 2 copy exch vpt sub exch vpt sub vpt2 vpt Rec fill
       2 copy exch vpt sub exch vpt Square fill Bsquare } bind def
/S15 { BL [] 0 setdash 2 copy Bsquare fill Bsquare } bind def
/D0 { gsave translate 45 rotate 0 0 S0 stroke grestore } bind def
/D1 { gsave translate 45 rotate 0 0 S1 stroke grestore } bind def
/D2 { gsave translate 45 rotate 0 0 S2 stroke grestore } bind def
/D3 { gsave translate 45 rotate 0 0 S3 stroke grestore } bind def
/D4 { gsave translate 45 rotate 0 0 S4 stroke grestore } bind def
/D5 { gsave translate 45 rotate 0 0 S5 stroke grestore } bind def
/D6 { gsave translate 45 rotate 0 0 S6 stroke grestore } bind def
/D7 { gsave translate 45 rotate 0 0 S7 stroke grestore } bind def
/D8 { gsave translate 45 rotate 0 0 S8 stroke grestore } bind def
/D9 { gsave translate 45 rotate 0 0 S9 stroke grestore } bind def
/D10 { gsave translate 45 rotate 0 0 S10 stroke grestore } bind def
/D11 { gsave translate 45 rotate 0 0 S11 stroke grestore } bind def
/D12 { gsave translate 45 rotate 0 0 S12 stroke grestore } bind def
/D13 { gsave translate 45 rotate 0 0 S13 stroke grestore } bind def
/D14 { gsave translate 45 rotate 0 0 S14 stroke grestore } bind def
/D15 { gsave translate 45 rotate 0 0 S15 stroke grestore } bind def
/DiaE { stroke [] 0 setdash vpt add M
  hpt neg vpt neg V hpt vpt neg V
  hpt vpt V hpt neg vpt V closepath stroke } def
/BoxE { stroke [] 0 setdash exch hpt sub exch vpt add M
  0 vpt2 neg V hpt2 0 V 0 vpt2 V
  hpt2 neg 0 V closepath stroke } def
/TriUE { stroke [] 0 setdash vpt 1.12 mul add M
  hpt neg vpt -1.62 mul V
  hpt 2 mul 0 V
  hpt neg vpt 1.62 mul V closepath stroke } def
/TriDE { stroke [] 0 setdash vpt 1.12 mul sub M
  hpt neg vpt 1.62 mul V
  hpt 2 mul 0 V
  hpt neg vpt -1.62 mul V closepath stroke } def
/PentE { stroke [] 0 setdash gsave
  translate 0 hpt M 4 {72 rotate 0 hpt L} repeat
  closepath stroke grestore } def
/CircE { stroke [] 0 setdash 
  hpt 0 360 arc stroke } def
/Opaque { gsave closepath 1 setgray fill grestore 0 setgray closepath } def
/DiaW { stroke [] 0 setdash vpt add M
  hpt neg vpt neg V hpt vpt neg V
  hpt vpt V hpt neg vpt V Opaque stroke } def
/BoxW { stroke [] 0 setdash exch hpt sub exch vpt add M
  0 vpt2 neg V hpt2 0 V 0 vpt2 V
  hpt2 neg 0 V Opaque stroke } def
/TriUW { stroke [] 0 setdash vpt 1.12 mul add M
  hpt neg vpt -1.62 mul V
  hpt 2 mul 0 V
  hpt neg vpt 1.62 mul V Opaque stroke } def
/TriDW { stroke [] 0 setdash vpt 1.12 mul sub M
  hpt neg vpt 1.62 mul V
  hpt 2 mul 0 V
  hpt neg vpt -1.62 mul V Opaque stroke } def
/PentW { stroke [] 0 setdash gsave
  translate 0 hpt M 4 {72 rotate 0 hpt L} repeat
  Opaque stroke grestore } def
/CircW { stroke [] 0 setdash 
  hpt 0 360 arc Opaque stroke } def
/BoxFill { gsave Rec 1 setgray fill grestore } def
/BoxColFill {
  gsave Rec
  /Fillden exch def
  currentrgbcolor
  /ColB exch def /ColG exch def /ColR exch def
  /ColR ColR Fillden mul Fillden sub 1 add def
  /ColG ColG Fillden mul Fillden sub 1 add def
  /ColB ColB Fillden mul Fillden sub 1 add def
  ColR ColG ColB setrgbcolor
  fill grestore } def
%
%
/PatternFill { gsave /PFa [ 9 2 roll ] def
    PFa 0 get PFa 2 get 2 div add PFa 1 get PFa 3 get 2 div add translate
    PFa 2 get -2 div PFa 3 get -2 div PFa 2 get PFa 3 get Rec
    gsave 1 setgray fill grestore clip
    currentlinewidth 0.5 mul setlinewidth
    /PFs PFa 2 get dup mul PFa 3 get dup mul add sqrt def
    0 0 M PFa 5 get rotate PFs -2 div dup translate
	0 1 PFs PFa 4 get div 1 add floor cvi
	{ PFa 4 get mul 0 M 0 PFs V } for
    0 PFa 6 get ne {
	0 1 PFs PFa 4 get div 1 add floor cvi
	{ PFa 4 get mul 0 2 1 roll M PFs 0 V } for
    } if
    stroke grestore } def
/Symbol-Oblique /Symbol findfont [1 0 .167 1 0 0] makefont
dup length dict begin {1 index /FID eq {pop pop} {def} ifelse} forall
currentdict end definefont pop
end
gnudict begin
gsave
0 0 translate
0.100 0.100 scale
0 setgray
newpath
0.500 UL
LTb
450 300 M
63 0 V
3317 0 R
-63 0 V
0.500 UL
LTb
450 469 M
31 0 V
3349 0 R
-31 0 V
450 638 M
31 0 V
3349 0 R
-31 0 V
450 807 M
31 0 V
3349 0 R
-31 0 V
450 976 M
31 0 V
3349 0 R
-31 0 V
450 1145 M
63 0 V
3317 0 R
-63 0 V
0.500 UL
LTb
450 1314 M
31 0 V
3349 0 R
-31 0 V
450 1483 M
31 0 V
3349 0 R
-31 0 V
450 1652 M
31 0 V
3349 0 R
-31 0 V
450 1821 M
31 0 V
3349 0 R
-31 0 V
450 1990 M
63 0 V
3317 0 R
-63 0 V
0.500 UL
LTb
450 2159 M
31 0 V
3349 0 R
-31 0 V
450 2328 M
31 0 V
3349 0 R
-31 0 V
450 2497 M
31 0 V
3349 0 R
-31 0 V
450 2666 M
31 0 V
3349 0 R
-31 0 V
450 2835 M
63 0 V
3317 0 R
-63 0 V
0.500 UL
LTb
450 3004 M
31 0 V
3349 0 R
-31 0 V
450 3173 M
31 0 V
3349 0 R
-31 0 V
450 3342 M
31 0 V
3349 0 R
-31 0 V
450 3511 M
31 0 V
3349 0 R
-31 0 V
450 3680 M
63 0 V
3317 0 R
-63 0 V
0.500 UL
LTb
450 300 M
0 63 V
0 3317 R
0 -63 V
0.500 UL
LTb
619 300 M
0 31 V
0 3349 R
0 -31 V
788 300 M
0 31 V
0 3349 R
0 -31 V
957 300 M
0 31 V
0 3349 R
0 -31 V
1126 300 M
0 31 V
0 3349 R
0 -31 V
1295 300 M
0 63 V
0 3317 R
0 -63 V
0.500 UL
LTb
1464 300 M
0 31 V
0 3349 R
0 -31 V
1633 300 M
0 31 V
0 3349 R
0 -31 V
1802 300 M
0 31 V
0 3349 R
0 -31 V
1971 300 M
0 31 V
0 3349 R
0 -31 V
2140 300 M
0 63 V
0 3317 R
0 -63 V
0.500 UL
LTb
2309 300 M
0 31 V
0 3349 R
0 -31 V
2478 300 M
0 31 V
0 3349 R
0 -31 V
2647 300 M
0 31 V
0 3349 R
0 -31 V
2816 300 M
0 31 V
0 3349 R
0 -31 V
2985 300 M
0 63 V
0 3317 R
0 -63 V
0.500 UL
LTb
3154 300 M
0 31 V
0 3349 R
0 -31 V
3323 300 M
0 31 V
0 3349 R
0 -31 V
3492 300 M
0 31 V
0 3349 R
0 -31 V
3661 300 M
0 31 V
0 3349 R
0 -31 V
3830 300 M
0 63 V
0 3317 R
0 -63 V
0.500 UL
LTb
0.500 UL
LTb
450 300 M
3380 0 V
0 3380 V
-3380 0 V
450 300 L
LTb
LTb
1.000 UP
1.000 UL
LT0
868 428 M
14 2 V
14 3 V
15 3 V
14 3 V
15 3 V
14 3 V
15 3 V
15 3 V
14 3 V
15 3 V
14 4 V
14 3 V
15 3 V
14 4 V
14 3 V
15 4 V
14 4 V
14 3 V
15 4 V
15 4 V
15 4 V
14 4 V
14 4 V
15 4 V
14 5 V
15 4 V
14 5 V
14 4 V
15 5 V
14 4 V
14 5 V
16 5 V
14 5 V
15 5 V
14 6 V
14 5 V
15 6 V
14 5 V
14 6 V
15 6 V
14 6 V
15 6 V
14 6 V
14 6 V
16 7 V
14 6 V
15 7 V
14 7 V
14 7 V
15 7 V
14 7 V
14 7 V
15 8 V
14 7 V
14 7 V
15 8 V
15 7 V
15 8 V
14 8 V
14 8 V
15 8 V
14 8 V
15 8 V
14 8 V
14 8 V
15 8 V
14 8 V
14 8 V
15 8 V
15 7 V
15 8 V
14 7 V
14 8 V
15 8 V
14 8 V
14 7 V
15 8 V
14 8 V
14 7 V
15 8 V
14 7 V
16 7 V
14 7 V
14 6 V
15 7 V
14 7 V
15 4 V
14 1 V
14 -1 V
15 -1 V
14 -3 V
15 -3 V
14 -4 V
14 -4 V
16 -5 V
14 -6 V
15 -5 V
14 -7 V
14 -7 V
15 -7 V
14 -7 V
14 -8 V
15 -9 V
14 -6 V
stroke
2371 849 M
14 -6 V
15 -8 V
15 -3 V
15 -6 V
14 -4 V
14 -7 V
15 -5 V
14 1 V
14 0 V
15 3 V
14 3 V
15 4 V
14 4 V
14 4 V
15 7 V
15 9 V
15 7 V
14 7 V
14 3 V
15 -1 V
14 5 V
14 9 V
15 9 V
14 2 V
14 -6 V
15 -21 V
14 -12 V
16 10 V
14 -42 V
14 -88 V
15 -9 V
14 -9 V
15 -9 V
14 -9 V
14 -9 V
15 -9 V
14 -9 V
14 -9 V
15 -9 V
14 -9 V
16 -7 V
14 -9 V
14 -9 V
15 -9 V
14 -10 V
14 -9 V
15 -11 V
14 -10 V
15 -9 V
14 -10 V
14 -12 V
15 -13 V
15 -9 V
15 -16 V
14 -24 V
14 -2 V
15 -2 V
14 -2 V
14 -1 V
15 -2 V
14 -2 V
14 -1 V
15 -2 V
14 -1 V
15 -2 V
15 -2 V
14 -1 V
15 -2 V
14 -1 V
15 -2 V
14 -1 V
14 -1 V
15 -2 V
14 -1 V
14 -1 V
15 -2 V
14 -1 V
16 -1 V
14 -2 V
14 -1 V
1.000 UL
LT1
868 353 M
14 -32 V
14 -3 V
15 65 V
14 -3 V
15 35 V
14 -50 V
15 -3 V
8 -62 V
84 0 R
9 61 V
14 2 V
15 -4 V
14 141 V
14 -4 V
15 164 V
15 -90 V
15 210 V
14 -4 V
14 179 V
15 62 V
14 -5 V
15 -21 V
14 -5 V
14 12 V
15 -5 V
14 25 V
14 -5 V
16 50 V
14 93 V
15 -5 V
14 132 V
14 -6 V
15 48 V
14 -5 V
14 69 V
15 -6 V
14 102 V
15 130 V
14 -6 V
14 130 V
16 -44 V
14 113 V
15 -7 V
14 112 V
14 -7 V
15 162 V
14 157 V
14 -8 V
15 187 V
14 -8 V
14 202 V
15 -8 V
15 176 V
15 -8 V
14 128 V
14 -8 V
15 116 V
14 76 V
15 -8 V
14 8 V
14 -8 V
15 -26 V
14 -8 V
14 -15 V
15 -7 V
15 30 V
15 -35 V
14 -7 V
14 -62 V
15 -8 V
14 -93 V
14 -7 V
15 -115 V
14 -8 V
14 -94 V
15 -106 V
14 -6 V
16 -88 V
14 -6 V
14 -124 V
15 -7 V
14 -105 V
15 -5 V
14 -82 V
14 -122 V
15 1 V
14 -106 V
15 3 V
14 -98 V
14 4 V
16 -42 V
14 5 V
15 -58 V
14 -67 V
14 7 V
15 -119 V
14 7 V
14 -103 V
15 10 V
14 -44 V
14 6 V
15 -10 V
15 31 V
15 7 V
stroke
2430 1390 M
14 52 V
14 7 V
15 -22 V
14 -2 V
14 -78 V
15 -2 V
14 -13 V
15 55 V
14 -3 V
14 70 V
15 -7 V
15 -65 V
15 -7 V
14 4 V
14 -2 V
15 -37 V
14 -3 V
14 -9 V
15 26 V
14 -2 V
14 85 V
15 21 V
14 77 V
16 -10 V
14 93 V
14 175 V
15 9 V
14 45 V
15 9 V
14 24 V
14 10 V
15 65 V
14 9 V
14 111 V
15 102 V
14 9 V
16 224 V
14 9 V
14 61 V
15 9 V
14 10 V
14 9 V
15 47 V
14 10 V
15 29 V
14 -103 V
14 13 V
15 -120 V
15 9 V
15 -61 V
14 24 V
14 -126 V
15 2 V
14 -213 V
14 -204 V
15 1 V
14 -226 V
14 2 V
15 -213 V
14 1 V
15 -256 V
15 65 V
14 -337 V
15 -124 V
14 2 V
15 -81 V
14 1 V
14 -128 V
15 1 V
14 -118 V
14 1 V
15 -68 V
7 -38 V
0.500 UL
LTb
450 300 M
3380 0 V
0 3380 V
-3380 0 V
450 300 L
1.000 UP
stroke
grestore
end
showpage
}}%
\put(2140,50){\makebox(0,0){$R\ [\mbox{kpc}]$}}%
\put(100,1990){%
\special{ps: gsave currentpoint currentpoint translate
270 rotate neg exch neg exch translate}%
\makebox(0,0)[b]{\shortstack{${\Sigma(R)}/{\Sigma_{c}}$}}%
\special{ps: currentpoint grestore moveto}%
}%
\put(3830,200){\makebox(0,0){ 1000}}%
\put(2985,200){\makebox(0,0){ 500}}%
\put(2140,200){\makebox(0,0){ 0}}%
\put(1295,200){\makebox(0,0){-500}}%
\put(450,200){\makebox(0,0){-1000}}%
\put(400,3680){\makebox(0,0)[r]{ 0.4}}%
\put(400,2835){\makebox(0,0)[r]{ 0.3}}%
\put(400,1990){\makebox(0,0)[r]{ 0.2}}%
\put(400,1145){\makebox(0,0)[r]{ 0.1}}%
\put(400,300){\makebox(0,0)[r]{ 0}}%
\end{picture}%
\endgroup
 